\pdfoutput=1

\documentclass[online,faculty=firw,department=elt,phddegree=elt]{adsphd}

\usepackage{nomencl}   
\usepackage[nonumberlist]{glossaries} 

\title{Cross-Layer Optimization for Power-Efficient and Robust Digital Circuits and Systems}

\author{Yanxiang}{Huang}

\supervisor{Prof.~dr.~ir.~Wim Dehaene}{}
\supervisor{Prof.~dr.~ir.~Liesbet Van der Perre}{}
\president{Prof.~dr.~ir.Yves Willems}
\jury{Prof.~dr.~ir.~Francky Catthoor\\
	  Prof.~dr.~ir.~Sofie Pollin\\
	  Prof.~dr.~ir.~Rudy Lauwereins\\
}

\externaljurymember{Prof.~dr.~ir. Andreas P. Burg}{École Polytechnique Fédérale de Lausanne}
\externaljurymember{Prof.~dr.~ir. Liang Liu}{Lund University}

\researchgroup{ESAT}
\website{} 
\email{} 

\address{Kasteelpark Arenberg 10}
\addresscity{B-3001 Leuven (Belgium)} 
\addresscity{B-3001 Leuven} 

\date{September 12, 2017}
\copyyear{2017}

\renewcommand{\nomname}{List of Symbols}
\makeatletter
\let\@printnomenclatureorig\@printnomenclature
\def\@printnomenclature[#1]{%
  \cleardoublepage%
  \chaptermark{\nomname}
  \@printnomenclatureorig[#1]
}
\makeatother
\makenomenclature

\newcommand{\glossname}{Abbreviations}
\makeglossaries

\let\printglossaryorig\printglossary
\renewcommand{\printglossary}{%
  \renewcommand{\glossaryname}{\glossname}
  \cleardoublepage%
  \printglossaryorig[title=\glossname]
  \chaptermark{\glossname}
}

\usepackage{textcomp} 
\usepackage{pifont}   
\usepackage{booktabs}
\usepackage{amssymb,amsthm}
\usepackage{amsmath}

\usepackage{subfig}

\usepackage{flushend} 

\usepackage{microtype} 
\usepackage{makecell}
\usepackage{gensymb} 

\usepackage{booktabs}
\usepackage{multirow}
\usepackage{rotating, graphicx} 
\usepackage{afterpage} 

\usepackage{textcomp} 

\usepackage{placeins} 

\usepackage{titlesec}
\newcommand\sectionbreak{\ifnum\value{section}>1\FloatBarrier\fi}
\newcommand\subsectionbreak{\ifnum\value{subsection}>1\FloatBarrier\fi}
\newcommand\subsubsectionbreak{\ifnum\value{subsubsection}>1\FloatBarrier\fi}

\usepackage{algorithm,algorithmicx,algpseudocode}

\newglossaryentry{EDA}{name={EDA}, description={Electronic Design Automation}}
\newglossaryentry{CPU}{name={CPU}, description={Central Processing Unit}}
\newglossaryentry{MCU}{name={MCU}, description={Micro-Controller Unit}}
\newglossaryentry{DSP}{name={DSP}, description={Digital Signal Processing/Processor}}
\newglossaryentry{ASIP}{name={ASIP}, description={Application-Specific Instruction set Processor}}
\newglossaryentry{CMOS}{name={CMOS}, description={Complementary Metal Oxide Semiconductor}}
\newglossaryentry{FPGA}{name={FPGA}, description={Field-Programmable Gate Array}}
\newglossaryentry{IC}{name={IC}, description={Integrated Circuit/Chip}}
\newglossaryentry{SRAM}{name={SRAM}, description={Static Random-Access Memory}}
\newglossaryentry{ASIC}{name={ASIC}, description={Application Specific Integrated Circuit}}
\newglossaryentry{RAM}{name={RAM}, description={Random Access Memory}}
\newglossaryentry{RTL}{name={RTL}, description={Register Transfer Level}}
\newglossaryentry{STA}{name={STA}, description={Static Timing Analysis}}
\newglossaryentry{VLSI}{name={VLSI}, description={Very Large Scale Integration}}
\newglossaryentry{IoT}{name={IoT}, description={Internet of Things}}

\newglossaryentry{DVFS}{name={DVFS}, description={Dynamic Voltage Frequency Scaling}}
\newglossaryentry{AVFS}{name={AVFS}, description={Adaptive Voltage Frequency Scaling}}
\newglossaryentry{IR-drop}{name={IR-drop}, description={dynamic voltage drop}}
\newglossaryentry{OCV}{name={OCV}, description={On-Chip Variation}}
\newglossaryentry{PVT}{name={PVT}, description={Process Voltage Temperature}}
\newglossaryentry{SEU}{name={SEU}, description={Single Event Upset}}

\newglossaryentry{VOS}{name={VOS}, description={Voltage-OverScaling}}
\newglossaryentry{FF}{name={FF}, description={Flip-Flop}}
\newglossaryentry{MSFF}{name={MSFF}, description={Master-Salve Flip-Flop}}
\newglossaryentry{PoFF}{name={PoFF}, description={Point of First Failure}}
\newglossaryentry{PoFW}{name={PoFW}, description={Point of First Warning}}
\newglossaryentry{PoFD}{name={PoFD}, description={Point of First noticeable Degradation}}
\newglossaryentry{CS}{name={CS}, description={Computation-Skip}}

\newglossaryentry{CPI}{name={CPI}, description={Cycles Per Instruction}}
\newglossaryentry{RISC}{name={RISC}, description={Reduced Instruction Set Computer}}

\newglossaryentry{FFT}{name={FFT}, description={Fast Fourier Transform}}
\newglossaryentry{FIR}{name={FIR}, description={Finite Impulse Response}}
\newglossaryentry{LDPC}{name={LDPC}, description={Low-Density Parity-Check code}}
\newglossaryentry{CORDIC}{name={CORDIC}, description={COordinate Rotation DIgital Computer}}
\newglossaryentry{ECC}{name={ECC}, description={Error Correction Code}}
\newglossaryentry{SDR}{name={SDR}, description={Software Defined Radio}}

\newglossaryentry{ANT}{name={ANT}, description={Algorithmic Noise Tolerance}}
\newglossaryentry{BER}{name={BER}, description={Bit Error Rate}}
\newglossaryentry{EVM}{name={EVM}, description={Error Vector Magnitude}}
\newglossaryentry{SNR}{name={SNR}, description={Signal to Noise Ratio}}
\newglossaryentry{SDDR}{name={SDDR}, description={Signal to Digital Distortion Ratio}}
\newglossaryentry{ENoB}{name={ENoB}, description={Effective Number of Bits}}

\newglossaryentry{SIMD}{name={SIMD}, description={Single Instruction, Multiple Data}}
\newglossaryentry{MAC}{name={MAC}, description={Multiply Accumulate}}
\newglossaryentry{MSB}{name={MSB}, description={Most Significant Bit}}
\newglossaryentry{LSB}{name={LSB}, description={Least Significant Bit}}
\newglossaryentry{LUT}{name={LUT}, description={Look-Up Table}}

\newglossaryentry{AWGN}{name={AWGN}, description={Additive White Gaussian Noise}}
\newglossaryentry{LTE}{name={LTE}, description={Long Term Evolution}}
\newglossaryentry{MIMO}{name={MIMO}, description={Multiple Input Multiple Output}}
\newglossaryentry{OFDM}{name={OFDM}, description={Orthogonal Frequency Division Multiplexing}}
\newglossaryentry{QAM}{name={QAM}, description={Quadrature Amplitude Modulation}}
\newglossaryentry{QPSK}{name={QPSK}, description={Quadrature Phase Shift Keying}}
\newglossaryentry{DL}{name={DL}, description={Down Link}}
\newglossaryentry{UL}{name={UL}, description={Up Link}}
\newglossaryentry{TDD}{name={TDD}, description={Time division duplex}}
\newglossaryentry{BS}{name={BS}, description={Base Station}}
\newglossaryentry{MMSE}{name={MMSE}, description={Minimum Mean Square Error}}
\newglossaryentry{ZF}{name={ZF}, description={Zero-Forcing}}

\newglossaryentry{DAC}{name={DAC}, description={Digital-to-Analog Converter}}
\newglossaryentry{DFE}{name={DFE}, description={Digital Front-End}}
\newglossaryentry{PA}{name={PA}, description={Power Amplifier}}

\makeglossaries

\setglossarypreamble{This list of acronyms only specifies the most important ones. The abbreviations lists below represent both the singular and the plural forms.}

\usepackage[backend=biber, url=false, style=alphabetic,maxalphanames=1]{biblatex}
\usepackage{threeparttable} 

\DeclareLabelalphaTemplate{
  \labelelement{
    \field[final]{shorthand}
    \field{labelname}
    \field{label}
  }
  \labelelement{
    \literal{,\addhighpenspace}
  }
  \labelelement{
    \field{year}
  }
}

\defbibenvironment{midbib}
  {\list
     {}
     {\setlength{\leftmargin}{\bibhang}%
      \setlength{\itemindent}{-\leftmargin}%
      \setlength{\itemsep}{\bibitemsep}%
      \setlength{\parsep}{\bibparsep}}}
  {\endlist}
  {\item}

\begin{document}


\makefrontcoverXII

\maketitle

\frontmatter 

\includepreface{preface}

\includeabstract{abstract}
\includeabstractnl{abstractnl}


\glsaddall[]
\printglossary


\tableofcontents
\listoffigures
\listoftables


\mainmatter 


\cleardoublepage


\chapter{Introduction}\label{sec:intro}

The groundbreaking digital revolution has refined the work and life styles of every individual. However, the gap between great application requirements and the enabling CMOS technology limitations is prominent. Therefore, design techniques to balance the quality and the power consumption of integrated circuits are demanded in the nano-CMOS era. This work promotes cross-layer optimizations for power and quality trade-off. It accepts errors that traditional designers advocated avoiding at all cost. The CMOS devices are working at their extremes with fewer safety margins. This enables extra power saving without noticeable quality degradations. This chapter introduces this work. 

The chapter is structured as follows: Section~\ref{sec:intro_motivation} describes the motivation of this thesis. It reviews the contradiction between the need and the reality of current digital circuits and systems. A cross-layer optimization approach is therefore promoted (Section~\ref{sec:intro_shape}). Section~\ref{sec:intro_scope} summarizes the scope of this thesis. Section~\ref{sec:intro_contribution} lists the main contributions of this work. Finally, Section~\ref{sec:intro_structure} presents the structure of this thesis.

\section{Context: motivation for power consumption and reliability trade-offs}\label{sec:intro_motivation}
\begin{figure}[H]
\centering
\includegraphics[width=0.85\textwidth]{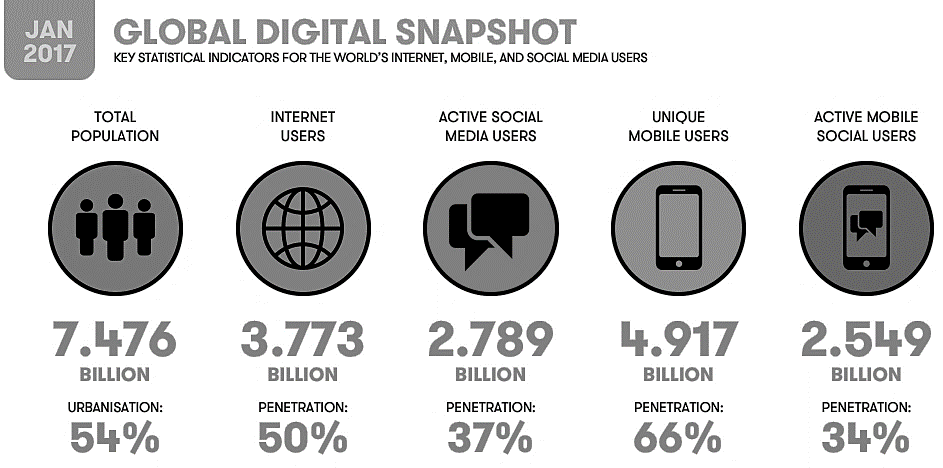}
\caption[The world is immersed in digital services.]{The world is immersed in digital services~(\cite{simon2017}).}
\label{fig:intro_usage}
\end{figure}

The ubiquitous digital infrastructure and services have totally changed our way of life, leisure and our means of communication and information. The new generation is enjoying the conveniences offered by the computers, phones, gadgets, and varieties of smart IoT devices that exceed the visions of last generation's craziest sci-fi movies. The immersion of digital world is playing a key role in human lives (Fig.~\ref{fig:intro_usage}). According to \cite{hoursonline}, adults are spending more than 12 hours accessing digital data in 2017.

The digital reality cannot be realized without the development of the enabling technologies. For the past few decades, we have seen marvelous breakthroughs in the domains such as personal computers, the internet, and mobile communications. Those breakthroughs, not only redefine the frontier of technologies but also expedite the mass adoption of them. For instance, almost three-quarters of the world’s population now use a mobile phone, with the total number of unique global mobile users rapidly approaching 5 billion~\cite{simon2017}. The advance of technology is now spreading to the applications as artificial intelligence, smart vehicle, and IoT devices.

\subsection{Demand: performance and power efficiency improvements}\label{sec:intro_demand}
Consumers are demanding higher data volume, in a faster rate. For instance, the world monthly mobile data traffic goes steeply from 3.7~EB (Exabytes) in 2015, to 9.9~EB in 2017, and will increase to 30.6~EB by 2020~\cite{index2016global}. The exponential growth of wireless data services driven by the mobile internet and smart devices has triggered the investigation of the 5G cellular network. The mobile phone of the future has to provide seamless connectivity anywhere and anytime. Around 2020, the new 5G mobile networks are expected to be deployed \cite{andrews2014will}. These networks will support multimedia applications with a wide variety of requirements, including higher peak and user data rates (more than 100 megabits per second for metropolitan areas \cite{6815890}), reduced latency (less than 1~ms \cite{best2014race}), enhanced indoor coverage, improved energy efficiency and so on.

Powering up those digital services requires a big amount of energy. Worldwide, data centers use about 400 terawatt-hours of electricity each year \cite{andrae2015global}. That's a little more electricity than all of the United Kingdom uses. Already, they have mushroomed from virtually nothing 10 years ago to accounting for about 2 per cent of total greenhouse gas emissions~\cite{datacenter}. That gives it the same carbon footprint as the airline industry. If left unchecked, they could use almost 8,000 terawatt-hours by 2030. That's about the amount of electricity all of Europe and Africa and much of Asia use today. Another optimistic estimation predicts that the global IT services will consume 15\% electricity production world-wise by 2025 \cite{greenit}. Considering the continuing demands for increased digital services, the energy efficiency for digital computation must be improved accordingly.

\begin{figure}[H]
\centering
\includegraphics[width=0.7\textwidth]{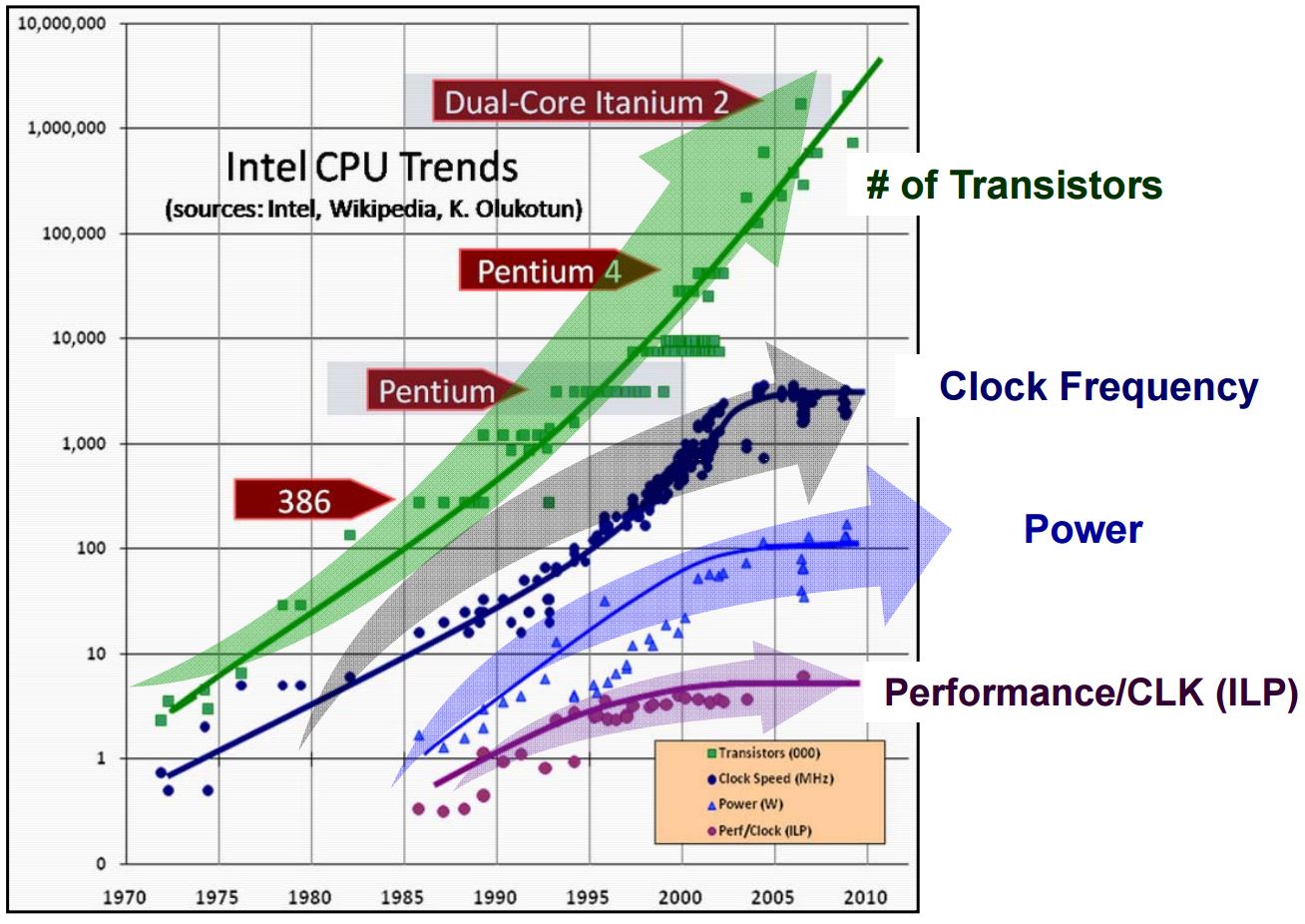}
\caption[The free lunch of CMOS scaling is over.]{The free lunch of CMOS scaling is over~\cite{sutter2005free}.}
\label{fig:intro_moore}
\end{figure}

In the past few decades, the demands of increasing performance and power efficiency were realized with the CMOS technology progress. The complexity of integrated circuits has approximately doubled every 18 months; the cost per function has decreased several thousand-fold. The exponential growth has fit to the well-known Moore's Law (prediction) \cite{moore1998cramming}. In the computer sector, the CPU had kept pace with the Dennard's scaling \cite{dennard1974design}, which suggests that the performance per watt also grows exponentially (Fig.~~\ref{fig:intro_moore}). However, CPU performance, which is largely coupled with the clock speed, has stalled at around 2005, since the 65nm CMOS process. This suggests the end of the free lunch brought by the CMOS scaling \cite{sutter2005free}. 

The power consumption (too high), heat (too much of it and too hard to dissipate), and current leakage problems are among the biggest challenges for the continuing of free-lunch scaling. Multicore parallelisms are employed to keep improving the performance, which eventually leads to new power challenges  (dark silicon) that may end the multicore era \cite{esmaeilzadeh2011dark}. Nowadays, power consumption has become an, arguably the most, important metric for digital computing devices.

Apart from saving energy, another motivation to minimize power consumption is to fit digital chips into ubiquitous IoT devices, much of which are powered by batteries or energy harvesters \cite{bravos2005energy, gyselinckx2005human++, rawat2014wireless}. Looking to the future, Cisco IBSG predicts there will be 25 billion devices connected to the Internet by 2015 and 50 billion by 2020 \cite{evans2011internet}. 

In summary, the increased IT service calls for, in addition to advanced CMOS technologies, successful enhancement in design of digital circuits and systems, to fulfill the performance and power efficiency requirements.

\subsection{Reality: device and application uncertainties endanger successful designs}\label{sec:intro_reality}
\begin{figure}[H]
\centering
\includegraphics[width=\textwidth]{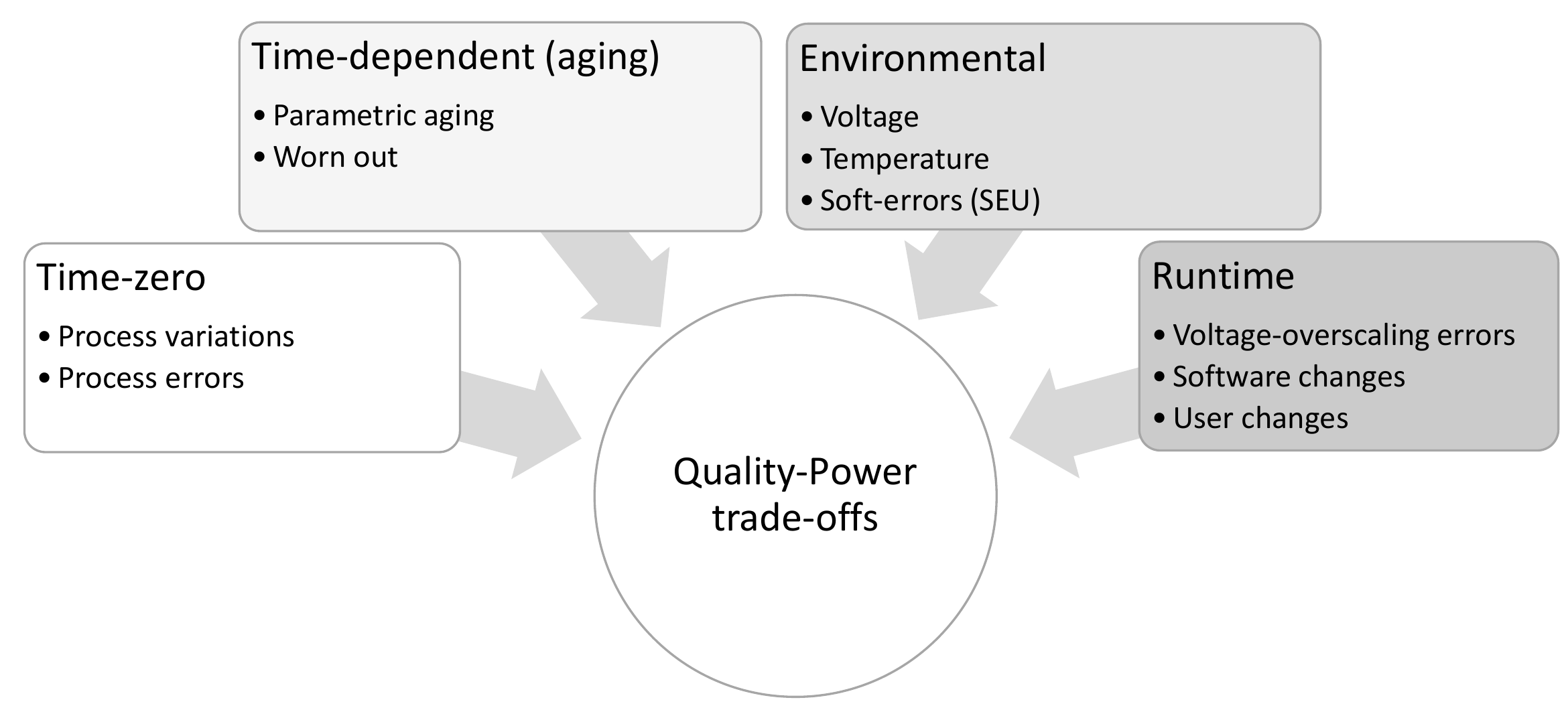}
\caption{Digital designs must cope with permanent (time-zero and time-dependent) and temporary (environmental and runtime) uncertainties (variations), to reach a balance between power consumption and output quality.}
\label{fig:intro_overview}
\end{figure}

The variability has become a major roadblock to CMOS scaling (Fig.~\ref{fig:intro_overview}). Below the sub-65nm regime, transistors no longer act deterministically as a consequence of fluctuations in device parameters. This phenomenon is caused by the process challenges (lithography, etching, chemical mechanical polishing, etc.) \cite{ghosh2010parameter}. These process challenges not only alters chip parameters (speed, area, power, etc.), they also result in functionality failures, e.g. stuck-in fault. The time-zero manufacturing process is not the only source of device uncertainties, chips also face aging problems during lifetimes. Worse still, the aging phenomena are uncertain in themselves. They heavily depend on the environment or workloads \cite{mintarno2013workload}.

Apart from these (semi-) permanent effects, environment \cite{unsal2006impact} (voltage, temperature, cosmic particle strikes, etc.)  and runtime (timing errors, software, and user) variations also add uncertainties to chip designs. For instance, high energy particle strikes can lead to random bit flippings on storage elements. VOS (Voltage Over-Scaling) circuits \cite{hegde2004voltage, jeon2012design} tries to save power by operates at a riskily-low supply voltage \footnote{It is different from the sub-threshold computing technique \cite{dreslinski2010near} where the $V_{dd}$ and the clock frequency are very low. In contrast, the VOS reduces $V_{dd}$, but not the clock frequency. The operating voltage of VOS is much higher than the sub-threshold region.}, which produces uncertain errors. In summary, these uncertainties share a similarity that they only affect the IC temporaries.

The time-zero and time-dependent challenges are often regarded as a yield problem, which is mainly tackled by the IC foundries \cite{tsai2004yield}. In contrast, the fast changing runtime uncertainties are so dynamical that they cannot be simplified solved by traditional post-silicon testing \cite{nithin2010dynamic}. Therefore, runtime uncertainties management is a domain that is still wide open and much profitable for digital circuits and systems designers. The fact that these uncertainties only leads to temporary effects makes trading quality for power savings possible.

The source and effects of them are further explained in Section~\ref{sec:variation_pvt}. To conclude, careful consideration of all those uncertainties is essential for successful digital circuit and system design.

\section{Calling for cross-layer optimizations}\label{sec:intro_shape}
\subsection{Traditional pyramid-shaped design}
\begin{figure}[H]
\centering
\includegraphics[width=0.5\textwidth]{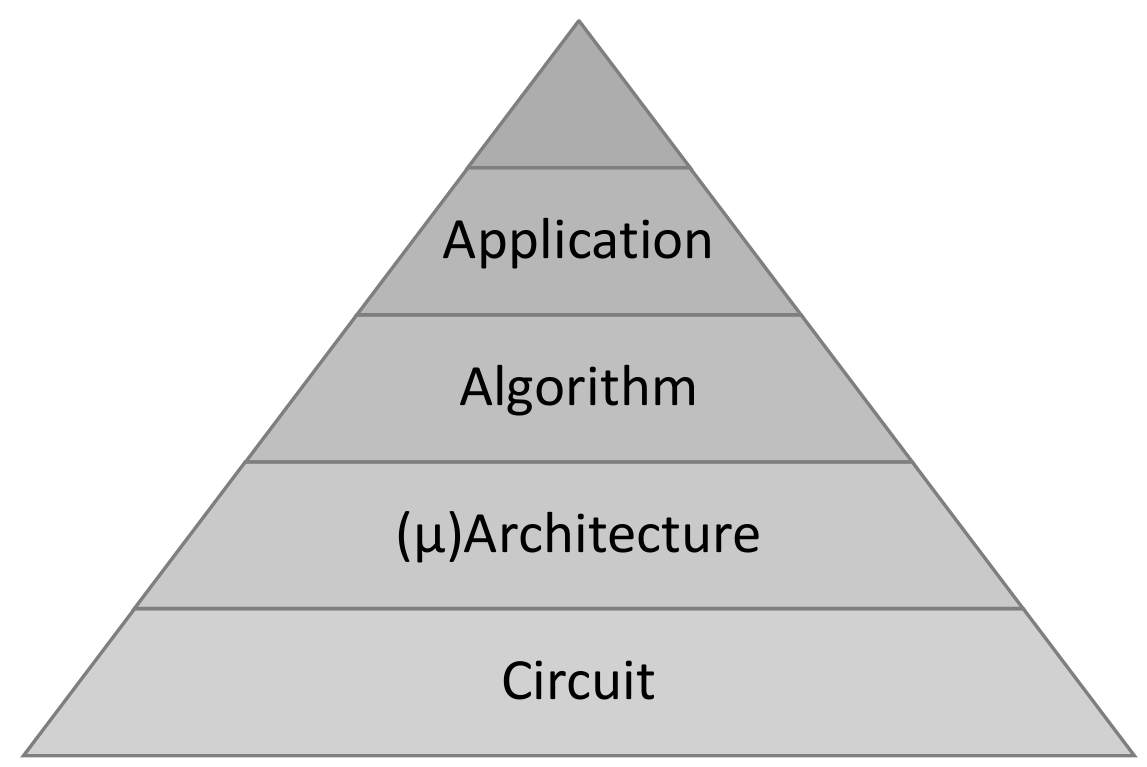}
\caption[The traditional pyramid-shaped design flow leads to over-designs at all levels.]{The traditional pyramid-shaped design flow leads to over-design at all levels (derived from \cite{hennessy2011computer}).}
\label{fig:intro_pyramid}
\end{figure}

Traditionally, the digital design flow follows a top-down procedure (Fig.~\ref{fig:intro_pyramid}). That is, the top-level design specs, e.g. quality, speed, power consumption, are assigned by system architects. The algorithm and circuit-level designers are forced to come up with solutions to fulfill these requirements. This rigid task dividing simplifies the design process, as fewer confusions during design are expected.

Following this design flow, the variations are traditionally tackled in particular levels. For instance, the process, voltage, and temperature (PVT) variations are packed up together and guaranteed with worst-corner safety margins at the circuit level (Section~\ref{sec:variation_wc}). Temporal degradation issues are hidden in the time-zero variations margins, albeit early works on designing with accurate time-dependent models \cite{liu2017comprehensive} start to gain popularity.

\begin{figure}[H]
\centering
\includegraphics[width=0.7\textwidth]{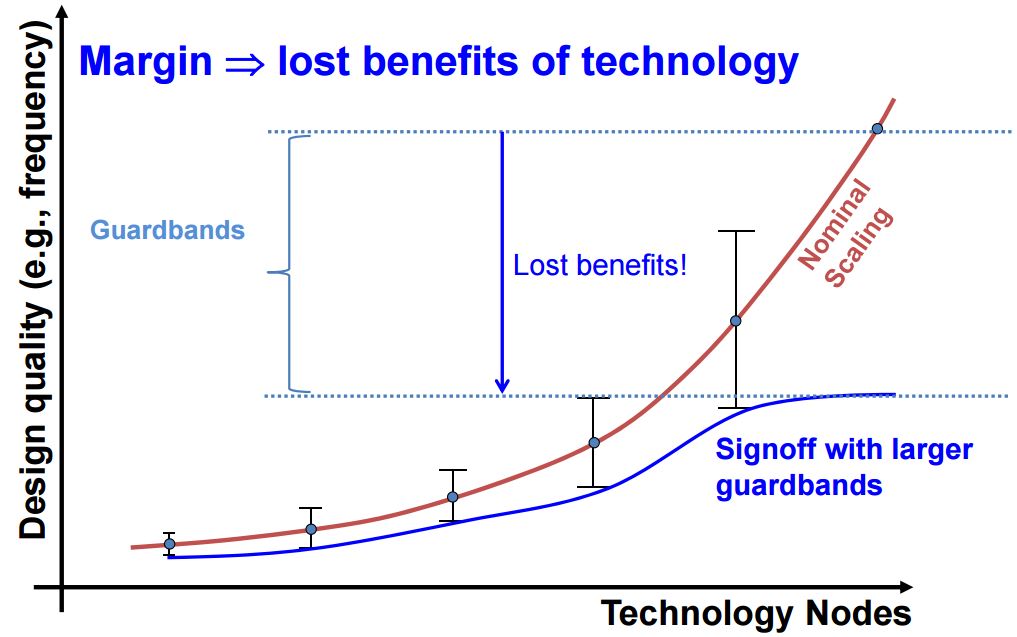}
\caption[Over-designed guardbands diminish the benefits of CMOS scaling.]{Over-designed guardbands diminish the benefits of CMOS scaling~\cite{kahngdesign}.}
\label{fig:variability_khang}
\end{figure}

A variety of safety margins are inserted in all levels of the design for manufacturing cost, reliable computing, acceptable device lifetime, device performance, etc \cite{austin2008reliable}. If keeps inserting margins and being pessimistic, the gain obtained by the CMOS scaling will reduce and might eventually diminish (Fig.~\ref{fig:variability_khang}).

The drawback is that, following the traditional pyramid-shaped design flow, smart engineering on quality and power consumption is usually limited to specific design levels, leading to only local optimal solutions. The application-level and the algorithm-level designs call for error-free results from lower levels, which makes lower level designs (e.g., at the architecture and circuit-level) unnecessarily complex sometimes. For example, chips should work at any temperature within the specification range, and thus transistors delays are pre-characterized on all temperatures. Circuits designers always try to avoid uncertainties and constrain chips to work equally on that (wide) delay range, which is not easy. Not many ``if-then-else'' cross-layer co-optimization are possible conventionally. The waste of resources of this worst-case guard-band is reviewed in Section~\ref{sec:variation_wc}.

Considering that the lower levels are utilizing too much resources to provide the higher level application service, the design flow is named as a pyramid-shaped method. In summary, the lack of information across design levels leads to wasted resources (in terms of power consumption and area cost). It is therefore calling for a new design and optimization paradigm.

\subsection{Cross-layer optimization}
\begin{figure}[H]
\centering
\includegraphics[width=0.7\textwidth]{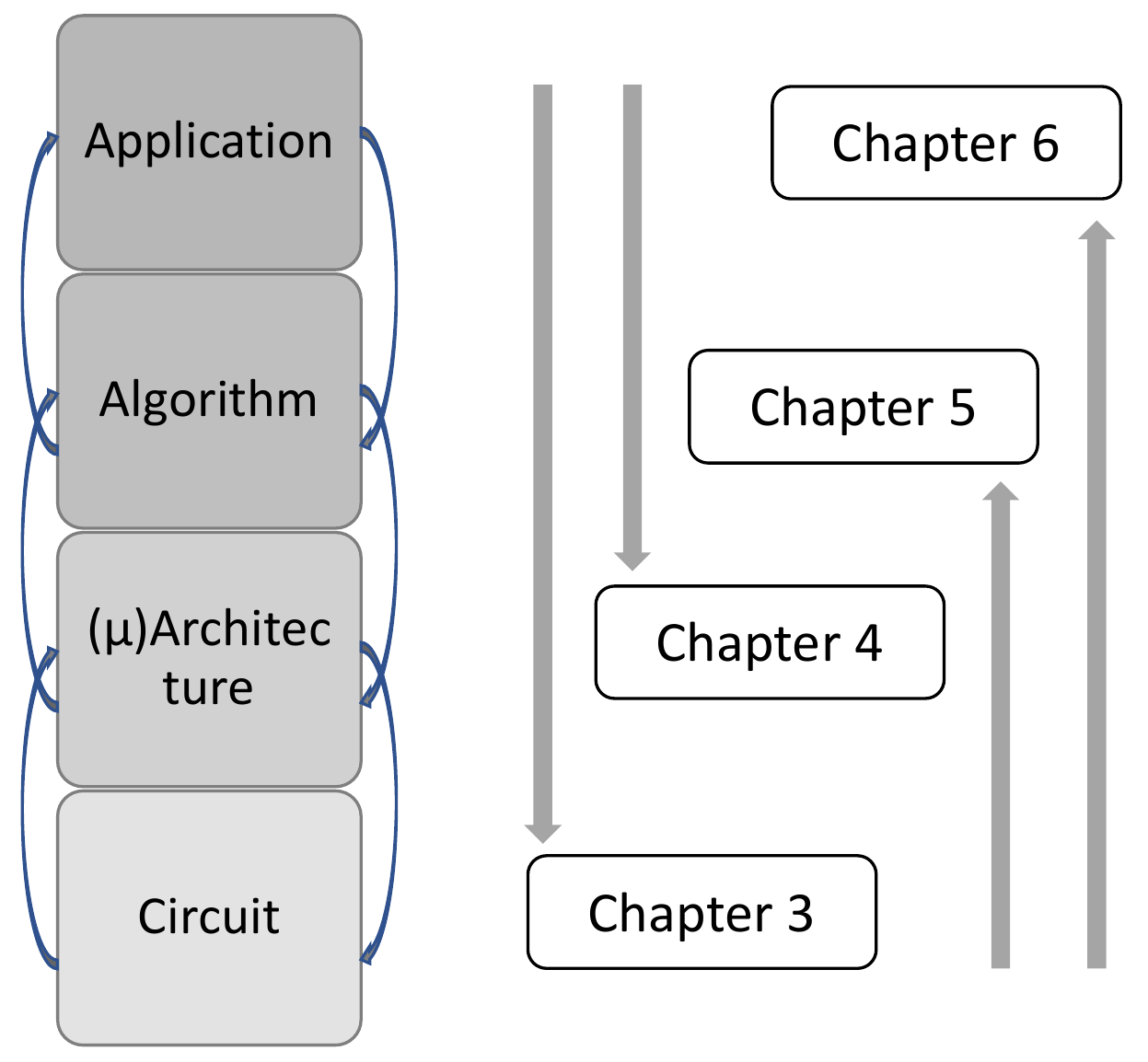}
\caption[Promoted tower-shaped cross-layer design approach reduces unnecessary power and area waste.]{Promoted tower-shaped cross-layer design approach reduces unnecessary power and area waste. This is achieved by exchanging information across design levels. Comparing with the pyramid-shaped cartoon (Fig.~\ref{fig:intro_pyramid}), the new paradigm requires less space at the low level, yet provides more rooms at the top of the tower (hence a tower rather than a pyramid). Chapter 3 and 4 transfers information from higher levels to circuit and micro-architecture levels. Chapter 5 and 6 mitigates lower level errors at algorithm and application levels.}
\label{fig:intro_tower}
\end{figure}

To exploit the resource waste in the pyramid-shaped design, this thesis promotes a cross-layer optimization design approach \footnote{Corss-layers in this work means to optimize across different design levels, which is different from the notion of ``layer'' from the Open Systems Interconnection (OSI) model.}. With this optimization paradigm, information are sharing across design layers for power savings or quality improvements. As will be demonstrated in the rest of this thesis, by extending optimization across different design levels, the over-design can be much reduced. This provides an advantage in conserving area and power consumption at the silicon foundation and increasing performance and quality at the application-level. This paradigm is named as a tower-shaped design flow (Fig.~\ref{fig:intro_tower}), reflecting the aim that each design layer is utilizing just-needed resources, which avoids over-designs.

In contrast of the tower-shaped design, errors are often permitted in the context of cross-layer optimizations. \cite{djahromi2007cross} showed that by allowing data bearing memories to have a controlled number of errors, significant gains in power consumption are possible. In fact, for a 3GPP modem, power savings of up to 17.5\% are achievable assuming a 32nm technology. The Razor \cite{razor03} and ANT \cite{ant99} techniques are excellent cross-layer optimization examples. They are discussed in details in Section~\ref{sec:variation_vos}.

\cite{carter2010design, mitra2010cross, gimmler2013cross} overviewed of the design techniques for cross-layer resilience. They highlights that distributing resilience and reliability across the system stack can improve performance and reduce power and area costs by taking advantage of the strengths of each layer and exploiting the characteristics of individual applications. 

Admittedly, the design effort of this new paradigm is larger because a cross-layer information should be shared and evaluated. Though, its benefits are worth the effort. This thesis demonstrates that the cross-layer approach brings considerable benefits for digital circuits and systems, with limited design effort increase. For example, lower-level designs assume (predicate) some properties of the higher-level, which reduce the power cost once predicated successfully; lower-level designs generate some errors under the worst situations, which will be handled by the higher-level designs.

\section{Thesis scope}\label{sec:intro_scope}
This thesis aims to reduce safety margins and saving power. Runtime adjusting approaches are employed to exploit those margins. More specifically, this work mainly handles the environmental and runtime uncertainties. The reason is that, these uncertainties are so dynamical that they cannot be simplified solved by traditional post-silicon testing \cite{tsai2004yield, nithin2010dynamic}. It not only responses to the environmental changes as the DVFS techniques (Section~\ref{sec:variation_adaptive}), but also allows occasional errors to propagate through different levels (e.g. Section~\ref{sec:algo}). Techniques in this work, however, also exploit margins for slowly changing process variations (time-zero and time-dependent), assuming that they are already packed in the safety margin. The permanent breakdowns are not covered in the thesis. This is because time-zero breakdowns are classified as yield problems and are mostly taken care of by semiconductor foundries; workload-dependent aging effects are considered by EDA tools recently \cite{karapetyan2015integrating}.

The targeted applications of this thesis are soft-quality requirement digital signal processing devices. These cross-layer optimizations are particularly beneficial in the context of soft-quality requirement systems. In these systems, errors are acceptable as long as the system output still meets the requirements. For instance, the transient error in a wireless communication application can easily be tolerated by the symbol detector, considering that the error vector magnitude is allowed \cite{karakonstantis2012exploitation}. This distinguishes from the general-purpose computer case \cite{carter2010design} where usually no errors are permitted. Another example is video processing applications, where skipping few erroneous pixels can be tolerated \cite{driscoll2003byzantine}. This leaves plenty of spaces for quality-power trade-offs. 

Therefore, in this work, the uncertainties incurred errors are allowed to propagate to the algorithmic and application-level. The end result is a good balance between the quality and power consumption. In specific, Chapter~\ref{sec:model} firstly models the algorithm-level impact of random errors, in the circuit-level, without digging into the algorithm. It then provides valuable guidelines to selectively harden designs against random errors at the circuit-level. Chapter~\ref{sec:arch} predicates programs' behavior, and modifies the microarchitectural structure, which leads to power savings for typical usages. Chapter~\ref{sec:algo} mitigates circuit-level errors at the microarchitecture and eventually the algorithm-level, leading to a graceful quality degradation. Chapter~\ref{sec:system} demonstrates that lower-level generated errors can be handled at the application-level. This provides opportunities to embrace hardware uncertainties for power saving. It takes the Massive MIMO wireless communication application as a case-study and demonstrates its resilience to lower-level errors. The chapter, therefore, encourages to use low-power yet erroneous components in the Massive MIMO.

\section{Main contributions}\label{sec:intro_contribution}
This work has contributed to better power solutions of digital designs in scaled CMOS, through a cross-layer optimization. These approaches were demonstrated in the content of wireless communication applications. The main research contributions of this thesis are summarized below. A more elaborate version of the main messages is provided in Chapter~\ref{sec:con}.

\begin{itemize}
\item[$\bullet$] Introduction of an analytical circuit-level random error effects model (Chapter \ref{sec:model}). This thesis proposed a graph travel approach that solves the model. It is shown how a graph based scheme can identify the sensitivity of indivisual Flip-Flops. This helps to selectively protect only those. It also demonstrated the scalability and effectiveness of the model on ISCAS and ITC benchmark circuits. Finally, this work validated the benefits of the model on an FFT processor design that reduces soft-error hardening overhead. This work was unveiled in \cite{huang16dac-serial}.

\item[$\bullet$] Proposed a novel fine-grain hardware-switch scheme to save power in embedded processors (Chapter \ref{sec:arch}). The thesis applied the proposed scheme to the multiplier unit of an OpenRISC processor. This technique competes with the idea of VLIW based SIMD instructions that requires large compiler modification. It demonstrated power savings on 11 typical signal processing applications, e.g. FFT, IIR, AES, JPEG. This work was published in \cite{huang16el-hardware-switch}.

\item[$\bullet$] Application of Razor circuit-level error detection techniques with error mitigation achieved through an algorithmic approach. Proposed a novel computation-skip scheme to mitigate errors for recursive applications (Chapter \ref{sec:algo}). It saves power saving by reducing the supply voltage, exploiting not only the error-free but also error resilient safety marigns.  This is achieved by skipping part of the computation and sacrificing some accuracy. The thesis implemented the scheme on a CORDIC hardware accelerator in 28nm CMOS technology with standard digital design flow. The work was published in \cite{huang14sips-error-resilient}, and elaborated in  \cite{huang16jsps-error-resilient}. 

\item[$\bullet$] Investigation of application-level error absorption and handling for Massive MIMO wireless communication applications (Chapter \ref{sec:system}). It demonstrated the error resiliency of Massive MIMO systems under hardware errors and even antenna outage. It also proposed a damage control strategy for Massive MIMO applications. The work is published in \cite{huang17icassp-mimo-dfe}.
\end{itemize}

\section*{List of publications}
The list of publications can be found in the attached Curriculum Vitae section.

\section{Thesis structure}\label{sec:intro_structure}
\begin{figure}[H]
\centering
\includegraphics[width=0.98\textwidth]{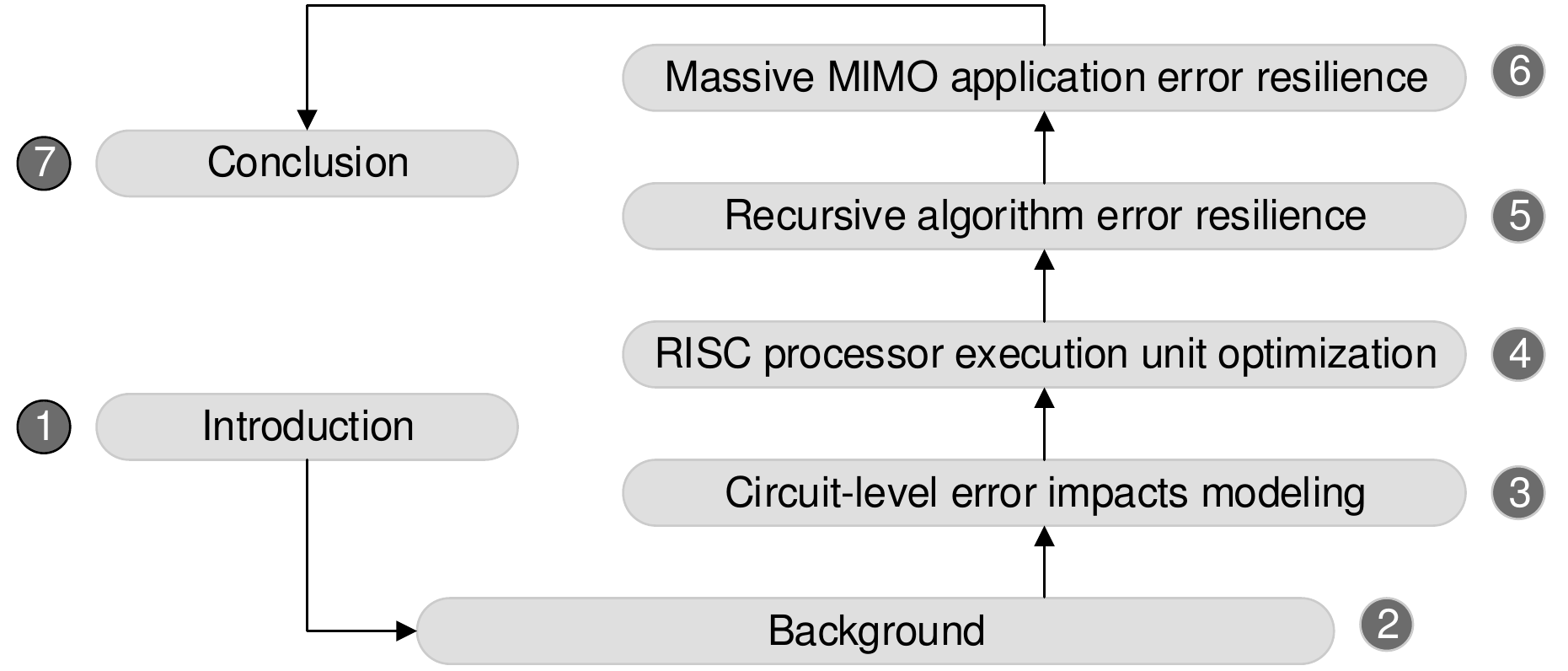}
\caption{Structure of the thesis and overview of the chapters.}
\label{fig:intro_structure}
\end{figure}

In this thesis, cross-layer optimization are performed for digital circuits and systems for power consumption and reliability trade-off (Fig.~\ref{fig:intro_structure}). 

Chapter~\ref{sec:variation} reviews techniques for power and quality tradeoffs. It starts with device-level phenomena of variations. The chapter then reviews the worst-case design and the adaptive scaling method that tunes supply voltage to save power. Finally, the benefit of the cross-layer VOS (Voltage Over-Scaling) approach is justified.

As shown in Fig.~\ref{fig:intro_tower}, Chapter~\ref{sec:model} develops a gate-level random error model, SERIAL. It models the importance of flip-flops regarding their impact on algorithm outputs. The efficiency and effectiveness of the model are shown in typical circuits, including ISCAS and ITC benchmarks, and an LDPC decoder. Finally, the model is applied to design a reliable FFT processor.

In Chapter~\ref{sec:arch}, a microarchitecture-level fine-grain hardware-switch scheme for embedded processors power savings is proposed. In specific, the chapter modifies the multiplier unit of the OpenRISC platform. It demonstrates power savings on typical signal processing applications.

Chapter~\ref{sec:algo} proposes a method for cross-layer error interplay between the circuit-level and the algorithm-level. It presents a computation-skip scheme to mitigate errors in recursive applications. The error mitigation scheme, together with the state-of-the-art timing error detection benchmark, are applied to a hardware CORDIC accelerator. The CORDIC accelerator is processed and verified in a standard 28nm CMOS process with only standard-cells.

Chapter~\ref{sec:system} presents the application-level error absorption and handling. It focuses on a Massive MIMO communication case-study. This chapter assesses hardware random errors (VOS) and antenna outage impacts. Finally, damage control strategies are proposed.

\cleardoublepage

\chapter{Background: pessimistic safety margins should be exploited}\label{sec:variation}


This chapter reviews techniques for power and quality trade-offs. It firstly introduces PVT (process, voltage, and temperature) variations and reliability threats. It then reviews the worst-case design and the adaptive scaling method that tunes supply voltage to save power. Finally, the benefit of the cross-level VOS (Voltage Over-Scaling) approach is discussed.

The rest of this chapter is structured as follows: Section~\ref{sec:variation_pvt} explains the variations from process, environment and runtime changes. Section~\ref{sec:variation_voltage} analyses two conventional methods, i.e. the worst-case approach, and dynamical voltage scaling, which handle variability. The limitation of the worst-case approach is pointed out. The benefits and disadvantages of adaptive scaling are also reviewed. Section~\ref{sec:variation_vos} discusses adventurous VOS methods that reach a balance between power consumption and output quality. Finally Section~\ref{sec:variation_con} concludes this chapter.

\section{Uncertainties in circuit paramaters}\label{sec:variation_pvt}
IC design has always been subject to variations which make it impossible at design time to determine exactly how a circuit will perform. Worse still, these uncertainties have become increasingly significant as a result of the scaling of technology. To cope with those variations, the concept of design margin has been introduced in the design process.

This section reviews the cause, and more importantly the effects, of the long-term and short-term variations, from a digital circuits and systems designer's perspective. This thesis does not try to model for these variations directly. Instead, it saves power by reducing over-pessimistic safety margins.

\subsection{Process variations and aging effects}\label{sec:variation_p}
Process variations and aging effects are long-term or permanent uncertainties for IC. They result to parametric variability (in area, speed, power consumption) and functional breakdowns. The parametric variability effects are reviewed in this subsection, because they usually lead to pessimistic parametric safety margins. Functional breakdowns, e.g., stuck-in errors, electrical stress, burn-in, \cite{veendrick2008nanometer}, are not covered in this work. The reason is that those extra threats have very specific characteristics that require individual studies. Another reason is that, most of them have already been addressed properly in the subjects of design-for-testability \cite{fujiwara1985logic} and design-for-manufacturability \cite{strojwas1989design, chiang2007design, orshansky2007design}.

\subsubsection{Time-zero process variations}
\begin{figure}[H]
\centering
\includegraphics[width=0.9\textwidth]{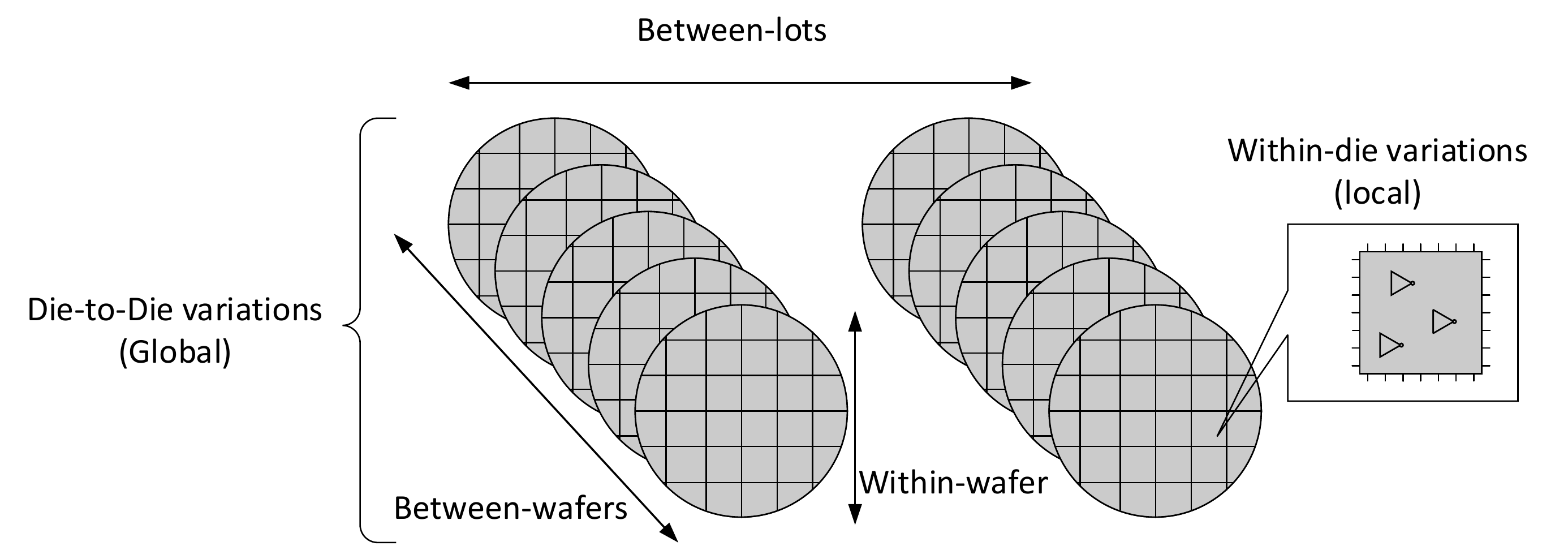}
\caption{Spatial process variations consist of global and local components.}
\label{fig:variation_variation}
\end{figure}

Spatial process variations are deviations of IC parameters compared to their targeted values at the design time. They are created by the limited controllability of a manufacturing process \cite{nourani2006testing}. The origins of these variations are categorized into inter-die (global) and intra-die (local) components \cite{ghosh2010parameter}. As plotted in Fig.~\ref{fig:variation_variation}, the global components consist of between-lots, between-wafers, and within-wafer variations. The local components are the within-die variations.

Global variability refers to the parameter changes for identical devices/interconnects separated by a longer distance, or fabricated at a different time, that result from factors such as processing temperature, equipment/tool properties, etc. between different runs, lots, wafers and dies \cite{saha2010modeling}. Therefore, they impact equally on all transistors and interconnects on a die \cite{bowman2009impact}. 

On the other hand, local variability causes parameter mismatch between identically designed devices/interconnects across a short distance within a die \cite{saha2010modeling, ohnari2013and}. Typical causes are fluctuations in length, width, oxide thickness, flat band control, and the number of dopants \cite{herr1986statistical}.

Conventionally, the main focus was on global variations. This is convenient for designers, as all transistors on the same IC can be modeled with an equal offset. However, local variations have recently entered into the interest zone, and have been pronounced more as a consequence of the aggressive technology scaling, increased chip area, clock frequency, and leakage power distributions \cite{boning1996statistical, tschanz2002adaptive, bowman2002impact}. Moreover, it was observed that local variations can have a much higher impact compared to global variations, e.g. lithography and etch technology can achieve 5\% mismatch of wafer-scale metal width uniformity, whereas within-die variations were reported on the order of 15\% \cite{dai2001timing}. 

The most prominent sources of the variabilities in nano-CMOS transistors are analyzed in \cite{wang2011statistical, asenov2007simulation}, which are Front-end-of-the-line (FEOL) random process variability sources, namely Random Discrete Doping (RDD), Line Edge Roughness (LER); PolySilicon (Poly-Si) Granularity (PSG), Metal Gate Granularity (MGG), and Oxide Thickness Variation (OTV).

\begin{figure}[H]
\centering
\includegraphics[width=.8\textwidth]{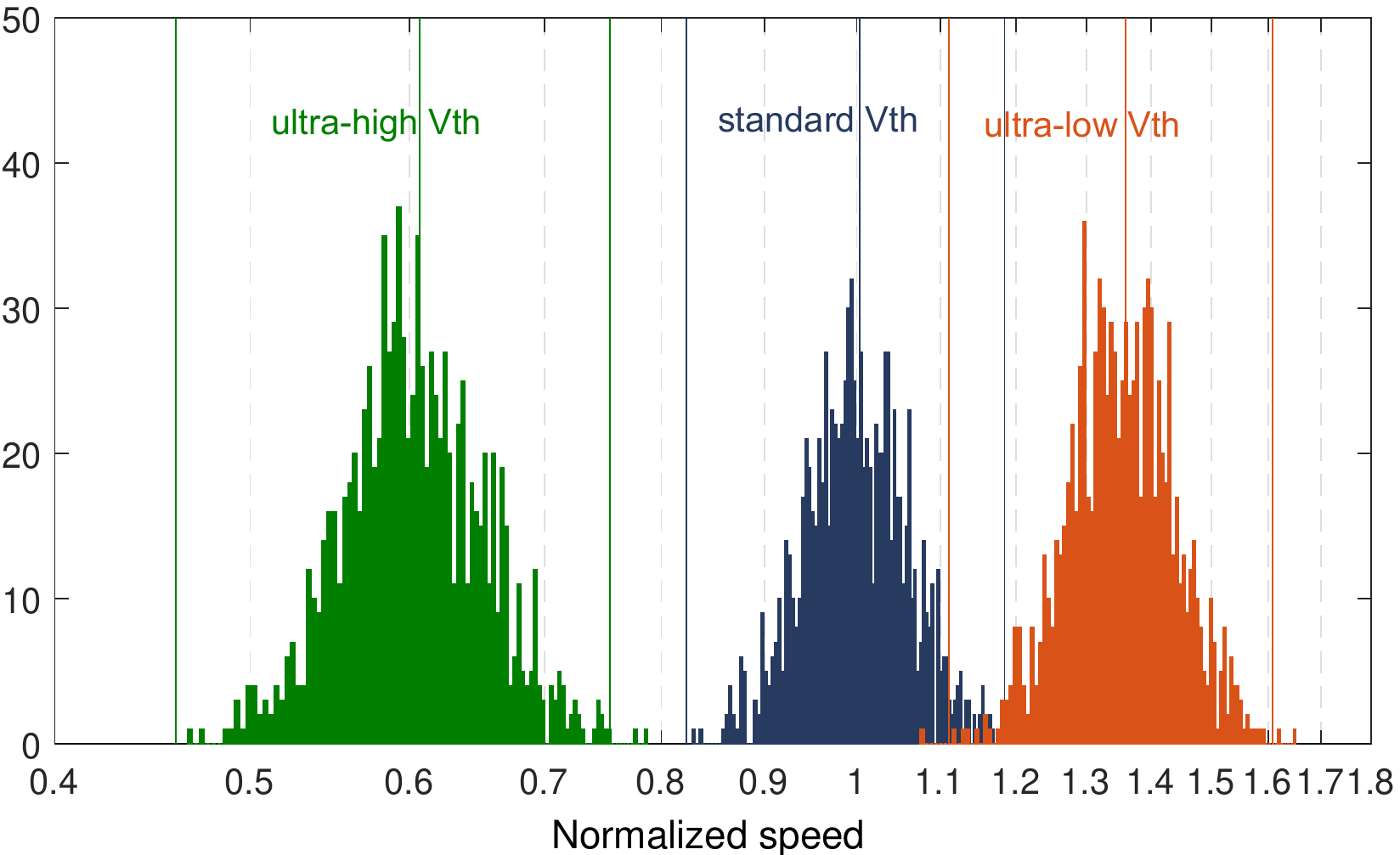}
\caption[The speed (inverse delay) of 28nm fan-out of 4 (FO4) ring oscillators (RO) follows a Gaussian distribution because of process variability.]{The speed (inverse delay) of 28nm fan-out of 4 (FO4) ring oscillators (RO) follows a Gaussian distribution because of process variability. Global and local variations are applied to the width and depth of transistors. The vertical lines indicate the mean speed and the $3\sigma$ speed.}
\label{fig:variation_distribution}
\end{figure}

The time-zero spatial process variations impact delays of a deeply scaled technology. Fig.~\ref{fig:variation_distribution} shows the circuit speed with randomized transistor width and depth, using the Monte-Carlo method. It demonstrates that transistor width and depth variations lead to 25\% speed difference ($3\sigma$) in the standard 28nm CMOS process.

\subsubsection{Time-dependent aging effects}\label{sec:variation_temporal}
A device will degrade over time, according to not only temperature impacts, but also workloads (e.g. voltage, frequency, duty-cycle) \cite{kukner2013impact, stamoulis2016capturing}. These problems are prominent than ever before in the sub-20nm technology nodes \cite{wilson2013international}. A throughout model that contains both reliability characteristics and system-level behavior modeling is thus required. An example is illustrated in \cite{chen2014system}.

As of today, the de-facto solution to the temporal degradation, during design, is to add another layer of margin. This margin is therefore stacked onto the existing process voltage and temperature (PVT) margins (discussed in Section~\ref{sec:variation_pvt}). While temporal degradations are not directly addressed in this thesis, the model introduced in Chapter~\ref{sec:model} can extend to temporal degradation impacts. Furthermore, the trade-offs between reliability (quality) and power savings (by margin shaving) in Chapter~\ref{sec:algo} and Chapter~\ref{sec:system} can also include temporal degradation margins.

\subsection{Environmental and runtime uncertainties}
In addition to (semi-) permanent variations, plenty of environmental and runtime threats also put the digital system quality in danger. These uncertainties are very difficult to model at design time. Therefore, they are assumed with the worst-case corner at design time. Luckily, they usually only affect IC behavior temporarily. Once the uncertainty resources are removed, the circuitry will return to normal condition. Therefore, those uncertainties only modify the circuit computation results, without permanent damages. Occasionally operating outside the specified region is acceptable, as long as the according errors are handled. Although some environmental or runtime conditions (e.g., permanent dose radiation error, mechanical vibrations and shocks, and electrostatic discharge) will fail chips irreversibly, those impacts are easier to deal with. Those permanent environmental impacts are beyond the scope of this work.

\subsubsection{Supply voltage variation}\label{sec:variation_v}
\begin{figure}[H]
\centering
\includegraphics[width=0.55\textwidth]{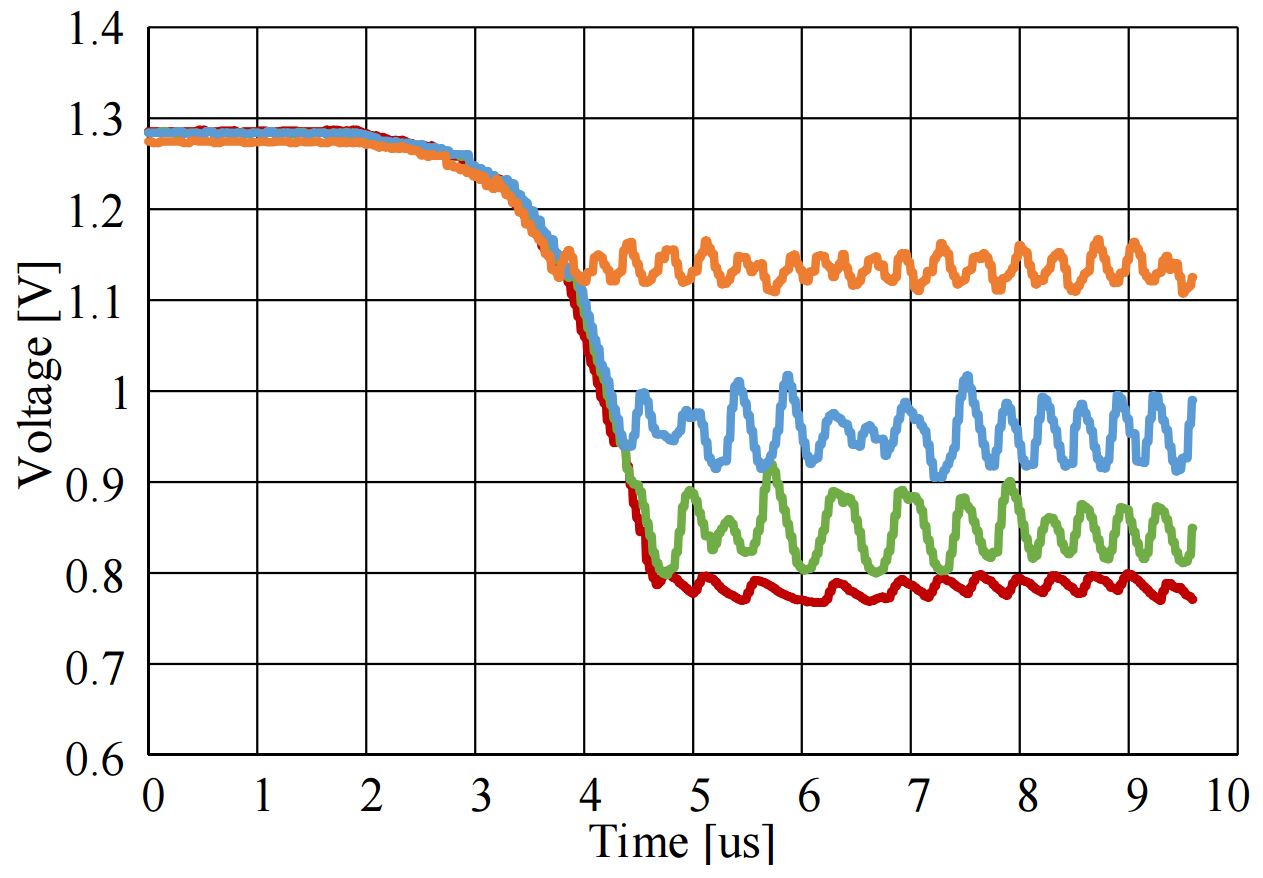}
\caption[The supply voltage of an IC is fluctuating.]{The supply voltage of an IC is fluctuating. The output of a state-of-the-art 28nm power regulator is shown \cite{rachala2016modeling}. Curves represent different output voltage targets.}
\label{fig:variation_regulator}
\end{figure}

Voltage variation is caused by non-idle power generators/regulators and IR-drop in the power delivery network. Fig.~\ref{fig:variation_regulator} shows that the voltage fluctuation from a 28nm power regulator can be as high as 100~mV (peak-to-peak).

In addition to the power source, the power delivery network also leads to voltage drop when currents are flowing through. This is called IR drop. This effect can be modeled with EDA tools so that the delay of each gate is modeled with an individual voltage. 


\subsubsection{Temperature variation}\label{sec:variation_t}
Integrated chips are also affected by the temperature. The usual temperature environment of an IC is -40 to 120 $^{\circ}C$. Furthermore, the circuit consumes energy and hence dissipates heat. This heats up the IC locally and temperature hotspots are produced. Counter-measures are needed for temperature variation, especially for complex multi-core processors \cite{chaparro2007understanding}. 

\begin{figure}[H]
\centering
\includegraphics[width=0.75\textwidth]{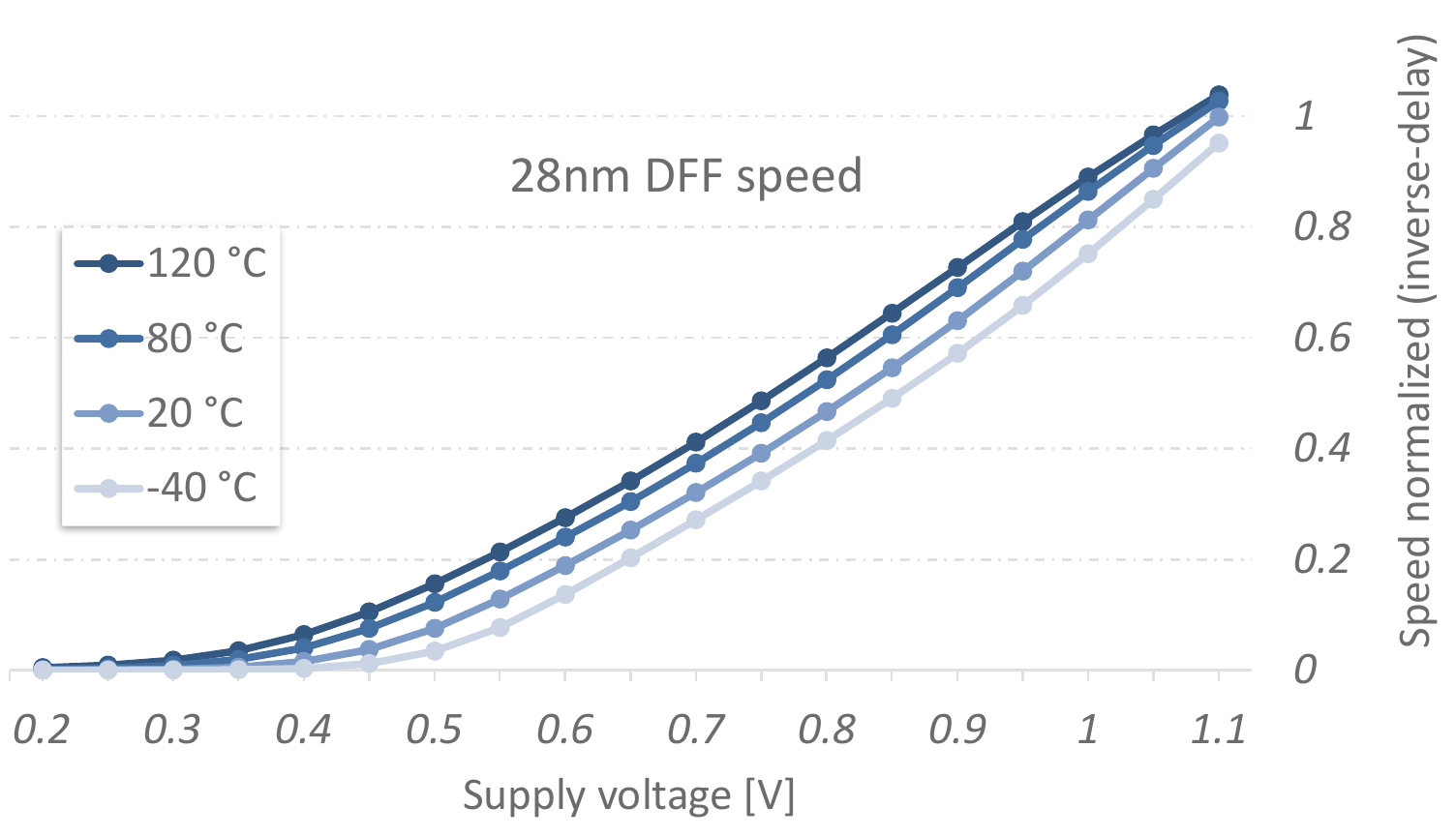}
\caption{The speed of a 28nm circuit changes along with temperature and supply voltage.}
\label{fig:variation_tempa}
\end{figure}

The temperature and voltage combined effects on 28nm circuit speed is plotted in Fig.~\ref{fig:variation_tempa}. In the regime of sub-1.1~V, higher temperature boosts up circuit speed. Reducing voltage decreases circuit speed linearly, down to threshold voltage ($V_{th}$).

\subsubsection{SEU (Single Error Upset)}\label{sec:variation_seu}
SEU is logic bits flipping triggered by external high-energy cosmic rays (e.g. protons and neutrons), and by $\alpha$-particles from package~\cite{pavlov2008cmos}. It is also called soft-errors. Fortunately, the earth magnetic fields shield a majority of the cosmic rays. Consequently, in the past, soft-errors are only relevant to space applications \cite{baumann2002impact, lesea2005rosetta}. Nonetheless, this view has been challenged, mainly due to the following two reasons. Firstly, with more and more data processed and stored in the cloud, the memory in these data centers easily exceeds hundreds of petabyte. Even though the individual error-rate is still negligible, the combined reliability threat should never be overlooked \cite{feng2010shoestring, kleeberger2013cross}. Secondly, the amount of mission-critical applications has or will increase substantially, spreading from the conventional niche industrial control market to daily usage scenarios. For the autonomous driving application, every decision in every car is critical \cite{marchio2014automotive}. In these systems, soft-errors become an issue that must be solved.

Soft-errors can be mitigated by using redundant circuits or systems, in the hope that they will not all fail at the same time. For instance, Chapter~\ref{sec:model} presents a model to help design systems with less redundancy. Note that high-energy particle rays also lead to permanent degradations. This degradation is not covered in this thesis because those degradations are mostly relevant to space applications, which is out of the scope of this work.

\subsubsection{VOS (Voltage Over-Scaling) hardware errors}\label{sec:variation_vos_error}
The other transient reliability threats are errors caused by hardware operating at risky situations. This category is different from the previous two in that, the reliability risks can be tackled at design time, but they are intentionally kept risky, knowing that they can be handled by higher level designs.

By applying VOS \cite{hegde2004voltage}, the energy consumption reduces quadratically according to the voltage, $V_{dd}$, while the delay only increases linearly. Although energy savings are achieved, the drawback lies in the mis-captured data for the memory elements that are caused by the reduced delay. For instance, designs with Razor FF error detectors \cite{razor03} produce sparse timing-errors that need to consider. The ANT techniques \cite{ant1999} intentionally introduce computation errors, in the hope that they will be removed at higher levels. 

The Razor FF and the ANT techniques are further investigated in Chapter~\ref{sec:algo}. In that chapter, error resilient VOS designs are employed to save power for a digital accelerator with algorithm-level optimization. Moreover, Chapter~\ref{sec:system} presents a power reduction potential in a Massive MIMO application, by embracing light-weight but erroneous hardware.

\section{Adjust supply voltage to trade quality for power savings}\label{sec:variation_voltage}

The dynamic power consumption of a digital IC scales with ${V_{dd}}^2$, where $V_{dd}$ is the supply voltage. Therefore, digital circuit designers usually reduce $V_{dd}$ (VOS) for power savings. The savings are error-free, as long as the signal setup timing constraint is satisfied \cite{7062936}. However, the critical (minimum) $V_{dd}$ that guarantees setup-timing closure cannot be determined at the design-time due to permanent and temporary variations. Consequently, hardware errors might be introduced: for logic components, the signal from the longest propagation paths are mis-captured \cite{6569370}; for memory components, this leads to incorrect write/read data/address or data loss \cite{karl2005timing}.

\begin{figure}[H]
\centering
\includegraphics[width=0.65\textwidth]{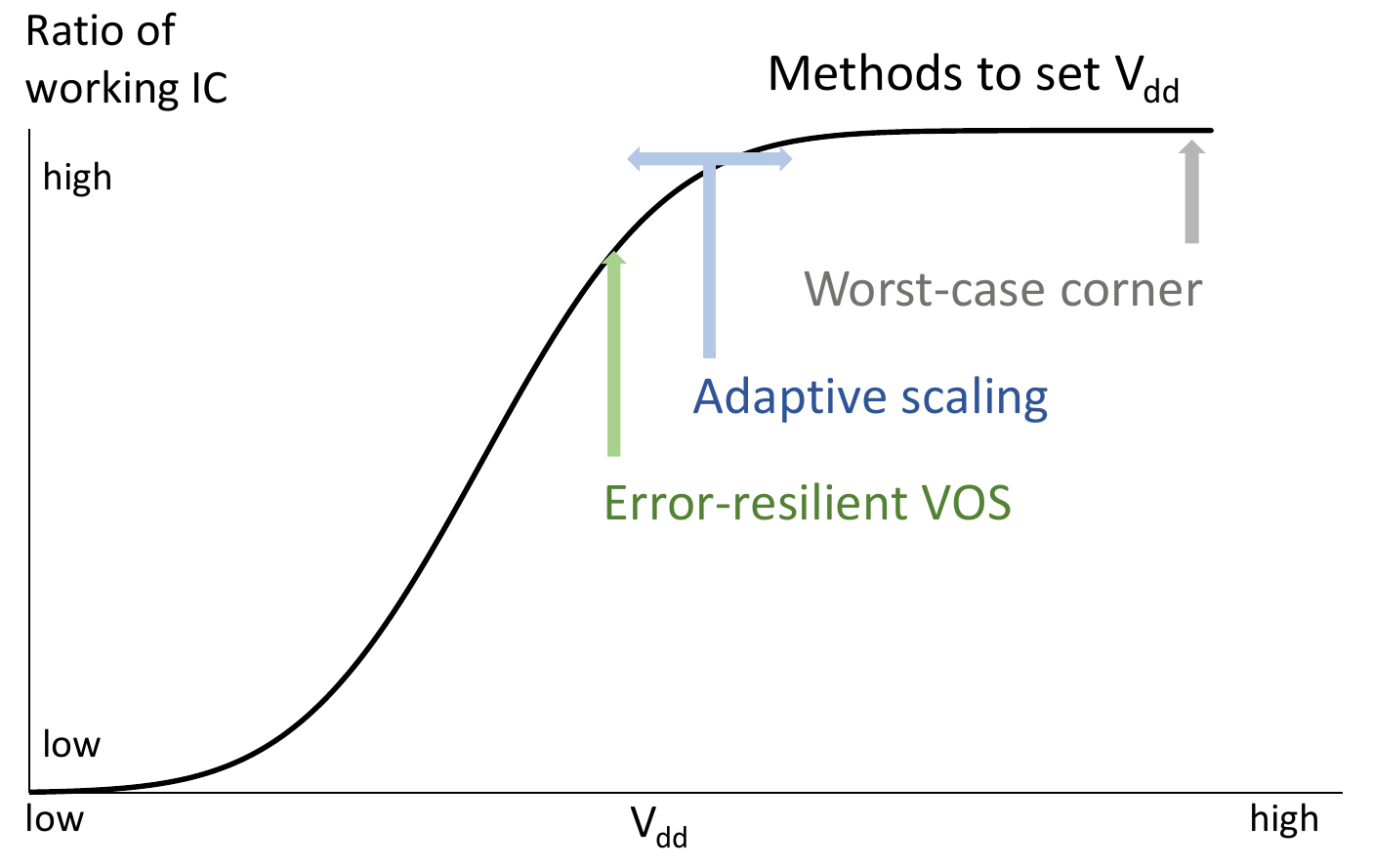}
\caption[Worst-case corner, adaptive scaling, and error resilient voltage over-scaling approaches all cope with speed variability.]{Worst-case corner, adaptive scaling, and error resilient VOS (Voltage Over-Scaling) approaches all cope with speed variability. The worst-case corner leaves energy savings on the table. The adaptive scaling (DVFS and AVFS) provides just needed $V_{dd}$ for error-free operation. Continuing to reduce $V_{dd}$ (VOS) saves more power, but errors will occur. This calls for error-resilient design techniques.}
\label{fig:variation_vdd}
\end{figure}

As shown in Fig.~\ref{fig:variation_vdd}, methods for selecting the $V_{dd}$ are summarized into three categories, i.e., the worst-case corner, adaptive scaling, and error-resilient VOS (Voltage Over-Scaling).

\subsection{Limitations to the worst-case corner approach}\label{sec:variation_wc}
\begin{figure}[H]
\centering
\includegraphics[width=0.55\textwidth]{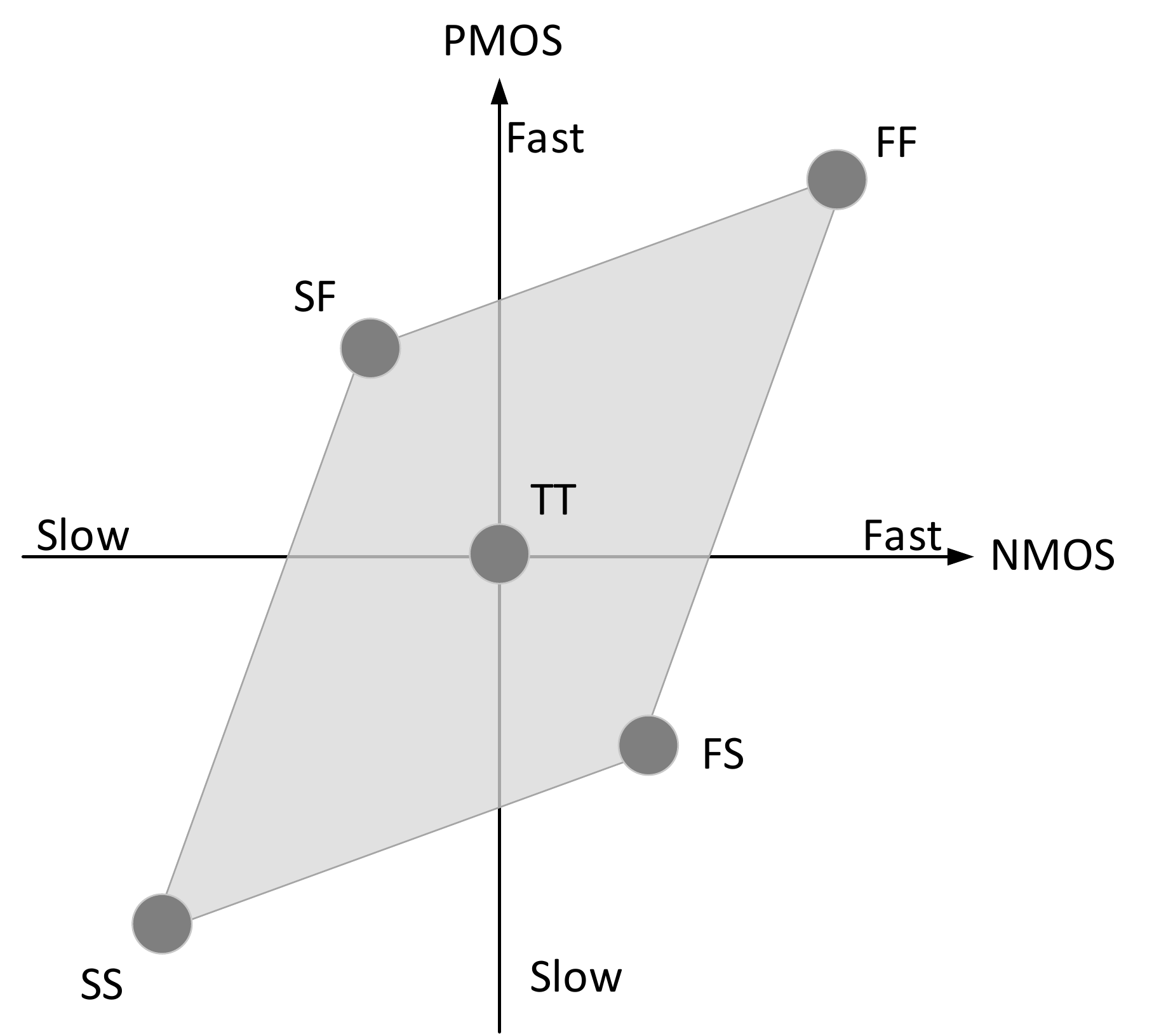}
\caption[Spatial process variability are described by slow and fast corner-case files.]{Spatial process variability are described by slow and fast corner-case files. This figure can be seen as a top-view of the pyramid in Fig.~\ref{fig:intro_pyramid}, where at the application-level TT is desired while all corners are ensured at the circuit-level.}
\label{fig:variation_corner}
\end{figure}

Conventionally, the worst-case corner approach is applied to manage the supply voltage. All chips are set to a fixed and more-than-enough $V_{dd}$, to meet the rarely occurring worst cases. Corner-files are usually used for circuit guard-banding. An example on describing the corner cases of process variations is drawn in Fig.~\ref{fig:variation_corner}. These files describe the worst-case, the typical-case, and the best-case delay values of standard-cells.

\begin{figure}[H]
\centering
\includegraphics[width=0.5\textwidth]{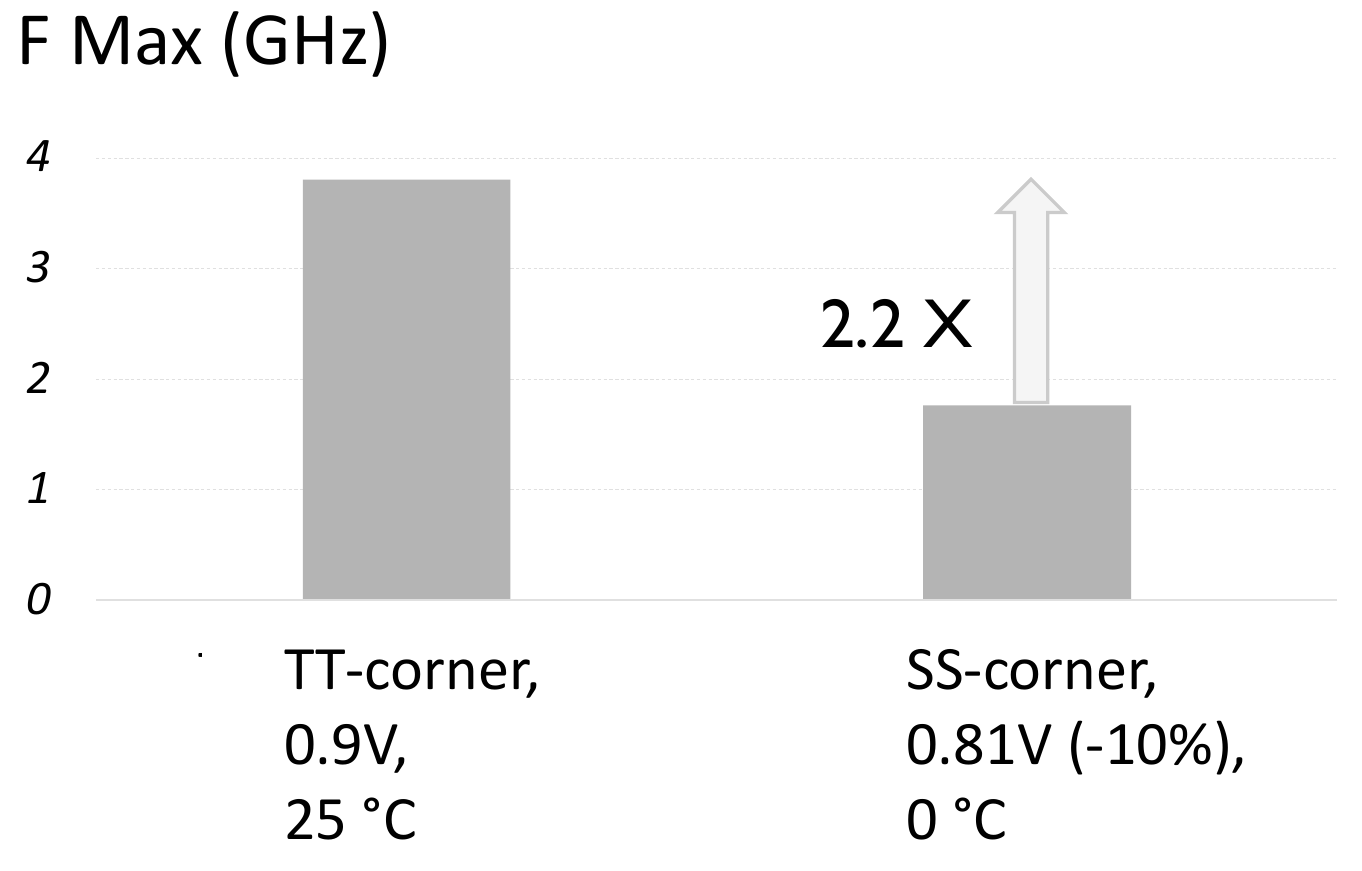}
\caption[The speed of a 28nm circuit exhibits a 2.2x difference between the typical-case and worst-case corners.]{The speed of a 28nm circuit exhibits a 2.2x difference between the typical-case and worst-case corners. Results are obtained by Spice simulation. The worst-case assumes the most pessimistic PVT variations.}
\label{fig:variation_famx}
\end{figure}

With the worst-case design approach, logic synthesis is performed for the slow-case process corner. However, these corner files lack detailed information on local within-die variations. Instead, global on-chip variation margins \cite{chang2012design,stine2007freepdk} are added for all transistors. The global timing margins assume pessimistically that all devices within a die are performing according to their worst-case process conditions. Another pessimistic assumption is that all chips always operate at the worst-case in terms of voltage and temperature. The gap between the worst-case and the typical case is large. For instance, Fig.~\ref{fig:variation_famx} shows that for a 28nm digital circuit, the performance difference (in terms of speed) is as large as 2.2x.

These worst-case conditions are often extremely rare combinations of complex interactions across an IC, which are almost impossible to predict at design time. Hence, as is often stated, the conventional worst-case design style leaves too much performance and power efficiency on the table.

\subsection{Adaptive supply voltage schemes}\label{sec:variation_adaptive}
Adaptive supply voltage schemes minimize the worst-case supply voltage margin used to account for PVT variations, using a suitable feedback signal to close the control loop. It finds the most optimal $V_{dd}$ for each chip, at chip setup stage (post-silicon) \cite{kulkarni2006statistical}, or periodically (runtime) adjust depending on the workload \cite{pillai2001real, martin2002combined, horvath2007dynamic}, and environments \cite{martin2002combined, das2006self, herbert2009variation, miro2014fine}. This method reduces the $V_{dd}$ and hence reduces the power consumption.

\begin{figure}[H]
\centering
\includegraphics[width=0.6\textwidth]{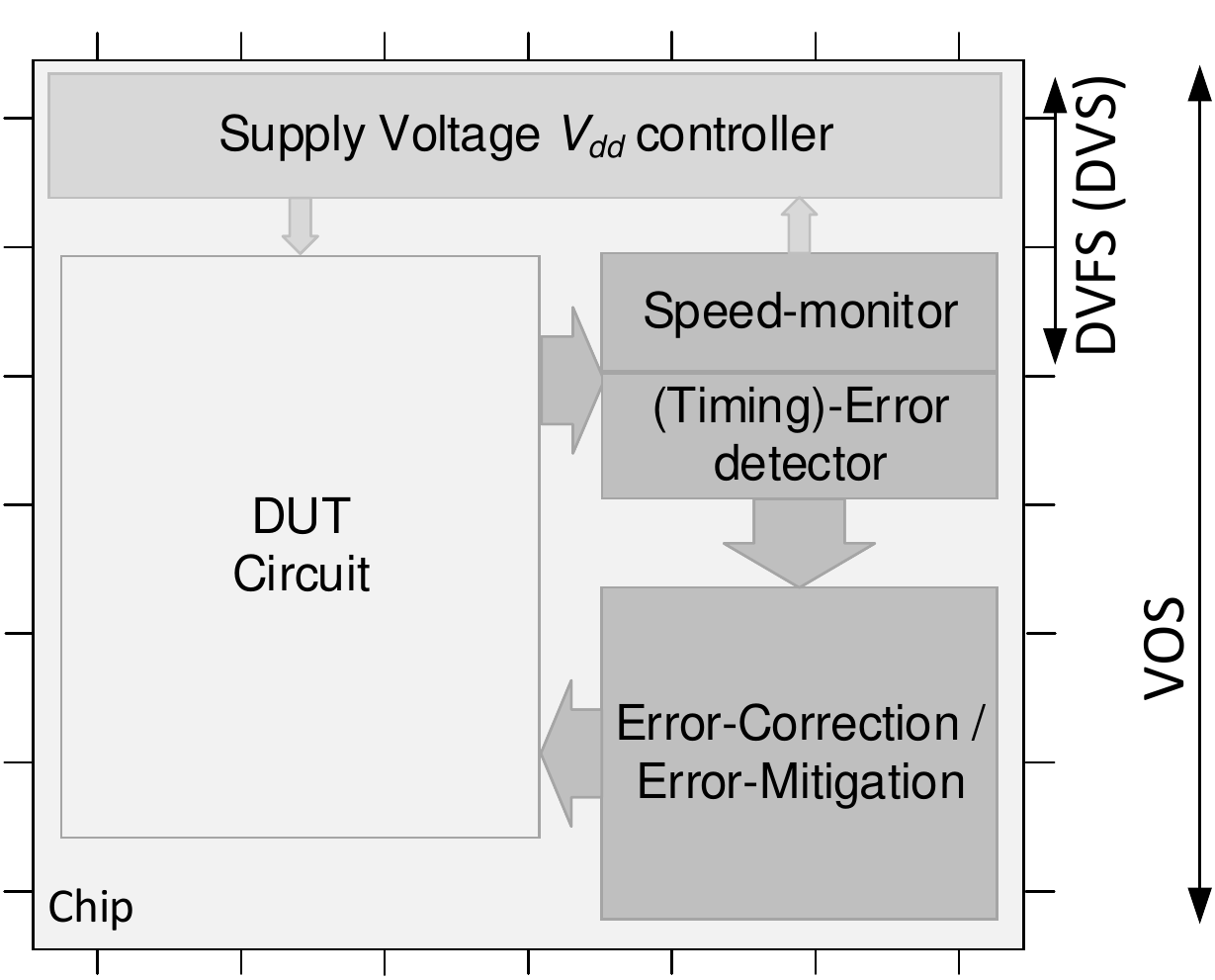}
\caption[The dynamical scaling method handles speed variability. Moreover, the error-resilient $V_{dd}$ scheme also handles circuit timing-errors.]{The dynamical scaling method handles speed variability. Moreover, the error-resilient $V_{dd}$ scheme also handles circuit timing-errors. The dynamical scaling approach utilizes speed monitors and $V_{dd}$ controllers. The VOS approach equips extra error detection and correction/mitigation units.}
\label{fig:variation_top}
\end{figure}

A stereotypical scaling scheme is illustrated in Fig.~\ref{fig:variation_top}. At run-time, the speed detector checks whether the DUT circuit has failed. This information is fed to a supply voltage control unit to adjust the $V_{dd}$. If no or very few errors are detected, the $V_{dd}$ will be scaled down to save power. The supply voltage controller adjusts the $V_{dd}$ slowly (in a coarse-grained temporal manner, e.g. every thousands/millions of cycles), for two main reasons: i) slowly adjusting saves power in the supply voltage controller itself; ii) the transition delay for modifying $V_{dd}$ is inherently much larger than the clock period of the core circuit.

The runtime adaptive schemes are divided into two sub-categories: dynamical voltage frequency scaling (DVFS) \cite{nowka200232, skadron2004hybrid, calhoun2006ultra} and adaptive voltage frequency scaling (AVFS) \cite{burd2000dynamic, das2006self, elgebaly2007variation, miro2014fine}. DVFS uses what’s called open-loop scaling. A system with DVFS schemes listens to the change of system requirements. The hardware vendor determines the optimal voltage for the chip based on the target application and frequency. DVFS is not calibrated to any specific chips. Instead, vendors create a statistical model that predicts what voltage level a chip that’s already verified as good will need to operate at a given frequency. For example, our LDPC decoder designs \cite{li15asscc-ldpc, li15el-ldpc} are characterized for different throughput-voltage combinations. Therefore, the system can opt to reduce $V_{dd}$, at run-time, if the required throughput is low. Note that these frequency-voltage sets are verified at the worst-case corner. So timing margins remain.

Although sometimes used interchangeably, AVFS, in contrast, uses a closed-loop system in which on-die hardware mechanisms manage the voltage — by taking real-time measurements of the junction temperature and current frequency, and adjusting the voltage to match them. This method eliminates the power waste discussed above by reducing the traditional guard bands that are required to ensure proper operation of every piece of silicon. AVFS can detect the circuit speed directly or indirectly. Indirect methods are observing the temperature and supply voltage that affects the circuit speed \cite{tschanz2007adaptive}. A replica circuit \cite{replica02} provides direct hints to the circuit speed. By monitoring the replica circuit, the actual speed of the circuit, which might be difficult to measure, can be guessed. However, these prediction schemes suffer from the delay mismatch between the replica and the actual critical path caused by within-die variations. In another word, the inaccuracy of the timing prediction limits the full exploitation of the variability design margin.

Therefore, in-situ timing-error detectors are proposed to measure the circuit speed. Canary FF \cite{canary04} compares the results on a flip-flop (FF) with a redundant FF that captures a delayed input. It warns the circuit when potential timing-errors are about to occur. The operating condition when the first warning is produced is called the point of first warning (PoFW). The circuit design of Canary FF is explained in Section~\ref{sec:algo_cscircuit}. Chapter~\ref{sec:algo} uses canary FF as the benchmark to demonstrate the power savings of the margin shaving. 

An example of potential adaptive scaling benefits is shown in the Appendix~A (published in \cite{huang16asscc-polar-dfe}). It presents a digital front-end for a polar RF transmitter, that demonstrates 70\% speed gain or 33\% power saving potential if the adaptive scaling method is applied.

\begin{figure}[H]
\centering
\includegraphics[width=0.8\textwidth]{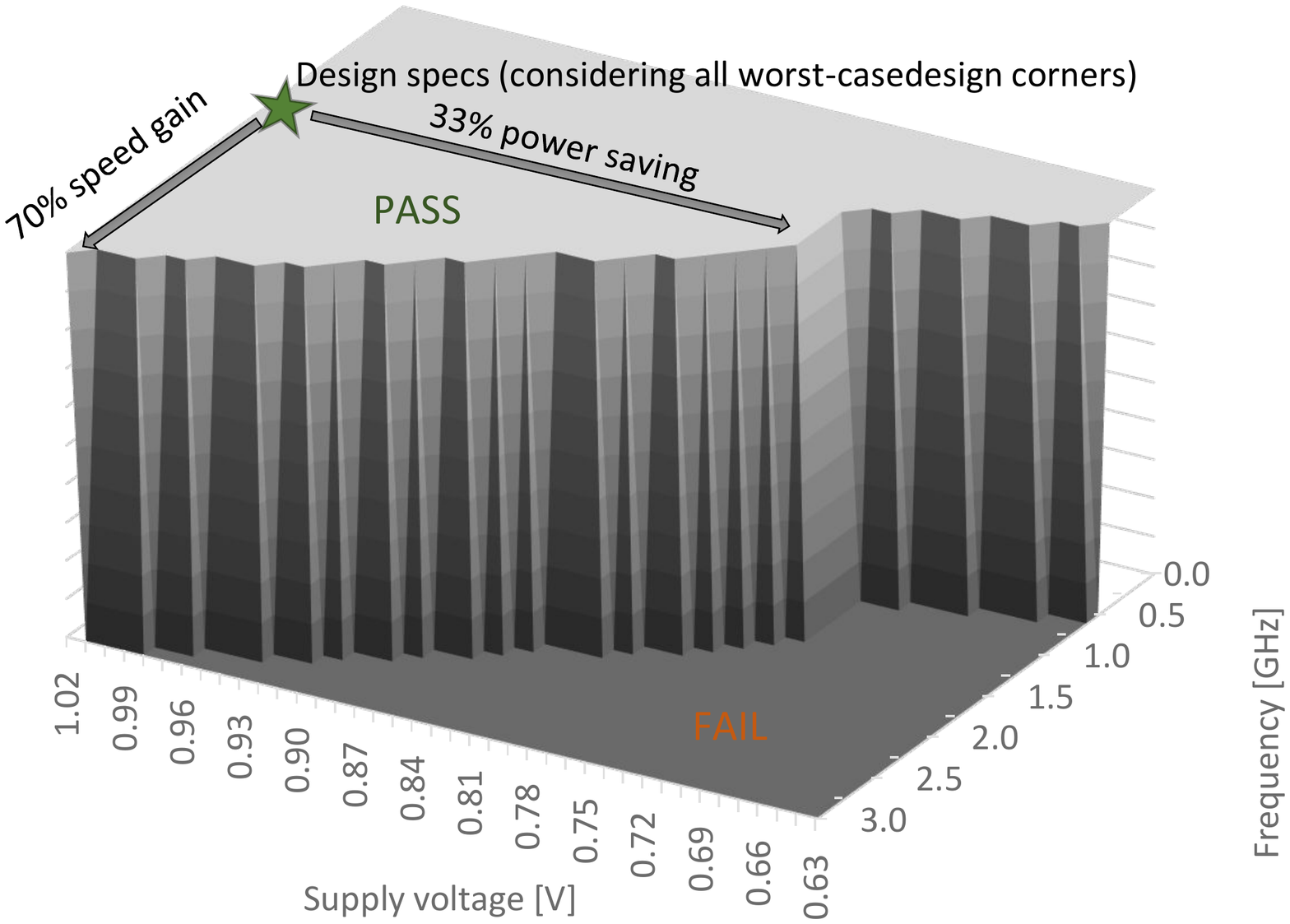}
\caption[The operation region of a digital front-end IC can be much wider than conservatively assumed.]{The operation region of a digital front-end IC can be much wider than conservatively assumed. The IC is measured at $25\celsius$. Orignal data is published in \cite{huang16asscc-polar-dfe}.}
\label{fig:variation_polar_yeild}
\end{figure}
 
These adaptive scaling schemes provide error-free power savings. However, because of transient degradation, e.g. supply voltage noises (IR drop), $V_{dd}$ will drop occasionally \cite{dietel2014compact}. To avoid errors, the circuit either boosts $V_{dd}$ up again quickly or guards with an additional margin from the PoFW. Both methods waste power. Therefore, a more aggressive version of adaptive scaling, VOS, is proposed. It shaves more design margins at the cost of infrequent timing violations.

\section{Cross-layer error-resilient voltage over-scaling schemes}\label{sec:variation_vos}
The third $V_{dd}$ adjustment method is VOS with error resilient designs. When $V_{dd}$ is further scaled down (lower than the safe operating condition), infrequent setup timing-errors occur. However, through adequate error resilient designs, these errors are be detected and corrected. In other words, the cross-layer optimizations are performed to handle errors. This goes beyond the dynamical scaling scheme where only environmental and runtime information is shared across design levels.

An error-resilient VOS scheme operates in a closed-loop as shown in Fig.~\ref{fig:variation_top}. In addition to $V_{dd}$ adjustment loop, it also utilizes an error detection and correction loop. The most popular error resilient approaches, i.e., in-situ error detection and corrections, and ANT techniques, are discussed in the following subsections. 

\begin{figure}[H]
\centering
\includegraphics[width=0.65\linewidth]{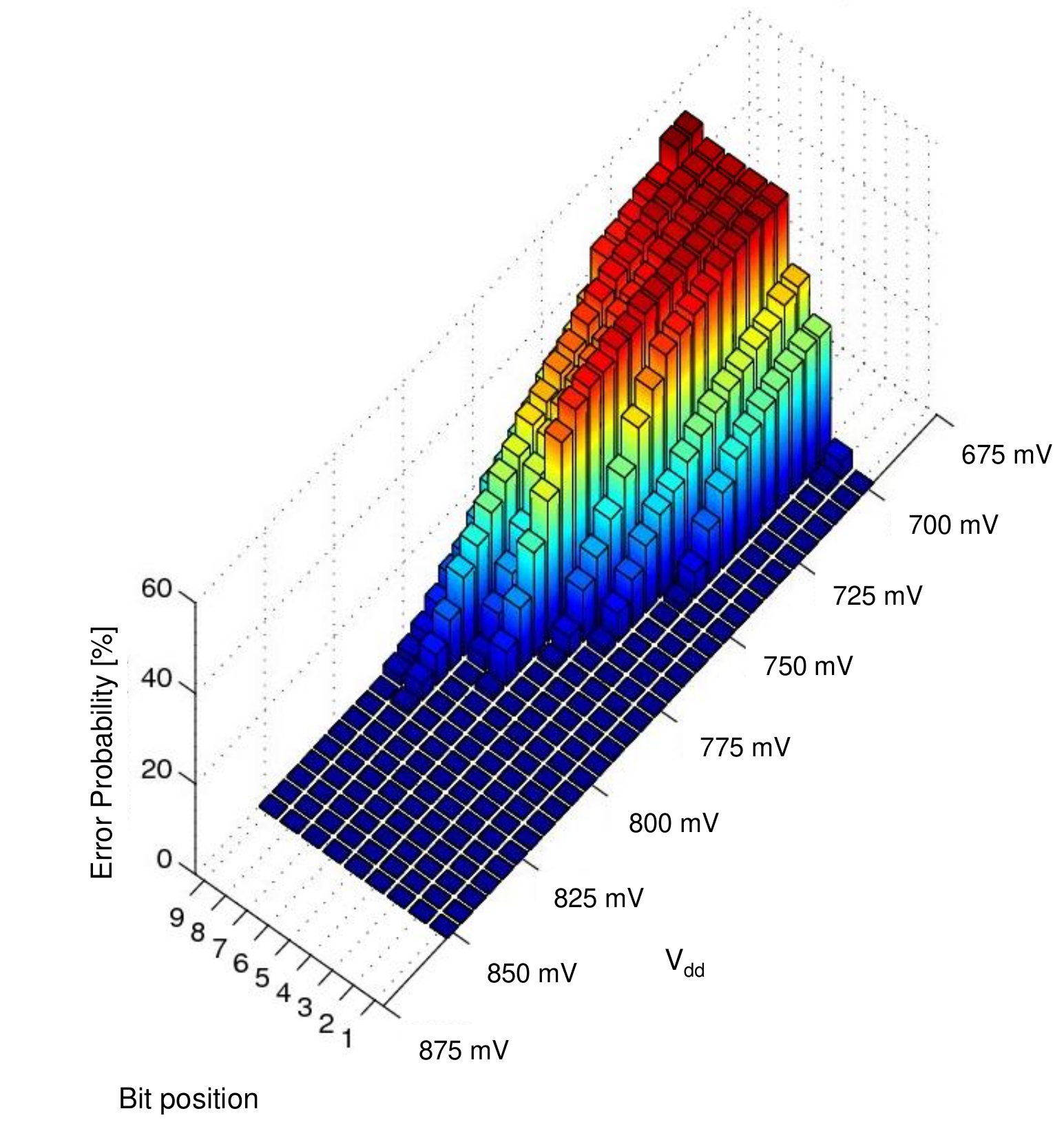}
\caption{With VOS, errors firstly occur at MSB for carry-ripple adders.}\label{fig:variation_error_position}
\end{figure}

Signals on timing critical paths take longer to propagate. Therefore, they call for more consideration during VOS, since they are more likely to fail during VOS. An example is shown in Fig.~\ref{fig:variation_error_position}, where the MSB usually has a higher error possibility than the LSB for a digital adder. This is especially true for carry-ripple adders, in which the carry propagates from the LSB to the MSB, making the delay of MSB longer \cite{liu2010computation}. 

\subsection{In-situ error detection and corrections}\label{sec:variation_detection}
For error detection, timing checks on flip-flops (FF) have been proposed and widely utilized. This scheme modifies the FF in the circuit. In-situ FF based schemes \cite{razor03, bowman09, nicolaidis2013adda, bubble13} are suggested to overcome this mismatch for microprocessors. Razor I \cite{razor03} detects a timing-error by employing an extra shadow latch. In contrast to Canary FF, the Razor can find the exact timing slack for design margin shaving. The details of Razor flip-flops and its variants are discussed in Chapter~\ref{sec:algo}.

For error correction, counter-flow \cite{razor03} and instruction-replay \cite{bowman09}, that issues extra clock cycles, are proposed. They result in multiple-cycle throughput penalties once a timing-error is detected. Bubble Razor \cite{bubble13} reduces the throughput penalty to 1 cycle. However, the design is based on two-phase latches, which is difficult to be incorporated into mainstream EDA tools. Global clock gating scheme \cite{razor03} also achieves one cycle penalty. Nevertheless, it is difficult to implement for large area high-speed circuits. Recently, a local 1-cycle error correction scheme is proposed \cite{shin13}. However, the timing constraint for its error flag signal becomes a challenge for multiple fan-in situations.

The Razor techniques \cite{razor03,5654663,bull2011power,fojtik2012bubble} predicts the PVT variation at circuit-level and exploits the design margin by providing minimum but sufficient $V_{dd}$ to the chip. \cite{bull2011power} reports 30\% and 52\%  power consumption saving on a typical die and a fast die, respectively; \cite{fojtik2012bubble} achieves 54\% saving on a typical die and 60\% saving on a fast one. A potential hazard of applying this technique is massive throughput reduction when $V_{dd}$ is dropped lower than the sufficient voltage. In this situation, the signal processor would terminate the processing to meet the throughput requirement, which leads to output quality degradation.

The infrequent errors, occurred in the time frame of nano seconds, are resolved by micro-architectural level error correction schemes. The supply voltage controller adjusts may  the $V_{dd}$ every tens of seconds,

\subsection{Arithmetic noise tolerance}\label{sec:variation_ant}
The arithmetic noise tolerance (ANT) techniques \cite{Shim2004,hegde2004voltage,narayanan2010scalable,karakonstantis2009system} save power in digital signal processors by gracefully sacrificing the signal-to-noise-ratio (SNR), admitting that a certain amount of errors might occur. They detect errors by algorithmic comparison. For example, \cite{hegde2004voltage} detects errors by observing the error-prone results, to check if it is within a reasonable range. The mitigation is accomplished by temporal or spatial redundancies. As a consequence, the setup timing-errors during $V_{dd}$ scaling are translated into graceful signal quality degradation.

ANT is applied to systems that deal with soft output requirements. That is, the output quality is not measured as a yes or no criteria. Instead, it is in the gray zone for quality. For instance, signals in a wireless communication system are measured by their SNR, EVM, or BER; signals in an audio player are measured by the PESQ value. In these soft output systems, a certain degradation can be acceptable. 

\begin{figure}[H]
\centering
\includegraphics[width=0.6\linewidth]{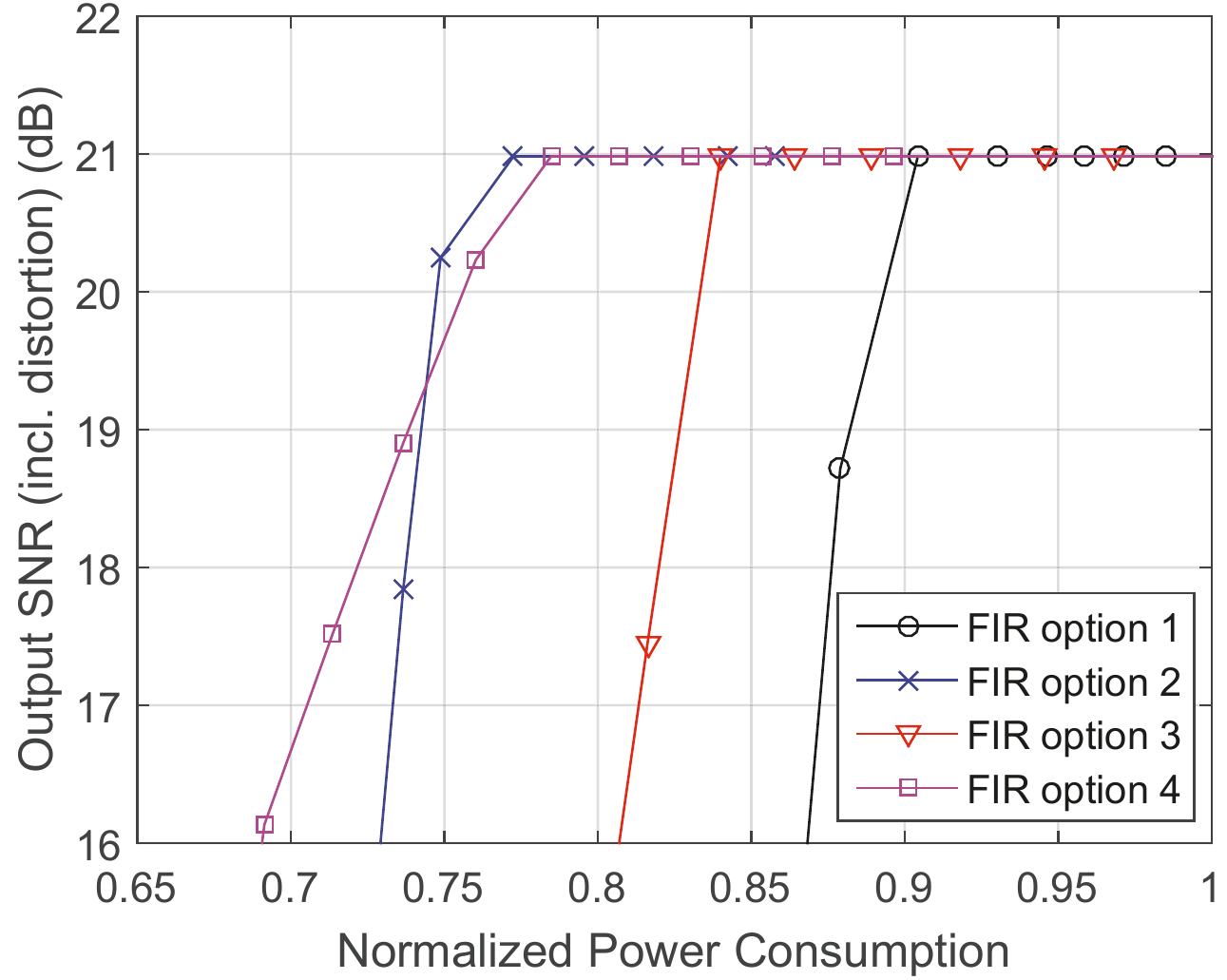}
\caption{VOS with error resilient techniques save power on FIR filter, at the cost of signal quality degradations \cite{liu2010computation}.}\label{fig:power_fir}
\end{figure}

Numerous designs \cite{hegde2004voltage,narayanan2010scalable,karakonstantis2009system,huang14sips-error-resilient,Shim2004,huang16jsps-error-resilient} are proposed to reduce the hardware errors at a given power budget. For instance, Fig.~\ref{fig:power_fir} provides an example of energy saving in 65nm COMS FIR filter brought by VOS, at the cost of SNR degradation. Besides, \cite{Shim2004} utilizes reduced precision redundancy to reduce the power consumption by 40\% on a digital FIR filter at the cost of slightly degrading the 23~dB SNR signal into 22~dB, and by 35\% for a 64-point FFT when lowering the SNR from 55.5~dB to 55~dB.

\section{Conclusion}\label{sec:variation_con}
PVT and other environmental and runtime variations challenge the quality of digital circuits and system. The traditional worst-case, which introduces big safety margin, waste power consumption. In 28nm CMOS technology, the safety margin is as high as 2.2x the speed potential. Consequently, careful trade-offs on power and quality that considers across all design levels are encouraged. 

Dynamical and adaptive scaling with on-chip timing-error monitors saves power by providing a just-needed supply voltage. The power saving will be maximized when the voltage is scaled down more aggressively where errors occur.  In these situations, error-resilient designs are crucial to mitigate errors and ensure quality.



\cleardoublepage

\chapter{Gate-level error impacts modeling analysis for random error mitigation}\label{sec:model}


This chapter presents a gate-level random error model. The model, named as SERIAL -- SignificancE RankIng ALgorithm, is an analytical approach that eliminates the pain of traditional time-costly Monte-Carlo simulation. It ranks the FF in a digital circuit according to their contribution to the outcome. The efficiency and effectiveness are shown on benchmark circuits. For instance, the computation of the algorithm can be finished within 30 seconds for a 64-point FFT accelerator (52k gates). With the ranking, circuit designers have the opportunity to selectively ensure the most important FF (e.g. FF hardening, VOS margin), without excessive hardening overheads. On an FFT circuitry, the algorithm helps to reduce the hardening overhead of 100\% FF into 45\%. The work in this chapter is unveiled in \cite{huang16dac-serial}.

The rest of this chapter is organized as follows: Section~\ref{sec:model_intro} presents the motivation. Section~\ref{sec:model_work} reviews pre-existing approaches and highlights the contribution of the proposed SERIAL model. Section~\ref{sec:model_serial} explains the model and the algorithm to calculate it. Section~\ref{sec:model_result} verifies the scalability and effectiveness of the model. Section~\ref{sec:model_apply} applies the model to facilitate the design of an FFT processor. Finally, Section~\ref{sec:model_con} concludes this chapter.

\section{Demands for a reliability model}\label{sec:model_intro}
System reliability has been a major concern since the beginning of electronic design age \cite{siewiorek2014reliable}. Traditional major threats are yield-related manufacturing faults (stuck-at) \cite{mei1974bridging}, space-radiation incurred soft-errors \cite{hazucha2000impact}, and wearing out/aging degradation \cite{yamabe1985time, vattikonda2006modeling}. Recently, VOS techniques introduce another reliability threat. These techniques save energy by design-margin shaving, e.g. Razor \cite{razor03} and ANT \cite{ant99}. However, errors are deliberately introduced during the margin shaving process. These traditional and newly introduced reliability threats need to be handled. Otherwise, system quality is at risk.

Circuit and system designers used to takes a `best effort' design flow. They apply reliability enhancement techniques with the best effort to ensure reliability. The most effective methods include ECC on memory \cite{slayman2005cache}, FF and SRAM hardening \cite{jahinuzzaman2009soft}, design rule check (DRC) modification and other safety margin insertion methods. The enhancement overhead ranges from as low as 12\% for a SECDED (Single Error Correction and Double Error Detection) Hamming code~\cite{richter2008new}, to as high as 200\% for TMR (Triple Module Redundancy). Designers usually verify the application-level reliability at the very end of the design process. Following this design flow, designers usually have no knowledge of which unit contributes the most to the system reliability, which is to say, we cannot distinguish which part is necessary and which part is over-design.

The `best effort' design flow, although easy to implement, either leaves safety margin wasted, or requires modifications at the very end (sometimes even leads to another round of IC process tape-out). A more efficient design flow is to model the reliability factor in every phase of the design process. As a consequence, methods to tackle these issues can be performed effectively and timely.

To tackle the reliability threat, a model to quantify impacts of reliability threats is needed. With the help of the model, RTL designers can provide just-needed reliability counter-measures, avoiding over-designing margin in chip power and area. The following section discusses prior modeling studies.

\section{Modeling and enhancement techniques for reliability}\label{sec:model_work}
Reliable system design depends on accurate and easy-to-use reliability models. Besides, it also requires effective and low-cost techniques to enhance reliability, once the system fails to meet the reliability target. This section discusses works on both aspects, as a background for this chapter.

\subsection{Error modeling}
\begin{figure}[H]
\centering
\includegraphics[width=1\textwidth]{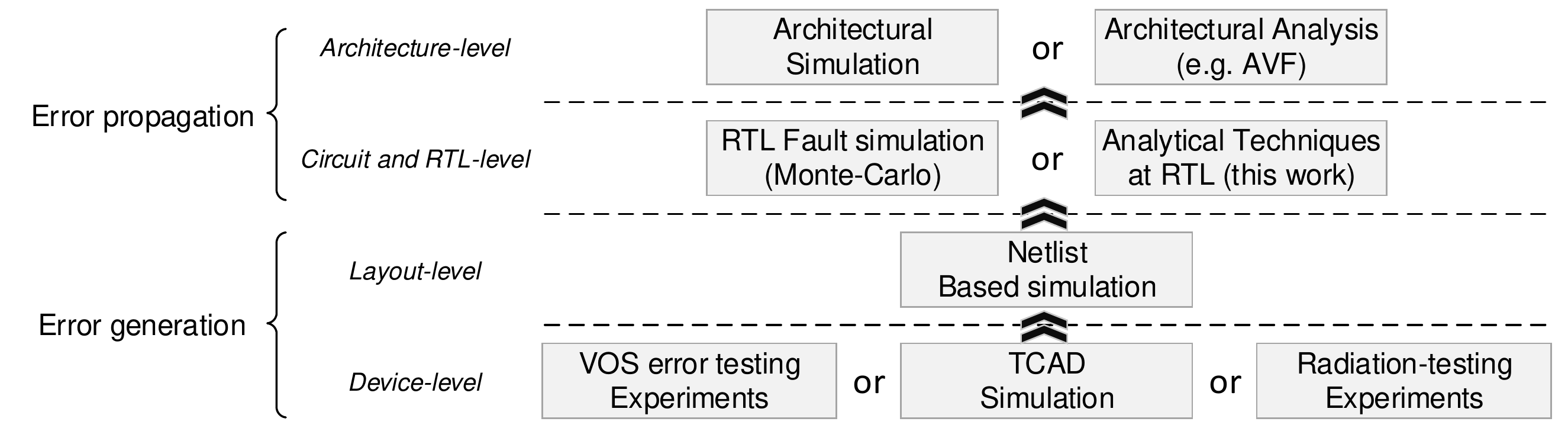}
\caption{Modeling process and structure for error impacts.}
\label{fig:model_rank_level}
\end{figure}

The works on reliability modeling consist of two categories: i) error generation and ii) error propagation (Fig.~\ref{fig:model_rank_level}). The partitioning is the same for both traditional errors (e.g. soft-error and stuck-at error) and VOS-induced errors.

\subsubsection{Error generation}

The principal difference among various error sources lays in the error generation. The error-generation model of soft-errors demands radiation-testing experiment and device-level simulation. It also requires standard-cell level analysis of error production. The model is finalized by the layout-level (e.g. Spice) single-cycle analysis to address effects of logical masking, electrical masking, and vulnerable window ~\cite{ebrahimi2015comprehensive}. 

In contrast, to model VOS-induced error generation, device-level reliability issues and timing (signal delay) degradation are analyzed at reduced $V_{dd}$ and IR-drop noise. The layout is simulated (e.g. using Spice), to provide error generation information.

\subsubsection{Error propagation}

The error propagation model describes the error impact on algorithm output, or on application-level performance. Therefore, the propagation model is uniform, regardless of the source of errors (soft-error and VOS error). 

Several estimation techniques for reliability have been proposed to investigate the behavior of circuits under faults. For instance, ad-hoc investigations into digital circuits are carried out for specific applications \cite{gaisler2002portable, karakonstantis2012exploitation, may2008case, murali2005analysis}. However, they require an in-depth understanding of both reliability and the circuit design, which are not always available. 

Therefore, a more generic design approach is coveted. Within this context, the Monte-Carlo simulation (e.g. \cite{holcomb2009design, clark1995fault, mirkhani2015depth}), taking the pre-computed error generation information, has become the most popular method for gate-level error propagation modeling. 

Whole system simulation on AVF (Architectural Vulnerability Factor) modeling~\cite{mukherjee2003systematic} are performed, taking the results from gate-level models. Recently, \cite{biswas2008computing,  wang2013accurate} discuss the vulnerability behavior of faulty structures at the architectural level. \cite{polian2011modeling} suggests a transient error model. It analyzes the possibility that soft-errors reflect on the output. Although it predicts the occurrence rate of an output error, it cannot model the consequent severity of errors, which however is essential.

\subsection{Reliability enhancement techniques}
This subsection discusses reliability enhancement techniques that will be utilized when the modeling is complete.

Many reliability enhancement techniques have been proposed \cite{6905763, 5236054}. At gate level, \cite{matush2010area, 5236054} introduced hardened flip-flops (FF hardening), which replace the traditional flip-flops, to reduce single-event upset without large area and power overhead. 

Various mitigation schemes to cope with the impact of soft-error have been investigated. For memories and communication systems, errors can be corrected by redundant data, e.g. error correction code \cite{4212048, mitra2005robust}. For digital circuits, selective triple modular redundancy \cite{samudrala2004selective, Bolchini2007} is effective in reducing the chance of transferring the circuit-level faults to micro-architectural level errors.

Besides the efforts in reducing the application-level occurrence of errors, designing systems that operate reliably even with the presence of errors is also of great interest. In these soft-quality requirements systems, errors are acceptable as long as the system output still meets the requirements. For instance, the transient error in a wireless communication system can easily be tolerated by the symbol detector, as long as the error vector magnitude is allowed \cite{karakonstantis2012exploitation}.

Despite the green-power benefit of VOS, the circuit reliability is at risk of timing-errors and noise contamination. Therefore, it is either suggested to apply an additional margin to guarantee reliability \cite{kunitake2011possibilities} or to implement error handling schemes \cite{razor03}. Both methods introduce overhead. 
   
Considering that all the error mitigation methods discussed above introduce area and power expenses for the system, circuit designers must carefully investigate the probability and characteristics of reliability risks, as well as their consequences, to provide just-needed reliability enhancement efforts.

Although methods for error mitigation are proposed, they do not provide systematically guideline on which part should harden. This chapter identifies the bottlenecks in random error hardening, and serves as a guideline for selectively hardening.

\section{Contributions of this chapter}
The error propagation model method is essential in judging the impact of reliability threats. It can analysis errors with various generation mechanism. However, the general Monte-Carlo method does not scale to large circuits because of runtime constraints, even with the help of grouping of similar gates. In summary, it is of immense interest to provide analytical approaches to investigate logic gate behavior under faults.

This chapter presents an analytical approach (SERIAL) for gate-level modeling. The approach efficiently identifies the most significant flip-flops that contribute mostly to the output quality. As a result, all flip-flops (FF) and input ports are sorted out based on their significance to the final output. It analyzes algorithmic effects of errors (error magnitude), in contrast to \cite{polian2011modeling} where only the chance of detectable errors 
were covered.

This chapter focuses only on finding out the most significant FF because they are the starting and end point of logic signals. Soft errors can be mostly eliminated locally by applying FF hardening \cite{ramanarayanan2003analysis}. Once the most significant FF are identified, circuit designers can exploit extra redundant FF or the corresponding logic (e.g. double/triple modular redundancy), to alleviate the reliability issue. From the VOS errors point of view, finding out the most significant FF helps to understand which FF need to be protected (from VOS errors). The only difference, compared to soft errors, is the error occurrence data, which can be obtained by Spice simulation.

This algorithm is challenging: errors might occur in random situations, and their impact is highly dependent on the circuit’s states. Therefore, this work assumes that the input data and circuit’s states are completely random. Another assumption is that signals coming from different FF to the same FF are independent. These assumptions make the model unrealistic at first sight. However, it still provides very useful information: the identified most significant FF are usually the clock-gating controllers, the state recording registers and loop controller, which are always essential for the system’s reliability, regardless of input data and states. By only hardening the most significant FF (a small percentage of FF), the system's reliability will improve remarkably.

For an absolute correct result, all FF should function correctly, which is difficult to guarantee due to area/power constraints. This work answers the question which FF to protect, when only a small portion of protected is allowed due to overhead restricts. Note that the result of SERIAL serves as a relevant guidance for circuit designers to choose which FF to harden, yet, it does not provide a guaranteed reliability model.

This approach focuses on the effects of faults in digital circuits, rather than specific consequences of soft-errors. Therefore, it can analyze the effect of soft errors, VOS errors, and even the conventional stuck-at faults, provided that the corresponding error generation model is given.

\section{SERIAL -- a SignificancE RankIng ALgorithm for error effects modeling}\label{sec:model_serial}

The proposed algorithm, SERIAL, takes the netlist of the digital circuits as the input, extracts the connection information between FF and in/out ports, and ranks all FF regarding their contribution to the final output. This section explains the algorithm in detail.

In this chapter, individual values are denoted by normal math symbols (e.g. $df$). The corresponding collections, e.g. sets, vectors, and matrices are denoted as upper case math symbols (e.g. $DF$).  

The \textbf{significance} of an FF or an in/out port is defined as the impact of its state change on the output. An FF with high significance implies radical system failure, once it is infected by an error. Therefore, to guarantee the system performance, circuit designers must identify the FF with high significance and then harden them. The significance on the output is initialized by the user of SERIAL users. The algorithm distributes the significance to all FF and input ports.

\begin{figure}[H]
\centering
\includegraphics[width=1\textwidth]{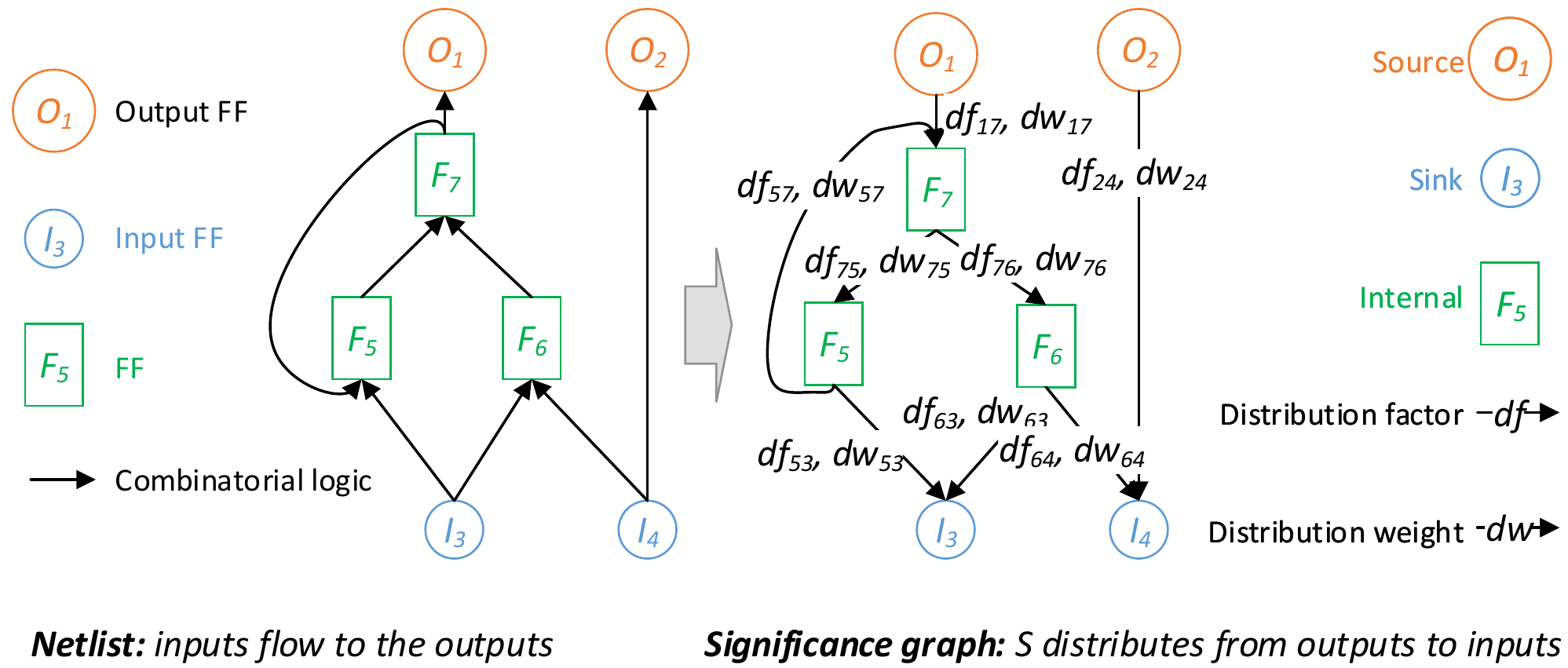}
\caption[A significance graph inverses the signal direction of the corresponding netlist.]{A significance graph inverses the signal direction of the corresponding netlist. Each node in the significance graph distribute its significance $S$ to its neighbors through the arrows.}
\label{fig:model_rank_netlist}
\end{figure}

To compute the significance, the notion of significance graph is introduced. Fig.~\ref{fig:model_rank_netlist} shows an example of converting a netlist into significance graph.

A \textbf{significance graph} is a weighted directed graph, which is formally modeled as a tuple <$O$, $I$, $F$, $S$, $DF$, $DW$>. The elements are explained as follows:
\begin{itemize}
\item[$\bullet$] \textit{Nodes}, which include:
    \begin{itemize}
        \item[$\cdot$] \textit{Source} ($O$): the output ports for the digital circuit are regarded as sources, since they distribute the significance to their neighbors;
        \item[$\cdot$] \textit{Sink} ($I$): the input ports; 
        \item[$\cdot$] \textit{Internal} ($F$): the FF, as the intermediate nodes for the significance graph.
    \end{itemize}
\item[$\bullet$] \textit{Significance} ($S$): The $s_i$ of each node represents its contribution to the final output.
\item[$\bullet$] \textit{Arrows} represent the $s$ distribution between nodes. Each arrow has two features:
    \begin{itemize}
        \item[$\cdot$] \textit{Distribution Factor} ($DF$): it represents the portion of $s$ distributed from one node to its adjacent tails. Therefore, the sum of the $df_{ij}$ that start from the same node is normalized to a unity 1, representing the significance of the node is distributed to its adjacent nodes. For instance, in Fig.~\ref{fig:model_rank_netlist}, $df_{75}+df_{76}=1$.
        \item[$\cdot$] \textit{Distribution Weight} ($DW$): it denotes the distributed $s$ in the arrow. Therefore, $DW = S \cdot DF$. 
    \end{itemize}
\end{itemize}

The $S$ of each node is determined by its adjacent heads, determined by the $DF$, which is shown as Equation~\ref{eq:model_s}:
\begin{equation}
\begin{bmatrix}
S_F \\
S_I
\end{bmatrix}
- DF *
\begin{bmatrix}
S_O \\
S_F
\end{bmatrix}
= \epsilon \approx 0,
\label{eq:model_s}
\end{equation}

where $S_F$ denotes the $S$ of the $FF$ ($n$ in total); $S_I$ represents the $S$ of the $m$ inputs, $S_O$ is the significance of the $k$ outputs; $\epsilon$ is the computational error (which is 0 for the exact solution); and $DF$ is the transfer matrix (size [$n$+$m$, $k$+$m$]) for $S$:
\begin{equation}
S_F =
\begin{bmatrix}
S_{f_1} \\
S_{f_2} \\
\vdots \\
S_{f_n}
\end{bmatrix};
S_I =
\begin{bmatrix}
S_{i_1} \\
S_{i_2} \\
\vdots \\
S_{i_m}
\end{bmatrix};
S_O =
\begin{bmatrix}
S_{o_1} \\
S_{o_2} \\
\vdots \\
S_{o_k}
\end{bmatrix};
\label{eq:model_s1}
\end{equation}

\begin{equation}
DF =
\begin{bmatrix}
DF_{O \rightarrow F} & DF_{F \rightarrow F} \\
DF_{O \rightarrow I} & DF_{F \rightarrow I}
\end{bmatrix}.
\label{eq:model_s2}
\end{equation}

The sum in each column of $DF$ is 1. The initial values of $S_O$ in Equation~\ref{eq:model_s} are initialized by users. By default, they are all set to 1, implying the equal importance of the output ports. 

The designer is free to set these initial values, according to the application requirement. For instance, if the output represents a 2's complementary binary number, it is advised to set the significance to $2^i$ for some applications, where $i$ is the bit position.

The matrix $DF$ is determined by the combinational logic between nodes in the significance graph, which is extracted from the netlist.

\begin{figure}[H]
\centering
\includegraphics[width=0.9\textwidth]{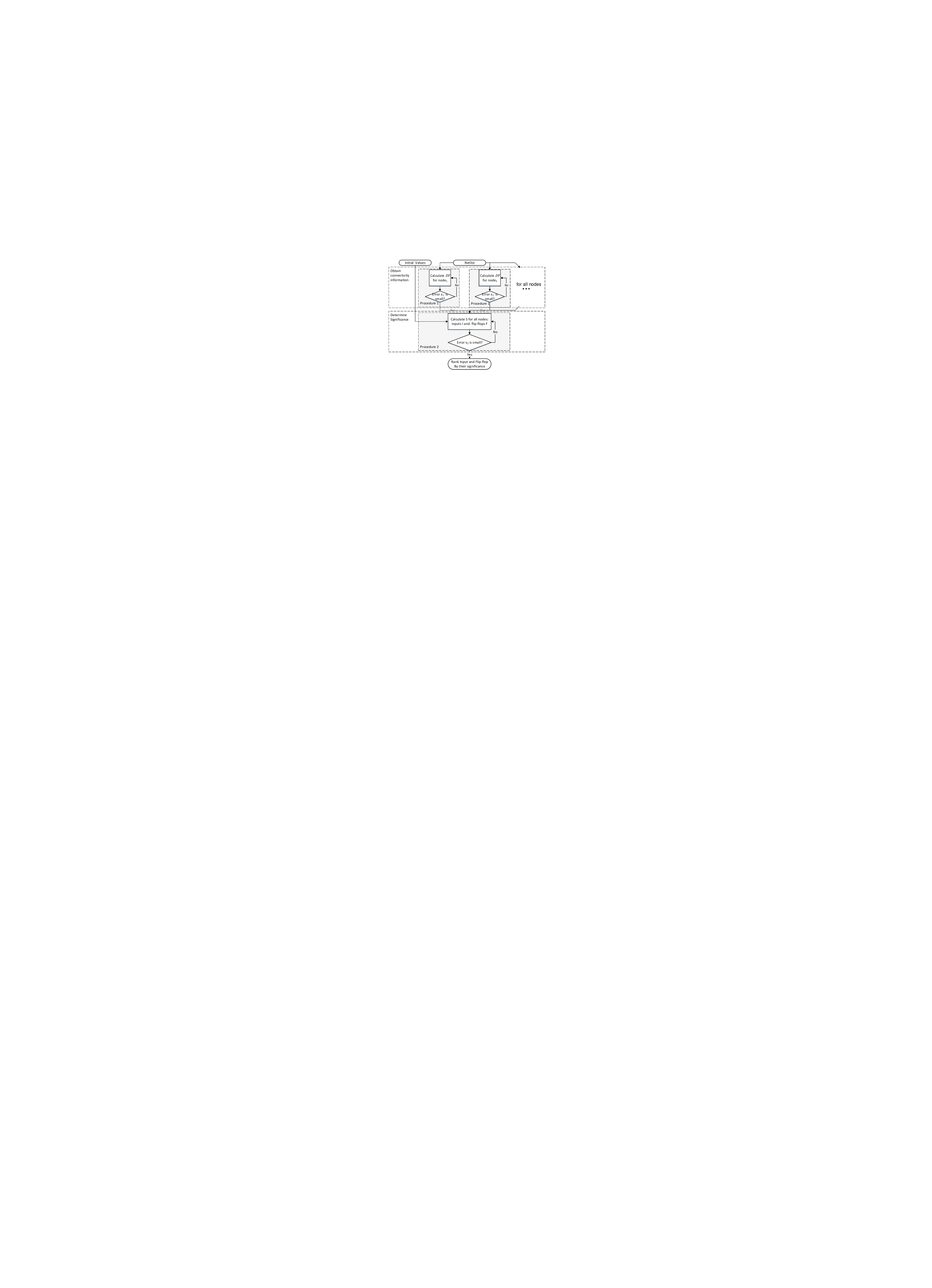}
\caption[The SERIAL model consist of two procedures.]{The SERIAL model consist of two procedures. It computes the connectivity before determining the significance.}
\label{fig:model_rank_system}
\end{figure}

Equation~\ref{eq:model_s} shows $n$+$m$ independent linear equations with $n$+$m$ unknowns ($S_F$ and $S_I$). The number of unknowns is huge for practical digital circuits. For example, there are more than 1M unknowns for a 1~$mm^2$ circuit in 28nm. This makes the exact solution ($\epsilon=0$) unrealistic. However, $DF$ is usually sparse, since not all FF are directly connected. This chapter proposes to use a heuristic algorithm to get an accurate-enough $S$ ($\epsilon$ close enough to 0) for all FF. The structure of the method is summarized in Fig.~\ref{fig:model_rank_system}. 

The solver of the SERIAL model consists of the following components:
\begin{itemize}
\item[1.] Initialize: set initial values $S$ for output ports ($S_O$);
\item[2.] Compute \textit{$DF$} (Procedure 1): determine the $df$ from each node to all adjacent tails of it;
\item[3.] Compute \textit{$S$} (Procedure 2): breadth-first search to update $DW$ for each arrow, and hence $S$ for each node ($S_I$ and $S_F$).
\end{itemize}

\subsection{Determine the distribution factor $DF$}\label{sec:df}
For one node, its adjacent tails inherit different portions of $S$, since the combinational logic is different. Procedure 1 computes $DF$ for one node (a row of $DF$). Therefore, it should be executed for every node to obtain the complete $DF$.

The notation of a \textbf{logic graph} is introduced. Similar to the significance graph, it also reverses the signal transferring direction of the netlist. The \textbf{logic graph} is modeled as a tuple <$O$, $I$, $G$, $LS$, $LDF$, $LDW$>:
\begin{itemize}
\item[$\bullet$] \textit{Nodes}, which include:
    \begin{itemize}
        \item[$\cdot$] \textit{Source} ($O$): the aim of this logic is to compute the $DF$ for this node;
        \item[$\cdot$] \textit{Sink} ($I$): all input ports and sequential FF; 
        \item[$\cdot$] \textit{Internal} ($G$): the combinational logic gates are the intermediate nodes for the graph.
    \end{itemize}
\item[$\bullet$] \textit{Logic Significance} ($LS$) for each node in the logic graph.
\item[$\bullet$] \textit{Arrows} denote the $LS$ distribution between nodes. Arrows have two features:
    \begin{itemize}
        \item[$\cdot$] \textit{Logic Distribution Factor} ($LDF$): the portion of $LS$ distributed from a node to its tails.
        \item[$\cdot$] \textit{Logic Distribution Weight} ($LDW$): the $LS$ on the arrow.
    \end{itemize}
\end{itemize}

The logic graph is a directed weighted graph. It contains branches and conjunctions, as the netlist might have branches and multi-port logic gates. The internal nodes are combinational logic gates, implying that the logic significance graph is non-cyclic (otherwise, the corresponding netlist contains cycles in combinational logics).

A head node distributes its $LS$ to its tail nodes along the arrows. The distribution factor $ldf$ is determined by the gate type: \textbf{$ldf$} is defined as the normalized possibility that an changed input data leads to changed output data. The input port corresponds to the tail node in the logic graph. The output port corresponds to the head node in the logic graph. Note that since the logic significance can only be distributed, the summation of $ldf$ from a node is a normalized 1.

For instance, the procedure of computing the $ldf$ for a 3-input AND gate (see Table~\ref{tab:model_truth_example}) is discussed with its truth table. For the input port A, if the signal is independently changed from 0 to 1, the state of the AND gate might change from state [1->5, 2->6, 3->7, 4->8], respectively, depending on the initial signal of port B and C. Of all four possible changes, only the state change 4->8 leads to a changed output Z. This also applies to changing port A  from 1 to 0. Therefore, for port A, the $ldf$ before normalization is 2/8, meaning that 2/8 of random changes in port A result to port Z value change. Considering that for input B and C the $ldf$ are both 2/8 (by symmetry), the normalized $ldf$ is therefore 1/3, since the sum of $ldf$ for any node is 1. In fact, for any symmetrical logic gates, the normalized $ldf$ is $1/n$, where $n$ is the number of input ports. This implies that all inputs contribute equally to the output.

\begin{table}[H]
\caption{Truth table of a 3-input AND gate.}\label{tab:model_truth_example}
\centering
\begin{tabular}{cccccccccc}
\toprule
\multicolumn{2}{c}{State} & 1 & 2 & 3 & 4 & 5 & 6 & 7 & 8 \\
\cmidrule(l){1-2} \cmidrule(l){3-10}
\multirow{3}{*}{in} & A & $0$ & $0$ & $0$ & $0$ & $1$ & $1$ & $1$ & $1$  \\
\cmidrule(l){2-2} \cmidrule(l){3-10}
 & B & $0$ & $0$ & $1$ & $1$ & $0$ & $0$ & $1$ & $1$  \\
\cmidrule(l){2-2} \cmidrule(l){3-10}
 & C & $0$ & $1$ & $0$ & $1$ & $0$ & $1$ & $0$ & $1$  \\
\cmidrule(l){1-2} \cmidrule(l){3-10}
out & Z & $0$ & $0$ & $0$ & $0$ & $0$ & $0$ & $0$ & $1$  \\
\bottomrule
\end{tabular}
\end{table}

One possible approach to computing $DF$ is enumerating the paths from the source node to each sink node. However, the number of unique paths increases exponentially with the number of branches and conjunctions, making the enumeration approach impractical. Therefore, this chapter proposes a heuristic breadth-first graph traversal algorithm to compute the $DF$ (see \textbf{Procedure~1}).  It records the $LS$ for each node and visits each node for only once during an iteration.

\clearpage
\begin{algorithm}
\caption{Procedure 1: Determine the $DF$ for all arrows that start from a node ($node_i$).}
\label{proc1}
\begin{algorithmic}[1]
\State \textbf{Input}: logic graph of $node_i$ \Comment{$node_i.ls$ are initialized to 1}
\State \textbf{Output}: $DF$  for all arrows of $node_i$
\Repeat     \Comment{loop until mismatch is small}
    \State push $node_i$ to the empty \textit{stack};
    \While{\textit{stack} not empty}
        \State pop a $node_j$ from \textit{stack};
        \State \textbf{visit}($node_j$);
        \For{each $node_k$ in <adjacent nodes of $node_j$>}
            \If{$node_k$ is a gate \textbf{AND}  $node_k$ not visited} \Comment{only when $node_k$ is not a end point of the logic graph}
                \State mark $node_k$ as visited;
                \State push $node_k$ to the \textit{stack};
            \EndIf
        \EndFor
    \EndWhile
\Until{error $\epsilon_1$ is small enough}
\For{each $node_j$ in <FF and inputs>} \Comment{end points of logic graph}
    \State return $node_i.node_j.df \gets node_j.ls$ \Comment{return the $ls$ as $df$}
\EndFor
\Statex
\Function{\textbf{visit}}{$node_j$}
\For{each $arrow_k$ that starts from $node_j$} \Comment{name the tail as $arrow_k$}
    \State $\Delta \gets node_j.ls * arrow_k.ldf - arrow_k.ldw$; \Comment{new vs. existing $ldw$}
    \State error $arrow_k.\epsilon_1 \gets \Delta / arrow_k.ldw * 100\%$; \Comment{error for proc.1}
    \State $arrow_k.ldw \gets arrow_k.ldw + \Delta$; \Comment{update $ldw$ of $arrow_k$}
    \State $node_k.ls \gets node_k.ls + \Delta$; \Comment{update $ls$ of $node_k$}
\EndFor
\EndFunction
\end{algorithmic}
\end{algorithm}

When visiting a node, the algorithm updates the $LS$ of its adjacent tails, as well as the weight for the adjacent arrows ($LDW$). Besides, the $LDW$ change rate ($\epsilon_1$) is recorded as the metric of computational error (see Procedure 1). $\epsilon_1$ is checked at the end of each iteration to determine whether additional iterations are required.

\begin{figure}[H]
\centering
\includegraphics[width=\textwidth]{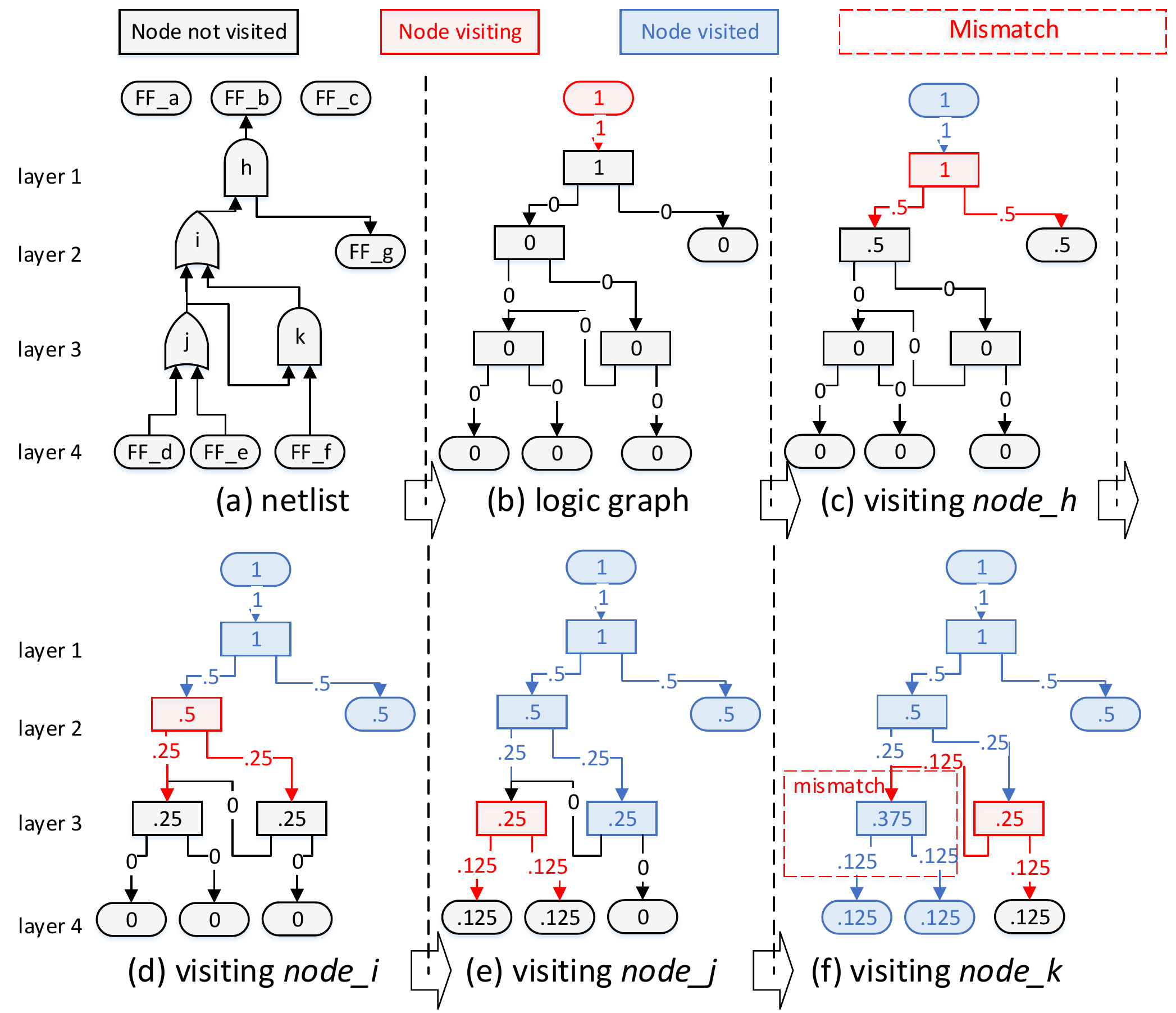}
\caption[An example of computing the distribution factor $DF$ (Procedure~1).]{An example of computing the distribution factor $DF$ (Procedure~1). (a): Netlist of the logic gates and FF within a pipeline stage; (b-f): Computing $DF$ using Procedure 1 in the first iteration. The procedure distributes the $S=1$ at $FF_b$ to $FF_d$-$FF_g$.}
\label{fig:model_rank_example}
\end{figure}

Fig.~\ref{fig:model_rank_example} shows an example of computing $DF$ for a specific FF ($node_b$). The netlist is converted to the logic graph in Fig.~\ref{fig:model_rank_example}(b) (Note the direction of the arrows). The number of the node indicates the $S$, and the number on the arrow denotes the $LDW$. Fig.~\ref{fig:model_rank_example}(c)-(f) shows the significance update in the first iteration.

A mismatch is observed in Fig.~\ref{fig:model_rank_example}(f), in which the logic significance of $node_j$, $ls_j$=0.375, is not equal to the sum of the $ldw$ of its adjacent arrows (0.125+0.125). This is because $node_j$ was visited in Fig.~\ref{fig:model_rank_example}(e), where its $ldw$ is updated. Therefore, $node_j$ will not be visited afterwards, even though its $ls$ is updated in Fig.~\ref{fig:model_rank_example}(f). Note that in the second iteration, the mismatch error on $node_j$ will be eliminated, since when visiting $node_j$, its $ls$ is already correct.

The mismatch is generated because of the conjunction of arrows from different layers (nodes in the same layer means they have the same logic depth). For example, $node_i$ (in $layer_2$) and $node_k$ (in $layer_3$) are both connected to $node_j$ (in $layer_4$), which results in the mismatch in Fig.~\ref{fig:model_rank_example}(f). The magnitude of the computation error $\epsilon_1$ is used to estimate the mismatch of the computation procedure.

To reach zero mismatches (zero error $\epsilon_1$) for the logic graph, the maximum number of iterations is equal to the maximum logic depth. This is because, after each iteration, the correct logic significance from nodes of the same layer will at least propagate to the nodes of the next layer. In reality, $\epsilon_1$ converges more quickly than the maximum bound, because, by the time a node is visited, it has probably got the correct update information ($ldw$) from all of its adjacent heads.

\subsection{Determine the significance $S$}
This section discusses the algorithm to solve Equation~\ref{eq:model_s}, which determines the significance for all nodes. \textbf{Procedure~2} shows the algorithm. 
\clearpage
\begin{algorithm}
\caption{Procedure~2: Determine the $S$ for all nodes.}
\label{proc2}
\begin{algorithmic}[1]
\State \textbf{Input}: the significance graph. \Comment{$node_i.ls$ are initialized to 1}
\State \textbf{Output}: $s$ for all nodes.
\Repeat     \Comment{loop until mismatch is small}
    \State push output ports (sources) to the empty \textit{stack};
    \While{\textit{stack} not empty}
        \State pop a $node_j$ from \textit{stack};
        \State \textbf{visit}($node_j$);
        \For{each $node_k$ in <adjacent nodes of $node_j$>}
            \If{$node_k$ is not a input port (sink) \textbf{AND} $node_k$ not visited} \Comment{only when $node_k$ is not a end point of the significance graph}
                \State mark $node_k$ as visited;
                \State push $node_k$ to the \textit{stack};
            \EndIf
        \EndFor
    \EndWhile
\Until{error $\epsilon_2$ is small enough}
\For{each $node_j$ in <FF and inputs>} \Comment{nodes in significance graph}
    \State return $node_j.s$ \Comment{return the $s$ for all nodes}
\EndFor
\Statex
\Function{\textbf{visit}}{$node_j$}
\For{each $arrow_k$ that starts from $node_j$} \Comment{name the tail as $arrow_k$}
    \State $\Delta \gets node_j.s * arrow_k.df - arrow_k.dw$; \Comment{new vs. existing $dw$}
    \State error $arrow_k.\epsilon_2 \gets \Delta / arrow_k.dw * 100\%$; \Comment{error for proc.2}
    \State $arrow_k.dw \gets arrow_k.dw + \Delta$; \Comment{update $dw$ of $arrow_k$}
    \State $node_k.s \gets node_k.s + \Delta$; \Comment{update $s$ of $node_k$}
\EndFor
\EndFunction
\end{algorithmic}
\end{algorithm}

Similar to Procedure 1, it is also a heuristic breadth-first algorithm. However, this problem is more complicated than Procedure 1, because the significance graph contains not only branches and conjunctions but also loops. By applying Procedure 2, each node is only visited once in an iteration, eliminating the situation of the algorithm being trapped in the loop. 

\begin{figure}[H]
\centering
\includegraphics[width=\textwidth]{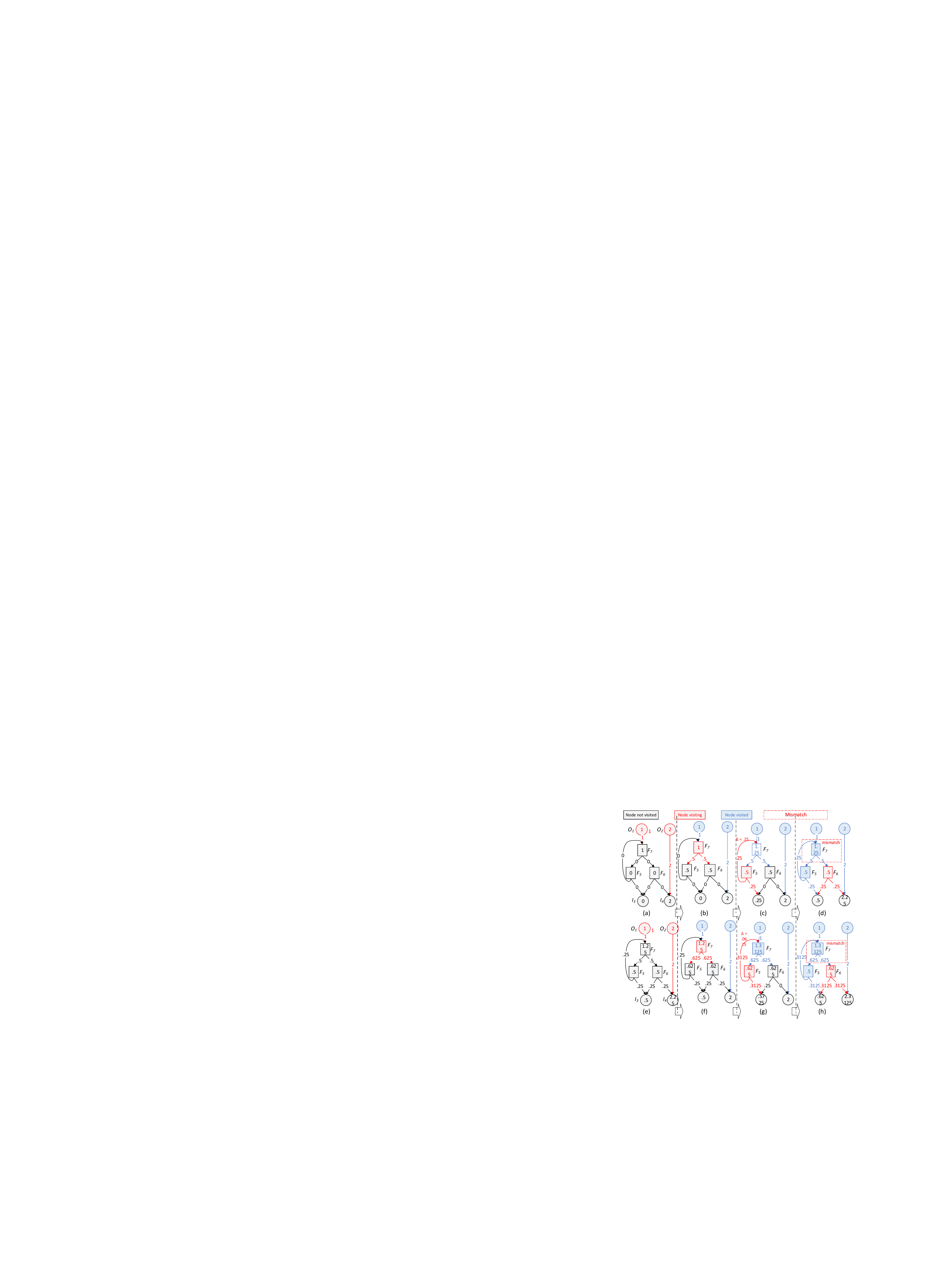}
\caption[An example of computing the significance (Procedure~2).]{An example of computing the significance (Procedure~2) for the netlist in Fig.~\ref{fig:model_rank_netlist}. (a-d): the first iteration; (e-h): the second iteration.}
\label{fig:model_rank_loop}
\end{figure}

Fig.~\ref{fig:model_rank_loop} shows the example of computing the significance during the first and the second iteration. The initial significance $s$ at $O_1$ is set as an unity 1, and the $s$ for $O_2$ is 2. In Fig.~\ref{fig:model_rank_loop}(d), a mismatch can be observed on $F_7$ due to loops. This mismatch is measured by computational error $\epsilon_2$, and it will reduce in future iterations: during the first iteration, $F_7$ obtains a $\Delta$ of 0.25 from $arrow_{F_5 \rightarrow F_7}$ (Fig.~\ref{fig:model_rank_loop}(c)):
\begin{equation}
S_{F_7} = 1 + \Delta = 1 + \frac{1}{4}.
\end{equation}
The same $\Delta$ $arrow_{F_5 \rightarrow F_7}$ becomes 0.25/$2^2$ in the second iteration (Fig.~\ref{fig:model_rank_loop}(g)):
\begin{equation}
S_{F_7} = 1 + \Delta_1 + \Delta_2  = 1 + \frac{1}{4} + \frac{1}{16}.
\end{equation}
This is because the loop $F_7$->$F_5$->$F_7$ contains 2 branches, i.e. $F_7$ and $F_5$ both have two adjacent tails. When the number of iterations goes to infinite, the $\Delta$ forms geometric progression with a common ratio of $1/4$, causing the final significance of $F_7$ to converge:
\begin{align}
S_{F_7} &= 1 + \Delta_1 + \Delta_2 + \Delta_3 + \cdots   \nonumber\\ 
        &= 1 + \frac{1}{4} + \frac{1}{16} + \frac{1}{64} + \cdots = 1.3\dot3.
\end{align}

Since every loop must contain branches (otherwise the corresponding loop in the netlist does not contain influx/conjunctions, as the graph reverses the direction of the netlist), the common ratio of $\Delta$ must be smaller than 1, preventing the significance to go to infinity. Therefore, the computational error $\epsilon_2$ is reduced with every iteration. Unlike Procedure 1 where a zero-$\epsilon_1$ result can be expected after a fixed amount of iterations, graphs with loops can only produce zero error after infinite iterations. However, in practical digital circuits, loops are usually very large (lots of branches in the loop, and hence the common ratio is very small), the breadth-first approach ensures that the algorithm converges faster. This is demonstrated with benchmark circuits in Section~\ref{sec:rank_speed}. Moreover, the most significant FF normally have very high fan-in rates in the significance graph. As a consequence, they emerge on the top of the significance ranking list after several iterations.

Loops in the graph might result in a larger significance value for internal nodes than that of sources, e.g. $SF_7$ is greater than $SO_1$. One example in digital circuits is the loop controller, which is more significant than the output ports, and thus should be protected more carefully.

Another important feature worth mentioning is the \textbf{confidence factor}. It is defined as the portion of significance each node absorbs from the arrow coming towards it. Typically it is fixed to 1, implying full absorbance from the head. If it is set to 0, the tail node of the arrow receives no significance (or logic significance) distribution from the corresponding head node. As a result, the significance only allocates to other interested FF. This is used for situations when circuit designers assure certain paths/components are secure (or important), where the significance is saved to the insecure components. In this work, the \textit{clk} port and the \textit{reset} port are assumed as hardened and thus are free from errors. Therefore, the confidences for the arrows connecting to them are fixed to 0.

\section{Experimental verification}\label{sec:model_result}
A proof-of-concept program written in Python was built to verify the proposed approach experimentally. It parses the input netlist into logic graphs and the significance graphs.

For benchmarking purpose, SERIAL is analyzed on a number of representative benchmark circuits: 6 ISCAS’89 circuits (e.g. s27) \cite{brglez1989combinational}, 3 ITC’99 benchmarks (microprocessors e.g. b14) \cite{corno2000rt}, 8-bit 32-point and 16-bit 64-point FFT processors (FFT32 and FFT64\footnote{FFT32: 8-bit resolution, 1 butterfly unit; FFT64: 16-bit resolution, 2 parallel butterfly units}) (developed from \cite{ShuoLi2015}), and a commercial-ready low density parity check (LDPC) decoder (developed from \cite{li15el-ldpc}). The ITC’99 circuits and LDPC are deliberately selected for their high complexity as each FF is connected to various FF.

\subsection{Scalability}\label{sec:rank_speed}
When performing Procedure 1 for one node, the visited gates are formed in tree-structure (see Fig.~\ref{fig:model_rank_example}). Therefore, 
\begin{equation}
\#gate\_visited \sim degree_{node}^2, 
\end{equation}
where $degree_{node}$ is the average degree of connectivity for the graph (it represents the number of nodes that one node, i.e. I/O and FF, connects to). Since the runtime for each gate is the product of mean iteration number ($iter_1$) and the $\#gate\_visited$, the overall complexity for all nodes of Procedure 1 ($T_1$) is summarized as:
\begin{equation}
T_1 \sim \#nodes * iter_1 * degree_{node}^2.
\end{equation}
Note that $\#nodes$ scales linearly with the circuit size, while $iter_1$ and $degree_{node}$ depend on the circuit characteristics.

The runtime for Procedure 2 is summarized as
\begin{equation}
T_2 \sim iter2 * \#node * degree_{node},
\end{equation}
where $degree_{node}$ implies the runtime for visiting one node, $\#node * degree_{node}$ suggests the runtime for one iteration, and $iter2$ is the number of total iterations for Procedure 2.

To measure the scalability of the algorithm, the SERIAL algorithm was applied to all benchmark circuits, on a desktop PC with Intel i7 CPU and 8 GB memory. The runtime analysis is shown in Table~\ref{tab:model_rank_runtime}. The benchmark complexity metrics that affect the algorithm execution time is listed.

\begin{table}[H]
\caption{SERIAL model solver runtime analysis for benchmark circuits.}\label{tab:model_rank_runtime}
\centering
\begin{tabular}{ccccccc}
\toprule
\multirow{2}{*}{Circuits} & \multicolumn{3}{c}{Circuit complexity} & \multicolumn{2}{c}{\makecell{Procedure~1\\$epsilon_1=0$}} &  Procedure~2 \\
\cmidrule(l){2-4} \cmidrule(l){5-6} \cmidrule(l){7-7}
     & $\#gate$ & \makecell{$\#node$\\(IO \& FF)} & $degree_{node}$ & \makecell{mean\\$iter_1$} & runtime & \makecell{runtime\\per iter.} \\
\cmidrule(l){1-1} \cmidrule(l){2-4} \cmidrule(l){5-6} \cmidrule(l){7-7}
s27        & 28    &8        &2.6    &2.25    &0.2 ms    &0.1 ms  \\
s510    & 315    &32        &3.3    &2.8    &2 ms    &0.8 ms \\
s641    & 392    &78        &6.6    &3.5    &9 ms    &3.4 ms \\
s5378    & 3.3k    &263    &9.3    &2.6    &39 ms    &18 ms \\
s13207    & 9.7k    &852    &6.3    &2.6    &97 ms    &44 ms \\
s38584    & 28k    &1.8k    &12.5    &2.8    &292 ms    &160 ms \\
\cmidrule(l){1-1} \cmidrule(l){2-4} \cmidrule(l){5-6} \cmidrule(l){7-7}
b14        & 9.6k    &233    &90.7    &14.4    &5.7 s    &158 ms \\
b20     & 21k    &546    &115    &14.0    &10 s    &325 ms \\
b21        & 22k    &546    &129    &14.6    &12 s    &369 ms \\
\cmidrule(l){1-1} \cmidrule(l){2-4} \cmidrule(l){5-6} \cmidrule(l){7-7}
FFT32     & 11k    &2.2k    &8.8    &2.8    &0.63 s    &188 ms \\
FFT64     & 52k    &8.6k    &12.6    &2.9    &6.1 s    &965 ms \\
\cmidrule(l){1-1} \cmidrule(l){2-4} \cmidrule(l){5-6} \cmidrule(l){7-7}
LDPC      & 778k    &59k    &136.9    &4.2    &1424 s    &40 s \\
\bottomrule
\end{tabular}
\end{table}

For Procedure 1, the runtimes for SERIAL on all ISCAS’89 circuits are all well below 1s. The runtime for the 52k-gate FFT64, comparing with FFT32, increases linearly with the node number, and quadratically with the $degree_{node}$. For LDPC and ITC’99 circuits, the runtime is relatively large, due to their high $degree_{node}$, which is the unique property of their memory addressing units. Note that the runtime for Procedure 1 can be substantially reduced by parallelism as the computing processes for all node are independent. For all circuits, the number of iterations to reach zero $\epsilon_1$ for Procedure 1 are all small, confirming our assertion in Chapter~\ref{sec:df}.
 
\begin{figure}[H]
\centering
\includegraphics[width=0.9\textwidth]{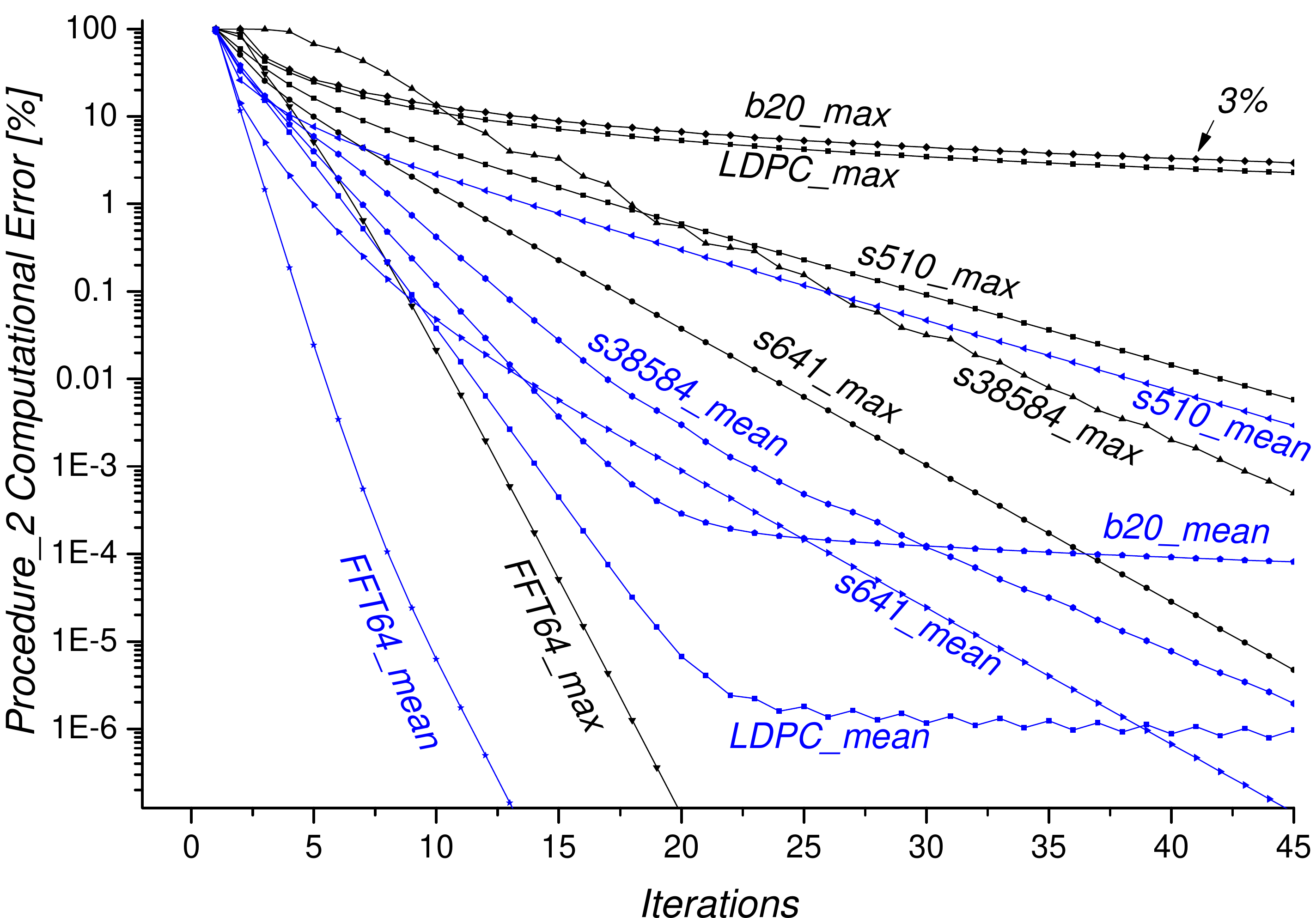}
\caption{In Procedure~2, the mismatch error $\epsilon_2$ converges rapidly towards 0.}
\label{fig:model_rank_delta}
\end{figure}

For Procedure 2, the runtime per iteration is reasonably small for all circuits. Fig.~\ref{fig:model_rank_delta} shows the computational error $\epsilon_2$ convergence w.r.t. iterations. Naturally, $\epsilon_2$ reduces with more iterations performed. After 40 iterations, even the most demanding circuit, LDPC, reaches a maximum of 3\% computational error $\epsilon_2$ for all arrows in the graph, which is very precise for the significance ranking. This also demonstrates the fast convergence of $\epsilon_2$, despite all the branches, conjunctions and cycles in the graph for practical circuits.

In summary, considering that $degree_{node}$ does not change with the circuit size, the SERIAL algorithm scales linearly with the circuit size ($\#nodes$). The only time-consuming case is the exceptional circuit, LDPC, of which the $degree_{node}$ is high. Despite that, the algorithm for all benchmark circuits finishes reasonably fast.

\subsection{Validation on an LDPC decoder}

\begin{figure}[H]
\centering
\includegraphics[width=1\textwidth]{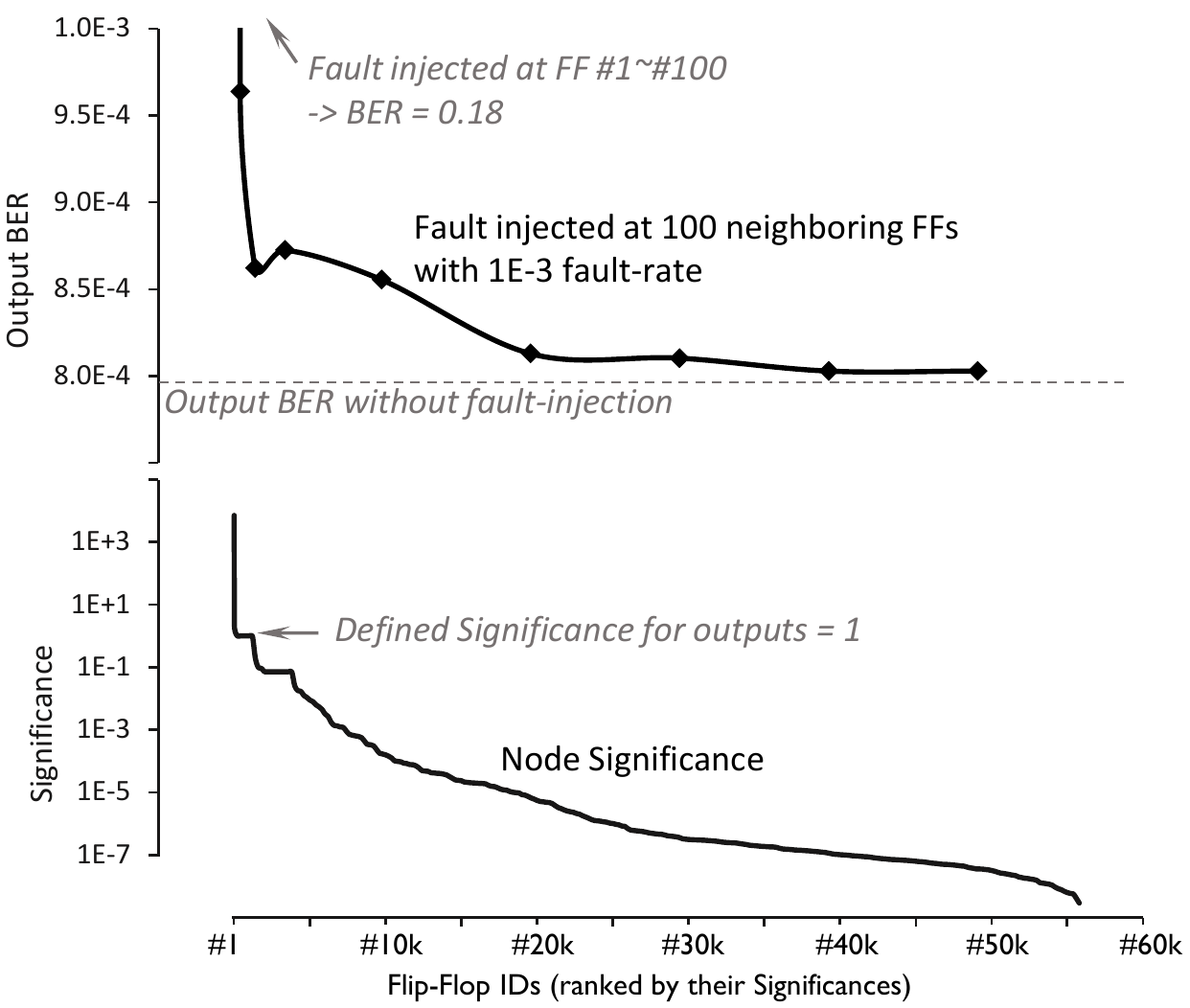}
\caption[In an LDPC decoder circuit, errors on FF with high significance values leads to worse BER.]{On an LDPC decoder circuit, errors on FF with high significance values leads to worse BER. The output quality is measured as BER, when random soft-errors are injected into FF. At each run, 100 neighboring FF with similar significance are under soft-errors, while the rest are free from soft-errors.}
\label{fig:model_rank_ldpc}
\end{figure}

After applying the SERIAL to the LDPC, each FF is labeled with its ranking order regarding significance. Fig.~\ref{fig:model_rank_ldpc}  (left y-axis) shows the significance distribution. A small portion (around 100) of FF has an extra-high significance value. These FF are mostly found in the control logic and clock-gating units. These high significance FF control lots of outputs. Moreover, the faults generated by these components cannot be flushed away easily.

For comparison, a Monte-Carlo method is applied to simulate the LDPC under soft-error (randomly fault injection). The simulator randomly flips the Q port for every selected FF at a chance of 0.1\% (1 in every thousand cycles) for each FF. For each run (denoted as a circle in Fig.~\ref{fig:model_rank_ldpc}), 100 FF with similar significance (e.g. $FF_{1000}$ to $FF_{1099}$, or $FF_{2000}$ to $FF_{2099}$) are selected for fault injection. Each chosen flip-flops is flipped during a randomly-chosen clock cycle with a uniform fault injection rate of 0.1\% (1 in every thousand cycles). The output decoded bit error rate (BER) is denoted as the system’s output quality.

If faults are injected to FF with significance ranking smaller than 300 (high significance), the output degrades significantly (BER = 0.18). This is because once errors are introduced in these most significant nodes, a huge amount of FF are affected, and the output results are gradually degraded. The significance of most FF are smaller than the defined output significance (defined as 1), suggesting that they are less sensitive to errors. This reflects the error absorbent capability of the LDPC circuit. Fig.~\ref{fig:model_rank_ldpc} shows a good coherence between the rank of the significance and the severity of soft-errors on the corresponding FF. 

Note that the BER curse is not monotonously decreasing around the Flip-Flop ID \# 3k. This shows that the ranking cannot guarantee a node with higher significance is more important than others. The inaccuracy can be because of a lack of statistical and run-time logic state information, and mismatches in the significance distribution, etc.  Despite that, the ranking serves as a guideline to distinguish which nodes are more important that should be protected, from the statistical point of view, as shown in Fig.~\ref{fig:model_rank_ldpc}.

\section{Application to harden an FFT design}\label{sec:model_apply}
\begin{figure}[H]
\centering
\includegraphics[width=1\textwidth]{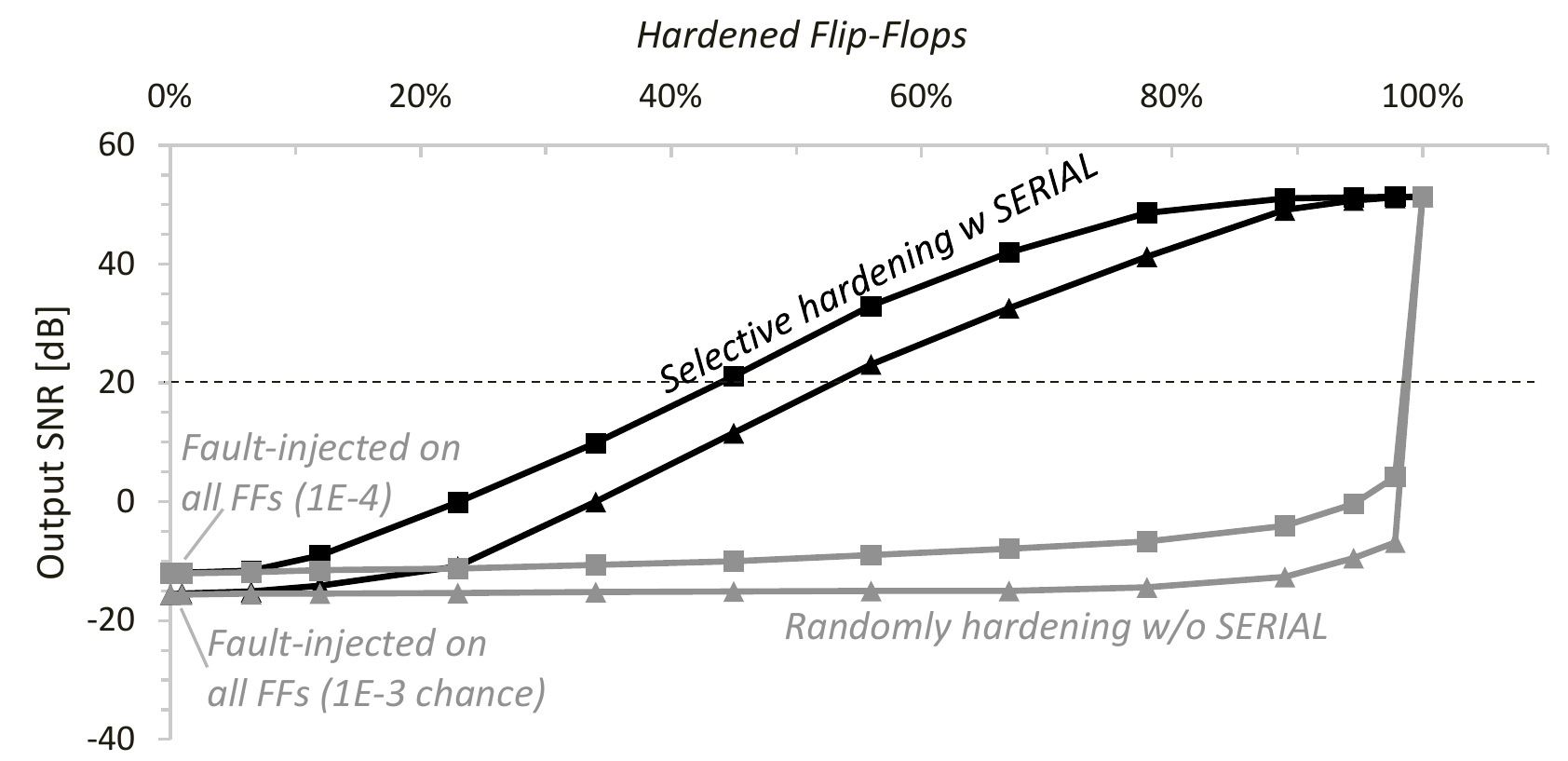}
\caption{Compared with hardening randomly, selective hardening with SERIAL increases the SNR, at the same hardening ratio.}
\label{fig:model_rank_fft}
\end{figure}

With the help of SERIAL, selective hardening can be performed on the FF with the highest significances. The FFT64 is simulated under random errors on FF, to capture its output SNR (Fig.~\ref{fig:model_rank_fft}).\footnote{In Chapter~\ref{sec:system}, the output SNR is named as the Signal-to-Digital-Distortion Ratio (SDDR), to avoid the confusion with the term SNR in wireless communication society to quantify the quality of channel.} It easily represents the scenario when chips are enduring soft-error when the error generation is uniform across all FF. It can also mimic the error of VOS, with some extra modeling of the error generation. 

An error rate of $10^{-3}$ and $10^{-4}$ is assumed for all basic FF, except for the hardened FF, where no error is assumed. A logical method to increase the reliability of a system under this threat is selective FF hardening, until the point that the reliability (in this case output quality) demand is met. 

Comparing with the hardening without the help of SERIAL, where randomly selecting FF for protection is the viable option, hardening with the guidance of SERIAL yields much better SNR even at a lower hardening coverage (smaller overhead). For instance, if the target SNR is 20dB under the fault rate of $10^{-4}$, only 45\% FF are required for hardening, compared with the almost 100\% hardening need for the random selection approach. Considering that hardening an FF doubles the FF area, the selective hardening saves more than half of the hardening overhead.

That proves that for the same reliability/quality target, the FFT that hardened with the help of SERIAL can be superior in power consumption and area cost.

\section{Conclusion}\label{sec:model_con}
In this chapter, an analytical approach to estimate the effect of error occurrence at gate-level is presented. The main novel idea in the proposed method is to pursue a heuristic approach to find out the controlling logics that is too important to fail. This serves as a guidance for circuit designers to selectively harden the design by methods such as FF hardening and design margin insertion, and consequently reduce the power overhead of error-hardening. 

The solving algorithm has shown a low execution runtime and good scalability of our method on various benchmark designs, i.e. ISCAS'89 and ITC'99 circuits, an FFT processor and an LDPC decoder. The coherence of the significance and the error effect, as well as the selective hardening benefits,  was verified. Using this model, an FFT processor is designed. It shows huge gains regarding output quality with the same error generation rate (power overhead). 

This work can further extend to analysis the impact when input data and circuit’s states are not completely random. Besides, the workload dependent error characteristic can also be introduced into this model to address aging-related errors \cite{hamdioui2013reliability, reddy2005impact}.



\cleardoublepage


\chapter{Microarchitecture-level power optimization exploiting application opportunities}\label{sec:arch}


This chapter performs microarchitecture-level power saving when considering the opportunities from applications. It proposes, in the context of general-purpose and domain-specific processors (embedded CPU, MCU or DSP), a fine-grained hardware switching scheme to select the proper device for low power computing. This scheme exploits word-length optimization opportunities for a multiplication unit. This scheme reduces the power of the multiplication unit from an OpenRISC processor by 23.7\%, which is equivalent to a 9.5\% power saving for the whole execution unit. The work in this chapter is published in \cite{huang16el-hardware-switch}.

The rest of this chapter is structured as follows: Section~\ref{sec:arch_intro} pinpoints the opportunity in the power-hungry multiplier for an embedded processor. Section~\ref{sec:arch_arch} reviewed the power consumption waste issue in an OpenRISC processor. Section~\ref{sec:arch_hs} proposes the fine-grained hardware switch scheme and applied it to the multiplier unit. Section~\ref{sec:arch_profile} profiles a variety of benchmark algorithms to find out the possible power benefit of this scheme. Section~\ref{sec:arch_verify} verifies the benefits in energy saving by netlist simulation. Section~\ref{sec:arch_con} concludes the chapter.

\section{Processor power waste in using unsuitable execution units}\label{sec:arch_intro}
Energy consumption is one of the most critical metrics for embedded signal processing systems. Traditionally, designers optimize the fixed-point word length that provides just-necessary precision for minimizing the power consumption. On the other hand, driven by the increasing demand for computing re-programmability, general-purpose computing devices, e.g. DSP, ASIP, and application processors, are becoming more favorable. 

In these systems, designers are constrained to perform the arbitrary word-length optimization, since processors typically sacrifice hardware costs to cater for the most complicated computing cases. In the meantime, a lot of lightweight computations that can be performed in low-energy operation devices, are executed on these over-complicated and power-hungry hardware, which leads to energy waste consequently.

In general-purpose processors, subword SIMD exploits the over-reserved word-length by applying parallelism in data-path processing. Employing SIMD reduces the number of operations, and hence decreases energy consumption. However, It requires dedicated hardware as well as software tuning to enable these SIMD intrinsic functions. Subword soft-SIMD \cite{kraemer2007interactive}, on the other hand, relies purely on software to exploit the sub-word parallelism. Nevertheless, in this scheme, guard bits are needed to be inserted, which is non-trivial for software developers \cite{fsoftsimd, catthoor2010exploiting}.

\section{Contribution of this chapter}\label{sec:arch_contribution}
This chapter introduces an alternative low precision computation unit besides the traditional full precision unit. A hardware word-length detector is used to switch the hardware units, in a fine-grained manner, to reduce the computational cost when the full precision computation is not necessary. 

Without degrading the output quality, this work detects small word-length computations and executes them in an extra reduced precision unit. This mechanism radically reduces the activation chance of the full precision unit. As a result, the dynamic power consumption decreases notably. 

The detection and execution units are both implemented with hardware at the microarchitecture level. Consequently, this technique requires neither modification on compiler nor software. The mechanism is applied to the multiplication unit. An alternative low-precision multiplier is therefore proposed. It leads to significant power saving, as the power consumption of a multiplier is $O(n^2)$ regarding the word-length $n$. 

This proposal is applied to OpenRISC, an open-source embedded microprocessor. This work implements the multiplier with proposed fine-grained hardware switch scheme. It verifies the energy improvement with algorithm profiling and gate-level simulation.

\section{Targeted embedded processing platform}\label{sec:arch_arch}
OpenRISC \cite{lampretopenrisc}, RISC-V \cite{waterman2011risc}, Sparc \cite{sparc1994sparc} are among the most famous open-source computing platforms. They are suitable to study power savings at the microarchitecture level. Without losing generality, this chapter adopts a simple 32-bit OpenRISC processor, called mor1kx (Cappuccino implementation) \cite{openrisc}.

\subsection{OpenRISC microarchitecture}\label{sec:arch_risc}
\begin{figure}[H]
\centering
\includegraphics[width=.9\textwidth]{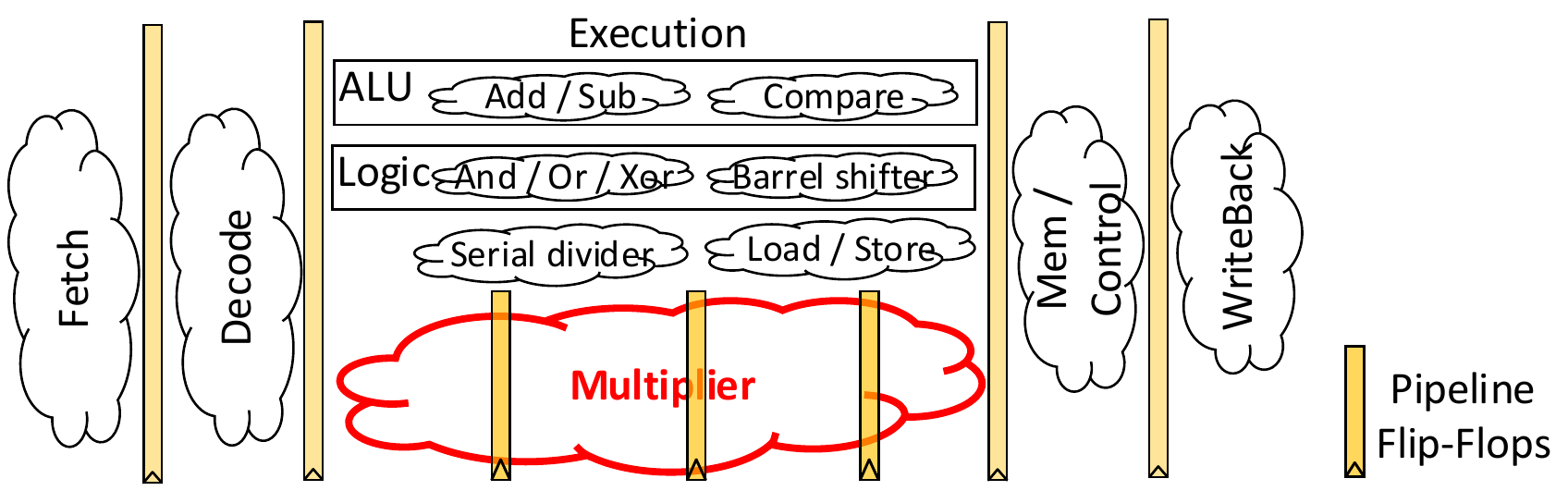}
\caption[A customized OpenRISC microarchitecture: the mor1kx Cappuccino flavor.]{A customized OpenRISC microarchitecture: the mor1kx Cappuccino flavor. It consists 5 pipeline stages. The 4-stage multiplier resides in the execution unit.}\label{fig:arch_top}
\end{figure}

The schematic of the processor is shown in Fig.~\ref{fig:arch_top}. The clock frequency is 1~GHz. The processor is realized in a standard 28nm CMOS technology. The execution stage consists of an ALU (Arithmetic Logic Unit), a logic computation unit, a Load/Store unit, a serial divider and a 4-cycle multiplier. 

\subsection{Power of the multiplication unit}\label{sec:arch_power}
\begin{figure}[H]
\centering
\includegraphics[width=.6\textwidth]{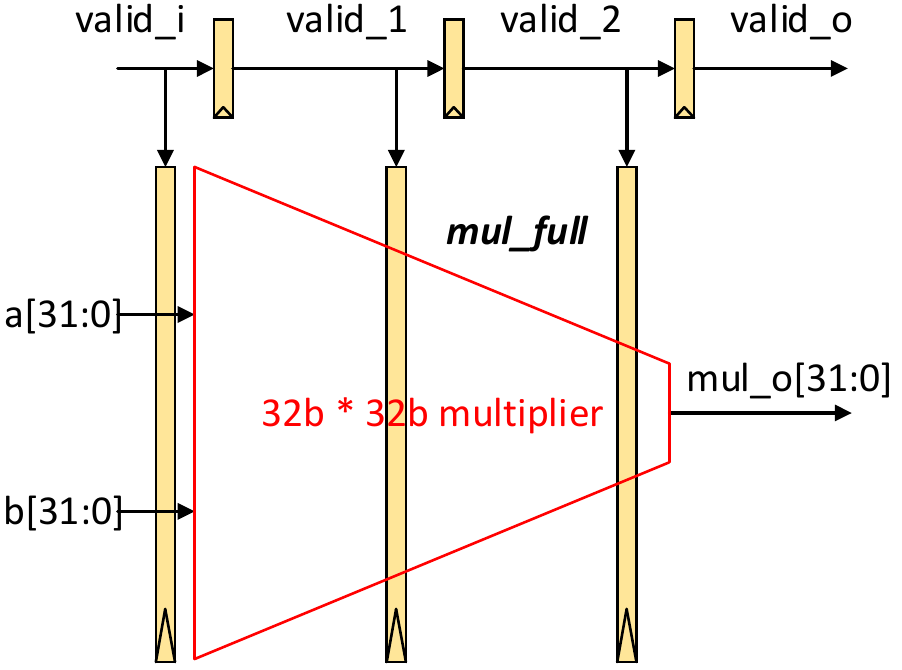}
\caption{The original multiplier in the Cappuccino implementation is 32~bits wide.}\label{fig:arch_implementation_ori}
\end{figure}

The circuit diagram of the multiplication unit (\textit{MUL}) of the default implementation is shown in Fig.~\ref{fig:arch_implementation_ori}. The \textit{MUL} contains four pipeline stages with clock-gating. Clock-gating helps to save energy, as it avoids signals from toggling when the \textit{MUL} is not in operation, for instance, when the processor is performing irrelevant instructions.

The area utilization and power consumption of the multiplier unit for each instruction are profiled in Table~\ref{tab:arch_mul_compare}. For the 32-bit multiplier, even if the word-length of multiplicands is much shorter than 32-bit, the power consumption is comparable with full-precision multiplication. The reason lies in the fact that multiplicands are represented in 2's complement form. In this form, the most significant bits (MSB) are filled with `1's or `0's when the number is short, which results in lots of toggling during positive-to-negative or negative-to-positive transitions. Nevertheless, if proper multiplier units are used, e.g. 8-bit multipliers for 8-bit multiplicand, the power consumption will be reduced accordingly.

\begin{table}[H]
\caption{Multiplier area and power consumption during each instruction.}\label{tab:arch_mul_compare}
    \begin{tabular}{ccccccc}
    \toprule
    Multiplier & Cell area & \multicolumn{5}{c}{Power during instructions [$\mu W$]} \\
    \cmidrule(l){3-7}
    size  & [$\mu mm^2$]    & NOP   & 4-bit & 8-bit & 16-bit & 32-bit \\
          &         &       & MUL   & MUL   & MUL    & MUL \\
    \cmidrule(l){1-1} \cmidrule(l){2-2} \cmidrule(l){3-7}
    4-bit & 109   & 12.152 & 48.022 & N/A   &  N/A  & N/A  \\
    8-bit & 289   & 16.497 & 100.417 & 111.18 & N/A   &  N/A \\
    16-bit &  1030 & 49.511 & 234.2 & 278.934 & 349.657 &  N/A \\
    32-bit &  1744 &  50.737 & 391.350  & 451.808  & 531.32  & 567.924  \\
    \bottomrule
    \end{tabular}
\end{table}%

This phenomenon provides an excellent opportunity for power optimization in processors, as the multiplicands type are not always declared as the full-size 32-bit long integer. Moreover, even if they are declared as 32-bit long integer, the actual value can be small, e.g. between -128 and 127 (which can be represented by an 8-bit number).

\section{Fine-grained hardware-switch scheme for multiplier (\textit{MUL})}\label{sec:arch_hs}
\begin{figure}[H]
\centering
\includegraphics[width=.9\textwidth]{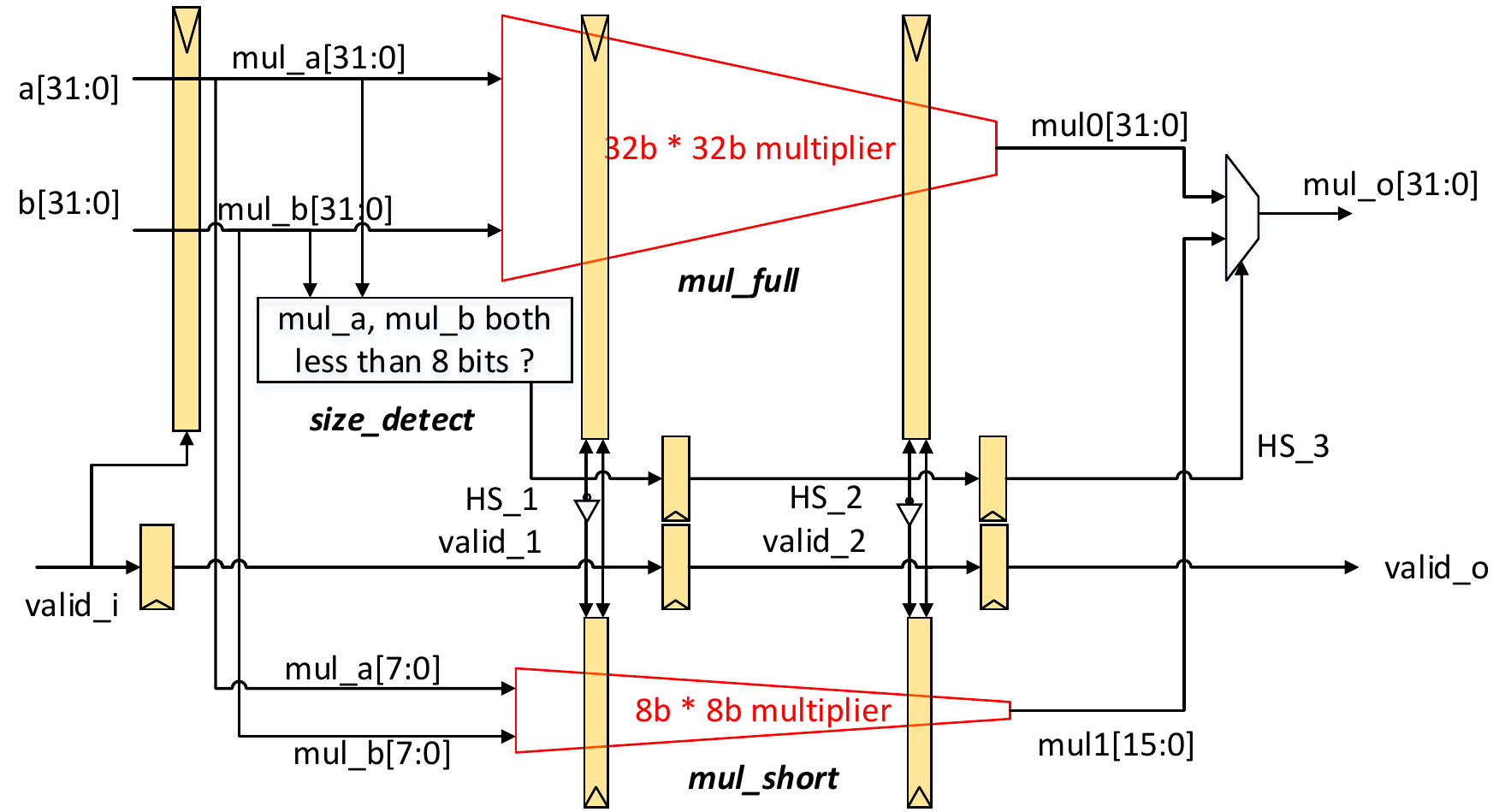}
\caption{\textit{MUL} with the proposed hardware-switching (HS) scheme selects between a 32-bit and an 8-bit multiplier, based on the data range.}\label{fig:arch_implementation_hs}
\end{figure}

Considering that there is huge power waste because of unnecessary gate toggling, this chapter introduces an alternative lower-cost multiplier to perform the computation for the cases when the word-length of multiplicands is short enough (see Fig.~\ref{fig:arch_implementation_hs}).

A simple size detecting unit (\textit{size\_detect}) is deployed to detect if both multiplicands are small, by checking if the MSBs (in this example from 8-bit on) is the same (all `1's or `0's). If both multiplicands are short, \textit{mul\_short}  will execute the operation while the \textit{mul\_full} is clock-gated, and vice versa. This hardware-switching (HS) scheme ensures that the signal only toggles in the proper multiplier unit, and the toggling in the other multiplier unit is minimized. The multipliers are divided into four stages by three sets of pipeline registers. Signals in the first stage always toggle even if the unit is not enabled since the logic inputs of the first stage are not clock-gated by the \textit{size\_detect}. 

The multiplier is retimed using a commercial RTL synthesizing compiler, which minimizes the power cost of the first pipeline stage. It moves more computations into the second stage, reducing the gate toggling incurred power consumption during irreverent operations in the first stage.

The cell area of the HS multiplier is 2053 $\mu mm^2$, which is 18\% higher than the Original multiplier. The area overhead is due to the introduction of the short multiplier and the corresponding MUX circuit. 

\begin{figure}[H]
\centering
\includegraphics[width=.98\columnwidth]{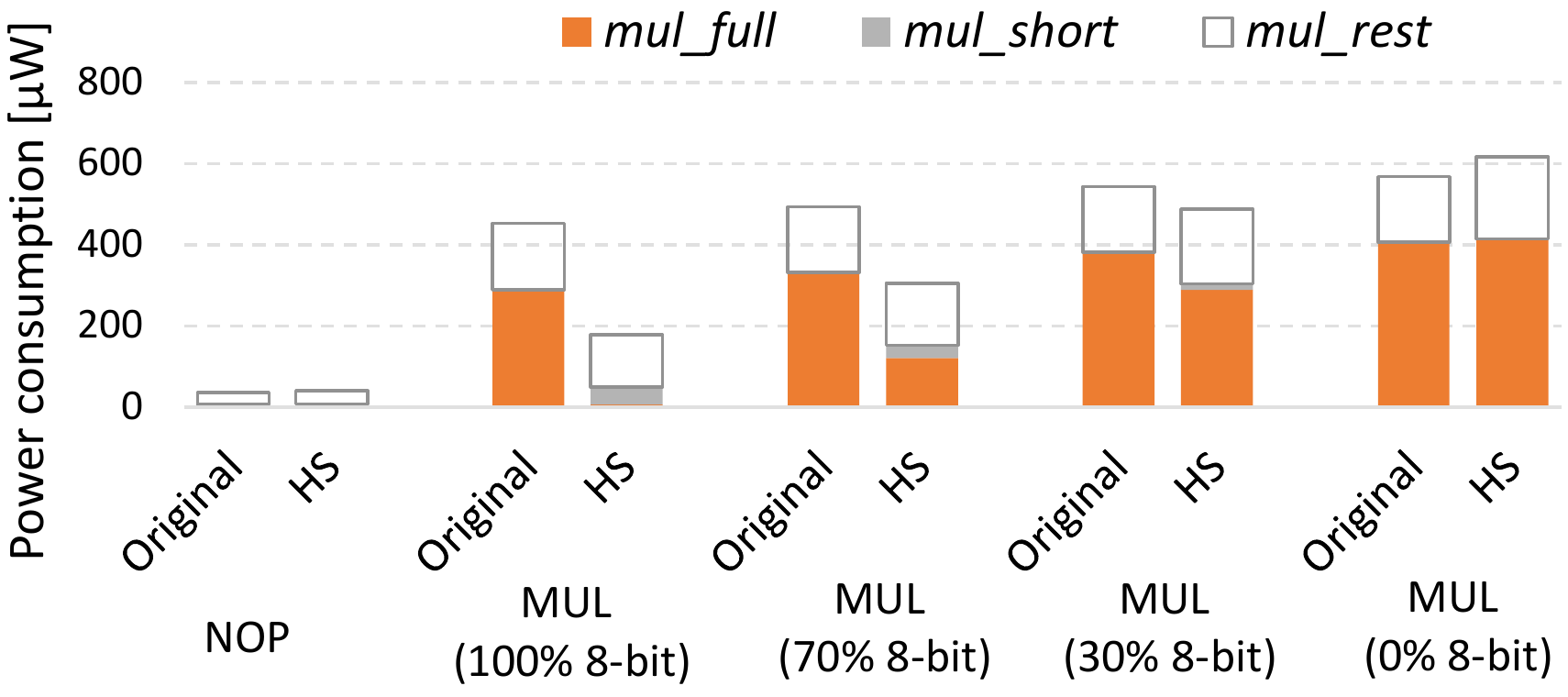}
\caption[\textit{HS} saves power when the multiplication inputs are small.]{\textit{HS} saves power when the multiplication inputs are small. During nops, the difference is marginal. The multiplicands are randomly generated to be either 8-bit or 32-bit, with the accordingly possibility.}\label{fig:arch_power}
\end{figure}

The power consumption of the \textit{Original} and the \textit{HS} multiplier is compared in Fig.~\ref{fig:arch_power}. It is broken down into 3 parts: \textit{mul\_full}, \textit{mul\_short}, and \textit{mul\_rest} (rest parts in the multiplier). During NOPs, both multipliers consume less than 40 $\mu W$, which mainly attributes to clock gating cells. For the HS multiplier, if the multiplicand is shorter than 8 bits, the \textit{mul\_full} unit is clock-gated, and the processing is assigned to the low-power \textit{mul\_short}. Therefore, the overall power consumption is significantly lower than the original multiplier. This advantage diminishes when all the multiplicands are larger than 8 bits. In that situation, the \textit{HS} multiplier suffers from the power penalty of the \textit{size\_detect} and the MUX unit.

\section{Algorithm profile}\label{sec:arch_profile}
The power savings only happen when using the low-precision multiplier. In order to measure the power consumption benefits of the HS multiplier, it is important to track the utilization frequency of the multiplication operation (\#multiplication/\#instructions), and the statistical chances that both multiplicands are short. These statistics depend heavily on application algorithms and the input data. 

For the benchmark, \textit{Cormark} 1.0 \cite{coremark} and ten other common algorithms for software radio and multimedia processing are profiled. \textit{Cormark} focuses on benchmarking CPU cores of embedded systems. The selected algorithms cover a broad range of typical applications in embedded processing, e.g. FFT, filtering, JPG decoding, cryptography, and error correction. The input data are set to represent the typical usage scenario.

\subsection{Utilization frequency of \textit{MUL}}

\begin{figure}[H]
\centering
\includegraphics[width=.9\columnwidth]{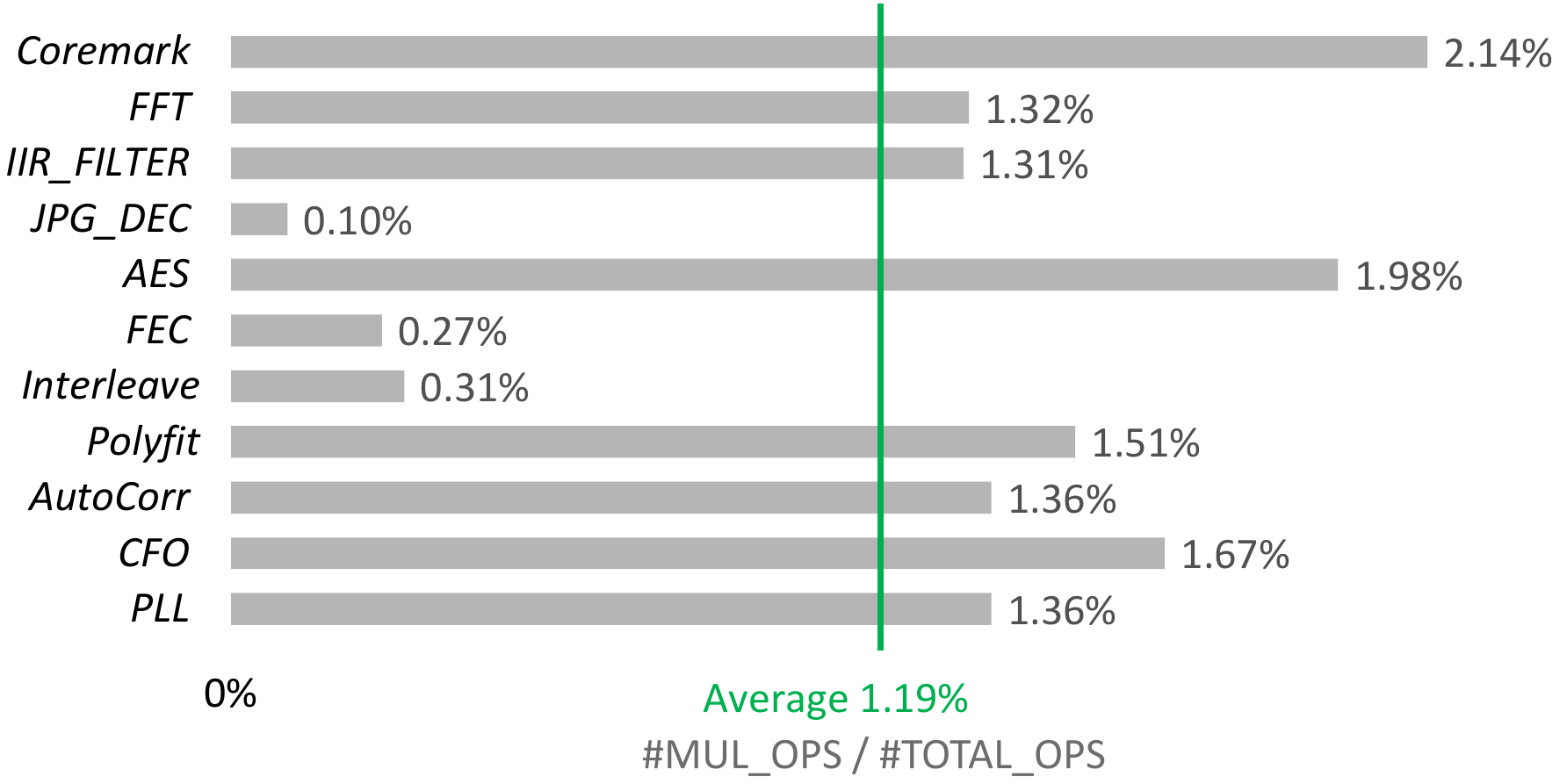}
\caption[Utilization frequency of the multiplier unit.]{The utilization frequency of the multiplier unit is around 1.2\%. So optimization on the multiplication unit is profitable.}\label{fig:arch_utilization}
\end{figure}

Fig.~\ref{fig:arch_utilization} shows the utilization frequency of the multiplier. In average, 1.2\% of the instructions is a multiplication. The processor actually takes more than 1.2\% of execution time in multiplication, as each multiplication takes four cycles.

Since each multiplication takes four cycles, the processor will take around 4.8\% of the cycles for multiplications.

\subsection{Word-length distribution for \textit{MUL}}

\begin{figure}[H]
\centering
\includegraphics[width=.98\columnwidth]{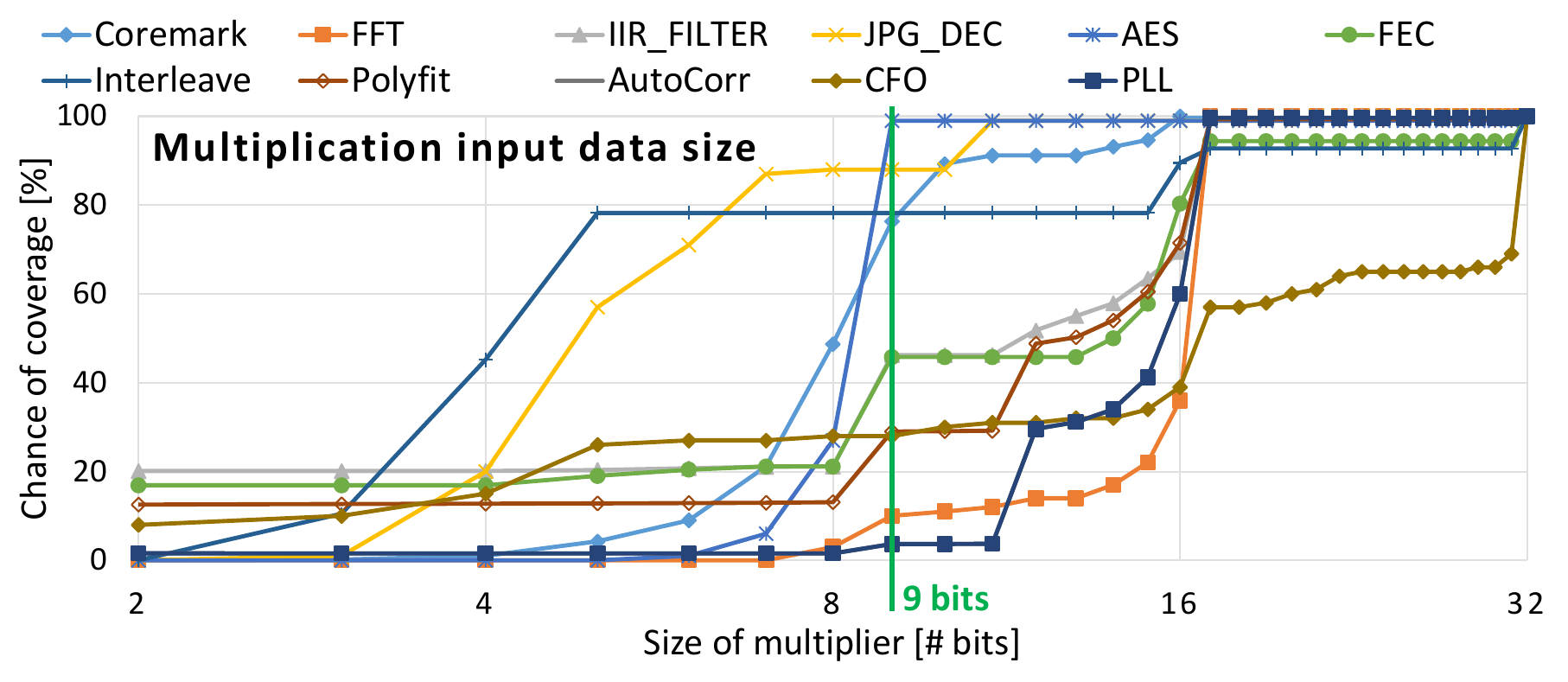}
\caption{The chance that both multiplicands are short increases when the criteria for `short' gets loose.}\label{fig:arch_chance}
\end{figure}

The word-length distribution of the multiplicands is illustrated in Fig.~\ref{fig:arch_chance}. The data is obtained by the cycle-accurate OpenRISC simulator. The multiplicands are recorded for each multiplication. There is a trade-off to choose how large the \textit{mul\_short} should be. If the criterion for short input is more strict, i.e. \# of bits is larger, the activation chance of the \textit{mul\_short} unit will increase, which leads to a lower power consequently. On the other hand, a larger \textit{mul\_short} unit itself consumes more power. Therefore, designers are suggested to profile the multiplication size coverage for typical applications and the corresponding power consumptions. 

Based on algorithm-level benchmark results (Fig.~\ref{fig:arch_chance}), a 9-bit multiplier is used for the \textit{mul\_short}. With this setting, the \textit{mul\_short} unit performs more than 80\% of the multiplication for \textit{Coremark}, \textit{JPG\_DEC}, \textit{AES}, and \textit{Interleave}; around 40\% multiplication for \textit{IIR\_FILTER}, \textit{FEC}, \textit{Polyfit}, and \textit{CFO}; around 5\% for \textit{FFT} and \textit{PLL}. 

This result shows that the HS scheme best fits algorithms that heavily use \textit{short\_integer} data-types for multiplications. In this scenario, the \textit{size\_detect} takes the role of the compiler to choose the suitable multiplication hardware. Moreover, for algorithms that use only full-width \textit{integer} data-type, e.g. \textit{IIR\_FILTER} and \textit{Polyfit}, the textit{mul\_short} still performs around 40\% of the multiplications. This is due to the fact that the varying input data has a very high tendency of falling into the short-size range, even though they are defined to be very wide to avoid the overflow in the worst case.

\section{Power saving validation of the proposed fine-grained hardware switch scheme}\label{sec:arch_verify}
The mor1kx is synthesized at 1GHz in a standard  28nm CMOS process. The derived netlist, together with its corresponding delay file (.sdf) and parasitic parameters, are simulated with the instructions from the most realistic and representative stimuli -- \textit{Cormark}.

\subsection{Area comparison}

\begin{figure}[H]
\centering
\includegraphics[width=.46\columnwidth]{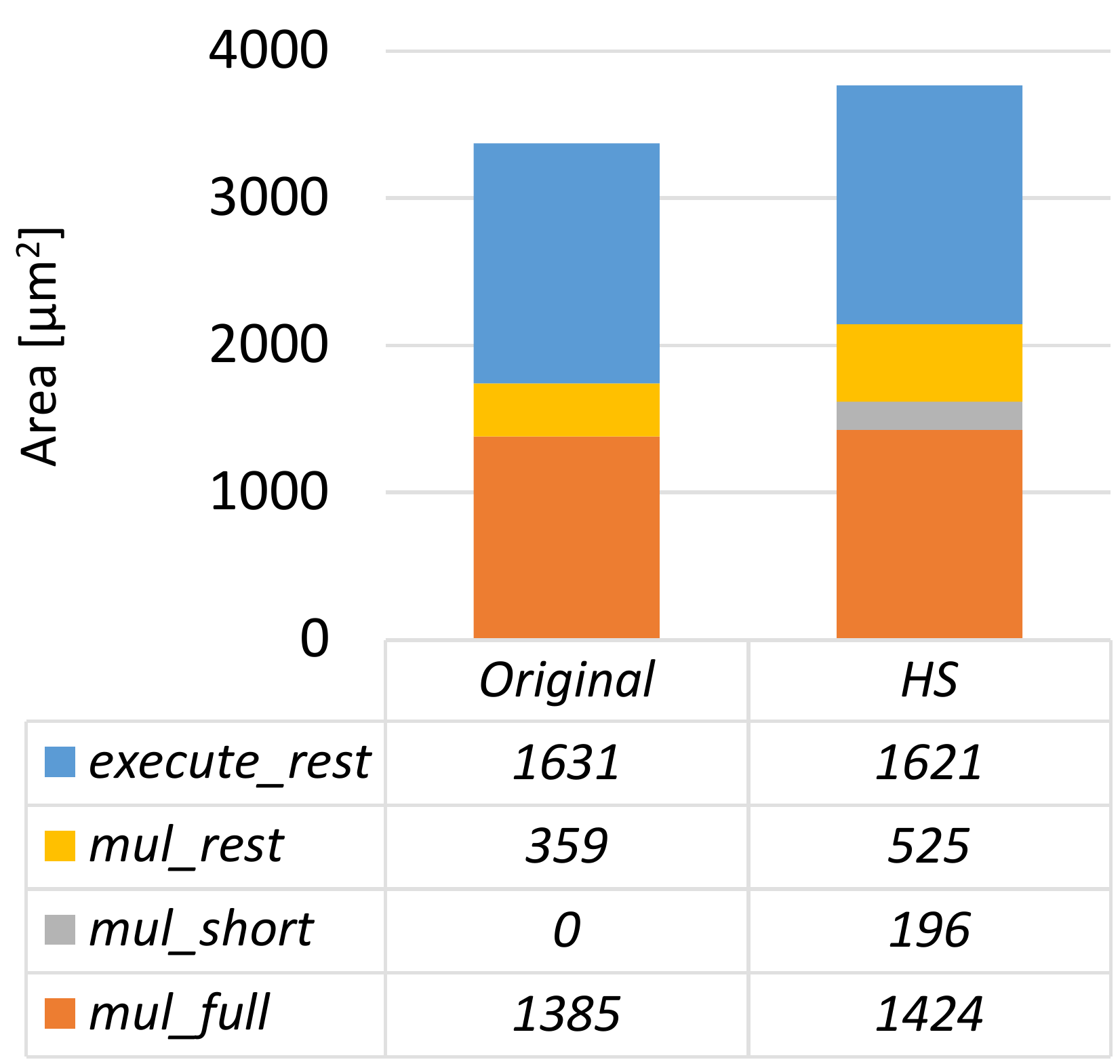}  
\caption{The proposed \textit{HS} scheme results to more area.}\label{fig:arch_metrics_area}
\end{figure}

\begin{figure}[H]
\centering
\includegraphics[width=.46\columnwidth]{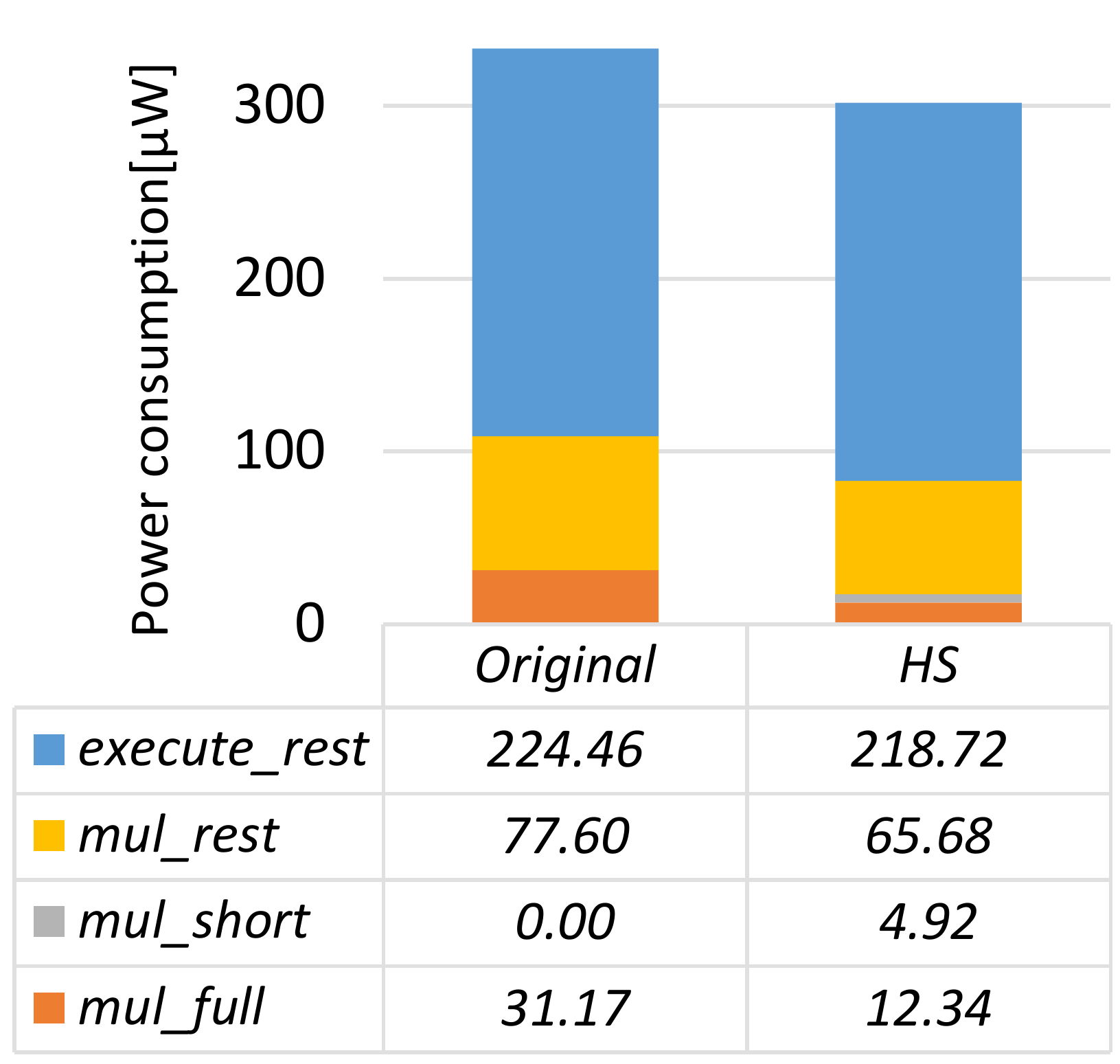}  
\caption[The power consumption of the \textit{HS}  reduces substantially.]{The power consumption of the \textit{HS}reduces substantially. The results are based on gate-level simulation with the Coremark benchmark.}\label{fig:arch_metrics_power}
\end{figure}

The area and power metrics with \textit{Original} or \textit{HS} (with 9-bit \textit{mul\_short}) schemes are shown in Fig.~\ref{fig:arch_metrics_area} and Fig.~\ref{fig:arch_metrics_power}. For the processor with HS scheme, the extra \textit{mul\_short} and \textit{size\_detect} results in 23.0\% area overhead for the multiplier unit, which is equivalent to 11.5\% area overhead for the whole execution unit.

\subsection{Power savings}
The power consumption of the \textit{mul\_full} is reduced from 31.167~$\mu W$ to 12.344~$\mu W$, since its execution ratio is greatly reduced. It accounts for a total of  23.7\% power saving for the multiplier unit and 9.5\% power saving for the execution unit. In summary, the fine-grained hardware switch scheme introduces redundant area, which saves execution power for low-precision computations.

\section{Conclusion}\label{sec:arch_con}
The HS scheme proposed in this chapter exploits word-length opportunities to reduce dynamic power consumption. It is achieved by utilizing an alternative short multiplier when the circuit detects the inputs are short enough.

The proposed scheme does not affect the software nor the compiler since the detection and switching are implemented at hardware level. It best fits processors which frequently perform short multiplications. In such processors, the multiplier unit power is significantly reduced. 

In this chapter, the hardware switch scheme does not alter the final output of the program, since the opportunistic hardware switch is only enabled if the input size is small enough to fit into the low-precision unit. In the future, for programs with soft quality requirements, the activation of the low-precision unit can be extended to scenarios that the data size is marginally larger than the low-precision unit. Therefore, the occurrence of low-precision activation is increased, and hence more power saving could be achieved, though at the cost of degraded output quality. 

In this regards, more cross-level optimization between the microarchitecture and the algorithm-level is encouraged. One such example of circuit and algorithm interplay is explained in Chapter~\ref{sec:algo}.



\cleardoublepage

\chapter{Algorithm-level error-resilient design mitigating circuit-level errors}\label{sec:algo}


This chapter proposes a method for cross-level error interplay between the circuit-level and the algorithm-level. Traditionally, circuit-level results need to be error-free. This confines the variability design margin shaving methods to error-free approaches, e.g. Canary FF. In contrast, this chapter promotes to embrace some circuit errors, for a more aggressive margin shaving. Errors produced at the circuit-level are mitigated at the algorithm-level. In particular, a \textit{computation-skip} scheme is discussed. The error mitigation scheme, together with the Canary FF that serves for benchmark purposes, is applied to a hardware CORDIC accelerator. The typical applications for CORDIC are QR decomposition and Cartesian to polar coordinate vector translation. The CORDIC accelerator is processed and verified in a standard 28nm CMOS process with only standard-cells. Using only standard-cells, this work eliminates the traditional semi- (or even fully-) customized design effort for in-situ error detection circuits. The work in this chapter is published in \cite{huang14sips-error-resilient, huang16jsps-error-resilient}.

The rest of this chapter is structured as follows: Section~\ref{sec:algo_intro} describes the background and the contributions. Section~\ref{sec:algo_implement} explains the design implementation trade-offs of the error-resilient circuits. It pinpoints the advantages of the proposed \textit{computation-skip} scheme. Section~\ref{sec:algo_case} takes a CORDIC hardware accelerator for case-study. Section~\ref{sec:algo_con} concludes the chapter.

\section{Motivatuion to algorithm-level error resilience}\label{sec:algo_intro}
Integrated circuits are designed with inherent guardbands, i.e. using worst PVT (Process, supply Voltage noise $V_{dd}$ and Temperature) corners, to ensure correct functionality for all chips with the presence of dynamic temperature and $V_{dd}$ noise fluctuation. However, typical usage patterns usually operate at the nominal PVT condition. The over-pessimistic margin leads to power waste. Nowadays, digital circuits aiming for low-energy consumption usually adjust the $V_{dd}$ ($V_{dd}$ scaling) to exploit the design margin. The energy consumption will reduce quadratically to the $V_{dd}$. Nevertheless, the circuit setup-timing constraints must be met, otherwise data errors will occur.

\subsection{Review of error-resilient techniques}
In-situ timing-error detection circuits have been proposed to detect setup timing-errors when reducing $V_{dd}$ ~\cite{razor03,canary04,bowman09,nicolaidis2013adda,bubble13}. A canary FF~\cite{canary04} generates a warning when the timing is critical. This enables dynamically adjusting the $V_{dd}$ and/or $f_{clk}$ for microprocessors. A circuit with canary FFs gradually reduces its $V_{dd}$, at the training phase, until the first occurrence of warning. This ensures correct functionality during operating, as no errors, but only warnings, occur. This scheme is regarded conservative, as a delay margin (between warnings and actual timing-errors) still exists. Razor like techniques, e.g. Razor \cite{razor03}, DSTB \& TDTB \cite{bowman09}, and Bubble Razor \cite{bubble13}, exploit the margin even further, by detecting actual timing-errors. This condition is called VOS (Voltage-OverScaling), as the voltage is scaled beyond the safety region. The corresponding error correction schemes, i.e. global clock gating \cite{razor03}, counter-flow \cite{razor03}, instruction-replay \cite{bowman09}, and Bubble Razor \cite{bubble13}, correct errors by issuing extra instructions. Therefore, the circuit can operate around the operating condition that produces sparse errors. These schemes, although applicable to general-purpose computers, result in throughput penalty when timing-errors are detected. Another common drawback of the previously proposed works in this class is the utilization of customized circuit design methodologies (mostly for error detection circuits), which is not a classical digital design flow.

TIMBER \cite{timber10, constantin2015exploiting} delays the clock for 1-phase to compensate the time borrowing for a timing-error at the circuit-level. However, this technique requires substantial effort in adjusting the clock phase for error recovery. This is very challenging for the timing closure with commercial EDA tools \cite{tam2000clock, bowman201616}.

Another class of methods to handle VOS errors is arithmetic noise tolerance (ANT) technique~\cite{Shim2004, hegde2004voltage, karakonstantis2009system, 6241554, 6636082}. ANT techniques detect errors by algorithmic comparison, and correct them without extra cycle penalty, by linear prediction \cite{hegde2004voltage}, reduced precision redundancy \cite{Shim2004} or adaptive error cancellation \cite{wang2003low}. However, a major drawback of regular ANT techniques is the non-generic algorithmic error detection. They require careful ad-hoc design. Another common problem of most ANT schemes is the requirements of dedicatedly selected data-path, e.g. specific adder micro-architectures~\cite{6636082}. This prevents data-path synthesis optimization that modern EDA offers.

\section{Contributions of this chapter}
A common drawback of the previously proposed work is the utilization of customized circuit design methodologies (mostly for error detection circuits), which is not a classical digital design flow. Besides, although extensive measurements are performed for Razor-like error recovery circuits, seldom applications address the relation between $V_{dd}$ drop and output quality for normal DSP blocks~\cite{6241554, 6636082}. Furthermore, the design margin reduction effectiveness is not verified with deeply scaled sub-28nm technologies.

In contrast, this work proposes a power reduction method (named as \textit{computation-skip} scheme) for a DSP accelerator. It demonstrates the power reduction and implementation feasibility of the mentioned techniques in deeply scaled technologies. The \textit{computation-skip} scheme handles errors at the algorithm-level that were created at the circuit-level. It mitigates the timing-error during $V_{dd}$ scaling for recursive application, at the cost of output SNR degradation without throughput drop. The proposed \textit{computation-skip} scheme can be applied to signal processing algorithms with a recursive structure. In these algorithms, signals will be processed by the same combinational logic for multiple times. Examples are CORDIC, Viterbi, LDPC decoding, loop counter and genetic algorithms. The \textit{computation-skip} scheme is compared with existing in-situ error correction schemes in Table~\ref{tab:algo_summary}.

\afterpage{
\begin{sidewaystable}[h]
\caption{In-situ timing-error detection and correction/mitigation schemes.}
\label{tab:algo_summary}
\scalebox{0.75}{
{\footnotesize
\begin{tabular}{lcccc}
\toprule
Features & FF based techniques & Temporal clock adjust & Algorithmic approximation & \textit{computation-skip} Error mitigation \\
\cmidrule(l){1-1} \cmidrule(l){2-5}
Examples  & \makecell{Razor~\cite{bowman09}; \\ DSTB and TDTB~\cite{bowman09}; \\ Bubble Razor~\cite{bubble13}; \\ One-cycle~\cite{shin13}}
          & TIMBER~\cite{timber10} & \makecell{ ANT \\ \cite{ant99,viterbi09}} & proposed \textit{computation-skip} \cite{huang16jsps-error-resilient}  \\
\cmidrule(l){1-1} \cmidrule(l){2-5}
Error Detection & \makecell{double sampling \/ \\ transition detection}  & \makecell{double sampling \/ \\ transition detection} & algorithmic inference &  \makecell{double sampling \/ \\ transition detection}  \\ \cmidrule(l){2-5}
~~Accurate detection? & yes      & yes    & no (inference)  & yes \\ \cmidrule(l){2-5}
~~When detector samples?      & after clock edge    & after clock edge  & at clock edge  & after clock edge \\ \cmidrule(l){1-1} \cmidrule(l){2-5}
Error correction                    & \makecell{counter-flow \\ instruction replay \\ bubble insertion}  & \makecell{delay clock \\ for one phase}    & use approximated $Q$    & \makecell{partial computation-skip \\ in the next cycle} \\ \cmidrule(l){2-5} 
~~Guaranteed quality?           & yes  & yes  & \makecell{no\\ (error accumulated)} & \makecell{no\\ (error not accumulated)}\\ \cmidrule(l){2-5}
~~Guaranteed throughput?               & \makecell{no\\ (1+ cycles penalty)}  & \makecell{no\\ (1 clock phase penalty)} & yes  & yes \\ \cmidrule(l){1-1} \cmidrule(l){2-5}
Clock overhead                     & moderate  & heavy    & none      & none  \\ \cmidrule(l){1-1} \cmidrule(l){2-5}
Logical gate overhead             & small     & small    & moderate  & small \\ \cmidrule(l){1-1} \cmidrule(l){2-5}
Time borrowing?                    & yes      & yes    & no  & yes \\ \cmidrule(l){1-1} \cmidrule(l){2-5}
Application independent?           & yes       & yes      & no        & \makecell{no \\ (recursive  applications only)}\\\bottomrule
\end{tabular}}}
\end{sidewaystable}
\clearpage
}

The major advantage of this scheme, compared with other FF based techniques and the temporal clock adjusting schemes, is the 0-cycle overhead during error correction. On the other hand, comparing with other ANT techniques, the \textit{computation-skip} scheme simplifies the error-detection design by adopting the FF based approach.

From the implementation point of view, the state-of-the-art canary FF error detection scheme, as well as the newly proposed \textit{computation-skip} scheme, were implemented and verified on silicon. Both circuits were processed in 28nm CMOS with standard digital design flow. This eliminates the conventional semi- (or even fully-) customized design effort for in-situ error detection circuits.

\section{Error resilient circuits implementations}\label{sec:algo_implement}
The circuit-level implementations of error resilient circuits are described in this section. This section describes the concept of Canary FF circuits. Moreover, it explains the design trade-offs of the computation-skip scheme. 

\begin{figure}[H]
\centering
\includegraphics[width=0.8\textwidth]{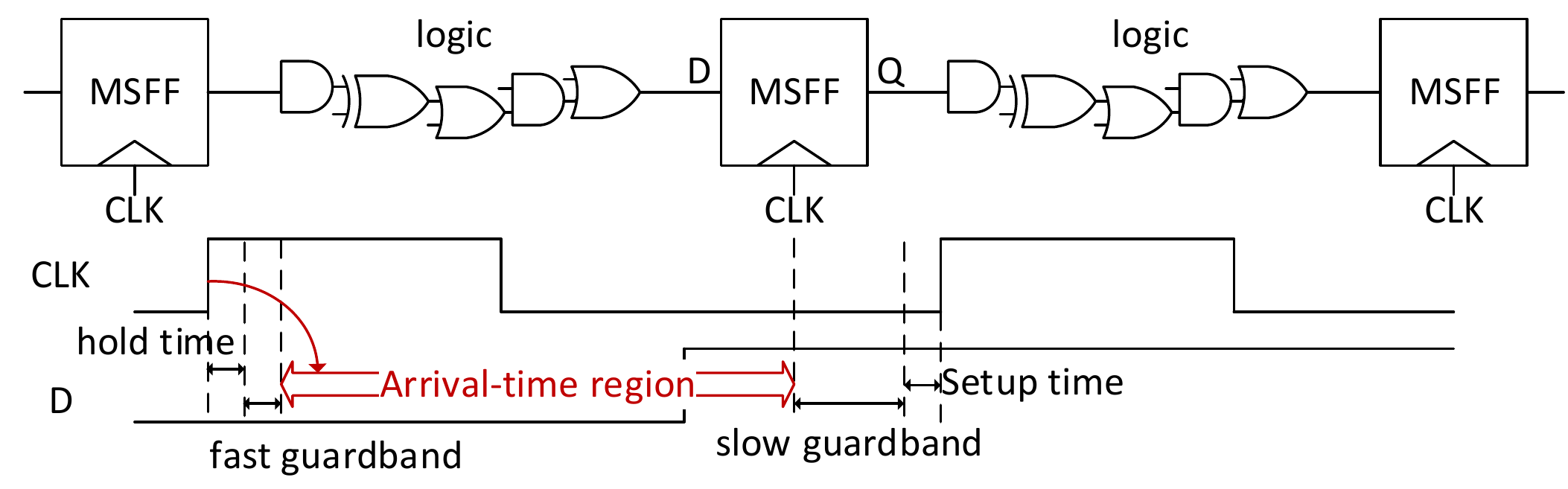}
\caption{The conventional pipeline scheme requires worst-case design margins.}
\label{fig:algo_conv}
\end{figure}

A conventional pipeline circuit diagram is outlined in Fig.~\ref{fig:algo_conv}. The signal arrival time for the $D$ port of an MSFF (Master-Slave Flip-Flop) is constrained by i) its hold-time plus the fast corner timing variation guard-band, and ii) its setup time plus the slow corner PVT guard-band. These corner-based guard-bands limit the capability of power saving.

\subsection{Circuits with Canary FF}
\begin{figure}[H]
\centering
\includegraphics[width=0.9\textwidth]{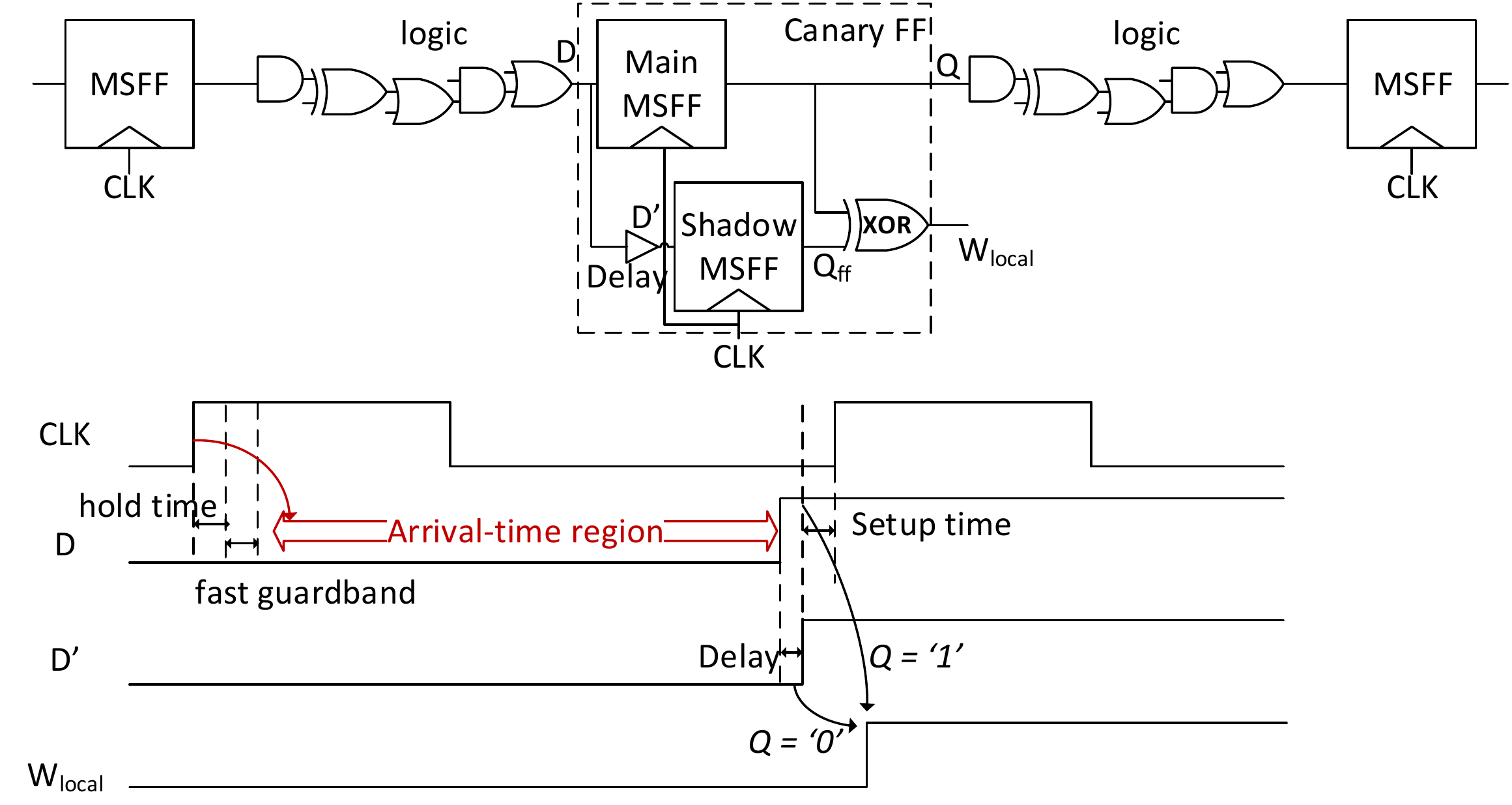}
\caption{Circuits with in-situ \textit{canary FF} that shave worst-case design margins.}
\label{fig:algo_canary}
\end{figure}

An in-situ \textit{canary FF} based circuit applies a second \textit{shadow MSFF} to detect dangerous (slow) timing at critical $V_{dd}$ or $f_{clk}$ (Fig.~\ref{fig:algo_canary}). When the signal arrival on $D$ is critical, its delayed signal $D\prime$ will violate its constraint. As a consequence, the \textit{main MSFF} and the \textit{shadow MSFF} capture different values. This triggers a local warning ($W_{local}$) since the \textit{shadow MSFF} has failed. Because the \textit{main MSFF} does not fail, the situation is only reported as a warning, rather than an error. This critical operation condition is hence called the PoFW (Point of the First Warning). However, if signal delays are enlarged under worse conditions, \textit{main MSFF} will fail, and functionality errors will occur. The operation condition that the first error emerges is called the PoFF (Point of the First Failure).

The width between the PoFW and the PoFF is the error detection window, which is tuned by the delay element. The circuit usually operates around the PoFW condition. The error detection window in \textit{canary FF} allows infrequent warnings and also ensures correct circuit functionality. The PVT slow guardband is exploited as each chip operates with just needed energy.

\subsection{Circuits with \textit{computation-skip} scheme}\label{sec:algo_cscircuit}
\begin{figure}[H]
\centering
\includegraphics[width=0.9\textwidth]{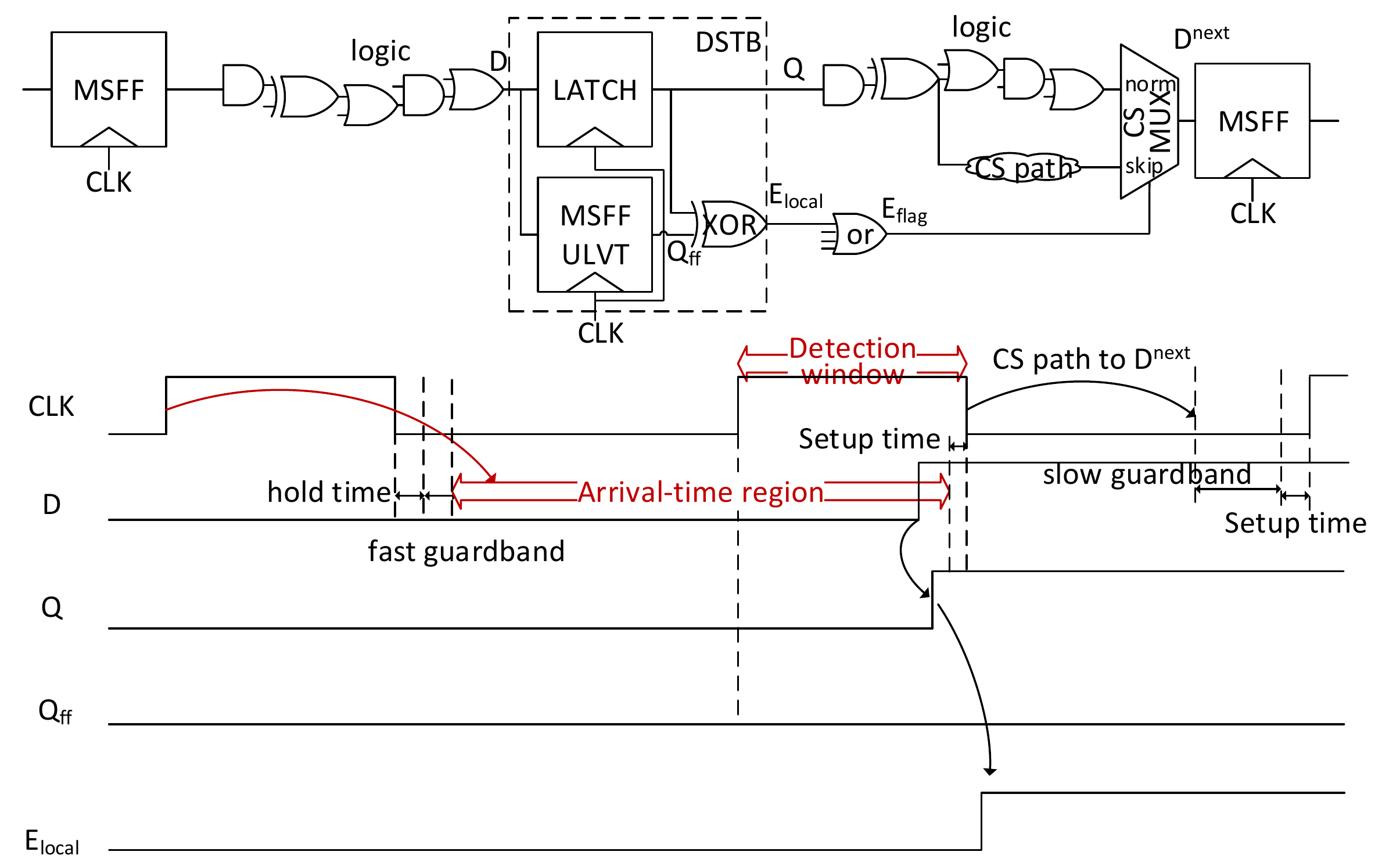}
\caption{Circuits with \textit{computation-skip} scheme, which utilizes \textit{DTSBs} for error-detection, eliminate worst-case design margins.}
\label{fig:algo_cs}
\end{figure}

The \textit{computation-skip} error resilient scheme \cite{huang16jsps-error-resilient} utilizes DSTB (Double Sampling with Time Borrowing)~\cite{bowman09} for error detection (Fig.~\ref{fig:algo_cs}). A DSTB~(Fig.~\ref{fig:algo_cs}) consists of an enable-high \textit{latch}, an \textit{shadow MSFF}, and an \textit{XOR comparator}. 

If a signal arrives late than the clock rising edge, the latch captures the correct signal (with time borrowing) while the shadow MSFF captures an incorrect one, an error flag $E_{local}$ is produced. The operating condition that the first error emerges is PoFF. If the chip only operates around the PoFF, the circuit behaves similarly to \textit{canary FF} based circuits. However, by investing the signal processing algorithms, a further design margin can be exploited. \cite{huang16jsps-error-resilient} proposed that for recursive applications, e.g. CORDIC and Viterbi, a part of computations can be approximated, or even totally skipped, under the worst-case condition. The final output is slightly degraded, which can often be tolerated or even compensated by the upper layer of the system, e.g. by using more quantization bits or stronger ECC. The system can thus decide whether to continue to work with the degraded circuit or to improve the operating condition (reducing $f_{clk}$ and increasing $V_{dd}$) to prevent further degradations.

Fig.~\ref{fig:algo_cs} shows the timing diagram of a circuit with \textit{computation-skip} scheme. Comparing with \textit{canary FF}, it eliminates not only the PVT slow guard-band but also the delay margin between PoFW and PoFF. More importantly, a path with long propagation delay can borrow time from the next upcoming clock cycle, relaxing its timing constraint by a duty cycle of the clock ($\tau \cdot t_{clk}$). To compensate for the time borrowing, a short version of the upcoming computational path, i.e. a \textit{computation-skip} path, is selected to meet the constraint. However, a disadvantage for circuits using DSTB or Razor is the minimum delay requirement. This requirement overburdens the hold time fixing during layout. Unfortunately, this disadvantage is also present in this \textit{computation-skip} scheme. Bubble Razor \cite{bubble13} eliminates the minimum delay issue by adopting a two-phase latch design. However, these two-phase latches are still not fully compatible to mainstream EDA tools.

Once a timing violation is detected by the DSTB, $Q$ is fed to the MUX by the \textit{computation-skip} path. This is because re-computing the next logic with the late-arrived correct $Q$ is impossible, due to the setup timing constraint for the following cycle. As a consequence, the correct signals from the previous clock are preserved. More importantly, further timing violation is eliminated by bypassing with the \textit{computation-skip} path. Note that only part of the \textit{logic} is skipped by the \textit{computation-skip} path. This avoids heavy quality degradations of skipping the whole logic. 

This mitigation can be regarded as a naive implementation for the approximated version of the logic. For recursive applications, the bypassing can be as simple as a direct copy of the previous signal. Another benefit of the bypassing is that no accumulating approximation errors are introduced. The skipping only leads to less performed iterations. The skipped computations are delayed into future cycles. If extra cycles are allowed in the future, errors can be totally eliminated. However, this work aims to maintain a constant throughput. No time penalties are permitted. Therefore, the \textit{computation-skip}  error mitigation scheme alleviates timing-errors into insufficient iteration errors.

\subsection{\textit{Computation-skip} scheme settings}

The \textit{computation-skip} scheme tolerates longer propagation delay in data-path than a nominal circuit. In other words, the circuit can even operate at $V_{dd}$ that lead to setup timing violations, or sub-critical $V_{dd}$, to save energy.

\begin{figure}[H]
\centering
\includegraphics[width=\linewidth]{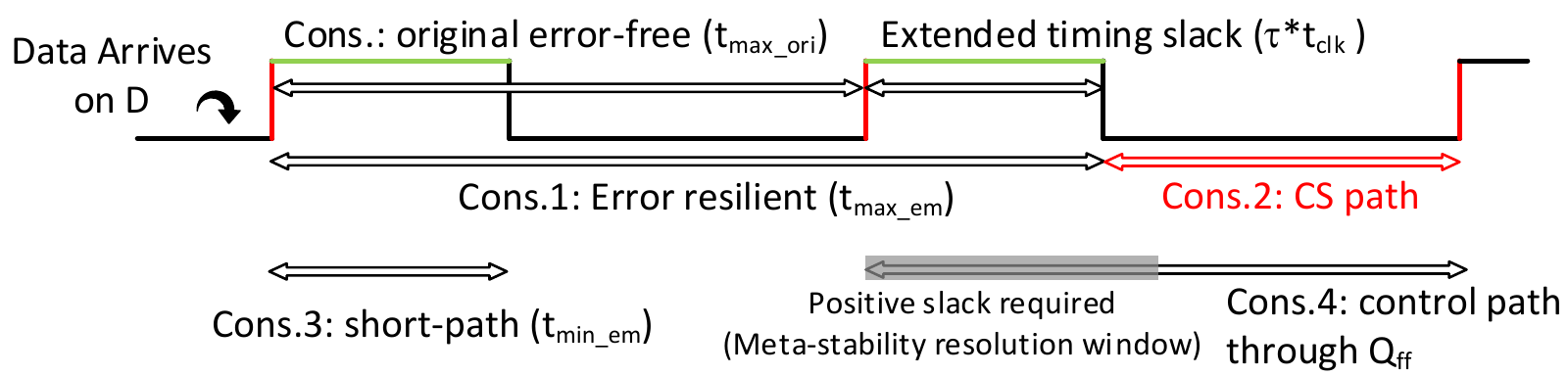}
\caption[Timing constraints for circuits with the \textit{computation-skip} scheme.]{Timing constraints for circuits with the \textit{computation-skip} scheme. The scheme relaxes the setup constraint by accepting late arrival signals.}
\label{fig:algo_timing}
\end{figure}

The detailed timing constraint is demonstrated in Fig.~\ref{fig:algo_timing}. By tuning the clock duty cycle factor $\tau$, digital circuit designers can extend the nominal delay constraint $t_{max\_orignal}$ to the error mitigation timing constraint $t_{max\_em}$:
\begin{align}\label{eq:algo_tmax}
t_{max\_ori} &= t_{clk}- t_{setup\_FF}  \nonumber\\
t_{max\_em} &= t_{clk} + \tau \cdot t_{clk} - t_{setup\_latch}
\end{align}
where $t_{clk}$ is the clock duration, $t_{setup\_FF}$ represents the setup time for a normal FF, and $t_{setup\_latch}$ represents the setup time for the main latch. During logic synthesis, the relaxed timing constraint will enable EDA tool to use smaller but slower gates, which reduces area and power consumption.

The speed constraint relaxation ratio $R$ is defined as:
\begin{equation}
R  \triangleq \frac{ 1 / t_{max\_em} }{ 1 / t_{max\_ori} }.
\end{equation}
Substituting $t_{max\_em}$ and $t_{max\_ori}$ from (\ref{eq:algo_tmax}) obtains
\begin{align}\label{eq:algo_rf}
R  & = \frac{ t_{clk}- t_{setup\_FF}  }{ t_{clk} + \tau \cdot t_{clk} - t_{setup\_latch} }\nonumber\\
 & \approx \frac{1}{ 1 + \tau}.
\end{align}
A lower $R$ represents loose constraint for data, and hence easier for setup timing closure.

At sub-critical $V_{dd}$ situations, circuits delay increase that the data input port of the shadow MSFF might change around clock rising edges. So these FF might become meta-stable. Therefore, paths starting from $Q_{ff}$ are guarded with positive slacks, serving as the resolution window for the FF to settle.

To find out the meta-stability resolution constant, Monte-Carlo simulation by Spice is performed on the ULVT FF. The meta-stability resolution is 20ps at nominal voltage. To mitigate the meta-stability issue, this chapter set an extra timing slack of 700ps. The worst case for MTBF estimation happens when exactly one signal at the D port of the FF changed close to the clock rising edge within setup time. Therefore, 
\begin{align}
MTBF &= \frac{e^{\frac{700~ps}{20~ps}}}{f_{clk}} \nonumber\\
    &\approx 41~days, ~~~when~f_{clk} = 450~MHz.
\end{align} 

It is sufficiently large to guarantee mean time before failure (MTBF) requirement for the system due to meta-stability. Note that the main latch is designed to never fail at even sub-critical situations, the data-path and \textit{computation-skip} path are immune from meta-stability, which is a big advantage for the DSTB. In reality, the system MTBF is much better than $41$ days because any other timing failure will set $E_{flag}$ to `1', masking the effect of the local meta-stability fault.

As the \textit{main latch} is still sensitive after the clock rising edge throughout the first half of the clock, if it captures the newly arrived signal too early, the signal from the previous cycle is flushed. In this situation, the error detection circuit might indicate a false error. Therefore, for the paths to the \textit{main latch}, a short-path timing constraint is required:
\begin{align}
t_{min\_em} = \tau \cdot t_{clk} + t_{hold\_latch},
\end{align}
where $t_{hold\_latch}$ is the hold time for the main latch. This short path constraint is guaranteed by inserting buffers during placement \& route.

The timing constraint is summarized in Tab.~\ref{tab:algo_timing}. With the \textit{computation-skip} scheme, the normal setup constraint is much relaxed with the ratio of $R$. This makes the timing much easier to meet, and hence EDA tools have the option to choose smaller and slower cells to save chip area and power consumption.

\begin{table}[H]
\caption{Timing constraints for the \textit{computation-skip} error mitigation scheme.}\label{tab:algo_timing}
\begin{tabular}{ll}
\toprule
Constraints & Path \\
\cmidrule(l){1-1} \cmidrule(l){2-2} 
error mitigation setup                 & $Q$ $\rightarrow$ Logic1 $\rightarrow$ Logic2 $\rightarrow$ $D$ \\
\textit{computation-skip} path setup   & $Q$ $\rightarrow$ CS path $\rightarrow$ $D$ \\
$Q_{ff}$ control path setup            & $Q_{ff}$ $\rightarrow$ $D$ \\
short-path hold                        & $Q$ $\rightarrow$ $D$ \\
\cmidrule(l){2-2} 
& Value \\ 
\cmidrule(l){2-2} 
error mitigation setup                 & $ \leq t_{clk} \cdot (1 + \tau)  - t_{setup\_FF}$ \\
\textit{computation-skip} path setup   & $\leq t_{clk} \cdot (1 - \tau) - t_{setup\_FF}$ \\
$Q_{ff}$ control path setup            & $\leq t_{clk} - t_{meta\_window} - t_{setup\_FF}$ \\
short-path hold                        & $\geq \tau \cdot t_{clk} + t_{hold\_latch}$ \\
\cmidrule(l){2-2} 
& Remarks \\
\cmidrule(l){2-2} 
error mitigation setup                 & The timing is relaxed for normal data-path \\
\textit{computation-skip} path setup   & Skip constraint for error recovery \\
$Q_{ff}$ control path setup            & Positive slack as meta-stability resolution \\
short-path hold                        & Short path constraint for latch \\
\bottomrule
\end{tabular}
\end{table}

\section{Case study on a CORDIC hardware accelerator}\label{sec:algo_case}
A CORDIC hardware accelerator is selected for a case-study to validate the error resilient design techniques. The canary FF case and the \textit{computation-skip} are applied to the \textit{core1 canary} and the \textit{core2 CS}, respectively. These two cores are synthesized and processed in a standard 28nm CMOS technology.

\subsection{CORDIC algorithm}
A CORDIC \cite{volder1959cordic} is a simple and efficient implementation to calculate trigonometric functions. A typical application is to compute the magnitude $M$ and initial angle $\phi$ of a complex input vector [$x_{0}$, $y_{0}$]:
\begin{align}
M & \triangleq \sqrt{x_{0}^2 + y_{0}^2} ;\nonumber\\
Cos(\phi) & \triangleq x_{0} / \sqrt{x_{0}^2 + y_{0}^2} ;\nonumber\\
Sin(\phi) & \triangleq y_{0} / \sqrt{x_{0}^2 + y_{0}^2} .
\end{align}

\begin{figure}[H]
\centering
\includegraphics[width=0.7\textwidth]{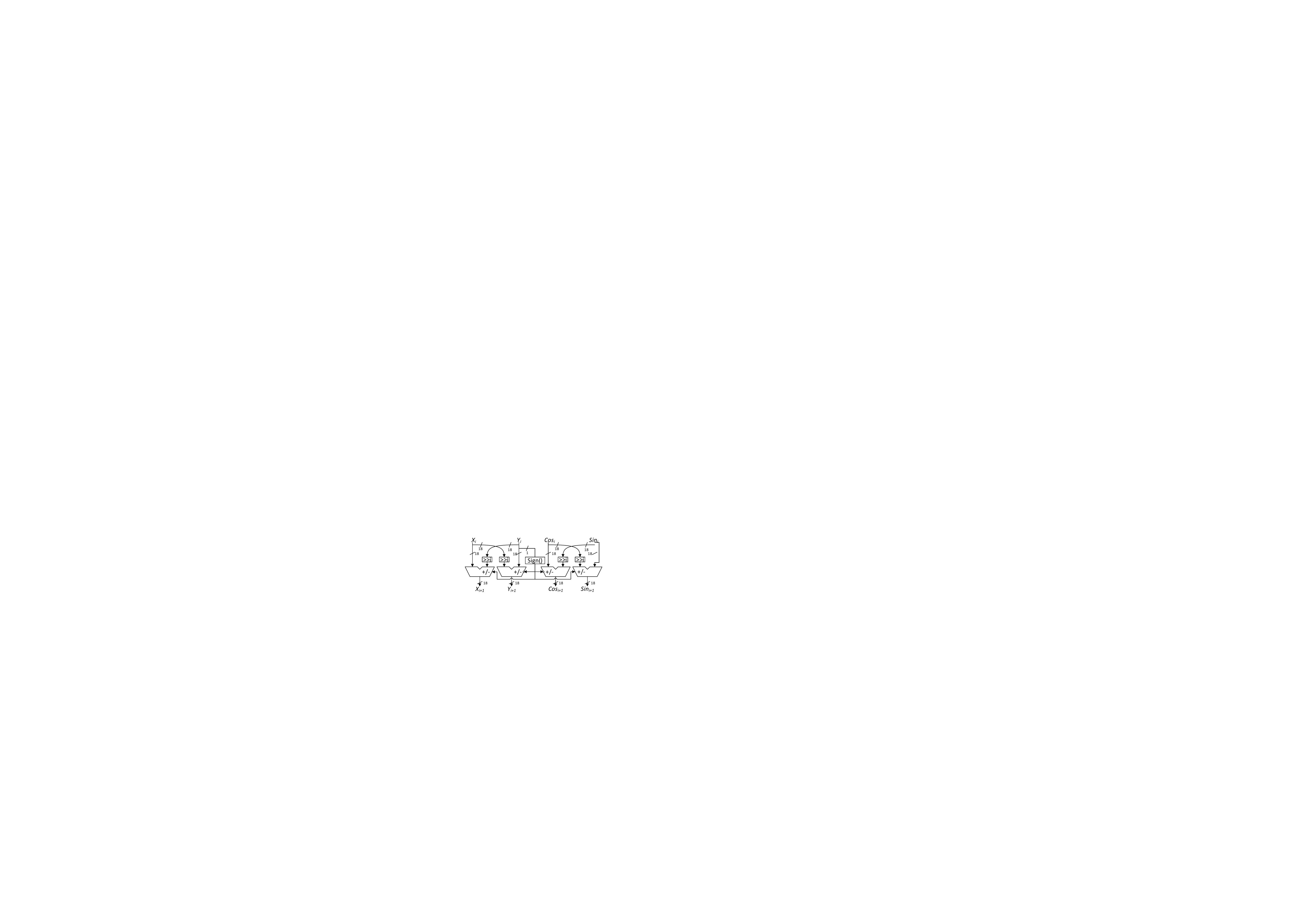}
\caption[]{A \textit{CORDIC cell} that performs a CORIC iteration. The example word-length is 18-bit.}
\label{fig:algo_cordic_cell}
\end{figure}

The CORDIC operation is an iterative process: the input vector [$x_{0}$, $y_{0}$] is rotated recursively by micro-rotations. The \textit{CORDIC cell} that performs one micro-rotation (iteration) is shown in Fig.~\ref{fig:algo_cordic_cell}.

The total number of iterations for a CORDIC operation is denoted as $n$. After $n$ iterations, $y_i$ converges to 0, and the resulted $x_{n}$ is the magnitude, with a scaling factor $K\approx 0.6072$. The angle ($\phi$) is represented in the sine ($Sin_n$) and cosine ($Cos_n$) format.

Because of non-ideal in the CORDIC algorithm, the resulted $x_{n}$, $Sin_n$, and $Cos_n$ can never reach the true values ($M$, $Cos(\phi)$, $Sin(\phi)$). The computation errors can be measured as the mean square error (MSE), the error vector magnitude (EVM), the signal-to-noise ratio (SNR), and the effective number of bits (ENOB). The computation procedures are given below.

The MSE directly depicts the magnitude of errors:
\begin{align}
MSE & = \frac{1}{T}\sum_{t=1}^{T}\left[ \left( x_{0}-x_{out} \right) ^2+ \left( y_{0}-x_{out} \right) ^2 \right] \\
& = \frac{1}{T}\sum_{t=1}^{T}\left[ \left( x_{0}- k \cdot x_{n} \cdot Cos_{n} \right) ^2+ \left( y_{0}- k \cdot x_{n} \cdot Sin_{n} \right) ^2 \right],
\end{align}
where $T$ is the number of samples for calculation. $x_{0}$ and $y_{0}$ are inputs for sample $t$. The $x_{out}$ and $y_{out}$ terms are the reverted format of CORDIC outputs ($x_{n}$, $Sin_n$, and $Cos_n$). The EVM compares the MSE with the signal power:
\begin{align}
EVM(dB) & = 10\log_{10}\frac{MSE}{P_{signal}} \\
  & = 10\log_{10}\frac{MSE}{ \frac{1}{T}\sum_{t=1}^{T} \left[ x_0^2 + y_o^2 \right] }.
\end{align}
The computation error is treated as noise. So the output quality can also be represented by SNR:
\begin{align}
SNR(dB) & = 10\log_{10}\frac{P_{signal} }{ MSE } \\
  & = 10\log_{10}\frac{ \frac{1}{T}\sum_{t=1}^{T} \left[ x_0^2 + y_o^2 \right] }{MSE}.
\end{align}
The SNR and EVM are in inverse linear relation, as illustrated in \cite{shafik2006extended}. In this chapter, the output quality of the CORDIC unit is denoted as ENOB \cite{geerts2006design}:
\begin{align}
ENOB = \frac{SNR-1.76dB}{6.02}.
\end{align}.

\begin{figure}[H]
\centering
\includegraphics[width=0.7\linewidth]{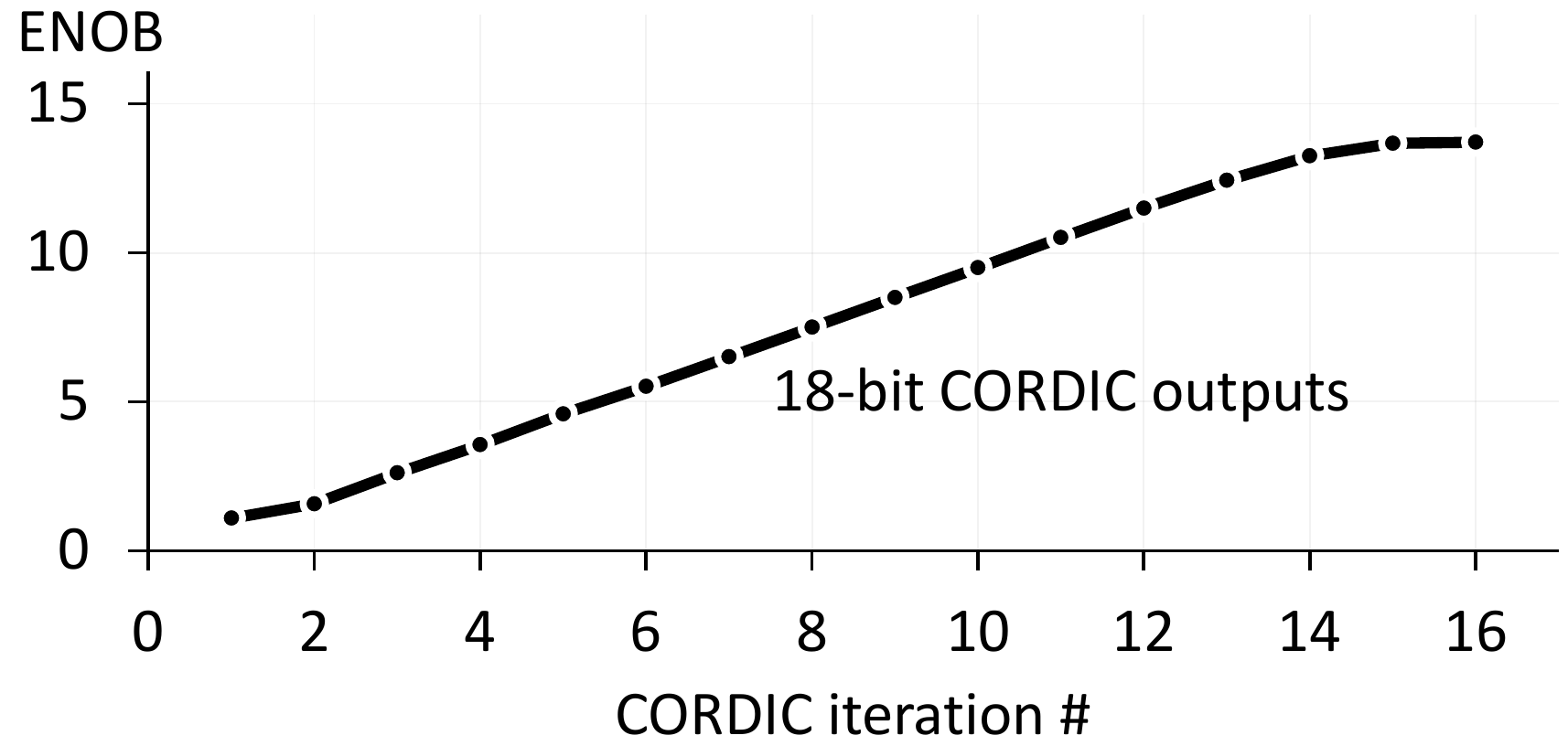}
\caption{The CORDIC output ENOB evolves with numbers of iterations $n$.}
\label{fig:algo_cordic_enob}
\end{figure}

Fig.\ref{fig:algo_cordic_enob} shows that the ENOB for the CORDIC evolves with each iteration. This evolution characteristic provides space for trading off the iteration number during VOS. Thus, this chapter proposes to apply the proposed \textit{computation-skip} scheme on the CORDIC application. Once a timing violation is detected, the chip skips part of the computation in the next cycle and adjusts the iteration counter. Due to the requirements of constant CPI, the final iterations, which contribute less to the EVM, are skipped to ensure that previous computations are guaranteed even in sub-critical situations. This is equivalent to computing with reduced iterations.

\subsection{State-of-the-art \textit{Core1} with Canary FF}
\begin{figure}[H]
\centering
\includegraphics[width=0.9\textwidth]{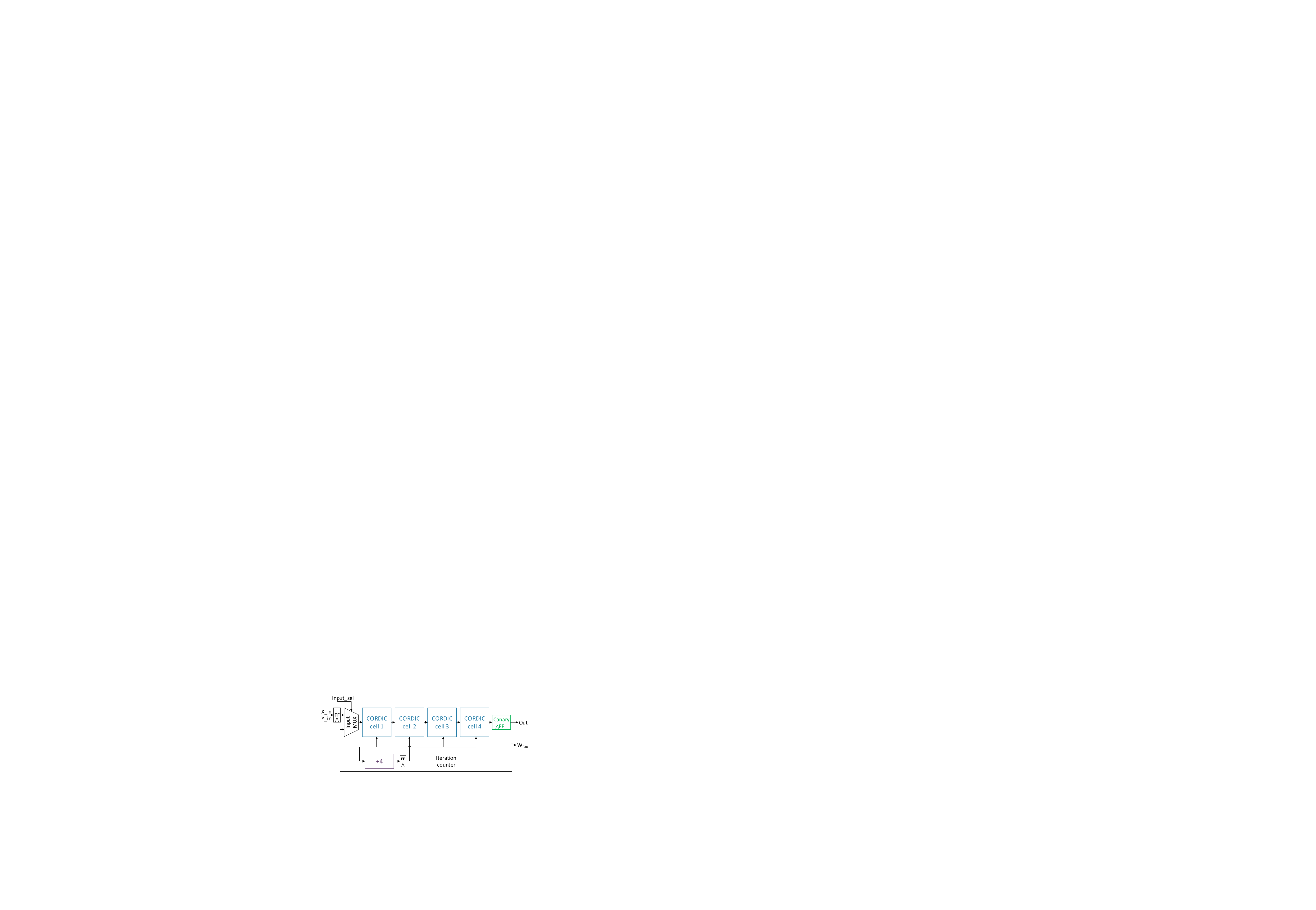}
\caption[]{\textit{Core1} hardware CORDIC accelerator replaces output FF with \textit{Canary FF}.}
\label{fig:algo_cordic_canary}
\end{figure}

The first implementation option (\textit{core1}) of the CORDIC accelerator is demonstrated in Fig.~\ref{fig:algo_cordic_canary}. The internal word size is 18-bit. 16 iterations are performed for each operation. The accelerator contains four \textit{CORDIC cells}; each performs a CORDIC iteration. Therefore, four clocks cycles are required to finish a CORDIC operation (CPI=4). Computation results are stored in the sequential cells (e.g. FF), whose outputs serve as the inputs for the next cycle. The iteration counter counts the CORDIC iterations and controls the barrel shifters in the CORDIC cells. For each cycle, the iteration counter is counted up by 4, meaning that 4 CORDIC iterations are finished in each cycle. Aiming for a high-performance design specification, the operating frequency is 450~MHz by design. This requires intensive gate up-sizing, as the speed target is challenging if worst-case design margins are considered.

The \textit{canary FF} were built out of two MSFF, a \textit{delay cell}, and a \textit{XOR gate} (see Fig.~\ref{fig:algo_canary}). The delay of the \textit{delay cell} is 150~ps, or 7\% of the clock period, under nominal conditions. EDA is disallowed from changing the \textit{canary FF}, otherwise the \textit{delay cells} will be deleted (optimized) for timing relaxation (and thus timing-error detection capabilities are lost). Besides, only the timing constraints for paths to \textit{main MSFF}, not for paths to \textit{shadow MSFF} (disallowing timing checks), are guaranteed, to avoid over-constraining.

The design was implemented in standard RTL. Afterward, the most timing-critical FF are replaced by \textit{canary FF}. Only the 9 MSBs from \textit{CORDIC cell 4} are substituted since they are associated with the longest delays, as suggested by the EDA tool (STA checks performed with Primetime). This reduces the overall area overhead (of applying \textit{canary FF}) to 1.3\%, compared with the alternative 7\% full substitute overhead.

\subsection{Proposed \textit{Core2} with \textit{computation-skip} scheme}
\begin{figure}[H]
\centering
\includegraphics[width=0.9\textwidth]{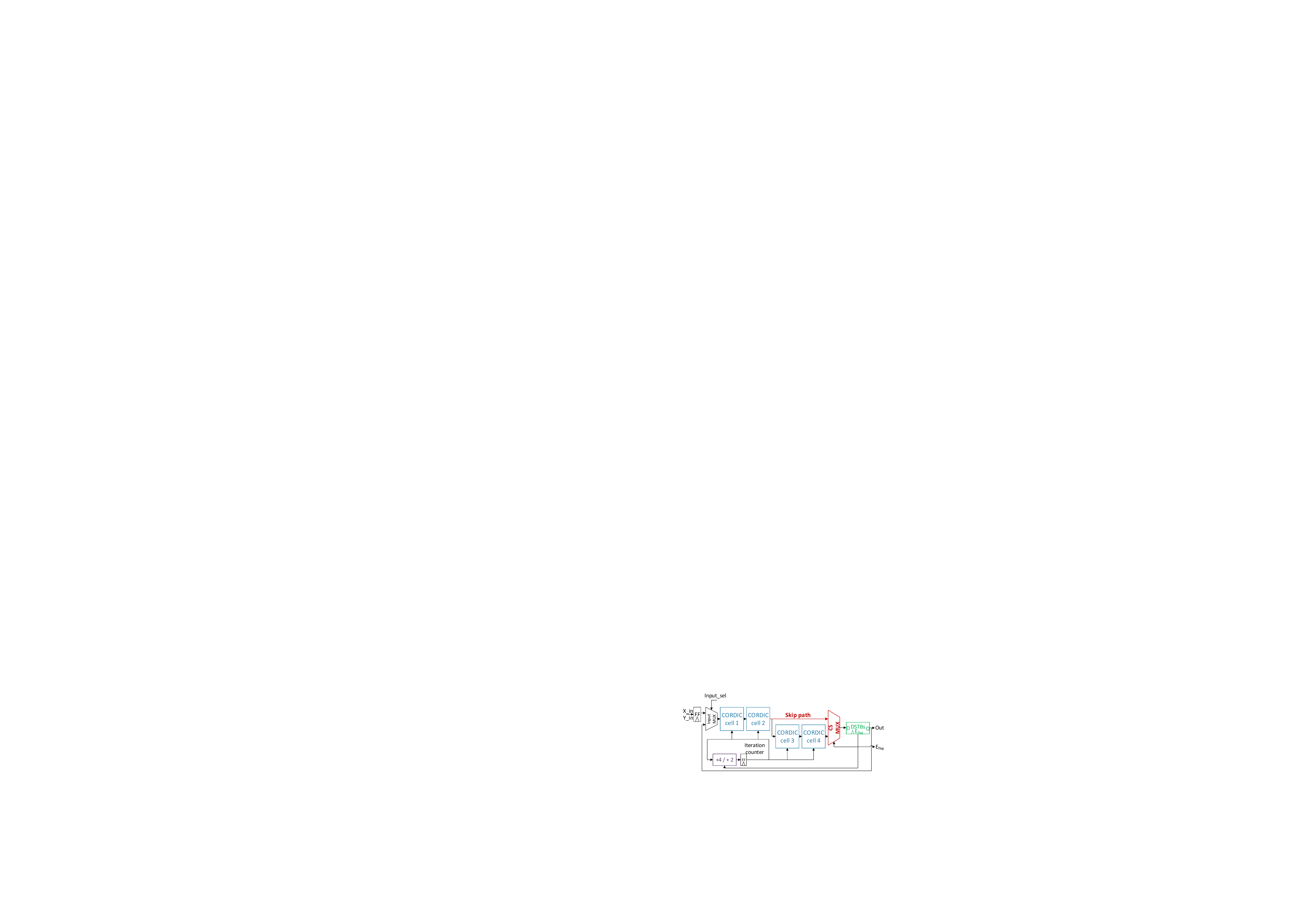}
\caption[]{\textit{Core2} features the proposed \textit{computation-skip} scheme.}
\label{fig:algo_cordic_cs}
\end{figure}

The second implementation option (\textit{core2}) is presented in Fig.~\ref{fig:algo_cordic_cs}. Similar to the \textit{core1}, FF that connect to \textit{CORDIC cell 4} are replaced by \textit{DSTB}. 


Knowing that the output quality depends on the \# of CORDIC iterations finished, the \textit{computation-skip} approach introduced a skipping path that shorts the \textit{CORDIC cell 3 and 4} when a short propagation delay is required. It is activated whenever the time is borrowed in the previous clock cycle (\textit{DSTB} will set a timing-error flag). Once the skip path is activated, the iteration counter is counted up by 2 (instead of 4), to provide correct right-shift commands for future iterations. The 9 MSBs of the \textit{CORDIC cell 4} were substituted by \textit{DSTB} to reduce substitution overhead. As the remaining 9 LSBs will fail under aggressive conditions, MSFF of these LSBs were replaced by latches to enable time borrowing. In summary, all outputs of \textit{CORDIC cell 4} allow time borrowing and only the 9 MSBs are responsible for timing-error detection. 

When a setup timing violation (for the previous cycle) is detected at the beginning of the current cycle, the circuit skips the last two CORDIC iterations (cell 3 and cell 4), due to insufficient processing time. Therefore, the iteration counter is counted up by two instead of by four. As a consequence, the intended four CORDIC iteration computation will eventually conduct two iterations. These skipped computations will be performed in later stages, as the iteration counter is only counted up by two. As a result, only the final computations are skipped because of the constraint of fixing the CPI to 4.

\begin{figure}[H]
\centering
\includegraphics[width=0.8\linewidth]{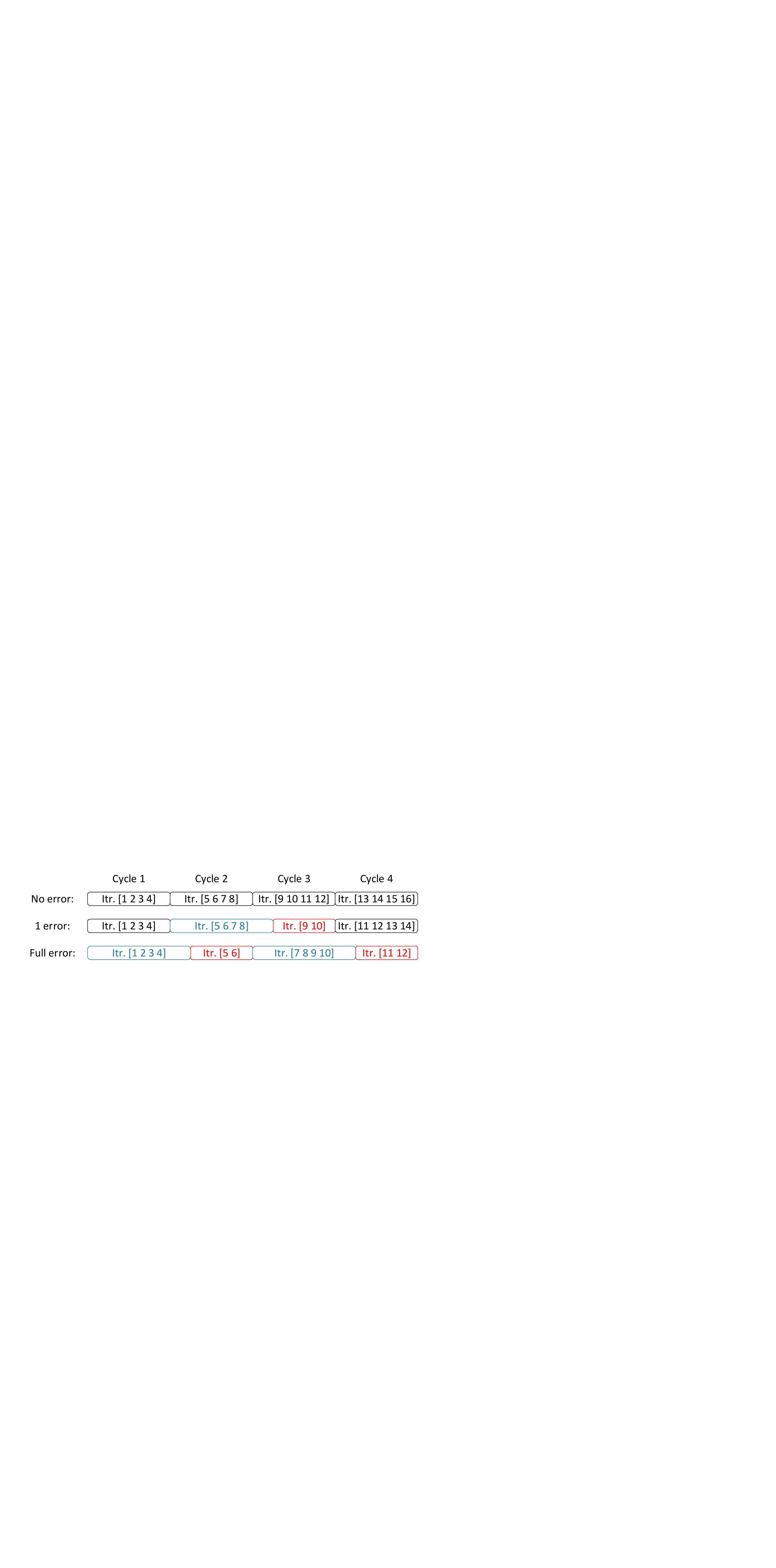}
\caption[Example timing diagrams of the proposed \textit{computation-skip} CORDIC.]{Example timing diagrams of the proposed \textit{computation-skip} CORDIC. It shows 3 cases: 1) no error detected; 2) error detected in one cycle (cycle 2); 3) every cycle triggers the $E_{flag}$. Timing-errors leads to reduced iterations. }\label{fig:algo_timing_cordic}
\end{figure}

Fig.\ref{fig:algo_timing_cordic} shows the timing diagram of the proposed CORDIC processor. When no error is detected, a complete CORDIC operation takes four cycles, computing 16 iterations. When one error is detected for cycle 2, the DSTB latches the late-arriving data and triggers the \textit{computation-skip} path in cycle 3. After four cycles, only 14 iterations instead of 16 are performed. This result serves as a reduced-quality output to maintain the throughput. When the timing situation is very severe that all cycles fail, the performed iteration number is 12. This is because the \textit{computation-skip} path (red in Fig.\ref{fig:algo_timing_cordic}) are constrained to never fail.

The timing constraints, as specified in Table~\ref{tab:algo_timing}, were applied during synthesis and layout. The following explains these constraints in the CORDIC context.
\begin{itemize}
\item[$\bullet$] The max time borrow constraint was set to 0 for the latches in \textit{DSTB}. This ensures no time borrowing under the nominal condition.

\item[$\bullet$] Max delay for paths through \textit{CORDIC cell 3 and 4} ($Q$ $\rightarrow$ \textit{cell\textunderscore 3} $\rightarrow$ $D$), which is also called error mitigation setup in Table~\ref{tab:algo_timing}, is  $t_{clk} \cdot (1 + \tau)  - t_{setup\_FF}$. The max delay is relaxed as they are designed for a better-than-worst-case situation, with the unlikely worst situations protected by time borrowing and the computation-skip. This enables area and power saving by gate-downsizing. The clock duty-cycle $\tau$  is set to 25\%, which results in 34\% area saving.

\item[$\bullet$] Max delay for \textit{computation-skip} paths ($Q$ $\rightarrow$ skip path $\rightarrow$ $D$) is $t_{clk} \cdot (1 - \tau)  - t_{setup\_FF}$. This constraint ensures computation skip during timing-borrowing.

\item[$\bullet$] Max delay for paths started from the \textit{MSFF} in \textit{DSTB} ($Q_{ff} \rightarrow D$) is $t_{clk} - t_{meta\_window} - t_{setup\_FF}$.  As the \textit{MSFF} in \textit{DSTB} might fail under aggressive operating conditions, $t_{meta\_window}$ = 700~ps is reserved for meta-stability resolution. To ensure stability resolution, ULVT MSFF are used instead of SVT ones, as they have higher loop gains and hence lower resolution time constant (20~ps).

\item[$\bullet$] Minimum delay for paths to \textit{DSTB} ($Q$ $\rightarrow$ $D$) is $\tau \cdot t_{clk} + t_{hold\_latch}$. This is achieved by automatic delay cell inserting during routing. This accounts for 8\% area overhead.
\end{itemize}

\subsection{Pre-silicon analysis}
To make a fair comparison between \textit{core1 canary} and \textit{core2 CS}, they are designed to fail (produce errors) at the same frequency and voltage condition. This is achieved by synthesizing the circuit delay of the CS core (Cons.1 in Fig.~\ref{fig:algo_timing}) the same as the clock delay of the Canary core.

\begin{figure}[H]
\centering
\includegraphics[width=0.85\linewidth]{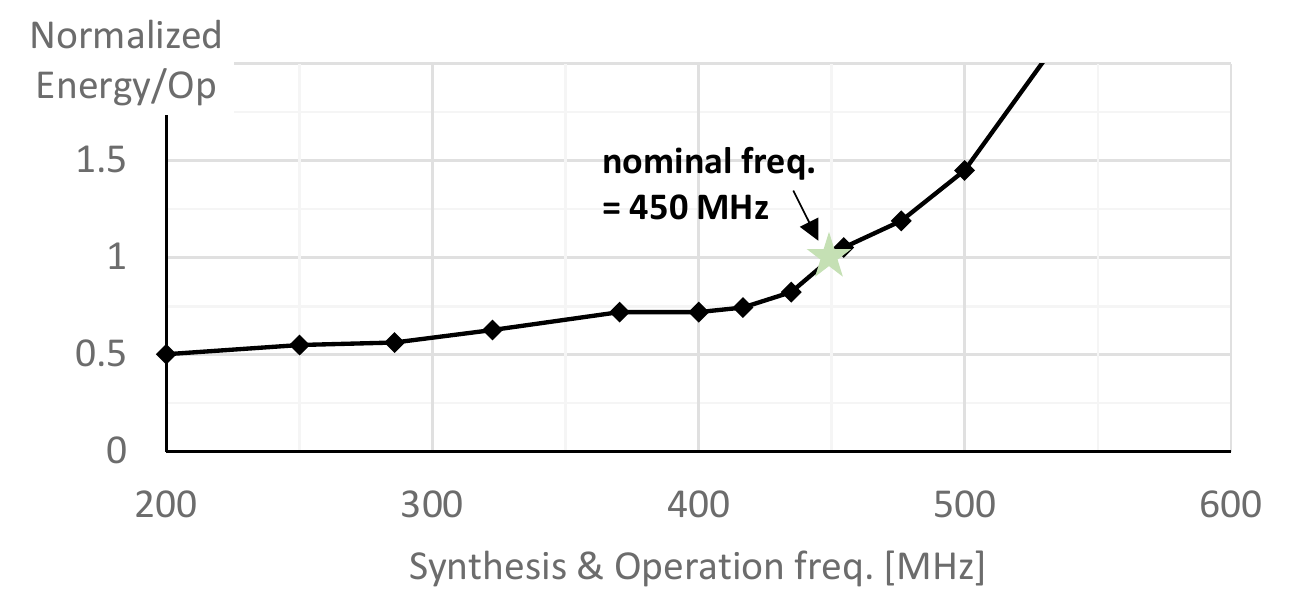}
\caption[450~MHz is selected for the operating condition.]{450~MHz is selected for the operating condition. The figure shows the normalized energy consumption per CORDIC operation for the conventional CORDIC. The energy is measured with simulated gate-level toggling information.}
\label{fig:algo_region}
\end{figure}

Fig.~\ref{fig:algo_region} shows the synthesizing frequency and energy consumption relation for the conventional CORDIC (without the modifications into $Canary$ nor $CS$). Without many surprises, higher clock frequency leads to automatically gate-upsizing (by the synthesizing EDA tools), and hence the power consumption is increased. 

This chapter chooses 450~MHz as the nominal operation frequency for both \textit{core1 canary} and \textit{core2 CS}, as shown in Fig.~\ref{fig:algo_region}. The reasons are i) the speed meets the high-speed requirement, and ii) the gate-upsizing is moderate, and hence the energy consumption is not increased dramatically, compared with over-relaxed situations (e.g., 200~MHz).

\begin{figure}[H]
\centering
\includegraphics[width=1\linewidth]{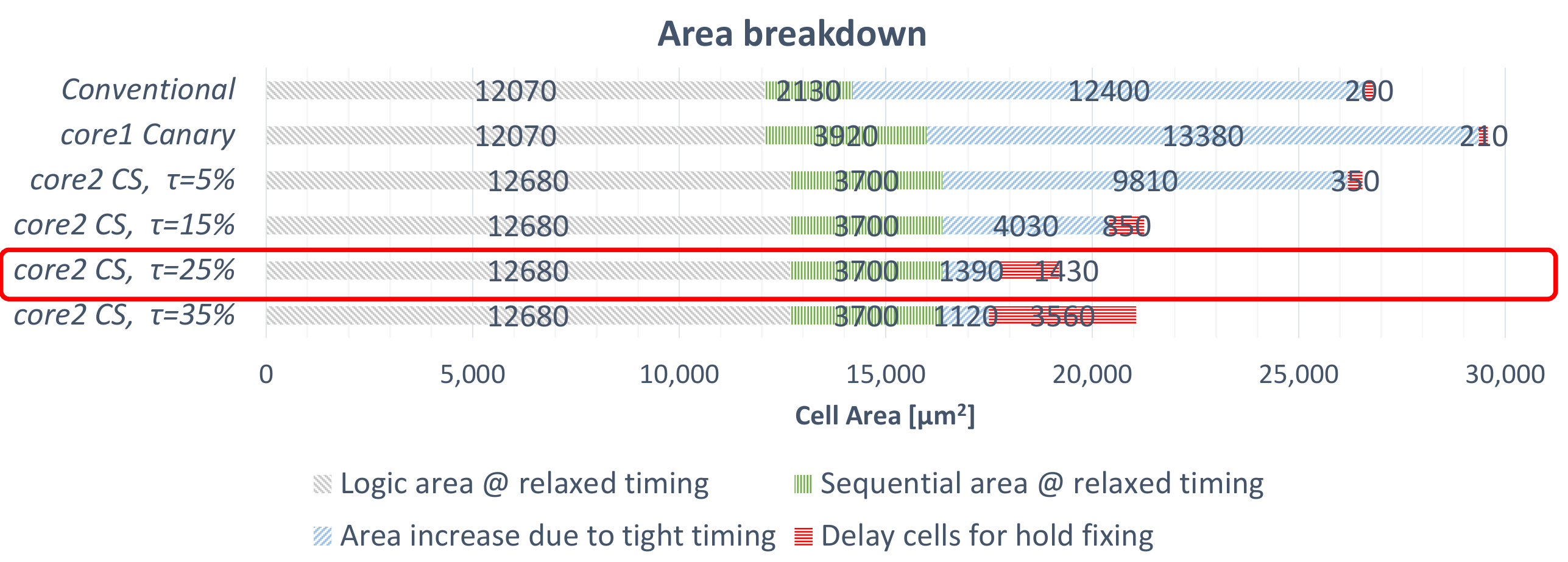}
\caption[Cell area breakdown of the conventional and the proposed hardware CORDIC accelerators designed at 450~MHz.]{Cell area breakdown of the conventional and the proposed hardware CORDIC accelerators designed at 450~MHz. The \textit{core2 CS} is designed with different duty cycle ($\tau$). The area increase due to tight timing is computed agaist the cell area under 1/10 clock frequency (45~MHz). \textit{cs} reduces area by loosen the timing constraint.}
\label{fig:algo_area}
\end{figure}

Both designs are synthesized at the frequency of 450~MHz. After synthesize, placement and routing, the cell areas of the original CORDIC and its variants, \textit{core1 canary} and \textit{core2 CS}, are shown in Fig.\ref{fig:algo_area}. The conventional CORDIC suffers from tight timing constraint to meet the worst-case corner, making the area very large. The \textit{core1 canary} CORDIC utilizes more area. This is because its sequential logic area is almost doubled by replacing normal MSFF with \textit{canary FF}.

For \textit{core2 CS}, when $\tau$ is large, the area increase because of timing constraint (light blue bar in Fig.~\ref{fig:algo_area}) is very small since the timing constraint is relaxed. This also exhibits a wider range of $V_{dd}$ drop monitoring. However, higher $\tau$ calls for more delay cells to fix the short path issue, which leads to more area cost. For a balance between area (and hence energy) and $V_{dd}$ drop monitoring, $\tau$=25\% is chosen in this chapter. $\tau$=25\% also makes the clock generation easy because it can be accomplished with a 4x frequency divider. According to Fig.\ref{fig:algo_area}, the delay cells account for 8\% of the total area when $\tau$=25\%.

\subsection{Post-silicon Measurement} 
\begin{figure}[H]
\centering
\includegraphics[width=0.7\linewidth]{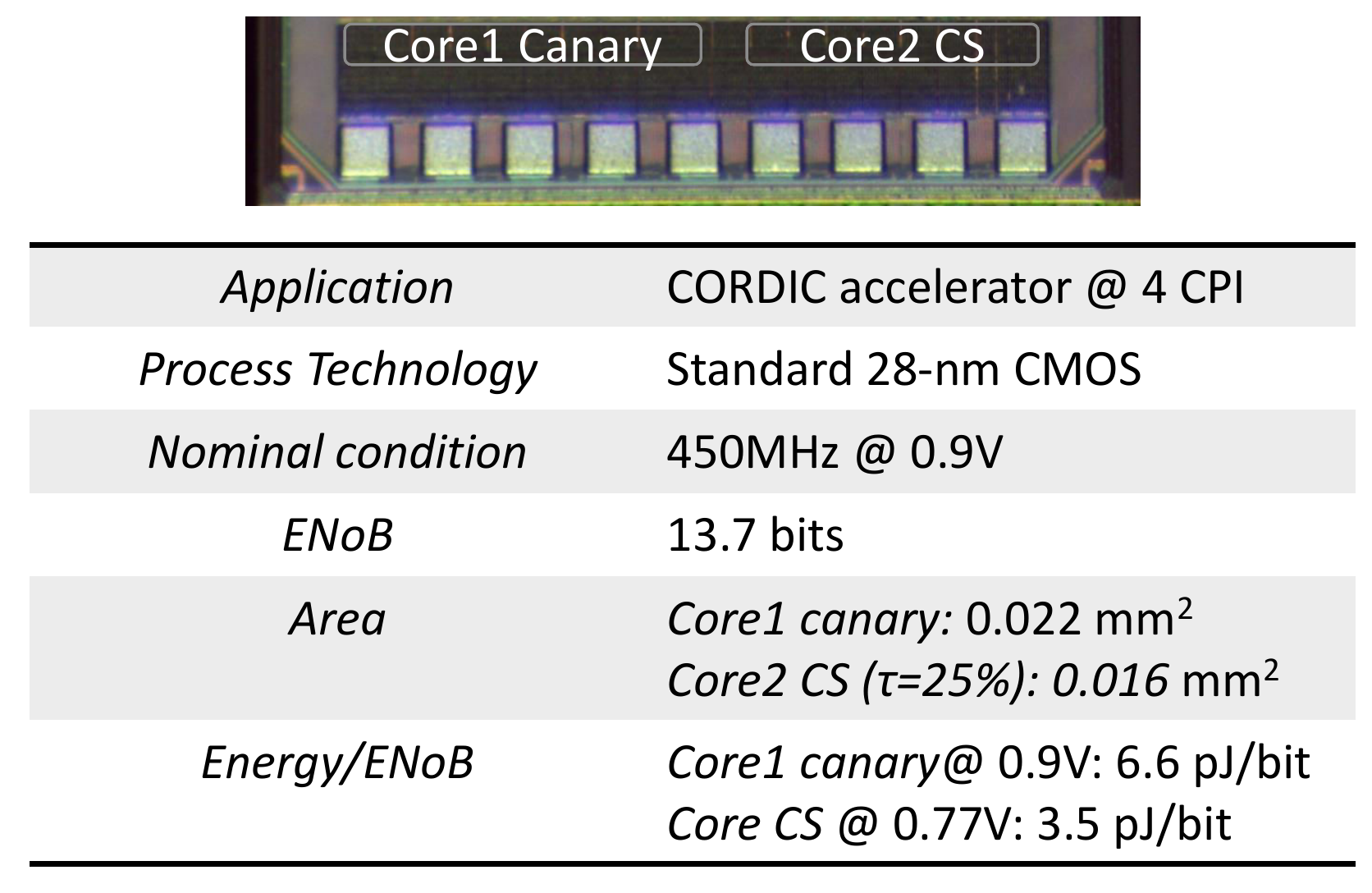}
\caption{Die photo and chip information.}\label{fig:algo_die}
\end{figure}

The CORDIC accelerators were processed in a standard 28nm CMOS technology. A micro-graph of the  chip is shown in Fig.~\ref{fig:algo_die}. The complete chip measures 1 * 0.2 $mm^2$. The area for the \textit{core1 canary} is 0.022 $mm^2$. The area for \textit{core2 CS} is 0.016 $mm^2$. The area reduction comparing to \textit{core1} is due to the relaxed timing constraints.

To measure the output SNR and ENOB, random stimuli are generated in a desktop computer and send to an FPGA controller. The FPGA then writes/reads the data in the on-chip memory serially, before/after a CORDIC execution. Matlab and Python scripts are used for ENOB computation. For energy consumption measurement, testing vectors are stored in the on-chip memory. So the chip is filled with data and will run continuously. This makes the energy consumption results realistic.

\begin{figure}[H]
\centering
\includegraphics[width=1\linewidth]{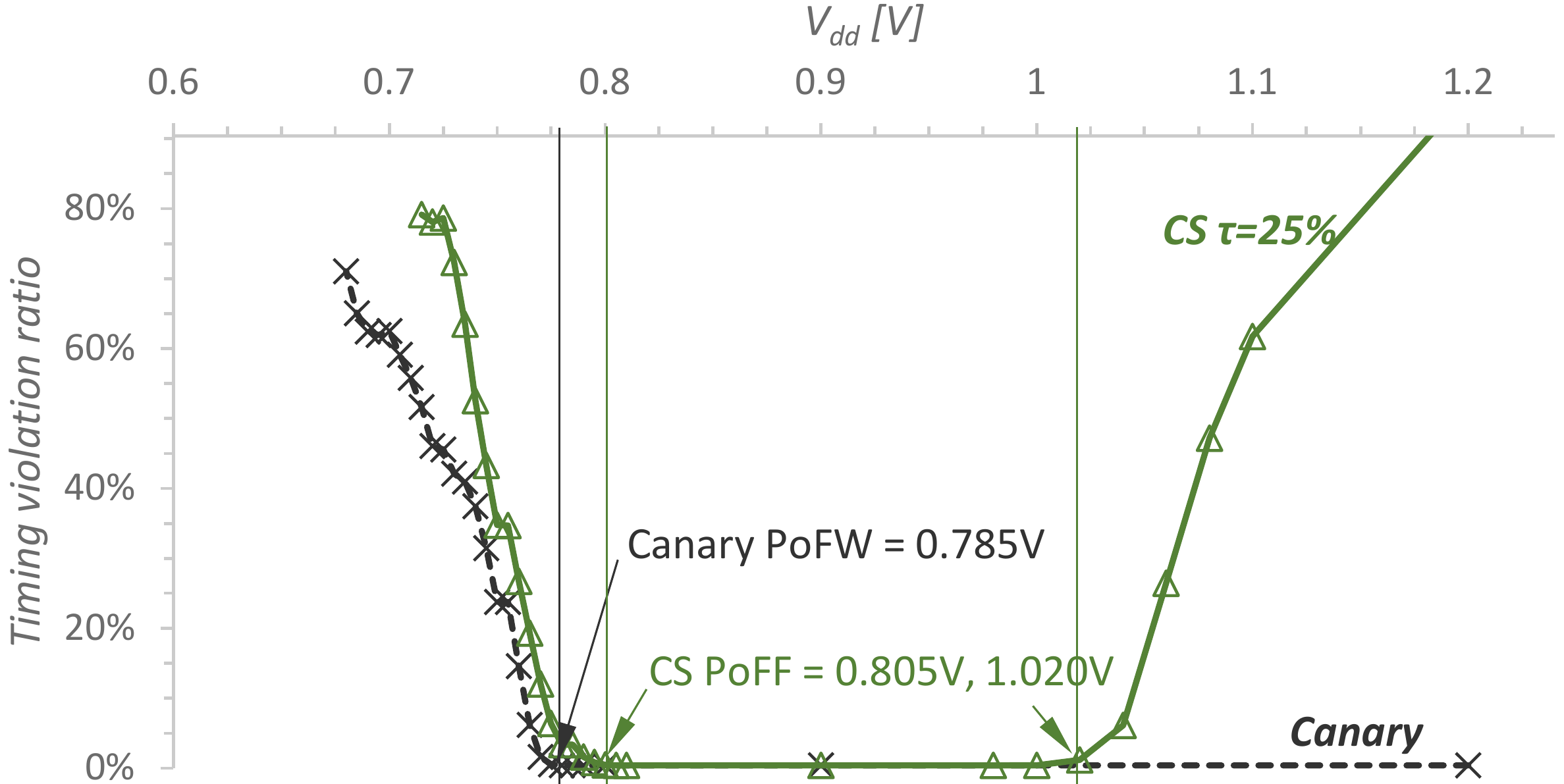}
\caption[Measured timing violation ratio.]{Measured timing violation ratio. A violation is observed if any path violates the timing. For \textit{canary}, the first timing violation condition is PoFW. For the proposed CS (\textit{computation-skip}), the first timing violation condition is PoFF.}\label{fig:algo_violation}
\end{figure}

The test chip was measured at 450 MHz. Fig.~\ref{fig:algo_violation} shows the timing violations for both cores. The violations are regards as warnings for \textit{core1 canary}. For \textit{core2 CS}, they are actual errors. The \textit{core1 canary} produces a warning when $V_{dd}$ is lower or equal to 0.785V (PoFW). For \textit{core2 CS}, when $V_{dd}$ is lower than 0.805V (PoFF) or higher than 1.020V (PoFF), timing violations are asserted. The prior case is because of timing failure in the error mitigation path. The later case is because of the short path failure.  

\begin{figure}[H]
\centering
\includegraphics[width=1\linewidth]{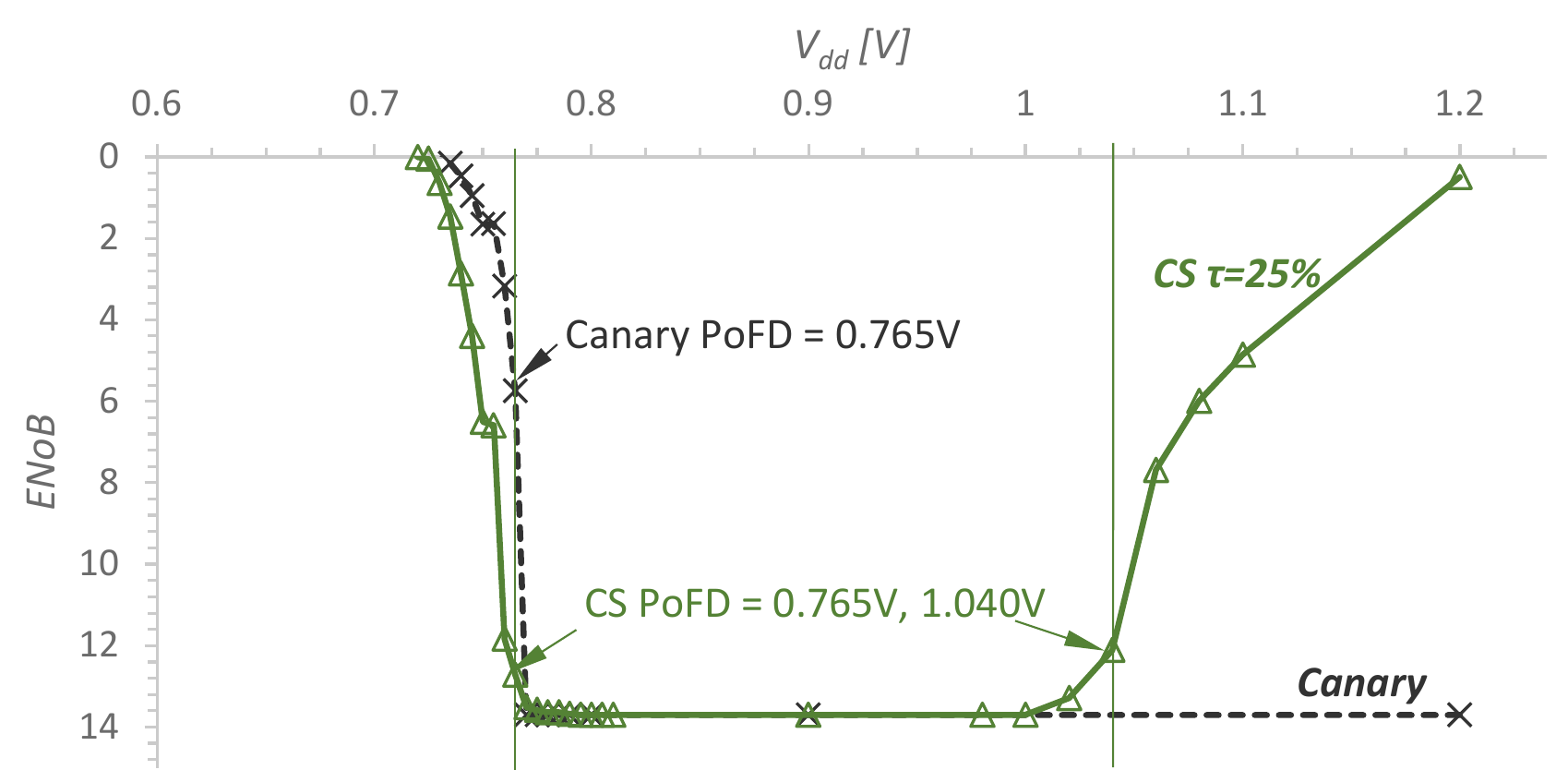}
\caption[Measured CORDIC ENOB.]{Measured CORDIC ENOB. For \textit{canary}, the PoFD coincides with the PoFF. For the proposed CS (\textit{computation-skip}), the PoFF does not result to PoFD, as the timing-errors are mitigated by the \textit{computation-skip} scheme.}\label{fig:algo_enob}
\end{figure}

To quantify the error degradation, the notion of PoFD (Point of the First noticeable Degradation) is introduced. It represents the critical situation when noticeable degradations are observed at outputs. Note that PoFD does not always coincide with the PoFF, as errors might be mitigated gracefully. Fig.~\ref{fig:algo_enob} shows the ENOB as a measure of the output SNR. The ENOB is unaltered until the 0.785V (PoFW), as no actual timing-error is introduced. The PoFD for \textit{core1 canary} is 0.765V. For \textit{core2 CS}, when $V_{dd}$ is lower than 0.805V (PoFF), \textit{computation-skip} paths are activated and thus the ENOB slightly drops. The PoFD for \textit{core2 CS} is marked on 0.765V. The PoFD for both cores are equal. This is because they are designed with the same frequency constraint, and are experiencing similar PVT conditions. Beyond the PoFD, the \textit{computation-skip} paths and control paths both fails and hence the error tolerance capability becomes invalid. If the $V_{dd}$ is higher than 1.040V, the ENOB decreases, due to minimum delay violations for DSTB. In summary, the ENOB performance is comparable for \textit{core1} and \textit{core2}, except for the high $V_{dd}$ situation.

\begin{figure}[H]
\centering
\includegraphics[width=1\linewidth]{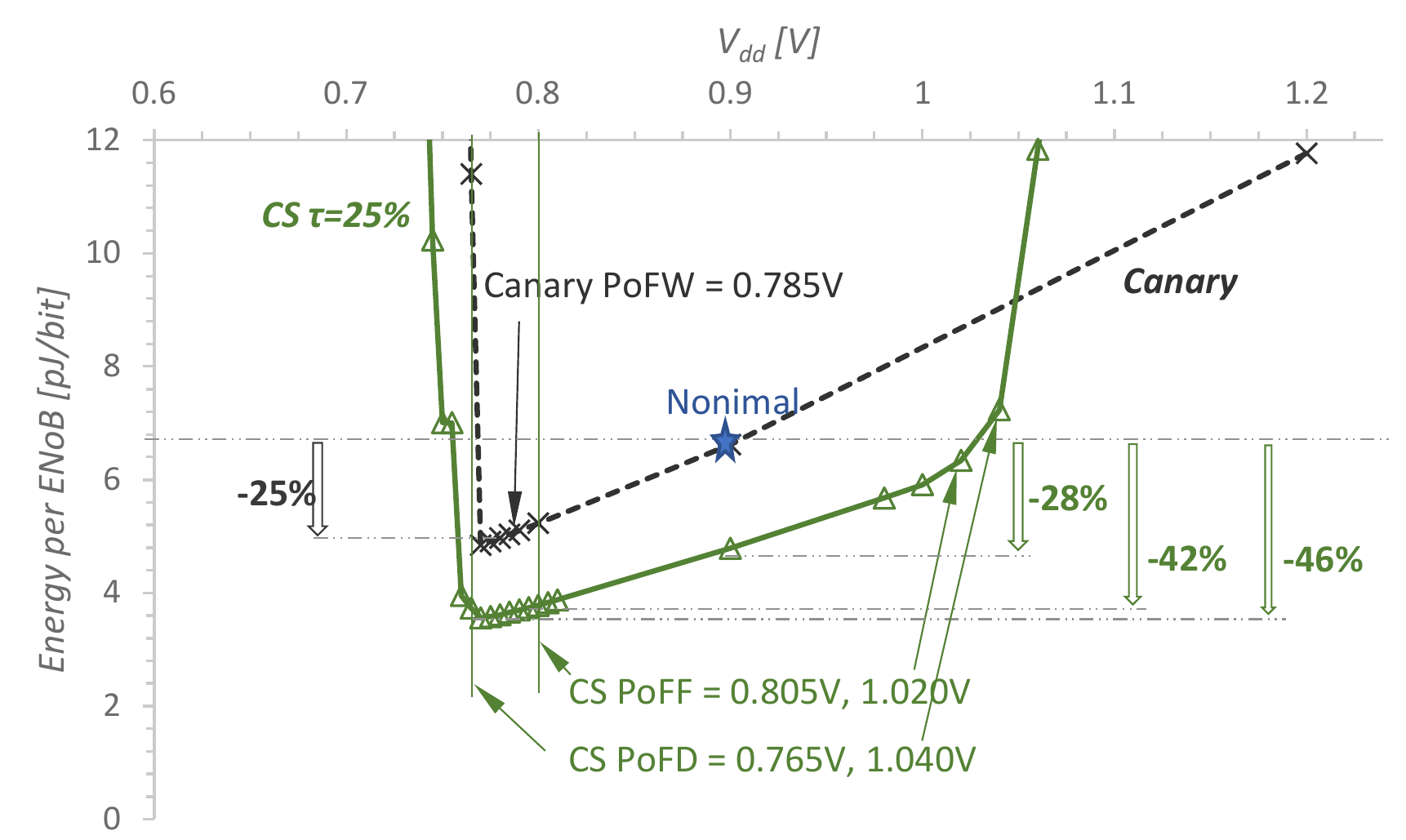}
\caption[Comparison of the \textit{core1 canary} (implementing the baseline state-of-the-art technique) and \textit{core2 CS} (implementing the proposed scheme) CORDIC accelerators during voltage scaling.]{Comparison of the \textit{core1 canary} (implementing the baseline state-of-the-art technique) and \textit{core2 CS} (implementing the proposed scheme) CORDIC accelerators during voltage scaling. \textit{core2} saves power by timing constraint relaxation, error-free adaptive scaling, and error-resilient VOS.}\label{fig:algo_energy}
\end{figure}

The energy consumption per CORDIC operation per effective bit is shown in Fig.~\ref{fig:algo_energy}. Under the nominal condition (0.9V), the \textit{core2 CS} reduces energy consumption per ENOB by 28\%, compared to the \textit{core1 canary} (6.6 pJ/bit), which is attributed to relaxed timing constraint. For both cores, reducing the $V_{dd}$ decreases the energy/ENOB. Going beyond the PoFDs leads to drastic energy/ENOB increase. The \textit{core1 canary} saves 25\% energy at the PoFW (0.785V), comparing with the 0.9V nominal case. This is error-free power saving that shaves the design margin. For the \textit{core2 CS}, the energy reduction is measured at 42\% at 0.805V (PoFF), where the ENOB is imperceptibly reduced. The energy/ENOB for \textit{core2 CS} keeps reducing when going beyond PoFF ($V_{dd}$ $\leq$ 0.805V), despite the fact that the ENOB is slightly reduced due to computation skips. The minimum energy/ENOB is 3.5~pJ/bit at 0.770V, which is 46\% lower than the nominal \textit{core1 canary}. The corresponding ENOB is reduced from 13.7 bits to 13.5 bits.  

\section{Conclusion}\label{sec:algo_con}
The CMOS variability margin leaves great room for energy savings. The timing-errors incurred by supply voltage over-scaling is most easily detected at the circuit-level. By providing this information to the algorithm-level, energy saving is achieved with gracefully output quality degradation.

In this chapter, error resilient techniques to set $V_{dd}$ for variability margin saving, i.e., Canary, Razor, TIMBER, ANT, and computation-skip, are analyzed. In particular, the \textit{computation-skip} scheme is discussed to mitigate timing-errors introduced by VOS. This scheme suits for evolutionary algorithms. It relaxes the timing constraint for a conventional circuit and hence saves power. The error flag can provide statistical timing-error rate, which indicates the output quality as well as the distance to the $V_{dd}$ scaling limit.

The canary and the \textit{computation-skip} schemes are applied to a recursive CORDIC processor. Effectively, the last CORDIC iterations are skipped once timing-errors are detected in this proposed \textit{computation-skip} CORDIC. A 28\% energy consumption per bit saving due to relaxed timing constraint (design margin shaving at design time) is observed. The energy/ENOB saving improves to 42\% because of adaptive scaling (error-free design margin shaving at run-time). Moreover, a total of 46\% saving is possible, with a 0.2-bit precision loss.



\cleardoublepage


\chapter{Application-level error-resilience on massive MIMO base stations}\label{sec:system}


This chapter investigates application-level error absorption and handling. The considered errors are generated at the circuit-level and the algorithm-level errors by hardware operating at risky conditions. This chapter focuses on a massive MIMO communication system case-study. It shows that the perceived performance will hardly be affected by sparse processing failures, while the power consumption can be considerably reduced as error resilient hardware are utilized. Furthermore, this chapter assesses antenna outage impacts and proposes damage control strategies. The work in this chapter is published in \cite{huang17icassp-mimo-dfe}.

The rest of this chapter is organized as follows: Section~\ref{sec:system_intro} introduces the opportunities brought by the massive MIMO system, as well as the cross-level optimization demands.  Section~\ref{sec:system_contri} highlights the contribution. The system description from a functional processing point of view is presented in Section~\ref{sec:system_system}. Section~\ref{sec:system_model} models the algorithm-level effects of circuit-level errors, or more specifically the VOS errors and the antenna outage error. The application-level impacts on the massive MIMO system are evaluated in Section~\ref{sec:system_exam}. Section~\ref{sec:system_improve} proposes an approach to enhance the massive MIMO system performance under errors. Finally, the major conclusions of this chapter are summarized and relevant directions for future elaboration of the proposed concept are outlined in Section~\ref{sec:system_con}.

\section{Opportunities and challenges of massive MIMO}\label{sec:system_intro}
\begin{figure}[H]
\centering
\includegraphics[width=0.8\linewidth]{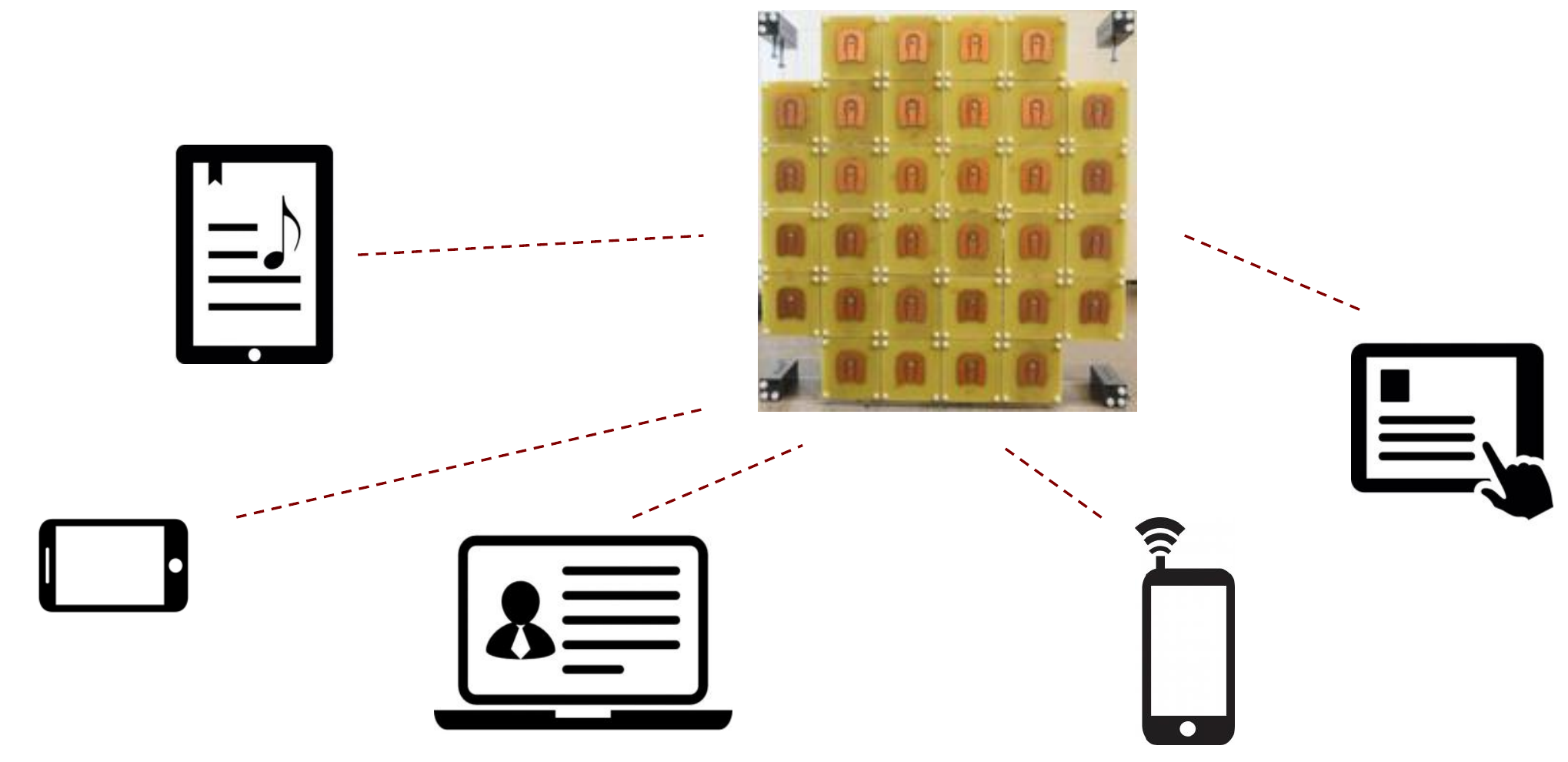}
\caption[Massive MIMO exploits large antenna arrays to spatially multiplex many terminals.]{Massive MIMO exploits large antenna arrays to spatially multiplex many terminals \cite{larsson2017massive}.}
\label{fig:system_mimo}
\end{figure}

Massive MIMO is the currently most compelling sub-6 GHz physical-layer technology for future wireless access. The main concept is to use large antenna arrays at base stations to simultaneously serve many autonomous terminals, as illustrated in Fig.~\ref{fig:system_mimo}.
  
Since its inception about a decade ago, the massive MIMO concept has evolved from a wild ``academic" idea to one of the hottest research topics in the wireless communications community, as well as the main work item in 5G standardization. The time for massive MIMO has come at this moment for two reasons: First, conventional technology has proven unable to deliver the spectral efficiencies that 5G applications are calling for. Second, the confidence in the exceptional value of the technology has spread rapidly since impressive real-life prototypes showed record spectral efficiencies, and the robust operation with low-complexity RF and baseband circuits has been substantiated \cite{prabhu20173}.

\subsection{Massive MIMO, the highly-demanded future technology}
Massive MIMO opens up a new dimension of wireless communications by using an excess of base station (BS) antennas, compared to the number of active terminals. This technique allows for very efficient spatial multiplexing, attainable using linear processing in a time-division duplex mode \cite{larsson2014massive}.

Conceptually, massive MIMO achieve a 10x or more increase in system capacity. What is even more important is the gain in reliability due to flattening out of deep fades, hardening of the channel, and array gain. This especially benefits the cell edge users and could be essential for low power terminals as in Machine Type Communications (MTC).

This stunning improvement of massive MIMO results from the fact that much less transmitted power is needed thanks to the array gain. It also benefits from the utilization of low complexity hardware \cite{gunnarsson17eucnc-loussy-processing}, as the individual antenna signals do not need to be of high precision \cite{7194033,gustavsson2014impact,bjornson2014massive}.

\subsection{Power consumption, a design challenge in massive MIMO digital processing}
However, an obvious concern is how a large number of antennas (and associated transceivers and signal processing) will affect the complexity and energy consumption of the BS. \cite{7114430} anticipates that the overall complexity and energy consumption in terms of J/bit can be lowered by a factor of 10 to 100 compared with current BS.

The energy consumption issue in digital systems is alleviated by the CMOS scaling. The scaling has brought steady power reduction for many generations, thanks to the decrease of the supply voltage that shows up as a quadratic factor in the dynamic power formula. However, Integrated circuits are facing ever increasing variability challenges in recent technology nodes (65nm and smaller). The process, voltage, and temperature (PVT) variability are considered as the three main contributors to circuit variability. Conventionally, to cope with this variability challenge, ICs are designed at the worst PVT corners, to ensure that they always operate correctly.

However, this approach brings considerable margins, leading to reduced peak performance and wasted power consumption. For instance, \cite{huang16jsps-error-resilient} shows for 28nm technology, the performance difference (in terms of speed) is as large as 2.2x between the typical case and the worst-case. To reduce the margins, dynamical scaling techniques manage power dissipation and temperature using variable $V_{dd}$. The most adventurous methods are the error resilient techniques. They scale down the $V_{dd}$ more aggressively (VOS) while accepting that errors might occur on individual chips. These methods have been proven to enable significant energy savings while maintaining excellent performance for wireless communication \cite{huang16jsps-error-resilient,hegde2004voltage}.

In this context, considerable energy reduction potential is expected for massive MIMO if its low accuracy need is extended to the CMOS implementation of digital signal processing -- to the point of sporadically processing distortion or even full failure of one or a few individual antenna signals. It, therefore, opens the door to much narrower design margins (comparing with the traditional semiconductor specification set at design-time). The circuits can operate at the lower supply voltage (and hence power).

\section{Contribution of this chapter}\label{sec:system_contri}
This work combine the results form error resilient hardware with the inherent antenna redundancy in massive MIMO. It focuses on the TDD option of 3GPP LTE in a massive MIMO context. \cite{bjornson2014massive} demonstrated the resilience to analog non-ideal hardware (e.g. nonlinearity). 

To embrace unreliable hardware, this work proposes to consider the digital computation as a faulty process. It demonstrates that if a limited number of circuit-level and algorithm-level computational errors can be tolerated at the application-level, the safety margins can be reduced significantly. This will bring considerable power saving with minor performance degradation. 

Apart from demonstrating the inherent error resilience, this work also proposed methods to detect extensive errors and adjust the massive MIMO application to prevent further quality loss.

\section{Masive MIMO system introduction}\label{sec:system_system}
In a massive MIMO system, the BS is equipped with $M$ antennas and serves $K$ single-stream users simultaneously, each equipped with a single antenna. Unless otherwise specified, $M$ = 100 and $K$ = 10 is set as typical values in this chapter. Fig.~\ref{fig:system_system} illustrates the BS architecture of a massive MIMO system. 

The BS consists of central digital modem functionality, the per-antenna processing including (I)FFT operations for OFDM (de)modulation, digital front-end (DFE), analog front-end (AFE) and power amplifier (PA). The signal processing complexity and power consumption of the inner-modem digital processing scale linearly with $K$, while the (I)FFT, the DFE, and the AFE complexity all scale linearly with $M$, and the pre-coder scales with $M*K$.

The massive MIMO digital processing complexity~\cite{7114430} is summarized by billion complex floating-point arithmetic operations per second in Table~\ref{tab:system_compare}. The data transfer overhead is included. For a typical massive MIMO system, the digital processing effort is dominated by the per-antenna functionality (mainly FFT and DFE filtering operations). This is due to the linear dependence of system power consumption to the massive BS antenna number $M$.

\afterpage{
\begin{sidewaysfigure}[h]
\centering
\includegraphics[width=1\textwidth]{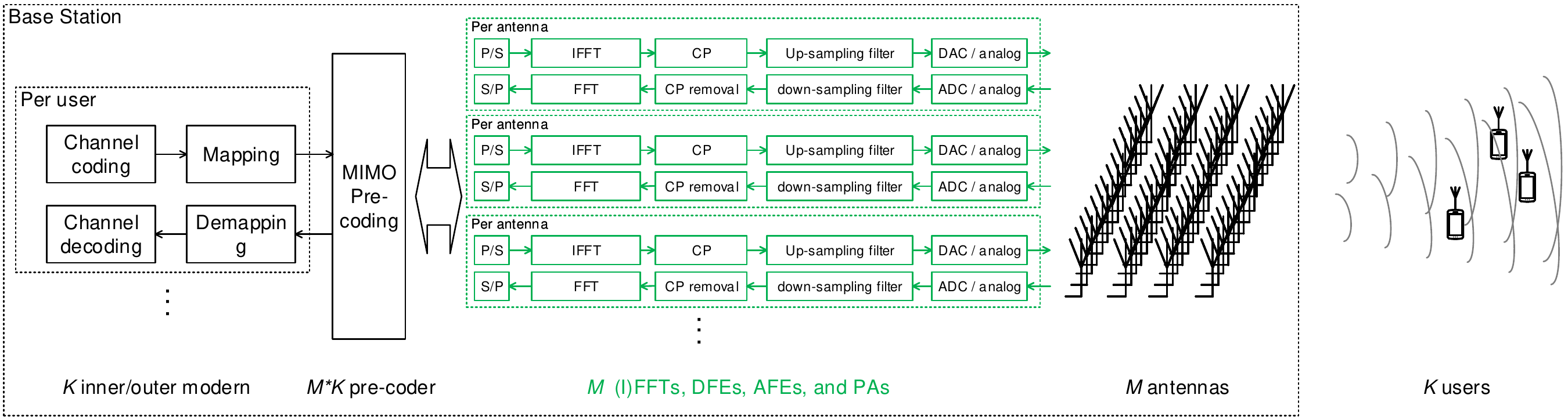}
\caption[A massive MIMO base station (BS) equips with $M$ antennas and serves $K$ users.]{A massive MIMO BS equips with $M$ antennas and serves $K$ users. Typically, $M$ = 100 and $K$ = 10.}\label{fig:system_system}
\end{sidewaysfigure}
\clearpage
}

The digital modem is split in the outer modem (processing information bits through channel coding/decoding) and the inner modem (in multi-carrier systems performing frequency-domain operations such as channel estimation and massive MIMO precoding). To determine the signal processing demands for a typical BS, the complexity of the digital components of massive MIMO \cite{7114430} is summarized in Table~\ref{tab:system_compare}. These complexity values estimate the number of billion complex floating-point arithmetic operations performed per second for each specific digital signal processing operation. They have been multiplied by an overhead factor (see \cite{7114430}) to take data transfers (in memories and registers) into account. Because they take a big portion of the power consumption of digital systems. 

\begin{table}[H]
\caption[Complexity of digital components for a 100x10 massive MIMO system.]{Complexity of digital components for a 100x10 massive MIMO system in each phase, with 20~MHz bandwidth, 3~bps/Hz (16-QAM, 3/4 coding rate, training not included)}\label{tab:system_compare}
\begin{tabular}{  c  c  c  c }
\toprule
\textbf{\textit{Subcomponent}}  & \textbf{\textit{Downlink (DL)}} & \textbf{\textit{Uplink (UL)}}  & \textbf{\textit{Training}} \\
 & \textbf{\textit{[GOPS]}}  & \textbf{\textit{[GOPS]}} & \textbf{\textit{[GOPS]}}     \\ \cmidrule(l){1-1} \cmidrule(l){2-4}
Inner modern &    175    & 520 &    290 \\
Outer Modern &    7    & 40  &    0     \\
DFE incl. (I)FFT &  920    & 920 &    920 \\\bottomrule
\end{tabular}
\end{table}

The workload of the massive MIMO is partitioned into downlink (DL), uplink (UL), and training phases. The training phase uses the UL signals to perform channel estimation. Therefore, its digital processing components are similar to the UL phase. For a typical massive MIMO system, the digital processing effort (see Table~\ref{tab:system_compare}) is dominated by the per-antenna functionality (mainly FFT and DFE filtering operations)~\cite{7114430}. Note that the downlink data traffic is usually larger (5 to 20 times more traffic than UL). In terms of overall complexity, a BS spends more effort on the UL phase.

To minimize the area cost and the energy budget of the BS, this chapter focuses on demonstrating the possibilities of accepting intermittent digital hardware errors in the DFE (incl. FFT). The hardware error effects are considered within the context of DL massive MIMO. The reasons are i) The DL data size is usually larger, which makes the DL the dominant power consumer. ii) The DFE for UL has a lot of similarities to the DL, and thus exhibits similar effects on errors.

\section{Digital hardware error impacts on signal quality}\label{sec:system_model}
This section illustrates the most common sources of errors in digital signal processing. The impact of these circuit errors on the (I)FFT and other DFE hardware are then modeled for the later application-level assessment. Accordingly, the unreliability of the circuits is becoming a non-negligible issue~\cite{rabeay}.

For the massive MIMO system, the digital hardware errors in (I)FFT \& DFE introduced by silicon unreliability and by adventurous design methodologies result in incorrect bit results during signal processing. This can be regarded as digital distortion noise. This work introduces a new metric to signal to digital distortion ratio (SDDR) to describe the quality of signal:
\begin{equation}
SDDR = 10 \cdot log \frac{{\sigma_s}^2}{ {\sigma_d}^2 },
\end{equation} 
where ${\sigma_s}^2$ and ${\sigma_d}^2$ are the powers of error-free DFE output, and the noise power of digital distortion due to circuit unreliability, respectively. The digital distortion noise results from circuit-level errors. Based on its mechanism, it is categorized into two class, i.e., VOS (temporary and local errors), and antenna outage (hard and full antenna errors).

The error-free output of a DFE, $y_s$, is contaminated by the VOS distortion $n_d$. Therefore, the final contaminated DFE output 
\begin{equation}
\tilde y=
    \begin{cases}
      y_s, & \text{if}\ error-free \\
      y_s + n_d, & \text{if}\ VOS~errors \\
      0~(fixed~value), & \text{if}\ antenna~outage
    \end{cases}
\end{equation}
The VOS errors $n_d$ of is modeled with a zero-mean Gaussian distribution \cite{liu2010computation} with ${\sigma_d}^2$ as the error power. The effect of VOS can thus be molded with SDDR. For the scenario of antenna outage, the DFE output stuck at a fixed value (1 or 0). The information from the input signal $y_s$ is completely lost. The rest of this section discuss the VOS and antenna outage effects at the algorithm-level.

Regardless of the DFE contamination class, the received signals at the MIMO receiver ($y$) are
\begin{equation}
\begin{bmatrix}
    y_1 \\
    \vdots \\
    y_K          
\end{bmatrix}
= 
\begin{bmatrix}
    h_{1,1}  & \cdots & h_{1,M} \\
    \vdots  & \ddots & \vdots \\
    h_{K,1}  & \cdots & h_{K,M}      
\end{bmatrix}
\begin{bmatrix}
    \tilde y_1 \\
    \vdots \\
    \tilde y_M          
\end{bmatrix}
+
\begin{bmatrix}
    n_1 \\
    \vdots \\
    n_N          
\end{bmatrix}
\end{equation}
\begin{equation}
y = H \tilde y + n,
\end{equation}
where $H$ denotes MIMO channel matrix. The vector of received symbols $y$ is distorted by the noise vector $n$. Note that the hardware distortion $n_d$ and $n$ are different in nature, as $n_d$ impacts individual transmitter antenna, $n$ suffers from the channel and receiver.

\subsection{VOS (Voltage-OverScaling) impacts}\label{sec:system_vos}
Section~\ref{sec:variation_voltage} reviewed the techniques to scale down the voltage to exploit design margins and save power consumption.  In this chapter, their impacts on SDDR is re-evaluated, in the context of the Massive MIMO digital front-end. 

The power saving consists of two part: error-free power saving and error-resilient power saving. The error-free part can be achieved usually by the Razor techniques. For instance, \cite{bull2011power} reports 30\% and 52\% power consumption saving on a typical die and a fast die, respectively; \cite{fojtik2012bubble} achieves 54\% saving on a typical die and 60\% saving on a fast one. The infrequent errors timing-errors are fully resolved by the micro-architectural level error correction schemes, and produce no errors to the signal output.

The error-resilient techniques, as reviewed in Section~\ref{sec:variation_ant}, reduces the power consumption of digital signal processors by gracefully sacrificing the SDDR, admitting that a certain amount of errors might occur. For example, Chapter~\ref{sec:algo} saves 45\% power consumption for CORDIC applications, at the cost of 1 ENoB degradation. However, if the supply voltage is further reduced for more aggressive power savings, the SDDR is reduced dramatically. e.g. lower than 0~dB, because a lot of setup paths are failed and the circuit cannot operate correctly.

In summary, state-of-the-art algorithm-level error-resilient techniques save around 40\% power, at the cost of potential sparse antenna processing distortion. The SDDR depends on the operating $V_{dd}$, the process variability and the environment temperature. This means that even with the same design, different (I)FFT \& DFEs of the massive MIMO might exhibit vastly different SDDR behavior. When the circuit is mainly subject to random SEU errors, designs can choose to either carry on using the erroneous signal, or selectively harden the most critical component using the knowledge presented in Chapter~\ref{sec:model}. In addition, error mitigation techniques, e.g. the recursive mitigation scheme presented in Chapter~\ref{sec:algo}, are encouraged in this work to avoid dramatic SDDR degradation.

If the SDDR of an individual functional block cannot be sustained in a cost efficient way, application-level redundancy is preferred. Section~\ref{sec:system_exam} analyzes the massive MIMO application-level effects of VOS errors and assures that circuit degradation on a small portion of antennas can be absorbed in the massive MIMO system.


\subsection{Antenna outage impacts}
Another hardware failure scenario for the DFE is the antenna outage (antenna is completely non-operational). This happens when the power supply systems are broken, or when a circuit controlling signal is corrupted, e.g. failure to wake-up the digital circuit. 
    
In an antenna outage scenario, the DFE output is stuck at a fixed value, which is assumed to be the maximum value (DFE output Y = maximum). The SDDR of the outage antenna is -$\infty$, as the signals from the victim antennas are completely lost. This model is regarded as one of the most pessimistic hardware failures. Note that the -$\infty$ SDDR does not imply infinite noise to the whole system, as only the victim antennas are affected and their PA powers are normalized among all antennas. Therefore, several antenna outages will not fail the system entirely.

\section{Random antenna error impacts assessment}\label{sec:system_exam}
Consider a TDD massive MIMO system in DL with $M$ = 100 and $K$ = 10, where the channel estimation and the minimum mean square
error (MMSE) MIMO pre-coding are free from digital hardware errors. The performances over a Rayleigh 20-tap i.i.d. channels are simulated. The system is OFDM-based according to LTE parameters, i.e., 1200 loaded subcarriers in a 20 MHz band. 

The channel is estimated through uplink pilots associated to the different user equipments (UE) in a round robin fashion, i.e., one pilot every 10 subcarriers for a given UE. Since the channel estimation is assumed to be perfect. The simulation in this work cannot take advantage of the MMSE pre-coding that would limit the digital distortion errors. This is because the digital distortion is not present in the channel training phase. SNR is defined based on a total transmit power normalized to 0~dB per user. The emitted power is normalized for each antenna. The simulations do not apply error correction coding (ECC), except for Fig.~\ref{fig:system_ldpc} where the effects of coding on digital hardware errors are studied.

\subsection{Error effects on uncoded QPSK 100x10 massive MIMO}
This subsection discusses the effect of digital hardware errors, when some per-antenna digital processing units are suffering from errors.

Assume that due to local PVT variation and semiconductor aging effects, a portion of the antenna IFFT \& DFE are suffering from slight VOS hardware errors, i.e. SDDR = 10~dB for victim antennas, and no digital hardware errors occur for the remaining antennas.

\begin{figure}[H]
\centering
\includegraphics[width=0.8\linewidth]{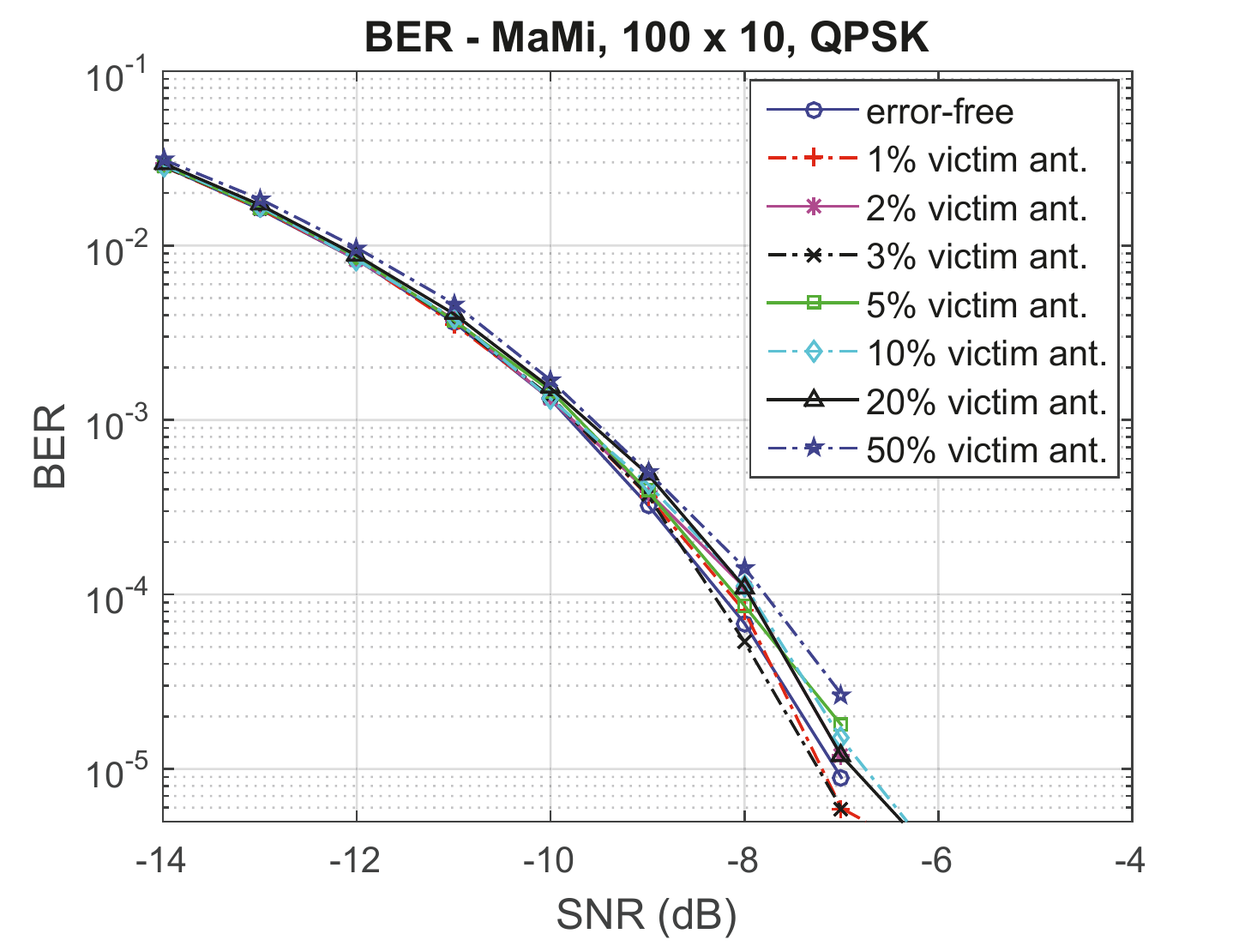}
\caption[Slight VOS digital distortion errors, i.e. SDDR = 10~dB, only degrades the system BER marginally.]{Slight VOS digital distortion errors, i.e. SDDR = 10~dB, only degrades the system BER marginally. This is observed by the massive MIMO system BER vs. channel SNR plot. Randomly chosen victim antennas are suffering from VOS errors. The remaining antennas are free from these errors. The pre-coding scheme is MMSE.}
\label{fig:system_10db}
\end{figure}

The system bit error rate (BER) degradation because of antenna errors is illustrated in Fig.~\ref{fig:system_10db}. The BER performance drops slightly as more antennas are affected. Nevertheless, the degradation is small even with 50\% antennas affected.

\begin{figure}[H]
\centering
\includegraphics[width=0.8\linewidth]{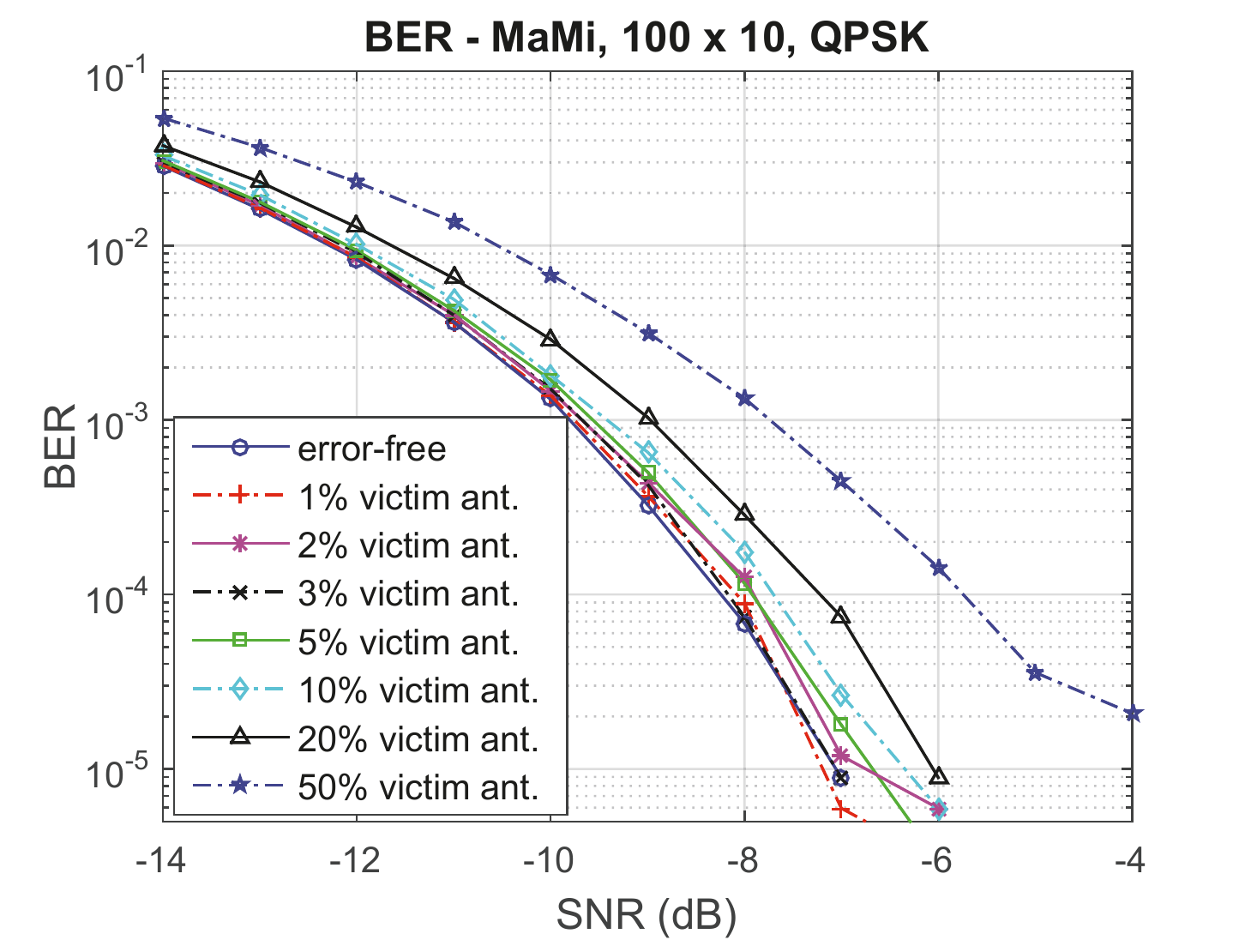}
\caption[If the massive MIMO system is suffering from extensive VOS digital errors on some antennas, i.e. SDDR = 0~dB, the output is still manageable.]{If the massive MIMO system is suffering from extensive VOS digital errors on some antennas, i.e. SDDR = 0~dB, the output is still manageable.}
\label{fig:system_0db}
\end{figure}

Supposing the VOS distortion noise is larger as designers further exploit the design margin, the digital hardware errors become more frequent and hence the SDDR is smaller for each given antenna, e.g. 0~dB. The resulting BER performance with the same settings is shown in Fig.~\ref{fig:system_0db}. For a target BER of $10^{-4}$, massive MIMO with 20\% antenna failing only requires a channel SNR of -7.4~dB, as opposed to -8.4~dB for the error-free case. This shows that the massive MIMO system still will operate correctly even if a noticeable amount of antennas suffer from digital hardware errors. 

\begin{figure}[H]
\centering
\includegraphics[width=0.8\linewidth]{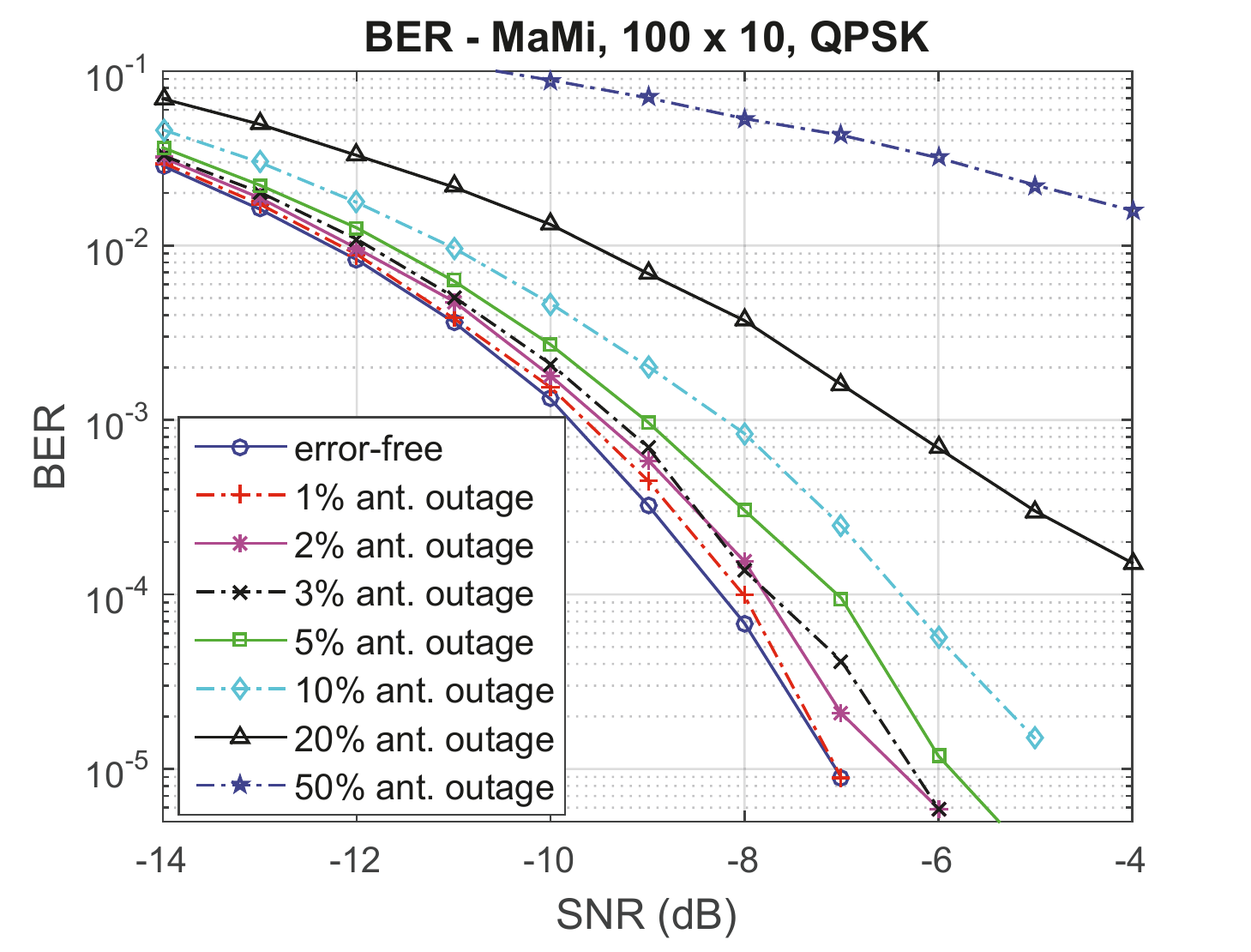}
\caption{The system can tolerates some antenna outage (DFE output stuck at a fixed value), i.e. SDDR = -$\infty$.}
\label{fig:system_out}
\end{figure}

In Fig.~\ref{fig:system_out}, the massive MIMO BER when applying the most pessimistic antenna outage model is shown. For the victim antennas, the useful signals are completely lost and a constant value is output from the DFE and emitted by the PA. This corresponds to an infinitely small SDDR for these victim antennas. The resulting BER performance shows larger SNR degradation for the same BER target. Nevertheless, the massive MIMO system can still cope with the antenna outage error thanks to the redundancy of antennas in the BS, at least for a failure rate up to 10\%.

\subsection{Antenna outage effects for other massive MIMO setups}
This subsection discusses the BER performance of massive MIMO for different settings. To analyze the most pessimistic situation, the antenna outage model is used.

\begin{figure}[H]
\centering
\includegraphics[width=0.8\linewidth]{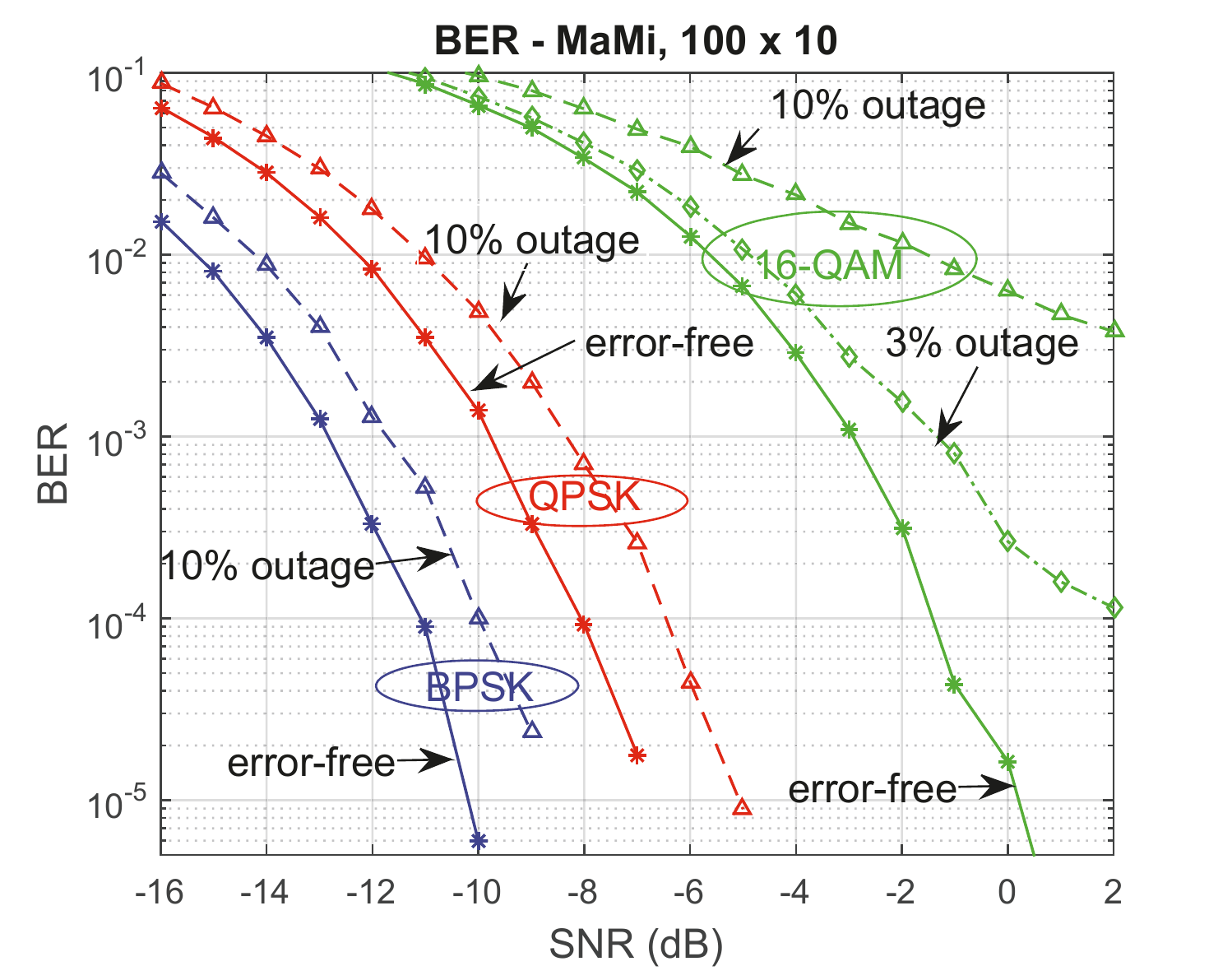}
\caption[Simple modulation schemes are more resilient to antenna outage.]{Massive MIMO system with random antenna outage errors for different modulation schemes, i.e. uncoded BPSK, QPSK and 16-QAM. Simple modulation schemes are more resilient to antenna outage.}
\label{fig:system_modulation}
\end{figure}

Fig.~\ref{fig:system_modulation} displays that for the 100x10 massive MIMO, 10\% antenna outage leads to slightly more BER degradation for QPSK, comparing with BPSK. This is due to the larger error margin for simpler modulation scheme. For the more sensitive 16-QAM modulation scheme, 10\% antenna outage leads to a huge degradation in DL BER. This implies that for communication systems where channel SNR is worse and simple modulation schemes are used, the reliability requirement of the antennas can be relaxed, to simplify the (I)FFT \& DFE design and reduce the power consumption budget.

\begin{figure}[H]
\centering
\includegraphics[width=0.8\linewidth]{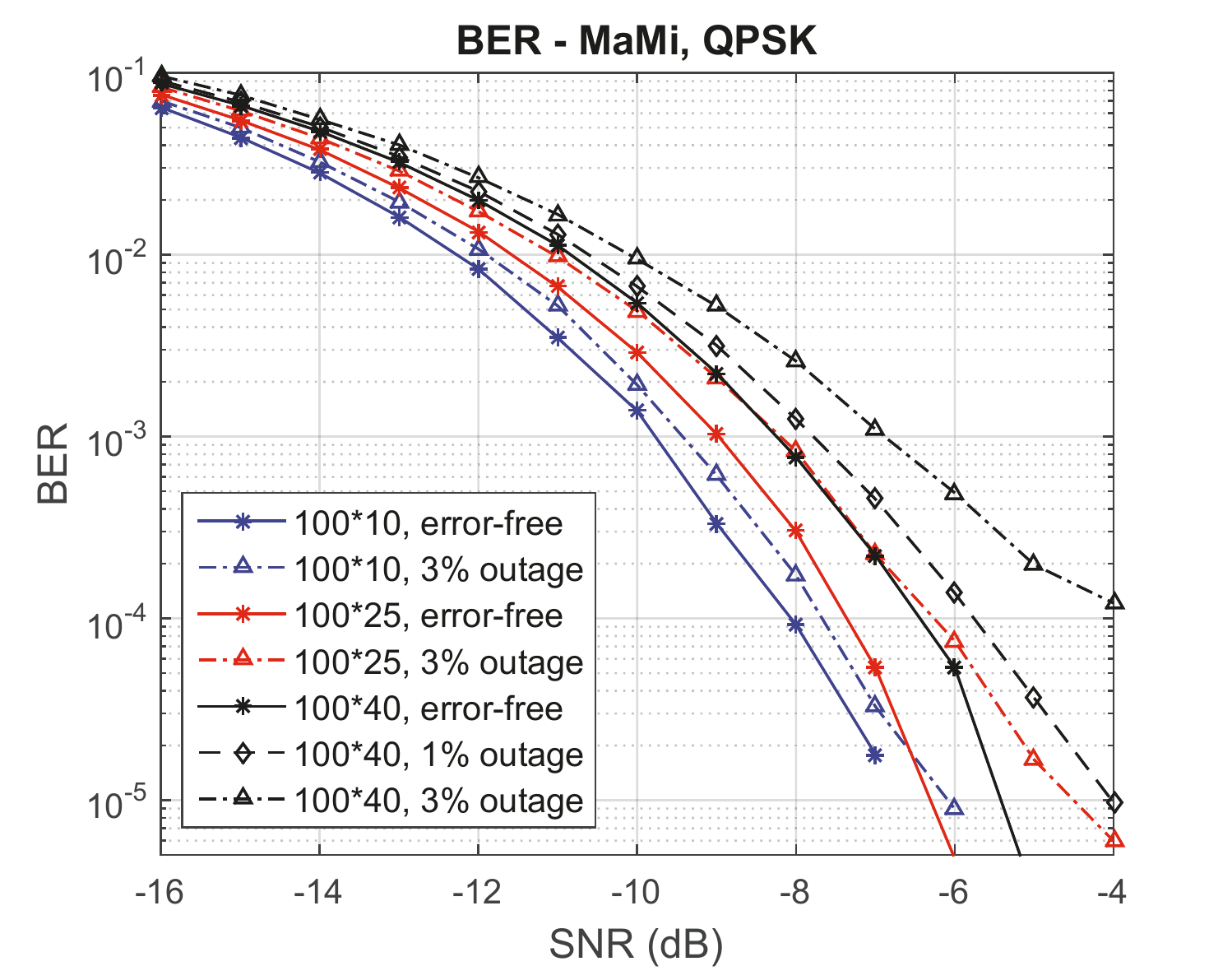}
\caption[Lower loads leave more spaces for error absorption.]{Massive MIMO system with random antenna outage errors for various loads, i.e. 100x10, 100x25, 100x40. Lower loads leave more spaces for error absorption.}
\label{fig:system_load}
\end{figure}

The BS antenna redundancy is reduced if the load of the massive MIMO system increases (the number $K$ of served users or streams is increased). In this scenario, the tolerance for antenna outage is decreased, compared to systems with small $K$ (Fig.~\ref{fig:system_load}).  Nevertheless, For massive MIMO systems where $M$ $>>$ $K$, the amount of antenna redundancy is sufficient to provide opportunities for antenna unreliability.

\begin{figure}[H]
\centering
\includegraphics[width=0.8\linewidth]{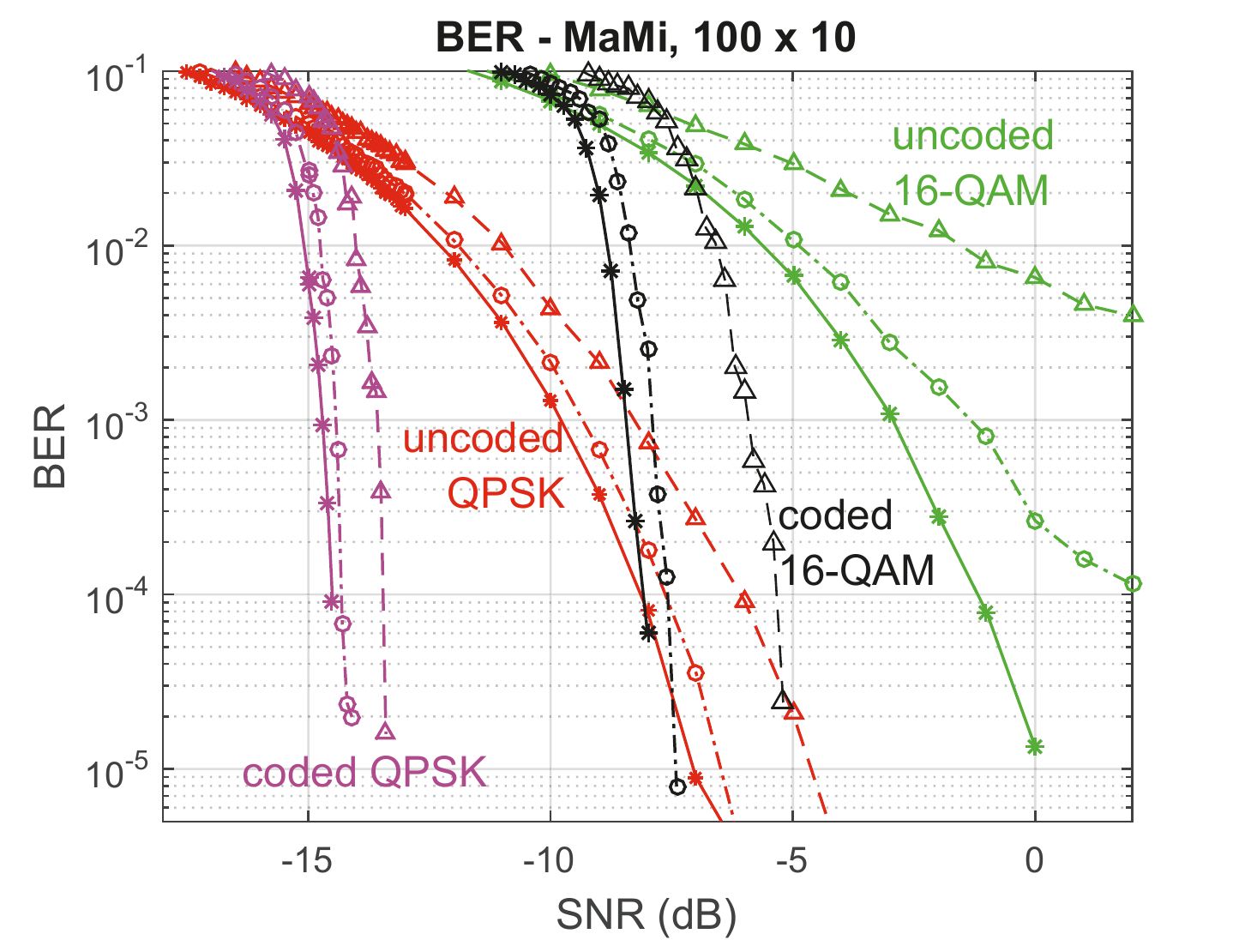}
\caption[Coding and uncoded degradations are similar.]{Massive MIMO system with random antenna outage errors for uncoded and coded (3/4 soft LDPC) QPSK, and uncoded and coded (3/4 soft LDPC) 16-QAM. The legend denotes: i) error-free (star shapes), ii) 3\% victim antennas (circle shapes), and iii) 10\% victim antennas (triangle shapes). Coding and uncoded degradations are similar.}
\label{fig:system_ldpc}
\end{figure}

So far uncoded results were presented. However, errors in massive MIMO systems can be mostly corrected by error correction codes, e.g. convolutional codes and LDPC codes. Fig.~\ref{fig:system_ldpc} shows the BER improvement when 3/4 soft decoded LDPC code is utilized in the massive MIMO system. At the targeted BER of $10^{-4}$, the SNR is 6~dB lower for the coded QPSK, compared with an uncoded case. For such BER, the SNR difference when considering antenna outage is smaller for the coded massive MIMO system, compared to the uncoded one, although a limited degradation always remains.

\section{Controlled antenna outage}\label{sec:system_improve}
According to Section~\ref{sec:system_vos}, VOS with error-resilient techniques bring up to 40\% power saving, at the risk of failures for few antennas. The simulation results from Section~\ref{sec:system_exam} illustrates performances when no error detectors are equipped. In other words, the massive MIMO system is operating as usual, regardless of hardware errors. In this situation, the massive MIMO manages to sustain system performance even if several antennas are non-operational (outage) due to aggressive VOS or completely DL failure. In order to improve the reliability of the system under hardware errors, this work proposes to firstly detect hardware errors, and then either correct errors, or circumvent the defective hardware if correction is not possible. It is worth noting that the distortion originating from digital circuit failures fundamentally differs from the random noise introduced in communication channels. While CMOS process variations may feature continuous random distributions, their effects typically lead to discrete antenna error events.

Dedicated monitoring circuit can be established to detect these errors and thus these erroneous bits can be labeled unreliable and potentially be corrected. Eventually, if some circuit errors get too large or systematic, measures at the system level can be taken to discard this hardware and increase the overall robustness.

If digital hardware designs provide monitors \cite{razor03,5654663,bull2011power,fojtik2012bubble} for each (I)FFT \& DFE, the massive MIMO system can equip a closed loop for error detection and correction. The error-resilient designs shown in Chapter~\ref{sec:algo} can detect and mitigate some errors at the algorithm level. Signals from erroneous antennas can thus be exploited, as the errors are small. The error effects are shown in Section~\ref{sec:system_exam}. 

\begin{figure}[H]
\centering
\includegraphics[width=0.8\linewidth]{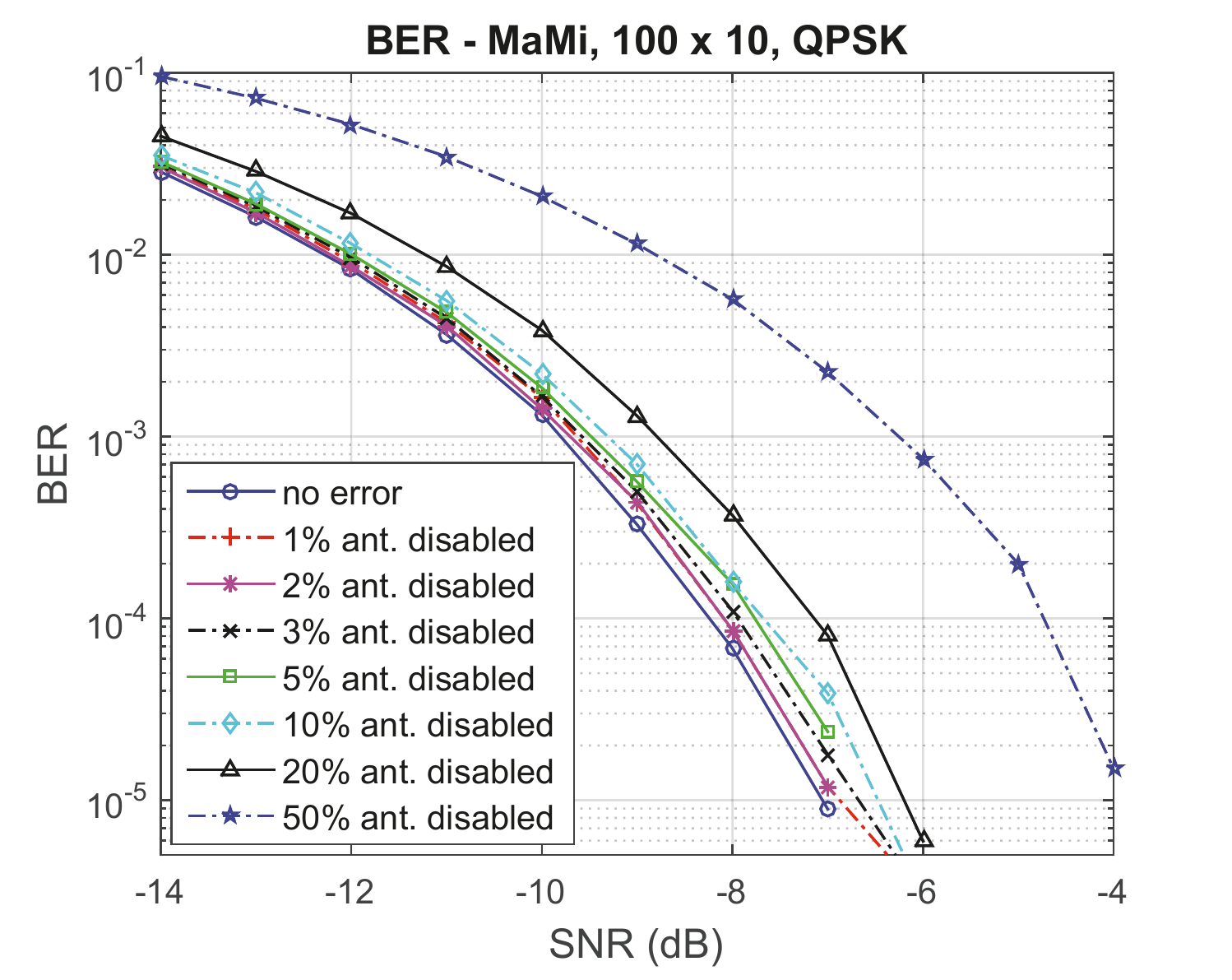}
\caption[Massive MIMO system performance when victim antennas are discarded and the channel estimation and DL is carried out by the remaining error-free antennas.]{Massive MIMO system performance when victim antennas are discarded and the channel estimation and DL is carried out by the remaining error-free antennas. This situation is better than antenna outage.}\label{fig:system_disable}
\end{figure}

Another countermeasure is to disable the victim antennas temporarily. Therefore, the channel estimation (and hence the precoding, and data transmission) are accomplished the remaining error-free antennas only. This method is equivalent to operating with a reduced number of error-free BS antennas $M$. For systems with large redundancy, e.g. using simple BPSK modulation, this method will hardly impact system quality. In Fig.~\ref{fig:system_disable}, erroneous antennas are taken out completely. This leads to BER worse than Fig.~\ref{fig:system_10db} and Fig.~\ref{fig:system_0db}, where antennas are affected with moderate digital hardware errors and the signal from these victim antennas are still exploited for communication.

This shows that by detecting the degree of antennas failing (noise power), designers have the option to determine whether to exploit the victim antennas or to discard them, for better performance. The noise power can be estimated from circuit-level or algorithm-level error monitors, or from the system level measurement that combines channel noise, e.g. by evaluating the measured channel information (CSI).

\begin{figure}[H]
\centering
\includegraphics[width=0.65\linewidth]{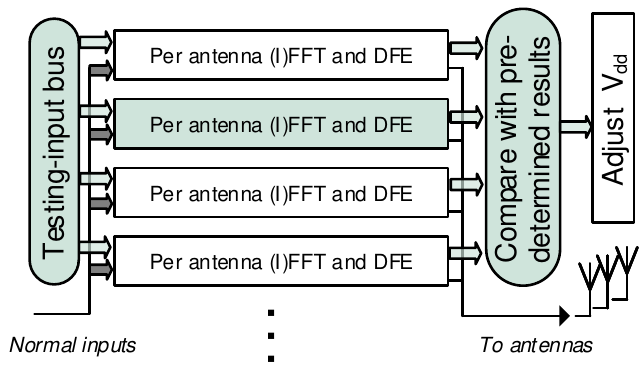}
\caption[An error resilient adaptive scaling technique to manage hardware errors.]{An error resilient adaptive scaling technique to manage hardware errors. This technique periodically checks DFE functionality, and adjusts $V_{dd}$ accordingly. In this figure, the second antenna is in testing-mode.}\label{fig:system_compare}
\end{figure}

This work proposes an error detection strategy to periodically check the antenna functionality by putting one antenna in the testing-mode at one time (Fig.~\ref{fig:system_compare}). During the testing mode, per antenna DSP are supplied with testing inputs. The outputs are compared with the pre-computed data. If the results are vastly different, the antenna is detected erroneous, and thus the $V_{dd}$ is increased to reduce errors and hence guarantee performance. Erroneous antennas are permanently disabled if they kept failing.

In this work, the periodical testing is scheduled when no data transmission is taken place. Moreover, it can even be performed on-the-fly during data transmission, since suppressing one (1\% for massive MIMO with 100 antennas) antenna during DL into the testing-mode would not introduce huge degradation (see Fig.~\ref{fig:system_disable}). This enables timely fine-grained $V_{dd}$ adjustment, which maximizes power savings. If, however, the antenna is permanently damaged and thus cannot recover by increasing $V_{dd}$, the antenna will then be labeled as defected.

\begin{figure}[H]
\centering
\includegraphics[width=0.98\linewidth]{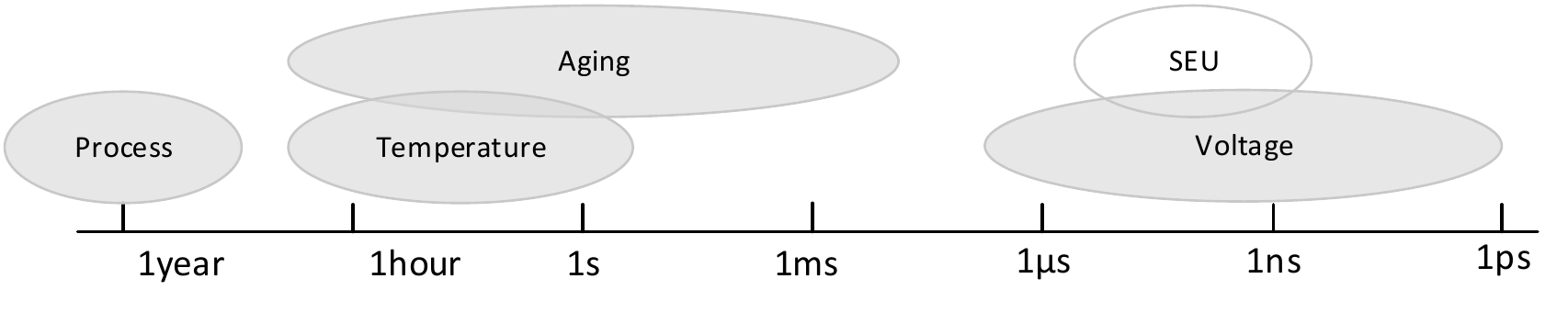}
\caption[The frequency of changing of the process, aging, temperature, and voltage noise that affect the timing errors of circuits.]{The frequency of changing of the process, aging, temperature, and voltage noise that affect the timing errors of circuits. Effects with slow changing rate can easily be resolved by periodical checking, e.g. process and temperature. Designs should either leave margins, or mitigate errors by resilient design toward slow changing effects, e.g. voltage noise and SEU.}\label{fig:system_duration}
\end{figure}

The time interval of the periodical testing depends on the nature of the error occurrence and the changing environment (Fig.~\ref{fig:system_duration}). The device process variability is usually determined after manufacturing. Thus, a pre-installation $V_{dd}$ adjustment is sufficient to account for this variability. The CMOS aging effects are becoming more evident in the scaled technology, they depend on the work loads (voltage, frequency, and the relaxation duty cycle). If long-term aging is the only concern, checking the correctness of every hour is sufficient. If, however, the relaxation factor is considered, which alters the device characteristics in several-hundreds cycles, periodically checking the results by every millisecond is recommended. The temperature, which is mainly subject to the heating and dissipation efficiency, usually changes in the ranges of seconds. The voltage noise occurs in the picosecond range. Thus, voltage-noise incurred errors, if the occurrence is rare, can be absorbed in the resilient computing, without counter-measures after periodical testing. If the occurrence becomes more frequent, e.g. once in every thousand cycles, errors will be captured by the periodical testing, and hence the system can opt to either disable the unit or to increase the voltage. The SEU occurrence rate for on-ground application is low. Therefore, communication systems usually do not address this issue.

\section{Conclusion}\label{sec:system_con}
This chapter examines the opportunity of using error-prone digital signal processing components in massive MIMO systems, and proposes a strategy to maximize power savings while still offering robust operation. The (I)FFTs \& DFEs in massive MIMO are the most critical digital components in terms of area and power consumption as they scale linearly with the massive BS antenna count $M$. Hardware errors in a number of antennas' (I)FFT \& DFE can be absorbed by the massive MIMO system thanks to the redundancy coming from the large antenna number. The massive MIMO system exhibits error resilience even for the worst-case antenna outage scenario. 

When the hardware error distortion power is low, e.g. lower than 0~dB, the massive MIMO system should continue using the erroneous signal. The errors can be corrected at the application-level by other redundant antennas. It is proposed for systems to equip with on-chip monitors, so that they can detect hardware errors on-the-fly, and discard the error-prone antennas when their SDDR is large, e.g. when antenna outage, or when components are suffering from severe aging effects.

This provides opportunities for the digital hardware designers to embrace cost-efficient and reduced-power digital components at the expense of sacrificing individual antenna reliability, yet maintaining overall systems performance. Up to 40\% power can be reduced for the considered digital processing components.



\cleardoublepage

\chapter{Conclusions and future work}\label{sec:con}


Digital circuit designs have benefited from the free lunch of technology scaling for many decades. Nevertheless, the free lunch starts to diminish as we are entering the deeply scaled era. Clever design optimizations are becoming more demanding than ever before. This work advocates a cross-layer to investigate through multiple design levels for power saving. This blurs the distinction between traditional design levels, especially in terms of handling variations, environmental and runtime uncertainties, and errors. This leads to global saving on power consumption as demonstrated in this work. This chapter concludes this thesis. Section~\ref{sec:con_con} highlights the results of the thesis. Section~\ref{sec:con_key} discusses the key messages from this work. Section~\ref{sec:con_future} lists some future work suggests interesting further extension of this work.

\section{Conclusions}\label{sec:con_con} 

In an era where performance and power are heavily constrained, the conventional worst-case design methodology no longer suffices. A cross-layer optimization is encouraged to quantify the actual need from all levels at design time to avoid over-design. Therefore, circuit and system designers should work more coherently to provide just-needed quality and minimize power consumption. This work advocated for a cross-layer optimization methodology for quality-power trade-offs.

In Chapter~\ref{sec:variation}, approaches for power and quality trade-offs were reviewed. It evaluated the device-level phenomena of PVT variations and reliability threats. It explained the limitation of the worst-case design approach. The chapter also reviewed the adaptive scaling method that uses replica circuits or in-situ error-detection flip-flops. The error-detection flip-flops offer large benefit, due to the capability of responding to fast variations and critical path activation differences. Finally, the motivation for cross-layer error resilience was presented. In a lot of applications, higher-level designs can easily absorb and handle errors that were seemingly inevitable at device and circuit-level.

\subsubsection*{Circuit-level: random error impact model}
In Chapter~\ref{sec:model}, a circuit-level random error model was presented. The model predicts the impact of device errors on algorithms. In contrast to conventional models that require time-costly Monte-Carlo simulation, the proposed model uses an analytical approach. It ranks the flip-flop nodes in a digital circuit according to their contribution to the outcome. The contribution is defined as its \textit{significance}. A flip-flop is significant if it tightly connects to a lot of significant flip-flops. The model is thus named as SERIAL, or SignificancE RankIng ALgorithm.

This automation eliminates the need for conventional trial-and-error searching for suitable error hardening. Circuit designers have the opportunity to selectively ensure the most important FF (e.g. FF hardening, VOS margin), without excessive hardening overheads. The efficiency and effectiveness were shown on benchmark circuits. The design principles were applied to the design of a reliable FFT processor, which cuts the hardening overhead by a half.

\subsubsection*{Microarchitecture-level: fine-grain hardware switch for power saving}
Chapter~\ref{sec:arch} proposed a microarchitecture-level fine-grain hardware-switch scheme to save power in embedded processors. This scheme exploits word-length optimization opportunities for a multiplication unit. The opportunities were justified on 11 typical signal processing applications. It is shown that half of the inputs of 32-bit multiplications are shorter than 8 bits.

Therefore, this work proposed a redundant short multiplier to perform these short applications. This leads to power saving as the toggle circuit complexity is reduced. The proposed hardware-switch scheme was validated on the OpenRISC platform. Without changing the software compiler, It brings 23.7\% power saving for the multiplication unit, which accounts for 9.5\% power saving for the whole execution unit.

\subsubsection*{Algorithm-level: computation-skip scheme to trade quality for power savings}
Chapter~\ref{sec:algo} proposed an algorithm-level error mitigation method, computation-skip scheme. It trades quality for power savings in recursive applications without throughput penalties. Errors produced at the circuit-level are mitigated, without error accumulation, at the algorithm-level. 

The chapter validated the power saving of the scheme on a CORDIC hardware accelerator. The accelerator is processed and verified in a standard 28nm CMOS process with only standard-cells. Using only standard-cells, this work eliminates the traditional semi- (or even fully-) customized design effort for in-situ error detection circuits. A 28\% energy consumption per bit saving due to relaxed timing constraint (design margin shaving at design time) is observed. The energy/ENOB saving improves to 42\% because of adaptive scaling (error-free design margin shaving at run-time). Moreover, a total of 46\% saving is possible, with a 0.2-bit precision loss.

\subsubsection*{Application-level: embracing erroneous hardware in Massive MIMO systems}
Finally, Chapter~\ref{sec:system} investigated application-level error absorption and handling. The considered errors are generated at the circuit-level and the algorithm-level errors by hardware. The chapter focuses on a Massive MIMO communication system case-study. It shows that the perceived performance will hardly be affected by sparse processing failures, while the power consumption can be considerably reduced as error resilient hardware are utilized. Furthermore, this work assesses antenna outage impacts and proposes damage control strategies.

When the hardware error distortion power is low, e.g. lower than 0~dB, the Massive MIMO system should continue using the erroneous signal. The errors can be corrected at the application-level by other redundant antennas. It is proposed for systems to equip with on-chip monitors, so that they can detect hardware errors on-the-fly, and discard the error-prone antennas when their SDDR is large, e.g. when antenna outage, or when components are suffering from severe aging effects. This provides opportunities for the digital hardware designers to embrace cost-efficient and reduced-power digital components at the expense of sacrificing individual antenna reliability, yet maintaining overall systems performance.

\section{Key messages}\label{sec:con_key} 
The traditional method to handle process variations, environmental and runtime uncertainties, where excessive safety margins are added, result to huge power efficiency loss. One benefit of this conventional approach is that it simplifies design -- if we are uncertain about some parameters, take a reasonable worst-case assumption and make sure that works. However, this must be changed as no precious power efficiency should be wasted because of a lack of engineering effort. 

There are in general two general directions to advance. First, an accurate model is desired. This model should eliminate the unnecessary pessimism. For instance, as the transistor ages according to the operating voltage and frequency, the worst-case settings for systems where the supply voltage is low should be less pessimistic \cite{kukner2013impact}. This enables easier timing closure and hence power saving. The more accurate the model, the more information need to consider during the modeling, this calls for cross-layer modeling. Eventually, the effort in modeling will outweigh the power gain at some point. Another roadblock for pursuing a perfectly accurate model is that the environmental and runtime uncertainties vary over time and changes very fast. An accurate model at one point becomes invalid at another time. Again, safety margins remain. 

Second, adjusting to the unpredictable changes when they arise is the advocated approach in this thesis. As it is fundamentally impossible to model precisely beforehand, the logical approach is dynamical changing according to the condition. Cross-layer information is encouraged as optimization locally within design levels are not enough.

The DVFS and the more aggressive AVFS falls in the second category. With speed detectors and fine-grained voltage regulator, the time-zero and time-dependent process variations can be largely eliminated. \cite{fojtik2012bubble} reports 54\% energy saving on an average die with this approach, without introducing any errors. Despite the remarkable benefits, the engineering challenges of i) the accuracy of speed monitor, ii) overhead of in-situ speed detection, iii) response time of the voltage-regulator are still the obstacles to mass adoption. Nevertheless, the huge energy saving has motivated engineers to exploit this opportunity, especially in the processor industry where a massive amount of chips are produced. The immense recursive benefit certainly outweighs the non-recursive engineering investment. The AVFS, realized by adaptive clocking distribution and power management, is becoming a common feature in modern processors, e.g. \cite{gonzalez20173}. The other demanding application for AVFS is the ultra-low power IoT terminal devices, where every microwatt matters. It is almost impossible to find a sub-threshold computing device that does not equip AVFS to manage variability.

Apart from error-free DVFS, the error is another factor for cross-layer optimization. It is a pity that engineers spend so many energy to optimize power with the error-free constraint while the constraint is not necessarily needed. In the end, the digital devices serve to provide service, not to compute correctly. This work demonstrates huge potential in the wireless communication system. In those systems, errors are corrected by the ECC as long as they are small. Traditionally, circuit-level reliability hardening (FF hardening) are employed in mission-critical circuits to ensure error-free circuit. This is however unnecessary.  In Chapter~\ref{sec:model}, a cross-layer optimization by selective FF hardening is performed on an FFT processor against random SEU. It increases reliability with much less hardening overhead. This methodology is advised in this work -- when trying to improve reliably, optimize with minimum overhead, and do not assume that reliable means error-free circuits.

Similar design approaches are carried out on algorithm (Chapter~\ref{sec:algo}) and application-level (Chapter~\ref{sec:system}) designs. The error-resilient approach advances from the canary FF method (an AVFS approach) in terms of power minimization, especially for high-speed circuits. This is because error-resilience relaxes the strict timing requirement in those circuits and uses slow but power-efficient digital cells. 

The massive MIMO and the 5G technology will certainly demand massive digital solutions in the future. A good news is that the massive MIMO tolerates plenty of errors thanks to the antenna redundancy. Errors include conventional channel errors, analog component non-ideality, quantization errors, VOS errors and even hardware failure. Its resilience motivates the popularity of massive MIMO technology. Moreover, it encourages low-power but erroneous hardware solutions. The DVFS with sparse errors and the algorithm-level solution in Chapter~\ref{sec:algo} fits into this picture perfectly.

Overall, device uncertainties, that used to be a silicon-only problem, should be solved by joint force of device and application designers. A cross-layer optimization mentality should be included in the digital circuit and system development. This called for compound knowledge from the device to the application during design.

\section{Future work}\label{sec:con_future}
Cross-layer optimization for quality and power remains a hot research domain. However, it should not stay in the research domain, yet actual industrial applications and deployments are anticipated. Fortunately, an early adoption of these techniques has been observed in the industry. The author firmly believes that the cross-layer methodology will become crucial to continue increasing the performance per watt for future digital circuits and systems. The following is a suggestion for future work in this area.

\begin{itemize}

\item[$\bullet$] \textbf{Extending the error effects model to non-uniform distributed errors.} The circuit-level model in this thesis covers the effects when errors are randomly generated. This is readily applicable to SEU effects where the error generation is uniform. For other non-uniform errors, especially time-dependent degradations, the toggling frequency, and operating voltage information, should never be dismissed. These non-uniformed error possibilities require special consideration not only in the error generation, but also error propagation. The significance factor can be extended to vulnerability factor, which is a product of the error generation possibility and error consequences. For instance, \cite{mukherjee2003systematic} proposes an architecture vulnerability factor, which is the product of the error generation possibility and the error propagation possibility (but not error severity).

\item[$\bullet$] \textbf{Complete hardware prototyping and system-level consideration of timing-error tolerant Massive MIMO systems.} The timing-error monitor on hardware is validated on recursive application CORDIC, and on non-recursive digital front-end. A complete system-level prototyping is possible. A key missing part is an integration with on-chip power regulators for autonomous voltage tuning. In addition, the proof of concept system-level demonstration, built with multiple chips, will allow consideration of realistic variations in environmental conditions. The considering of errors in uplink and channel estimations can also be researched. If that is considered, the co-optimization of digital front-end and channel pre-coder is much needed. It is believed by the author that the MMSE pre-coder might suppress the digital distortion errors more effectively than the ZF pre-coder. The increased channel estimation and pre-coding complexity should be carefully checked. 

\item[$\bullet$] \textbf{Considering workload-related CMOS aging effects.} Although the methods in this thesis can cope with the slow-changing CMOS aging effects, they cannot fully exploit the workload-related CMOS aging. Modern CMOS device is allocating more safety margins to the aging effects \cite{stamoulis2016capturing}. Therefore, designing circuits with workload-related aging models will promote new proposals to exploit these margins, and saves power consumption.

\item[$\bullet$] \textbf{Approximate computing.} The hardware switch in this work is only activated when no arithmetic errors will be produced. This can extend to situations when some small amount of errors occurs. For instance, when the input is slightly larger than the short multiplier size, saturation or truncation can be performed. This, however, needs compiler interplay to ensure application-level correctness. Moreiver, this thesis has not covered the approximate computing hardware. An approximate computing device produces errors by design, which simplies the computation process in returen. The comparisons between the VOS and approximate multiplier \cite{liu2014low} can be further investigated. The VOS and approximate hardware can also work together, its impact on application-level quality should be studied.

\item[$\bullet$] \textbf{Stochastic computing.} Known for its low-complexity, stochastic computing devices represents and processes information in the form of digitized probabilities. However, it was seen as impractical because of very long computation times and relatively low accuracy. However, if future technologies continue to increase uncertainty in circuit behavior, it will imply a need to better understand, and perhaps exploit, probability in computation.

\item[$\bullet$] \textbf{Systematical cross-layer design methodology.} This thesis performed ad-hoc cross-layer optimizations to digital circuits and systems, which spot and optimize the power / quality bottlenecks. Ideally, a more systematical cross-layer design methodology is welcomed. This task is not easy, as changing the whole design flow requires huge interplay from IC foundries to EDA vendors, designers, and system integrator. However, if the current design flow, combined with cross-layer optimizations, can not sustain the development of digital technologies, the systematical methodology is definitely one of the most promising solutions.

\item[$\bullet$] \textbf{Impacts on future semiconductor devices.} The device uncertainties and management techniques in sub-5nm novel device architecture (e.g. gate-all-around transistors, carbon nanotubes), novel materials (e.g. GaN, GaAs), novel integration (3-D chip and package) and memory technologies (e.g. M-RAM, R-RAM) can be studied. Technology engineering is trying to enhance the reliability of those new technologies. However, the progress is not always satisfying. The new device parameters in variations might demand new design methodologies.

\item[$\bullet$] \textbf{Applications to other soft-output systems.} The case-study in this thesis is around the wireless communication systems. In these systems, the quality of service is never strict, as long as SNR, BER, throughput requirements (among others) are fulfilled. The knowledge gained in this thesis can also apply to the power and quality trade-offs in other soft quality systems, e.g., heuristic searching problems, approximate simulation of supercomputing tasks, and training and inference in artificial intelligence applications. In the deeply scaled semiconductor era, the advance depends more and more on the application. The digital neural network is well believed to the next killer application. The convolutional neural network, especially in computer vision applications, does not require precise solutions. The data-path error resilience power minimization can be substantial, considering the massive amount of processing carried out.

\end{itemize}



\cleardoublepage




\appendix

\chapter{A digital front-end processor for 60 GHz polar transmitter}\label{ch:myappendix}




\section*{Abstract}
A complete Digital Front-End (DFE) processor for the 60~GHz polar transmitter is presented. It avoids supply modulating, RF limiters, and AM detection circuits, compared to traditional analog-centric polar transmitter architectures.

The front-end processor consists of i) a poly-phase Cascaded Integrator-Comb (CIC) filter for spectrum shaping, ii) parallel COordinate Rotation DIgital Computers (CORDIC) for rectangular-to-polar conversion, and iii) Power Amplifier (PA) non-linearities pre-distortion units using Look-Up Tables (LUTs). It is designed in the two-phase latch-based pipeline to achieve a throughput of 4x1.76~Gsps. Implemented in a standard 28nm CMOS technology, the DFE processor occupies 0.031~$mm^2$ and consumes 39mW from 0.9V supply. This result outperforms previously reported architectures.


\section{Introduction}
In contrast with the scarcely available spectrum in the sub-10~GHz range, the 60~GHz frequency band provides 4 channels of 2.1~GHz bandwidth each, as specified by the IEEE802.11ad standard~\cite{11ad}. This provides up to 6.75~Gbps data rate in Wireless Personal Area Network (WPAN). ~\cite{11ad}

However, due to the high free-space path loss, transmission at 60~GHz covers much less distance for a given power budget. This can be alleviated by employing phased array antennas.~\cite{Ref10} Nevertheless, as the number of Power Amplifiers (PAs) increases, the power consumption grows drastically. The power issue is more severe given the fact that, the 60~GHz class-A linear mode PAs usually provide less than 5\% efficiency.

Therefore, the polar architecture is proposed, which allows the PA to operate in the saturation region. In a polar transmission, the PHase (PH) and the AMplitude (AM) signals have separate paths before being combined by the PA. Conventional analog-centric polar modulation scheme suffers from several challenges, e.g. supply voltage linearity of the PA, AM-AM distortion, AM-PM distortion, nonlinearity of the envelope detector, and AM-PM distortion.

To cope with that, a digital-intensive transmitter architecture with the polar concept is explored at mm-waves high-bandwidth transmitters. Moreover, the polar concept is expanded to the whole transmitter, rather than only in the RF domain. The AM signal can then digitally modulate a variable-size PA. This avoids modulating the supply and also eliminates the need for an additional RF limiter and AM detection circuits, which would introduce extra nonlinearity and bandwidth limitations. Despite many advantages, the design of the digital front-end (DFE) processor is very challenging. For the 60~GHz application, the DFE processor mostly needed to work at a very high speed depending on the required oversampling factor. \cite{li15icassp-polar-dfe} discussed several design considerations for the polar conversion unit, without implementation.

This work presents the first DFE processor for such polar transmitter working in the 60~GHz band. It enables high-bandwidth data transmission with an output throughput of 7.04~Gsps (4x1.74~Gsps). The extensive measurement confirms the great potential of the polar architecture in an actual design. Section~II discusses the system architecture. The implementation details are illustrated in Section~III. In Section~IV, the chip measurement results are presented.

\section{System architecture}
\begin{figure}[H]
\centering
  \includegraphics[width = .9\linewidth]{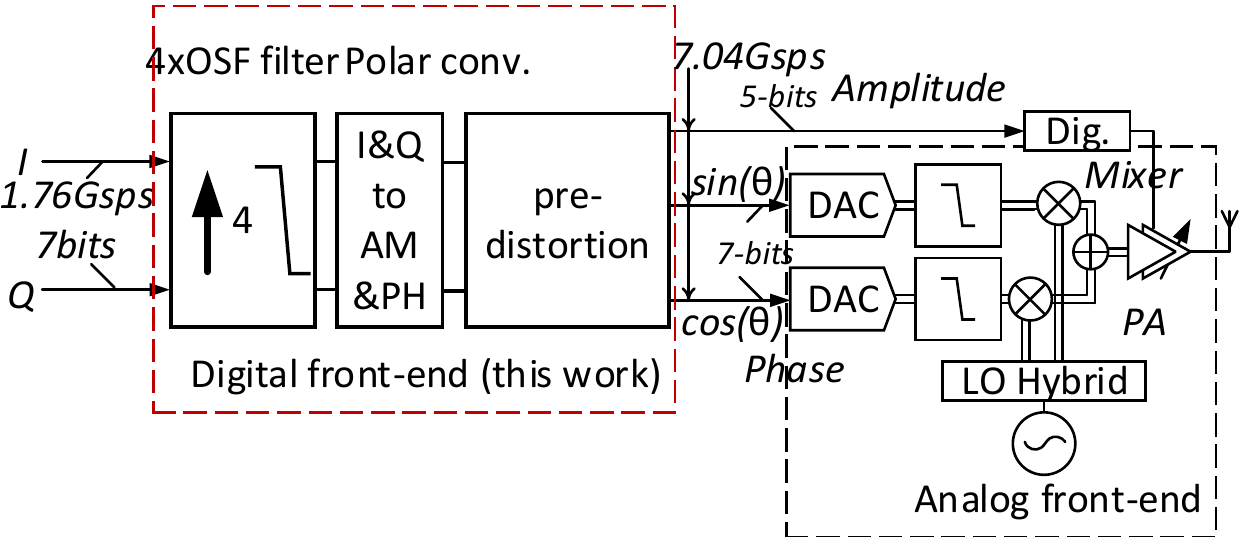}
\caption{Block diagram of the digital intensive 60~GHz polar transmitter.}
\label{fig:polar_system}
\end{figure}

Fig.~\ref{fig:polar_system} shows the high-level architecture of a polar transmitter system. The system consists of a DFE processor (which is presented in this work) and an analog front-end (described in details in~\cite{Ref10}). The DFE comprises DSP for upsampling the rectangular ($I$ \& $Q$) signals, for $I$ \& $Q$ signals to AM and PH signals conversion, and for pre-distortion compensation. The 802.11ad standard specifies -21 dB EVM for single carrier QAM-16 modulation. Considering the variations in this deeply scaled 28nm CMOS technology, -31dB is taken as the design goal with a design margin of 10 dB. 

\subsection{Polar conversion}
The rectangular to polar conversion takes in-phase $I$ and quadrature $Q$ signals, and provides the corresponding AM (A) and PH (in the form of $sin(\theta(t))$ and $cos(\theta(t))$):
\begin{eqnarray}\label{dirconv}
\nonumber
    A(t)&=&\sqrt{I(t)^2+Q(t^2)}\\
    \nonumber
    sin(\theta(t))&=&sin(arctan(\frac{Q(t)}{I(t)}))\\
    cos(\theta(t))&=&cos(arctan(\frac{Q(t)}{I(t)}))
\end{eqnarray}
\normalsize

This conversion involves multiple complex computations, e.g., square root, trigonometric and division computations. With the aim of energy efficient processing, this is achieved by deep-pipelined COordinate Rotation DIgital Computers (CORDIC)~\cite{li15icassp-polar-dfe}. Each CORDIC rotates the vector of the $I$ \& $Q$ signals iteratively until the vector angle reaches zero. The resulting vector amplitude and the rotated angle are recorded as the AM and PH signals.

\subsection{Phase shaping filter}
The polar conversion is a nonlinear computation, which broadens the spectrum. To avoid error vector magnitude (EVM) degradation from the spectrum overlap, signals are oversampled and digitally filtered before aliasing. Another reason for oversampling is to overcome alias generated by the RF-DAC in the analog stage.~\cite{van2007digital} To suppress the alias below the spectrum mask, a 4xOSF is investigated in~\cite{li15icassp-polar-dfe}, in combination with an analog Butterworth baseband filter in the PH path.

\begin{figure}[H]
\centering
\includegraphics[width =.75\linewidth]{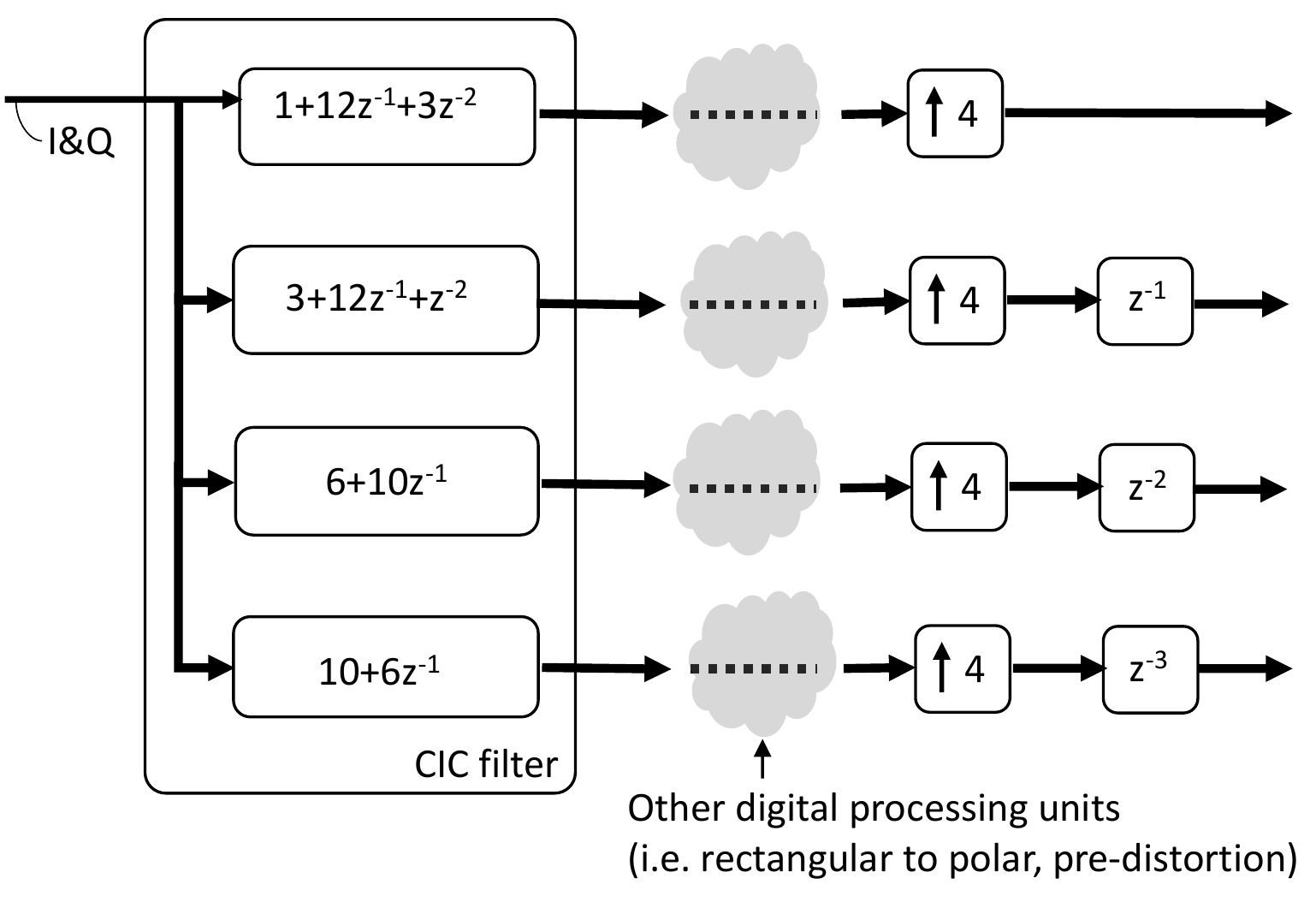}
\caption{Poly-phase implementation of the phase shaping filter.}
\label{fig:polar_cic}
\end{figure}

For pulse shaping, we utilized the Cascaded Integrator-Comb (CIC) filter, rather than the computation-intensive raised cosine filter. The structure and the transfer function of the CIC filter are illustrated in Fig.~\ref{fig:polar_cic}. The oversampling is performed after the CIC filter, rather than before it, to reduce the operating frequency of the CIC filter. 

\begin{figure}[H]
  \centering
  \includegraphics[width = .6\linewidth]{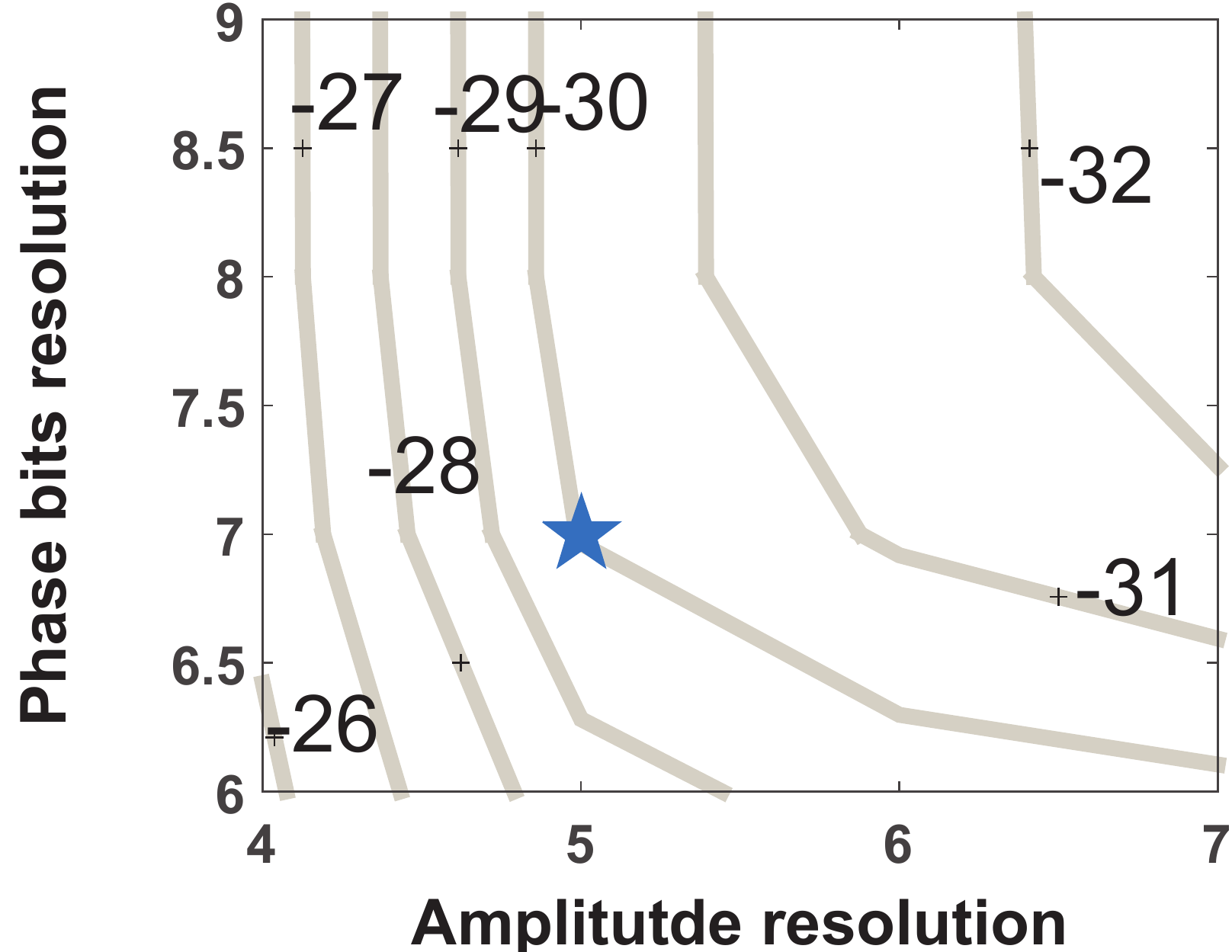}
  \caption[EVM vs. PH/AM resolution.]{EVM (in dB) vs. PH/AM resolution (in \# of bits).}
  \label{fig:polar_quantization}
\end{figure}

With a targeted EVM of -30 dB, the PH and AM resolutions for the combination of the phase shaping filter and the polar conversion are traded-off in Matlab (Fig.~\ref{fig:polar_quantization}), which decides 5 bits for AM and 7 bits for PH.

\subsection{Pre-distortion}

\begin{figure}[H]
\centering
\includegraphics[width =.65\linewidth]{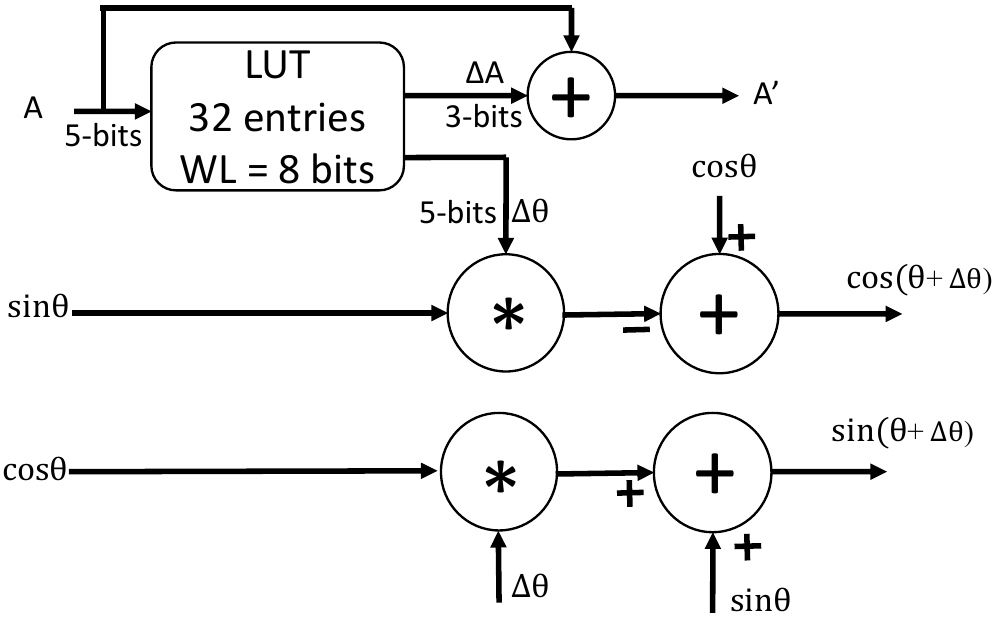}
\caption{Pre-distortion circuits for PA non-linearities.}
\label{fig:polar_pre}
\end{figure}

The DSP output signals are distorted by the analog processing functions. This is due to i) nonidealities such as bandwidth limitations of the analog components in the amplitude and phase paths, ii) delay mismatch between the amplitude and phase paths, and iii) the RF-DAC non-idealities. These distortions cause spectral regrowth and devastate the constellation diagram. Therefore, a Pre-Distortion (PD) circuit is provided.

As shown in Fig.~\ref{fig:polar_pre}, the pre-distortion unit is built with a lookup table (LUT), where the AM serves as the addressing index. The AM and PH signals are compensated with the derived $\Delta$ values from the LUT:
\begin{eqnarray}\label{eq:pre}
\nonumber
    A(t)' &=& A(t) + \Delta A \\  \nonumber
    sin(\theta'(t))&=&sin(\theta(t)+\Delta\theta(t))\\    \nonumber
     &=&sin(\theta(t))cos(\Delta\theta(t))+cos(\theta(t))sin(\Delta\theta(t))\\    \nonumber
     &\sim&sin(\theta(t))+cos(\theta(t))\Delta\theta(t)\\    \nonumber
    cos(\theta'(t))&=&cos(\theta(t)+\Delta\theta(t))\\    \nonumber
     &=&cos(\theta(t))cos(\Delta\theta(t))-sin(\theta(t))sin(\Delta\theta(t))\\
     &\sim&cos(\theta(t))-sin(\theta(t))\Delta\theta(t)
\end{eqnarray}
\normalsize

The pre-distorted AM is created by summing up the AM and the derived $\Delta A(t)$  from the LUT. Similarly, the pre-distorted PH signals are obtained by operating on respective PH signals with and the $\Delta\theta(t)$. To avoid heavy computations, the PH pre-distortion is approximated in Equation~\ref{eq:pre}, provided that the $\Delta\theta(t)$ is small. An LUT consists of 32 entries (because the AM signal is 5-bit width). Each entry is of 8 bits width. The 3 most significant bits are assigned to the $\Delta A$, and the 5 least significant bits to the $\Delta \theta(t)$.

\section{Implementation details}\label{sec:im}
Fig.~\ref{fig:polar_pipeline} shows the overall pipeline scheme. The DFE input throughput is 1.76~Gsps. The system comprises 4 parallel pipelined signal paths (each has a separate CIC filter, a CORDIC, and a pre-distortion unit), as the OSF is 4. Therefore, the output throughput is 7.04~Gsps (4x1.76~Gsps). The speed requirement is challenging, even with the 4x parallelism. Therefore, the DFE processor was implemented with a deep pipeline structure, i.e. pipelined after each addition.

\afterpage{
\begin{sidewaysfigure}[h]
\centering
\includegraphics[width = \linewidth]{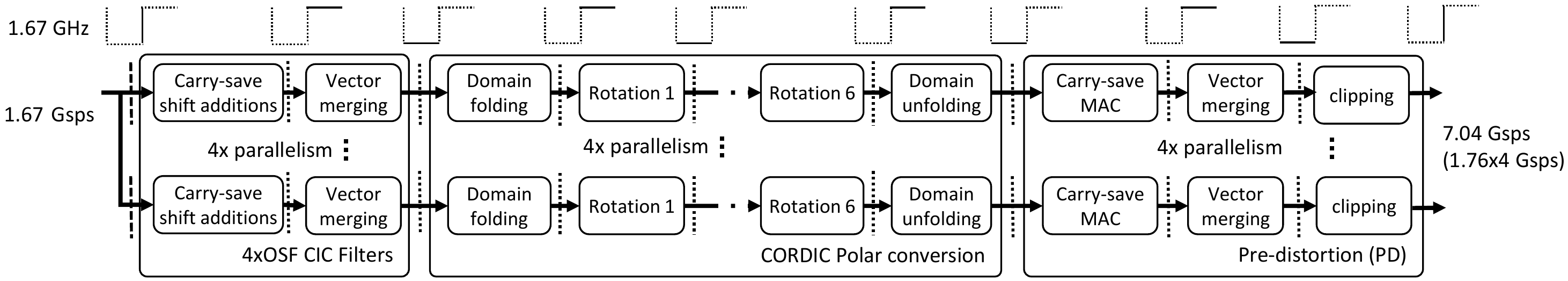}
  \caption{Pipeline scheme of the proposed DFE processor.}
  \label{fig:polar_pipeline}
\end{sidewaysfigure}
\clearpage
}

Since the coefficients of the CIC filters are pre-determined, the multiplications in those filters were accomplished by shift-additions for power saving. Moreover, signals are added by customized carry-save adders and finally adding up by the vector merging adders (adding up the vectors of sums and carries)~\cite{huang15micpro-deci-filter}. The pipeline breaks the CORDIC after each CORDIC rotation. As the AM resolution (5-bit) is less than the PH resolution (7-bit), the AM signals require fewer rotations. Therefore, they are ready before PH signals are produced. This is advantageous for the pre-distortion: by the time the PH signals are computed, the $\Delta A$ and the $\Delta \theta$ are already fetched by the AM signals from the LUT. The LUT is implemented with single-port RAMs. 

To reduce the power consumption, level-triggered two-phase latches were chosen as the sequential component. The input sequential elements are rising-edge enabled flip-flops. The enabling signals for the latches are indicated by a solid line. For instance, the first latch in the pipeline is active-high, while the second latch is active-low. This complementary two-phase latch methodology eliminates often-encountered hold time problem in latch based designs. Advantageously, the proposed latch scheme allows time borrowing. For instance, the data can arrive later than the rising edge for an active-high latch, and thus borrows time from the next pipeline stage. The advantage of time-borrowing is two-fold. Firstly, it eases timing closure, because it can perform stage balancing automatically. This eliminates manually moving computation and/or logic elements from one stage to another. For the exampled 1.76~GHz high-speed, it is especially beneficial. Secondly, the opportunistic time-borrowing principle addresses process and environmental variations. Due to such variations, even if the pipeline is carefully equalized at design time, the delay of each computation stage can vary in the fabricated chip, the effect of which becomes even more severe with technology scaling. In the DFE, time-borrowing allows for a slower computation stage to opportunistically borrow time from faster ones, which averages out some of the variations. 

Even with the techniques mentioned above, the speed requirement cannot be achieved for the standard 28nm technology, with 0.9V as the standard $V_{dd}$. Accordingly, we applied the following modifications during library selection. i) For the CIC filter and the CORDIC polar converter: we utilized fast but leaky Low $V_{th}$ (LVT) cells, rather than the Standard $V_{th}$ (SVT). This leakage increase should not change the overall power consumption, as the circuit power is dominated by the dynamical part (since the clock frequency is very high). Moreover, the designed $V_{dd}$ was increased to 1V for a higher speed. ii) For the pre-distortion unit: as the setup-timing requirement is not as difficult as the rest units, they were accomplished by SVT cells under 0.9V.   

Therefore, we divided the design into two power domains: 1V for the CIC filter and the CORDIC polar conversion, and 0.9V for the pre-distortion unit. The pre-distortion unit can be switched off and by-passed for scenarios where linearity is sufficient, for energy saving.

\section{Measurement results}
\begin{figure}[H]
  \centering
  \includegraphics[width = .8\linewidth]{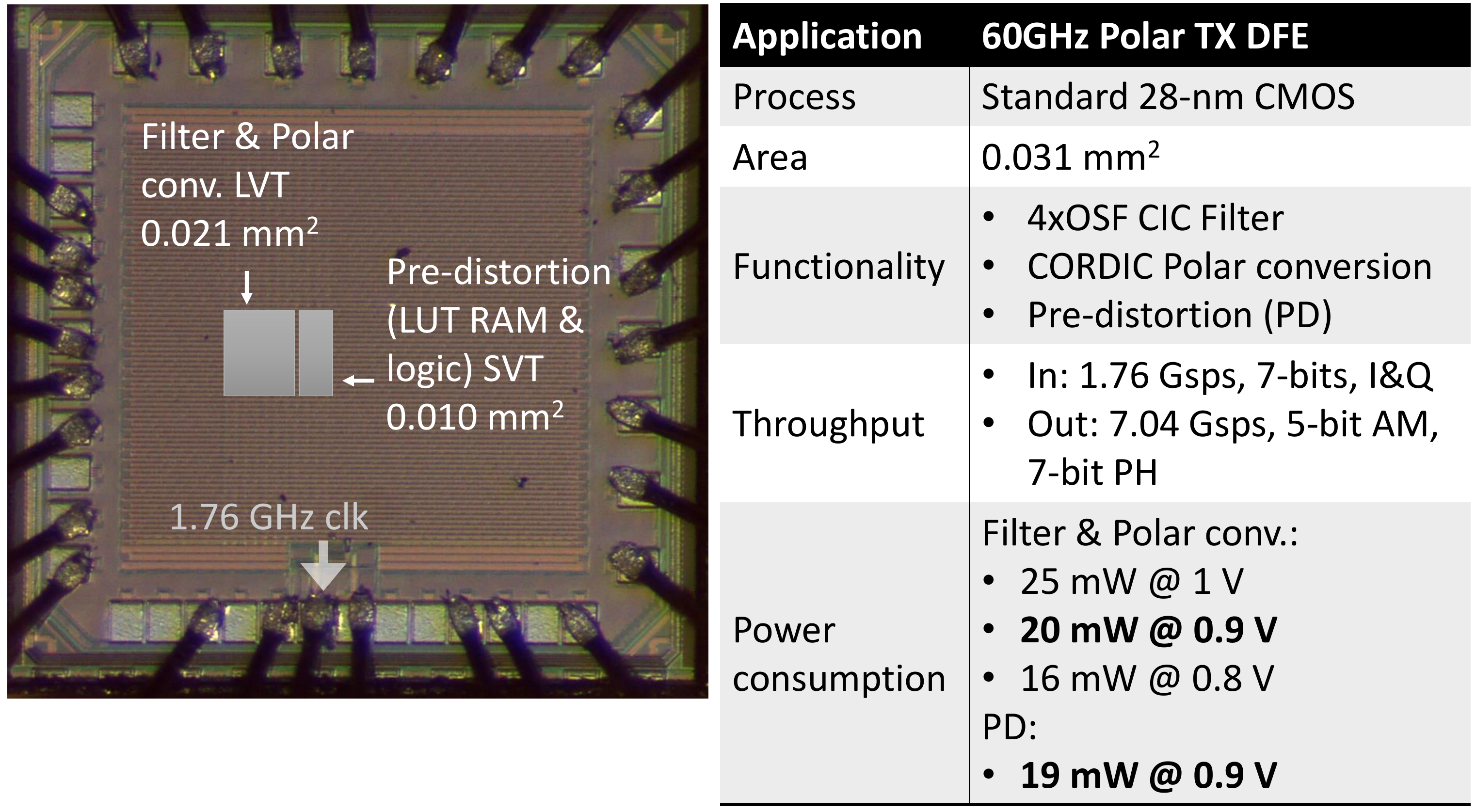}
  \caption{Die photo and chip information of the proposed digital polar DFE processor.}
  \label{fig:polar_shot}
\end{figure}

\begin{figure}[H]
  \centering
  \includegraphics[width=0.4\linewidth]{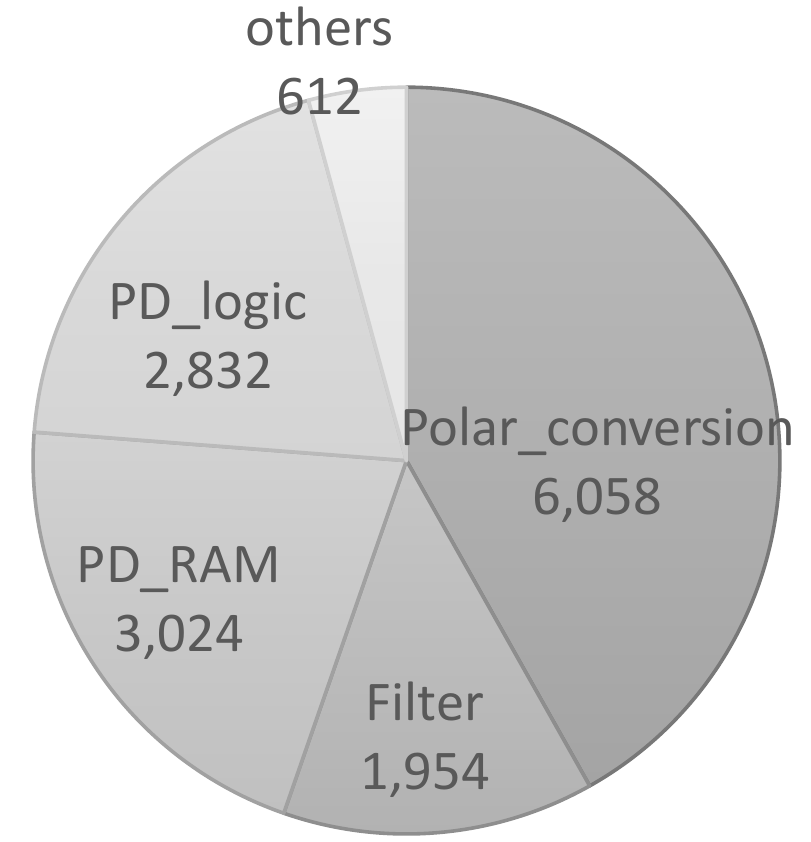} 
  \caption[Cell area breakdown of the proposed DFE processor.]{Cell area breakdown (in $\mu m^2$) of the proposed DFE processor. N.B.: Routing spacing not included.}\label{fig:polar_area}
\end{figure}

The DFE processor was processed in a standard 28 $nm$ CMOS technology. A micrograph of the chip is shown in Fig.~\ref{fig:polar_shot}. The complete design area is as small as 0.036$mm^2$, of which the switchable pre-distortion unit utilizes 0.015$mm^2$. The cell area breakdown is illustrated in Fig.~\ref{fig:polar_area}.

\begin{figure}[H]
  \centering
  \includegraphics[width=0.8\linewidth]{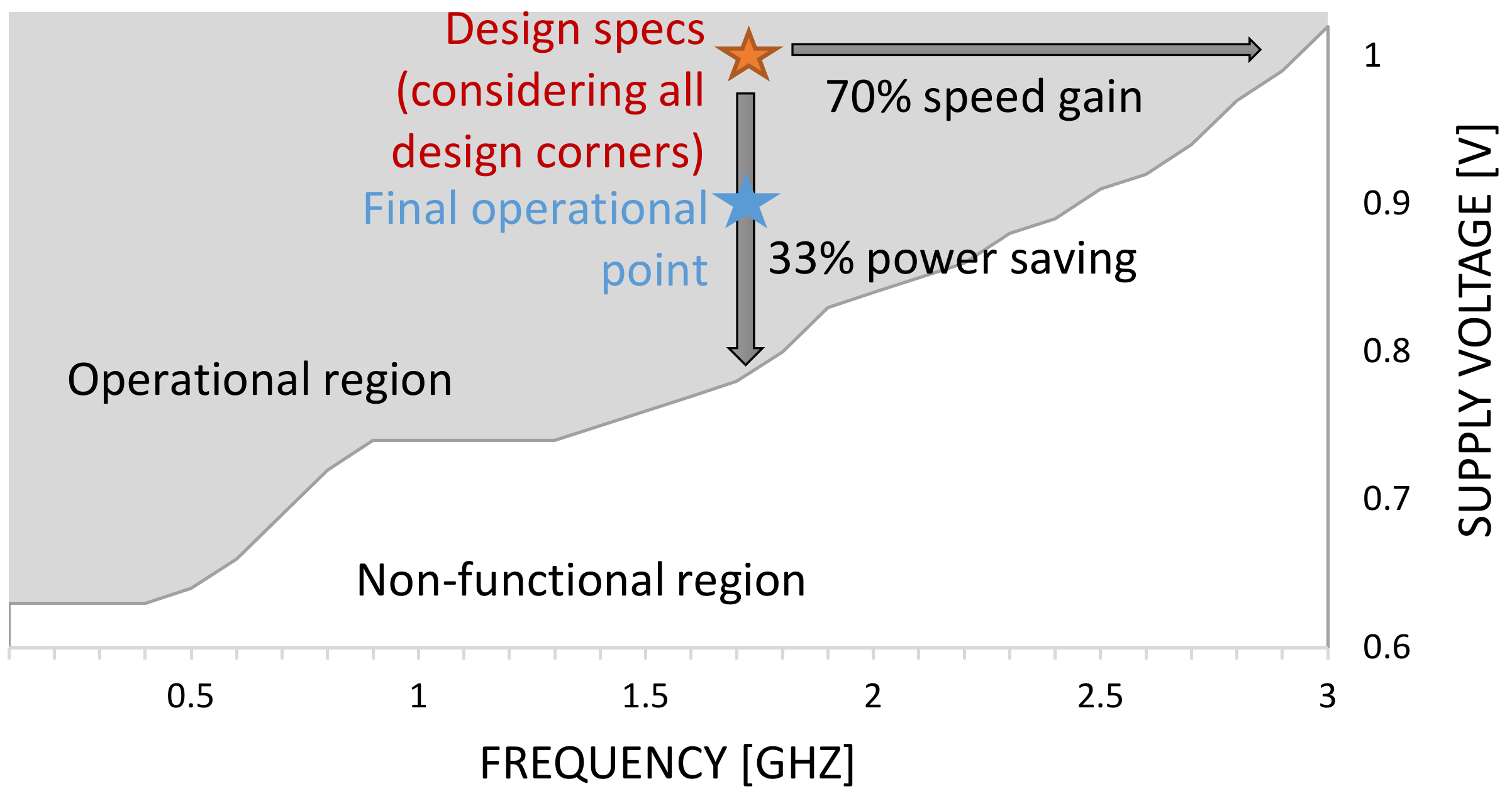} 
  \caption{Typical chip speed vs. $V_{dd}$ @$25^{\circ}C$.}\label{fig:polar_speed}
\end{figure}

The circuit speed was measured for a typical-case chip at 25$^{\circ}C$ room temperature (Fig.~\ref{fig:polar_speed}). Due to the inserted design margin, although the chip is designed to be 1.76~GHz @ 1V, a typical chip can operate correctly @ 3~GHz with the same $V_{dd}$, implying a throughput of 4x3~GHz. Alternately, it can also operate @ 0.8V $V_{dd}$ with a fixed frequency of 1.76~GHz, which brings 33\% power saving. The CIC filter and the CORDIC polar conversion consume 25mW @ 1V $V_{dd}$, 20mW @ 0.9V, or 16mW @ 0.8V. The pre-distortion unit consumes 19mW @ 0.9V. The overall leakage consumption power is less than 1mW at room temperature.

\begin{figure}[H]
  \centering
  \includegraphics[width=0.5\linewidth]{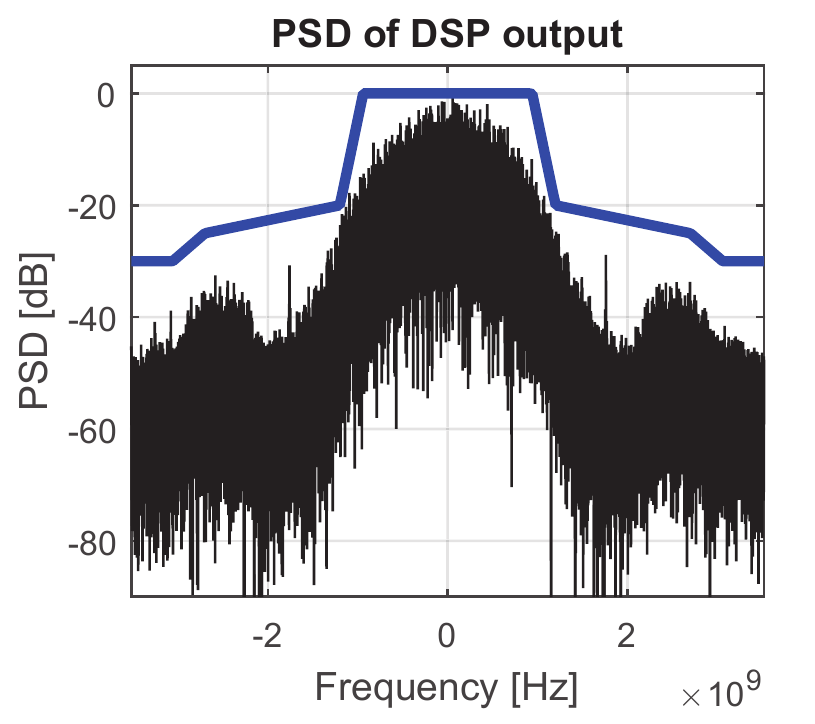} 
  \caption{Measurement DFE outputs PSD.}\label{fig:polar_psd}
\end{figure}

\begin{figure}[H]
  \centering
  \includegraphics[width=0.4\linewidth]{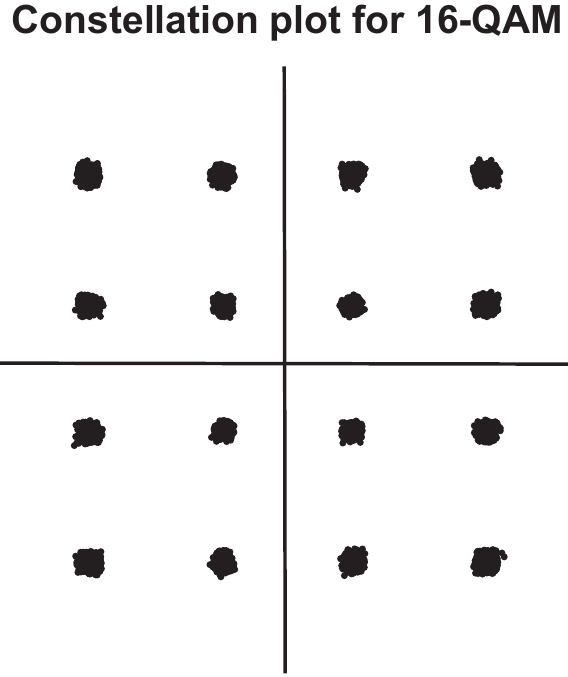} 
  \caption{Measurement DFE output 16-QAM constellation, with a corresponding EVM of -30.5~dB.}\label{fig:polar_constellation}
\end{figure}

The power spectrum density (PSD) of the DFE outputs is shown in Fig.~\ref{fig:polar_psd}. Both the in-band PSD and the alias rejection are confirmed to be compliant with the spectrum mask. The EVM of the produced signal is measured to be -30.5~dB. Fig.~\ref{fig:polar_constellation} plots the constellation for 16-QAM signals. It demonstrated clearly the nice purity of the signals.

As this work is the first polar transmitter for high bandwidth 60~GHz system, benchmarking can be difficult. Nevertheless, Table~\ref{tab:polar_comp} compares state-of-the-art digital signal processors \cite{polarhwang,polarstrollo,us4,mehta20100} that has similar functionalities. The only work with comparable speed is \cite{us4}, where merely the 4xOSF filtering function is provided. Even with the perfect scaling normalization, the only state-of-the-art work with comparable energy consumption is \cite{polarstrollo}, which only performs CORDIC polar conversion. In summary, the comparisons show the proposed design has significant advantages.

\begin{table*}[t]
\caption{Performance comparisons of digital polar front-end systems.}
\label{tab:polar_comp}
\includegraphics[width = 1\linewidth]{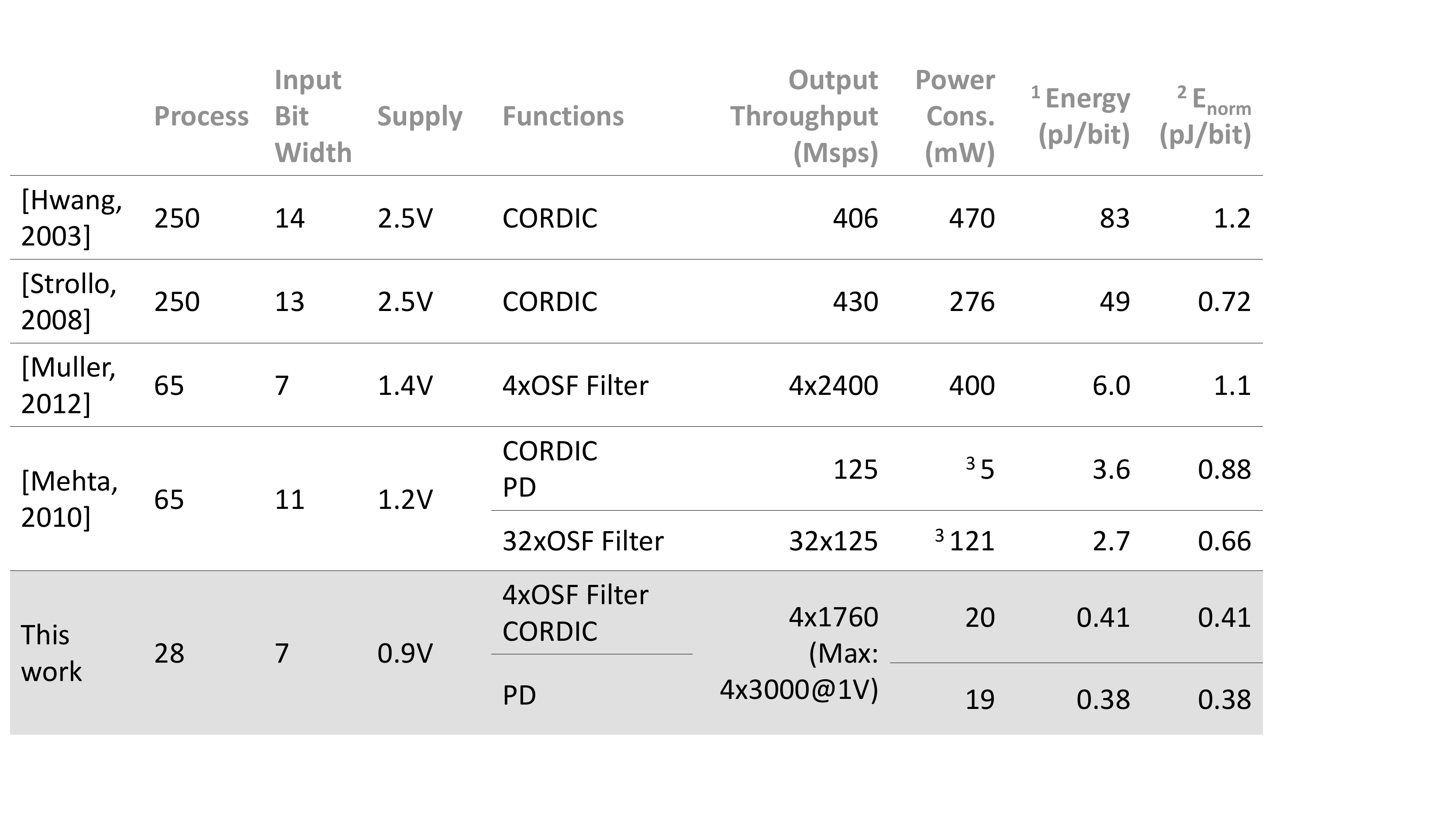}\centering
\begin{tablenotes}
        \footnotesize
        \item[] ~~$^1$ Energy = Power / Frequency / Bits.
        \item[] ~~$^2$ $E_{norm}$ =  Energy x $(V_{dd}/1V)^2$ x (tech/28$nm$).
        \item[] ~~$^3$ Estimated from a total current of 105mA.
\end{tablenotes}
\end{table*}

\section{Conclusions}
This paper presents the first DFE processor for polar transmitter working in the 60~GHz band. It enables digital-intensive transmitter architecture with polar concept expanded to the whole transmitter, rather than conventionally only in radio frequency domain. 

The DFE processor was processed in a standard 28~$nm$ technology with 0.036~$mm^2$ area. The processor provides -30.5~dB EVM, with 4x1.76~Gsps output throughput. The throughput can reach to 4x3~Gsps when 1V $V_{dd}$ is supplied. It consumes 39mW from 0.9V supply.



\cleardoublepage





\printbibliography[title=\bibname]


\includecv{curriculum}

\includepublications{publications}

\makebackcoverXII

\end{document}